%% file: Multilevel_Monte_Carlo_for_uncertainty_quantification_in_structural_engineering.tex
\documentclass[review,3p]{elsarticle}

\usepackage{lineno,hyperref}
\modulolinenumbers[1]

\usepackage{graphicx}
\usepackage{amsmath,amssymb,empheq,amsfonts,amsthm,bm}
\usepackage{booktabs}
\DeclareMathOperator{\E}{\mathbb{E}}
\DeclareMathOperator{\V}{\mathbb{V}}
\newtheorem{theorem}{Theorem}
\newcommand\norm[1]{\left\lVert#1\right\rVert}

\usepackage{floatrow}
\usepackage{caption}
\usepackage{subcaption}
\usepackage[usenames,dvipsnames]{pstricks}
\usepackage{booktabs}
\usepackage{caption}
\usepackage{subcaption}
\usepackage{multirow}
\usepackage[justification=centering]{caption}
\usepackage{physics}

\journal{Comput. Struct.}
\begin{document}

\begin{frontmatter}

\title{Multilevel Monte Carlo for uncertainty quantification in structural engineering}

\author[CW]{P. Blondeel\corref{cor1}} 
\ead{philippe.blondeel@kuleuven.be}

\author[CW]{P. Robbe} 

\author[BOKU]{C. Van hoorickx} 

\author[BOKU]{G. Lombaert} 

\author[CW]{S. Vandewalle} 

\address[CW]{KU Leuven, Department of Computer Science, NUMA Section, Celestijnenlaan 200A, 3001 Leuven, Belgium}
\address[BOKU]{KU Leuven, Department of Civil Engineering,  Kasteelpark Arenberg 40, 3001 Leuven, Belgium}

\cortext[cor1]{Corresponding author}

\begin{abstract}
Practical structural engineering problems  often exhibit a significant degree of uncertainty in the material properties being used,  the dimensions of the modeled structures,  the magnitude of loading forces, etc. In this paper, we consider two beam models: a cantilever beam clamped at one end and a beam clamped at both ends. We consider a static and a dynamic load. The material uncertainty resides in the Young's modulus, which is either modeled by means of one random variable sampled from a univariate Gamma distribution or with multiple random variables sampled from a Gamma random field. The Gamma random field is generated starting from a truncated Karhunen-Lo\`eve expansion of a Gaussian random field, followed by a transformation. Three different responses are considered: the static elastic response, the dynamic elastic response and the static elastoplastic response. The first two respectively simulate the spatial displacement of a concrete beam and its frequency response function in the elastic domain. The third one simulates the spatial displacement of a steel beam in the elastic as well as in the plastic domain.  The plastic region is governed by the von Mises yield criterion with linear isotropic hardening. In order to compute the statistical quantities of the static deflection and frequency response function, Multilevel Monte Carlo (MLMC) is combined with a Finite Element solver. This recent sampling method is based on the idea of variance reduction, and employs a hierarchy of finite element discretizations of the structural engineering model. The good performance of MLMC arises from its ability to take many computationally cheap samples on the coarser meshes of the hierarchy, and only few computationally expensive samples on the finer meshes. In this paper, the computational costs and run times of the MLMC method are compared with those of the classical Monte Carlo method, demonstrating a significant speedup of up to several orders of magnitude for the studied cases. For the static elastic response,  the deflection of the beam including uncertainty bounds in the spatial domain is visualized. For the static  elastoplastic response, the focus lies on the visualization  of a force deflection curve including uncertainty bounds. For the dynamic elastic  response, the visualizations consist of a frequency response function, including uncertainty bounds. \end{abstract}

%\maketitle


\end{frontmatter}

%\linenumbers
\nolinenumbers

\section{Introduction}

There is an increasing need to accurately simulate and compute solutions to engineering problems whilst taking into account model uncertainties. Methods for such uncertainty quantification and propagation in structural engineering can roughly be categorized into two groups: non-sampling methods and sampling methods. Examples of non-sampling methods are the perturbation method  and the Stochastic Galerkin Finite Element method. The perturbation method is based on a Taylor series expansion approximating the mean and variance of the solution \cite{Kleiber}.
The method is quite effective, but its use is restricted to models with a limited number of relatively small uncertainties.
The Stochastic Galerkin method, first proposed by Ghanem and Spanos \cite{GhanemAndSpanos}, is based on a spectral representation in the stochastic space. It transforms the uncertain coefficient partial differential equation (PDE) problem by means of a Galerkin projection technique into a large coupled system of deterministic PDEs. This method allows for somewhat larger numbers of uncertainties and is quite accurate. However, it is highly intrusive and memory demanding, making its implementation cumbersome and restricting its use to rather low stochastic dimensions.

Sampling methods, on the other hand, are typically non-intrusive. Each sample corresponds to a deterministic solve for a set of specified parameter values. Two particularly popular examples are the Stochastic Collocation method \cite{Bab} and the Monte Carlo (MC) method \cite{Fishman}.
The former samples a stochastic PDE at a carefully selected multidimensional set of collocation points. After this sampling, a Lagrange interpolation is performed leading to a polynomial response surface. From this, the relevant stochastic characteristics can easily be computed in a post-processing step. However, as is also the case for Stochastic Galerkin, the Stochastic Collocation method suffers from the curse of dimensionality; the computational cost grows exponentially with the number of random variables considered in the problem.
The MC method on the other hand, selects its samples randomly and does not suffer from the curse of dimensionality. A drawback is its slow convergence as a function of the number of samples taken.
The convergence of Monte Carlo can be accelerated in a variety of ways. For example, alternative non-random selections of sampling points can be used, as in Quasi-Monte Carlo \cite{Caflish,Niederreiter}, and Latin Hypercube \cite{Loh} sampling methods. Also, variance reduction techniques such as Multilevel Monte Carlo (MLMC) \cite{Giles} and its generalizations, see, e.g., \cite{PJ, PJ2}, can speed up the method.
Note that there also exist hybrid variants which exhibit both a sampling and non-sampling character. This type of methods combines, for example, the Stochastic Finite Element methodology with Monte Carlo sampling or a multi-dimensional cubature method, see, e.g., \cite{Ghanem_Hyrbid,Acharjee}.

Monte Carlo methods have since long been used in the field of structural engineering, for example in problems of structural dynamics \cite{Masanobu} or in elastoplastic problems where the structure's reliability is assessed \cite{Pulido}. The focus of this paper is to apply the MLMC method to a structural engineering problem, discretized by means of the Finite Element method. A comparison with the standard MC method will show a significant reduction in computational cost. 

The structure of the paper is as follows. In section 2, we formulate the mathematical model, introduce the problem statement and describe how the uncertainty is modeled. Section 3 recalls the MLMC method, and provides some additional algorithmic implementation details. In Section 4, numerical results are presented. First, a performance comparison is made between standard MC and MLMC for a static elastic and elastoplastic load case. Next, results showing the uncertainty propagation for the static elastic, static  elastoplastic and the dynamic elastic response are shown for when the Young's modulus is modeled by means of a random variable coming from a univariate Gamma distribution and as multiple random variables sampled from a Gamma random field. The fifth and last section offers concluding remarks and details some paths for further research.

\section{The mathematical model}
\subsection{Beam models and material parameters}
\begin{figure}[H]
\begin{subfigure}[b]{0.54\textwidth}
\centering
\includegraphics[height=1.9cm]{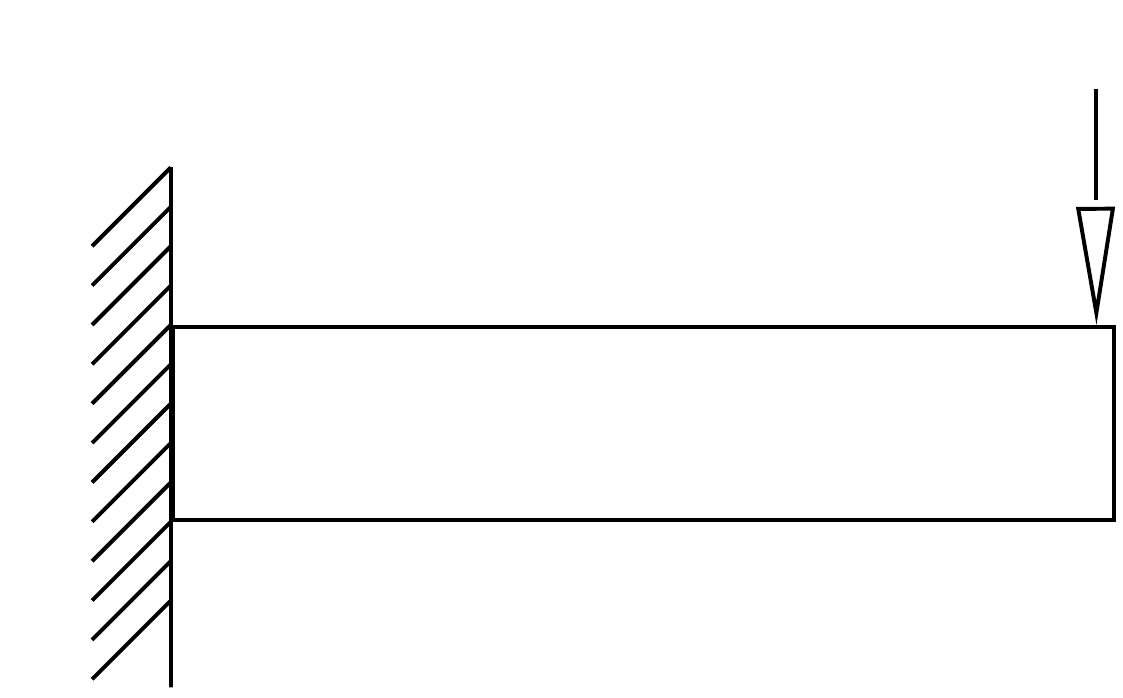}
\end{subfigure}
\begin{subfigure}[b]{0.45\textwidth}
\centering
\includegraphics[height=1.9cm]{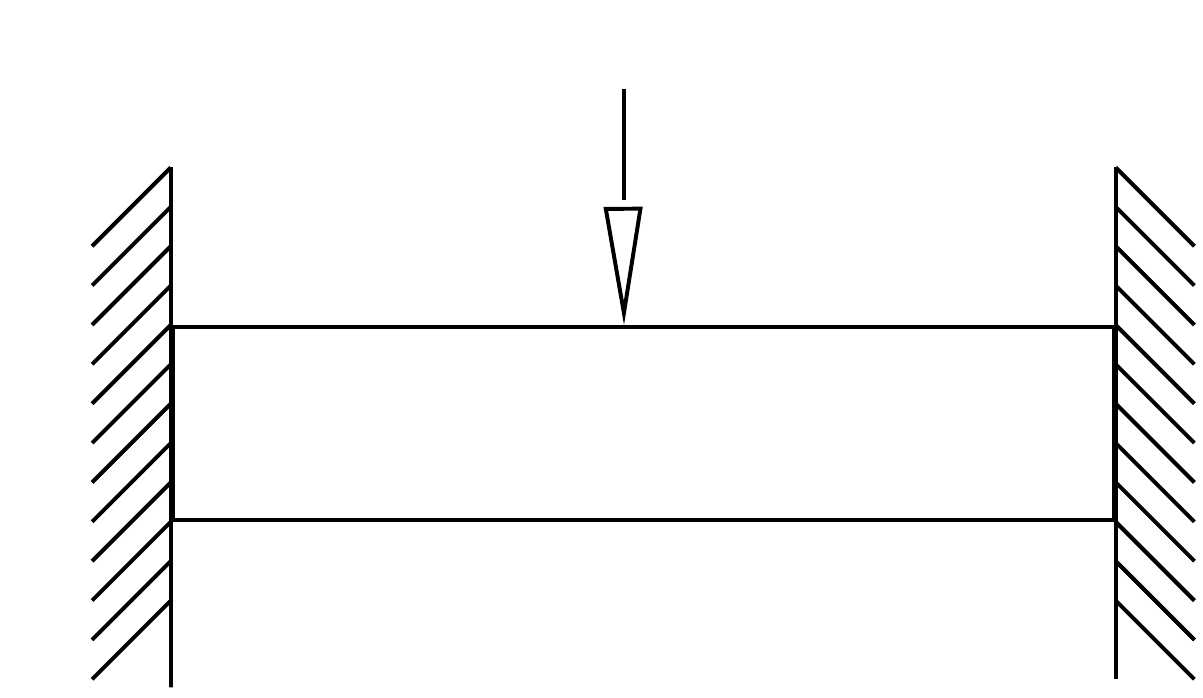}
\end{subfigure}
\caption{Cantilever beam loaded on its right end (left) and beam clamped at both sides loaded in the middle (right).}
\label{fig:bean_configurations}
\end{figure}

The considered engineering problem is the response of a cantilever beam clamped at one side and a beam clamped at both sides  as seen in Fig.\ref{fig:bean_configurations}, assuming plane stress. The following responses are considered: the static elastic response, the dynamic elastic response and the static elastoplastic response. The first two respectively simulate the spatial displacement and the frequency response function of a concrete beam model solely in the elastic region. The third one simulates the spatial displacement of a steel beam in the elastic as well as in the plastic region. The dimensions of the beam are $2.5\,\mathrm{m}$ (length) by $0.25\,\mathrm{m}$ (height) by $1\,\mathrm{m}$ (width)  for the elastic responses and $10^{-3}\,\mathrm{m}$ (width) for the elastoplastic response. The material parameters of the concrete are as follows: a mass density of $2500\,\mathrm{kg/m^3}$, a Poisson ratio of $0.15$ and a Young's modulus subjected to some uncertainty with a mean value of $30\,\mathrm{GPa}$. The material parameters of the steel are as follows: a yield strength of $240\,\mathrm{MPa}$, a Poisson ratio of $0.25$ and a Young's modulus subjected to a certain degree of uncertainty with a mean value of $200\,\mathrm{GPa}$. In order to model the material uncertainty, two uncertainty models will be considered. The first model consists of a homogeneous Young's modulus characterized by means of a single random variable. The second  model is a heterogeneous Young's modulus represented as a random field. Both uncertainty models will be used to compute the stochastic characteristics of the static elastic and elastoplastic response of a beam clamped at both sides as seen in Fig.\ref{fig:bean_configurations} (right) and the dynamic response of a cantilever beam clamped at one side as seen in  Fig.\ref{fig:bean_configurations} (left). 

\subsubsection{The homogeneous model}
Following \cite{YongLiu}, we opt to describe the Young's modulus in the homogeneous model by means of a univariate Gamma distribution. This distribution is characterized by a shape parameter $\alpha$ and a scale parameter $\beta$:
\begin{linenomath*}
\begin{equation}
f(x|\alpha,\beta) = \dfrac{1}{\beta^{\alpha} \Gamma(\alpha)} x^{\alpha-1} e^{\left(-\dfrac{x}{\beta}\right)}.
\end{equation}
\end{linenomath*}

\begin{figure}[H]
\centering
\begin{subfigure}[b]{0.54\textwidth}
\scalebox{0.45}{
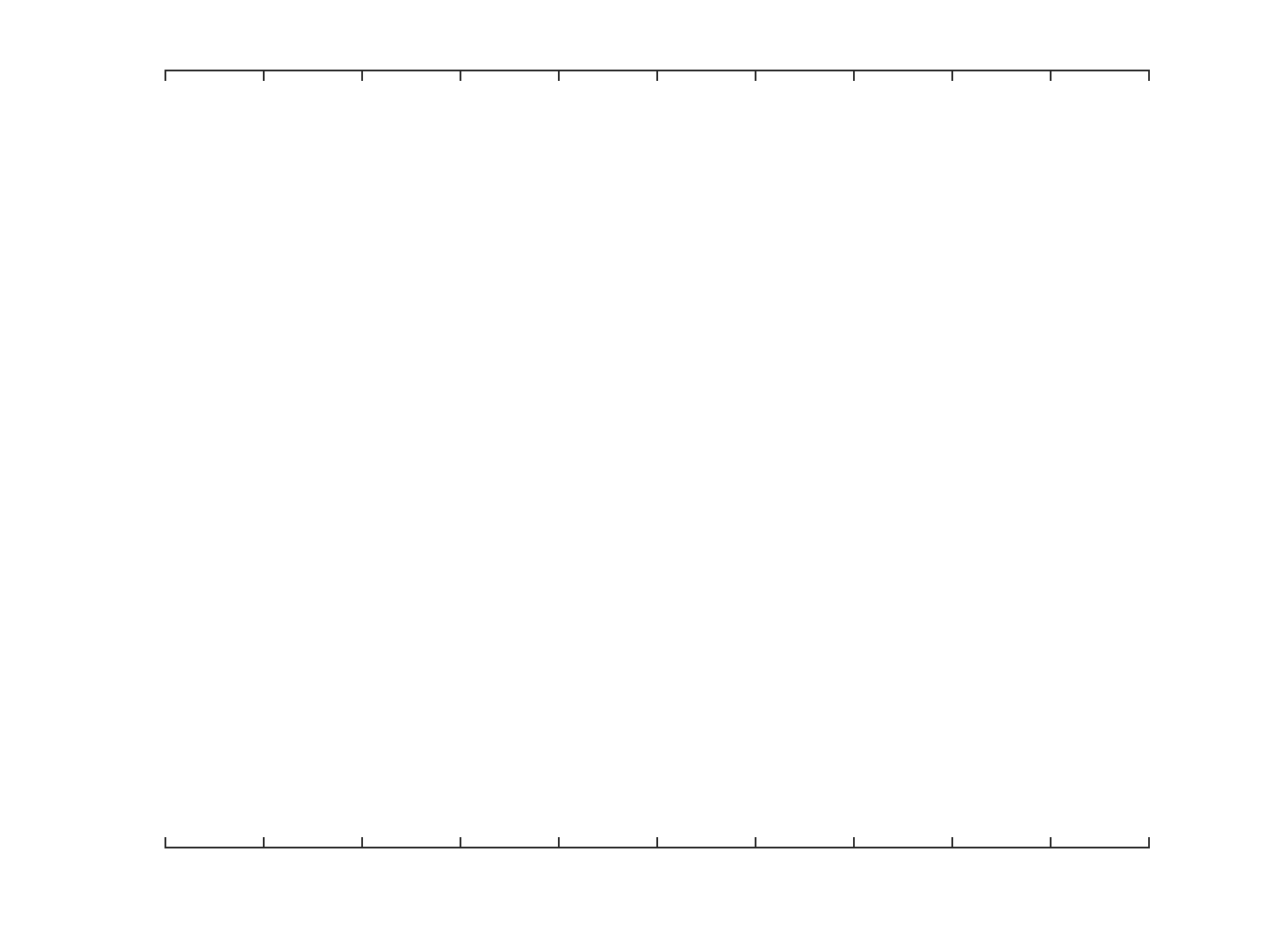}
\end{subfigure}
\begin{subfigure}[b]{0.45\textwidth}
\scalebox{0.45}{
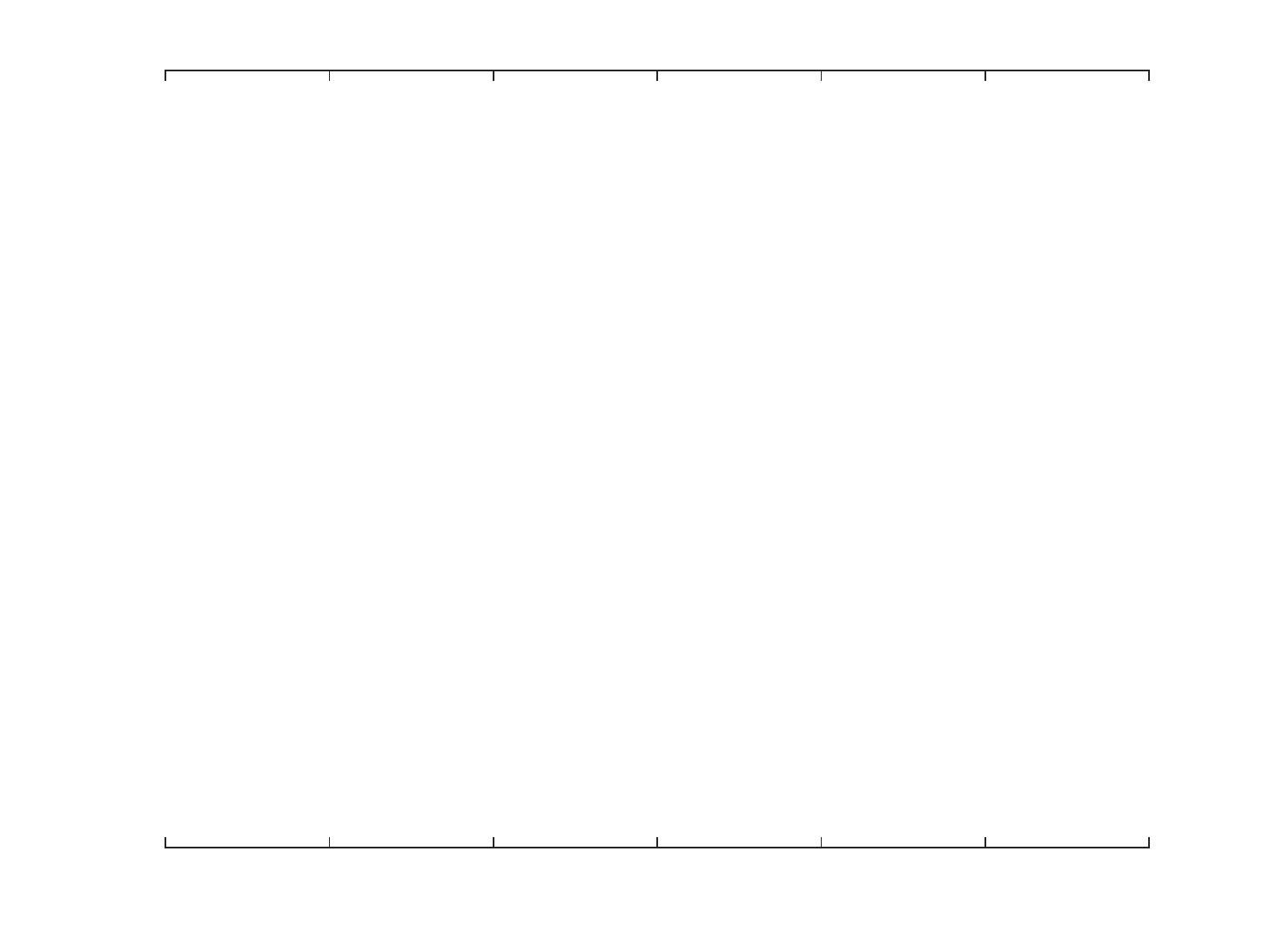}
\end{subfigure}
\centering
\caption{Probability density function for Young's modulus as a univariate distribution for concrete (left) and steel (right). Shown also is the mean $\mu$  and the standard deviation $\sigma$.}
\label{fig:pdf_gam_prior}
\end{figure}

The corresponding mean value and variance can be computed respectively as $\mu=\alpha \beta$ and
$\sigma^2=\alpha \beta^2$. In this paper, we will select $\alpha\!=\!7.1633$ and $\beta\!=\!4.1880 \times 10^9$ in order to model the material uncertainty for the concrete beam, which are based on values coming from \cite{Ellen}. This leads to a mean of $30\,\mathrm{GPa}$ and a standard deviation of $11.2\,\mathrm{GPa}$. For modeling the material uncertainty in the steel beam, we select $\alpha\!=\!934.2$ and $\beta\!=\!0.214\times 10^9$, see \cite{Hess}. This gives a mean of $200\,\mathrm{GPa}$ and a standard deviation of $6.543\,\mathrm{GPa}$. The Gamma distribution for both materials is plotted in Fig.\,\ref{fig:pdf_gam_prior}.

\subsubsection{The heterogeneous model}
The Young's modulus with spatially varying uncertainty will be represented by means of a (truncated) Gamma random field. The construction of this random field is done by means of a classical, two step process. First, a (truncated) Gaussian random field is generated, using a Karhunen-Lo\`eve (KL) expansion \cite{Loeve}. Next, this Gaussian random field is transformed into a Gamma random field with a memoryless transformation \cite{Grigoriu}.

Consider a Gaussian random field $Z(\mathbf{x},\omega)$, where $\omega$ is a random variable, with exponential covariance kernel,
\begin{linenomath*}
\begin{equation}\label{eq:covariance_kernel}
C(\mathbf{x},\mathbf{y}):= \sigma^2 \exp\left(-\dfrac{\norm{\mathbf{x}-\mathbf{y}}_p}{\lambda}\right)\,.
\end{equation}
\end{linenomath*}
We selected the 1-norm ($p\!=\!1$), a correlation length $\lambda\!=\!0.3$ and a standard deviation $\sigma \!=\!1.0$. The corresponding KL expansion can then be formulated as follows:
\begin{linenomath*}
\begin{equation}
Z(\mathbf{x},\omega)=\overline{Z}(\mathbf{x},.)+\sum_{n=1}^{\infty}  \sqrt{\theta_n} \xi_n(\omega) b_n(\mathbf{x})\,.
\label{eq:KLExpansion}
\end{equation}
\end{linenomath*}
 $\overline{Z}(\mathbf{x},.)$ denotes the mean of the field, and is set to zero. The $\xi_n(\omega)$ denote i.i.d.\,standard normal random variables.
The symbols $\theta_n$  and $b_n(\mathbf{x})$ denote, respectively, the eigenvalues and eigenfunctions of the covariance kernel \eqref{eq:covariance_kernel}. For the parameter values selected above, those eigenvalues and eigenfunctions can be computed analytically, see \cite{GhanemAndSpanos,Cliffe,Ghanem_prob,Zhang}. For a one-dimensional domain $\left[0 , 1\right]$, they are given by
\begin{linenomath*}
\begin{equation}\label{eq:1D_eigVec}%\label{eq:1D_eigVal}
 \theta_n^{1\text{D}}= \dfrac{2 \lambda}{\lambda^2 w_n^2 +1}
~~~~\mbox{and}~~~
b^{1\text{D}}_n(x) = A_n \left(\sin(w_n x) +\lambda w_n \cos(w_n x)\right)\,.
\end{equation}
\end{linenomath*}
The normalizing constants $A_n$ are chosen such that $ \norm{b_n}_2 = 1$.
The constants $w_n$ represent the real solutions, in increasing order, of the transcendental equation
\begin{linenomath*}
\begin{equation}\label{eq:transcendental}
\tan(w)=\dfrac{2 \lambda w}{\lambda^2 w^2 -1}\,.
\end{equation}
\end{linenomath*}
For the two-dimensional case, the eigenvalues and functions are obtained in a tensorproduct way,
\begin{linenomath*}
\begin{equation}\label{eq:2d_eigenVal}%\label{eq:2d_eigenVec}
\theta_n^{2\text{D}} = \theta_{i_{n}}^{1\text{D}} \theta_{j_n}^{1\text{D}}
~~~\mbox{and}~~
 b_n^{2\text{D}} (\mathbf{x}) = b_{i_n}^{1\text{D}} (x_1) b_{j_n}^{1\text{D}} (x_2)~~ \mbox{with}~~n = (i_n, j_n)\,.
\end{equation}
\end{linenomath*}

\begin{figure}[H]
\centering
\begin{subfigure}[b]{0.54\textwidth}
\scalebox{0.45}{
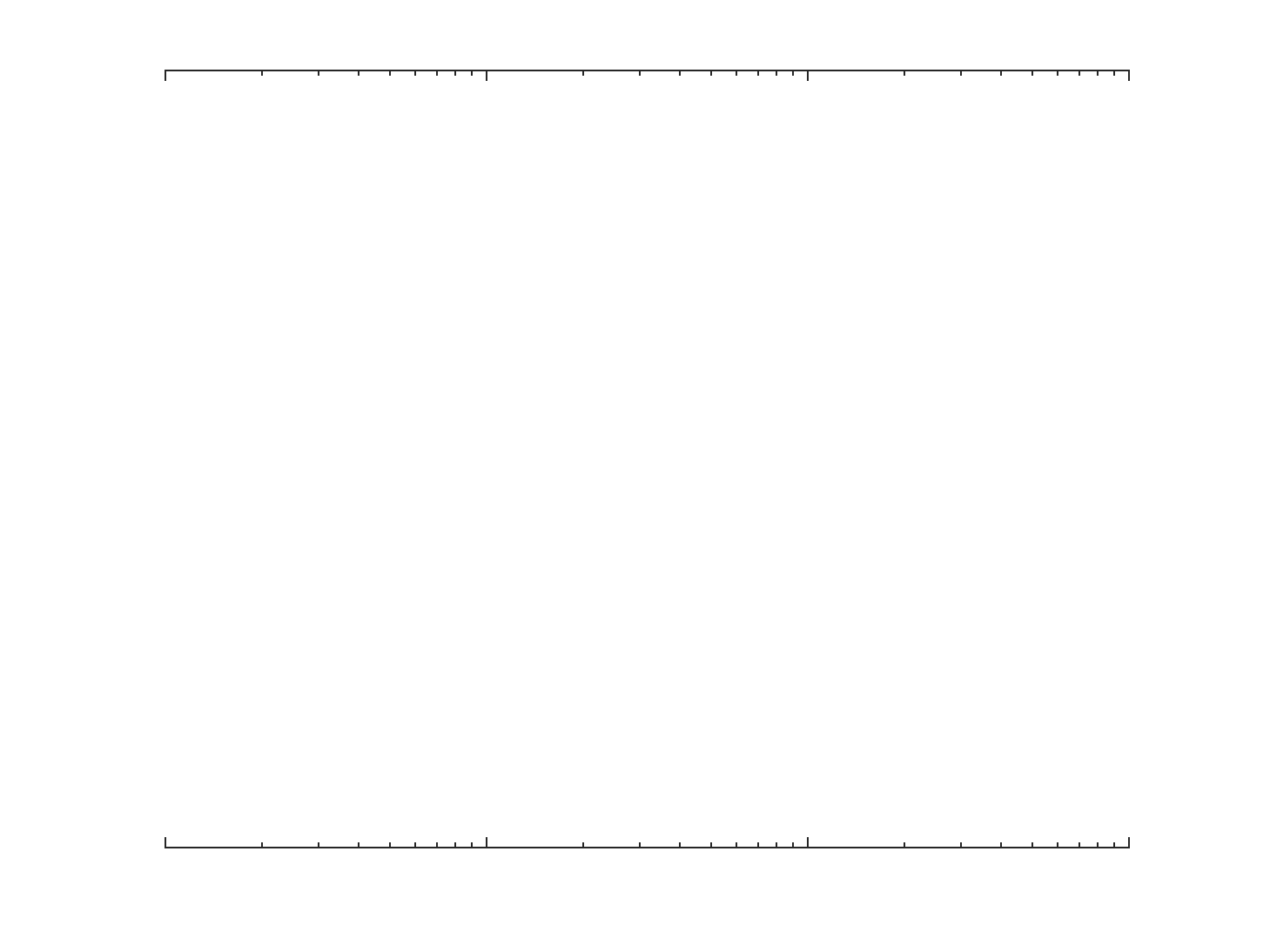}
%\hspace{1cm}
\end{subfigure}
\begin{subfigure}[b]{0.45\linewidth}
\scalebox{0.45}{
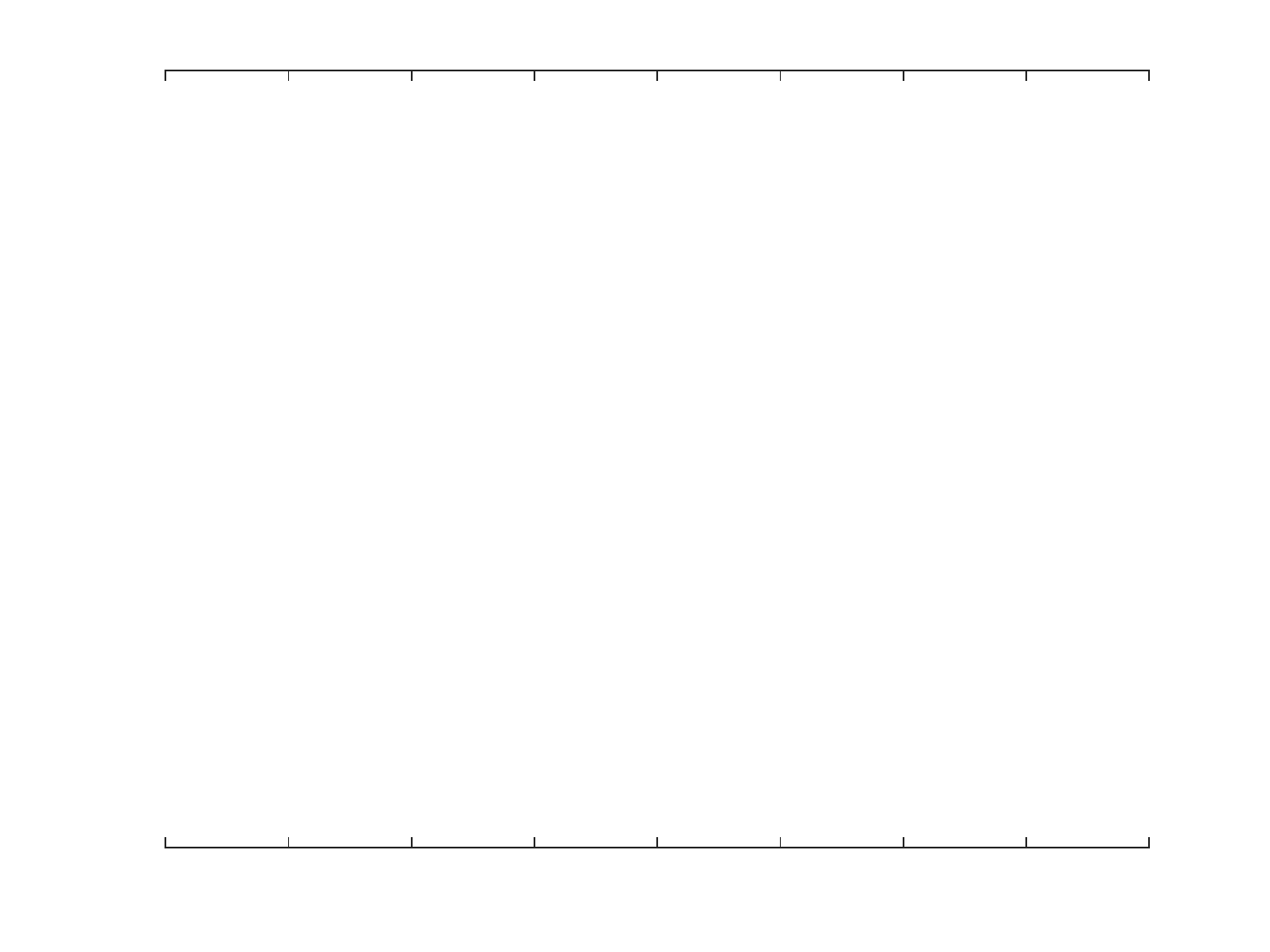}
\end{subfigure}
\caption{Magnitude of the eigenvalues and their cumulative sum (left) and memoryless transformation used to generate the Gamma random field (right).}
\label{fig:Eigenvalf}
\end{figure}

In an actual implementation, the number of KL-terms in \eqref{eq:KLExpansion} is truncated to a finite value. This value depends on the magnitude and, more precisely, on the decay of successive eigenvalues.
Those eigenvalues are plotted in Fig.\,\ref{fig:Eigenvalf} (left). Inclusion of the first 101 KL-terms is  sufficient in order to represent 90\% of the variance of the random field.
The percentage of the variance that is accounted for as a function of the number of included eigenvalues, corresponds to the cumulative sum of the eigenvalues, and is also illustrated in the figure. A cumulative sum of 1.0 corresponds to 100\% of the variance of the field being accounted for.

\begin{figure}[H]
\begin{subfigure}[b]{0.54\textwidth}
\centering
\scalebox{0.45}{
	\makebox[\textwidth]{
	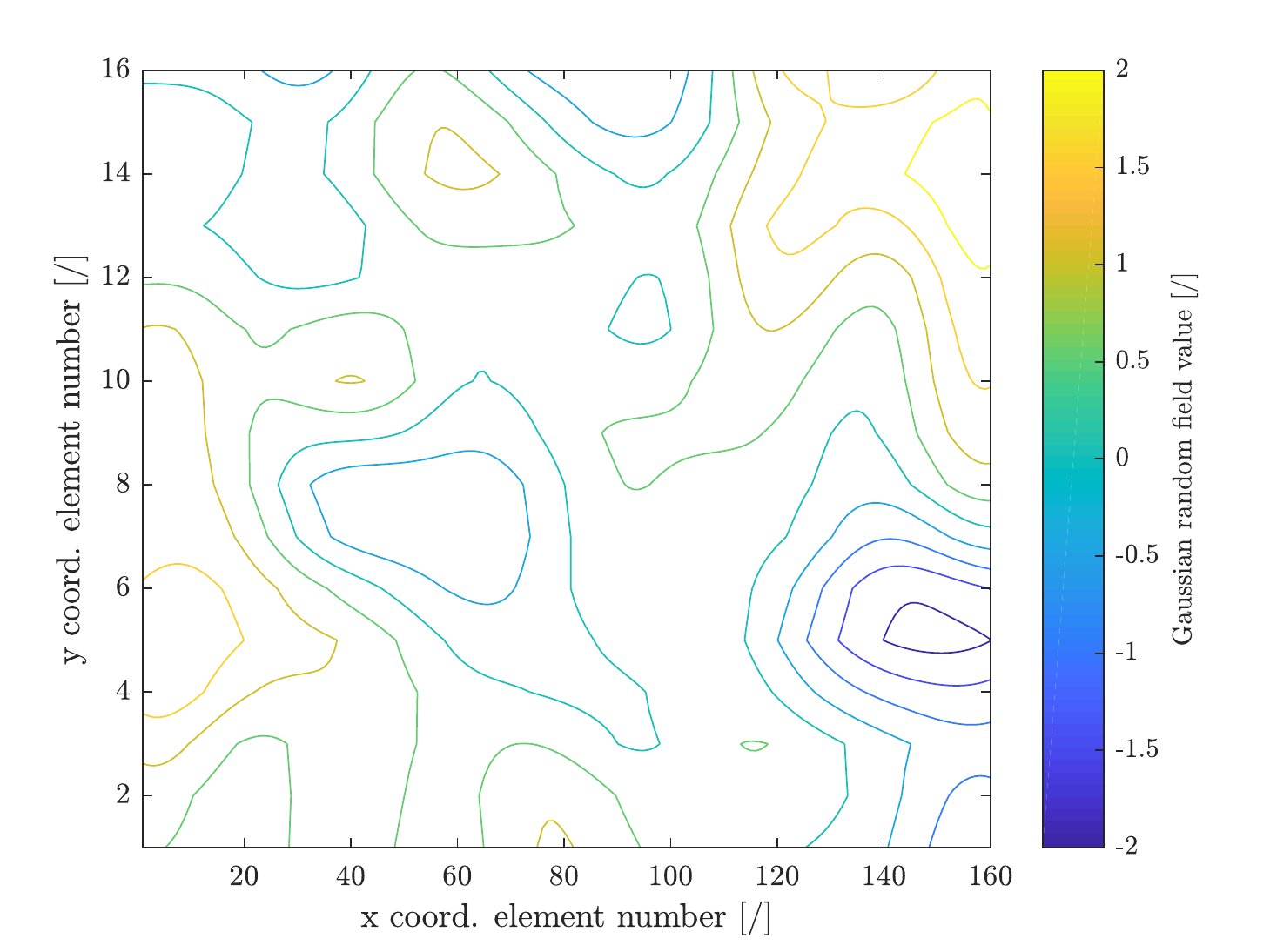}}
%	\caption{Guassian random field}
\label{fig:Gaussian Random field}
\end{subfigure}
%\centering
\begin{subfigure}[b]{0.45\linewidth}
\centering
	\scalebox{0.45}{
		\makebox[\textwidth]{
	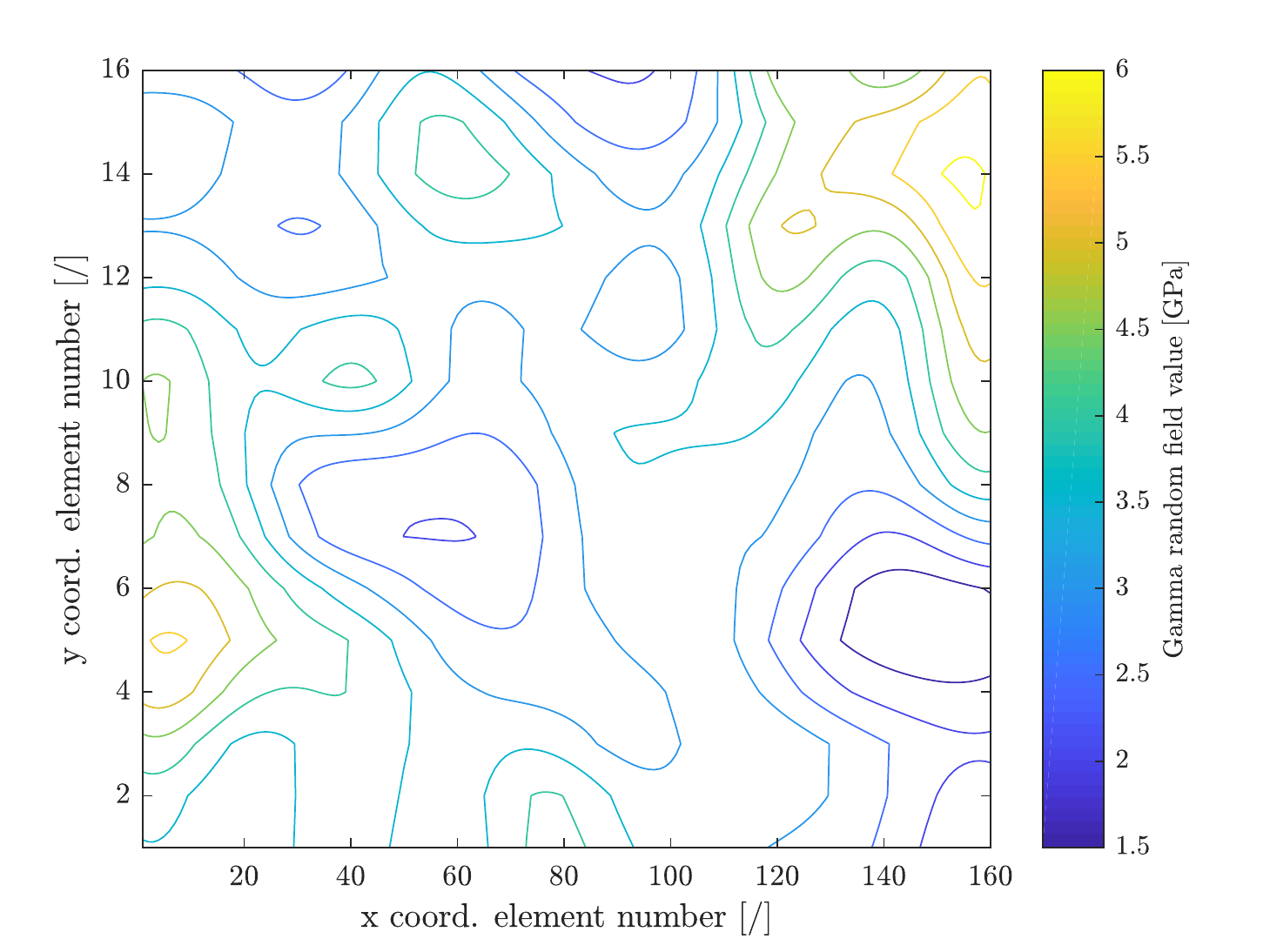}}
%	\caption{Gamma Random Field}
\label{fig:Gamma random field}
\end{subfigure}
\caption{Gaussian random field (left) and the corresponding Gamma random field (right).}
\label{fig:contour}
\end{figure}

Once the Gaussian field has been generated, a memoryless transformation is applied pointwise,
\begin{linenomath*}
\begin{equation}\label{eq:MemTrans}
g(y) = F^{-1}\left[\Phi(y)\right],
\end{equation}
\end{linenomath*}
in order to obtain the Gamma random field \cite{Grigoriu}.
Here, $F$ denotes the marginal cumulative density function (CDF) of the target distribution and $\Phi$ the marginal CDF of the standard normal distribution. This transformation is depicted in Fig.\,\ref{fig:Eigenvalf} (right) with the full line representing a realization of $F$, and the dashed line representing a realization of $\Phi$.
Contour plots of a realization of a Gaussian random field and the corresponding Gamma random field are presented for illustration purposes in Fig.\,\ref{fig:contour}.

\subsection{The Finite Element method}
The Finite Element method will be used to compute the responses of the beam assuming plane stress. An equidistant, regular rectangular mesh is applied with bilinear quadrilateral elements. The implemented solver for the elastic and the elastoplastic response is different. The underlying equations and solution methods are reviewed hereunder.
\\
\\
For the static elastic response, the system equation is of the form \label{Displacement_eq}
\begin{linenomath*}
\begin{equation}
\mathbf{K} \mathbf{\underline{u}} = \mathbf{\underline{f}},
\end{equation}
\end{linenomath*}
with $\mathbf{K}$ the global stiffness matrix, $\mathbf{\underline{f}}$ the global nodal force vector and $\mathbf{\underline{u}}$ the displacement. The global stiffness matrix and nodal force vector are obtained from the element stiffness matrices $\mathbf{K^e}$ and the element force vectors $\mathbf{\underline{f}^e}$. These are computed analytically by evaluation of the following integrals:
\begin{linenomath*}
\begin{equation}\label{eq:K}%\label{eq:f}
\mathbf{K^e} = \int_{\Omega} \mathbf{B}^T \mathbf{D} \mathbf{B} d \Omega
~~~~\mbox{and}~~~
\mathbf{\underline{f}^e} = \int_{\Gamma_t} \mathbf{N}^T \mathbf{\overline{t}}_n d \Gamma_t.
\end{equation}
\end{linenomath*}

The element nodal force vector $\mathbf{\underline{f}^e}$ is modeled as a Neumann boundary condition, where  $\mathbf{\overline{t}}_n$ stands for the surface traction specified as a force per unit area  and $\mathbf{N}$ the element shape function matrix, integrated over the free element boundary $\Gamma_t$. The element stiffness matrix $\mathbf{K^e}$ is obtained by integrating the matrix  $\mathbf{B^TDB}$ over the element's surface $\Omega$. Matrix $\mathbf{B}$ is defined as $\mathbf{LN}$ with $\mathbf{L}$ the derivative matrix specified below, and with $\mathbf{D}$ the elastic constitutive matrix for plane stress, containing the element wise material parameters,
\begin{linenomath*}
\begin{equation}
\mathbf{L} = 
\begin{bmatrix}
\pdv{}{x} & 0  \\
0 & \pdv{}{y}  \\
\pdv{}{y} & \pdv{}{x}  \\
\end{bmatrix}\,
~~~~\mbox{and}~~~
\mathbf{D} = \dfrac{E}{1-\nu^2}
\begin{bmatrix}
1 & \nu & 0 \\
\nu & 1 & 0 \\
0 & 0 & \dfrac{1-\nu}{2} \\
\end{bmatrix}\,.
\label{eq:BDmatrix}
\end{equation}
\end{linenomath*}

For the dynamic response case the following equation is obtained,
\begin{linenomath*}
\begin{equation}\label{eq:dyn}
\left(\mathbf{K}(1+\imath \,\eta) - (2 \pi f)^2 \mathbf{M}\right)\mathbf{\underline{u}} = \mathbf{\underline{f}}
~~~\mbox{with}~~\mathbf{M}^e= \int_{\Omega} \mathbf{N}^T \rho \mathbf{N} d \Omega\,.
\end{equation}
\end{linenomath*}
Matrix $\mathbf{M}$ denotes the system mass matrix obtained from the assembly of the element mass matrices $\mathbf{M^e}$. $f$ denotes the frequency  and $\rho$ the volumetric mass density of the material. The multiplication of the system stiffness matrix $\mathbf{K}$ with the imaginary unit $\imath$ and the constant $\eta$, denotes the damping matrix.
\\
\\
The approach for solving the static elastoplastic case differs due to  the nonlinear stress-strain relation in the plastic domain. The plastic region is governed by the von Mises yield criterion with isotropic linear hardening. An incremental load approach is used starting with a force of $0\,\mathrm{N}$. The methods used to solve the elastoplastic problem are based on Chapter 2 $\S$4 and Chapter 7 $\S$3 and $\S$4 of \cite{Borst}. For this case, the system equation takes the following form:
\begin{linenomath*}
\begin{equation}\label{Displacement_eq_plast}
\mathbf{K} \Delta\mathbf{\underline{u}} = \mathbf{\underline{r}},
\end{equation}
\end{linenomath*}
where  $\Delta\mathbf{\underline{u}}$ stands for the resulting displacement increment. The vector $\mathbf{\underline{r}}$ is the residual, 
\begin{linenomath*}
\begin{equation}
\mathbf{\underline{r}}=\mathbf{\underline{f}}+\Delta\mathbf{\underline{f}}-\mathbf{\underline{q}},
\end{equation}
\end{linenomath*}
where  $\mathbf{\underline{f}}$ stands for the sum of the external force increments applied in the previous steps, $\Delta\mathbf{\underline{f}}$  for the applied load increment of the current step and   $\mathbf{\underline{q}}$  for the internal force resulting from the stresses
\begin{linenomath*}
\begin{equation}
\mathbf{\underline{q}}=\int_{\Omega} \mathbf{B}^T \bm{\sigma} d\Omega.
\end{equation}
\end{linenomath*}

First the displacement increment of all the nodes is computed according to \eqref{Displacement_eq_plast}, with an initial system stiffness matrix $\mathbf{K}$ resulting from the assembly of the element stiffness matrix $\mathbf{K^e}$,  computed by means of a Gauss quadrature
\begin{linenomath*}
\begin{equation} \label{K_el_plast}
\mathbf{K^e} = \int_{\Omega} \mathbf{B}^T \mathbf{D}^{ep} \mathbf{B} d \Omega,
\end{equation}
\end{linenomath*}
where $\mathbf{D}^{ep}$ denotes the elastoplastic constitutive matrix. The initial state of $\mathbf{D}^{ep}$ is the elastic constitutive matrix from \eqref{eq:BDmatrix}. Secondly, the strain increment $\Delta \varepsilon$ is computed,
\begin{linenomath*}
\begin{equation}
\Delta \varepsilon = \mathbf{B}\Delta\mathbf{\underline{u}}.
\end{equation}
\end{linenomath*}
Thirdly, the nonlinear stress-strain relationship,
\begin{linenomath*}
\begin{equation}
d\bm{\sigma} = \mathbf{D}^{ep} d\varepsilon,
\end{equation}
\end{linenomath*}
is integrated by means of a backward Euler method. The backward Euler method essentially acts as an elastic predictor-plastic corrector; an initial stress state that is purely elastic is computed and then projected in the direction of the yield surface as to obtain the plastic stress state. Due to the implicit nature of the integrated stress-strain relation, this equation must be supplemented with the integrated form of the hardening rule and the yield condition. This system of nonlinear equations is then solved with an iterative Newton-Raphson method.
Afterwards, the consistent tangent stiffness matrix is computed \cite{Agusti}. This matrix is then used to compute the updated element stiffness matrix \eqref{K_el_plast}, resulting in an updated system stiffness matrix $\mathbf{K}$. 

The inner iteration step of solving the stress-strain relation and the updated system stiffness matrix is repeated for each outer iteration step which solves equation \eqref{Displacement_eq_plast}. The outer step  consists in balancing the internal forces with the external ones as to satisfy the residual, which in our case equals $10^{-4}$ times the load increment. The procedure used is incremental-iterative, relying on the iterative Newton-Raphson method.  This process is repeated for each load increment.

 Each finite element is assigned a value of the Young's modulus. This is accomplished by means of the midpoint approach, i.e., the value is taken constant within each individual element and equal to the value of the realization of the random field at the center point of the element \cite{jie_lie}.

\section{The Multilevel Monte Carlo method}
\subsection{Method overview}
\label{sec:MethodOverview}
\begin{figure}[H]
\centering
\includegraphics[height=1.9cm]{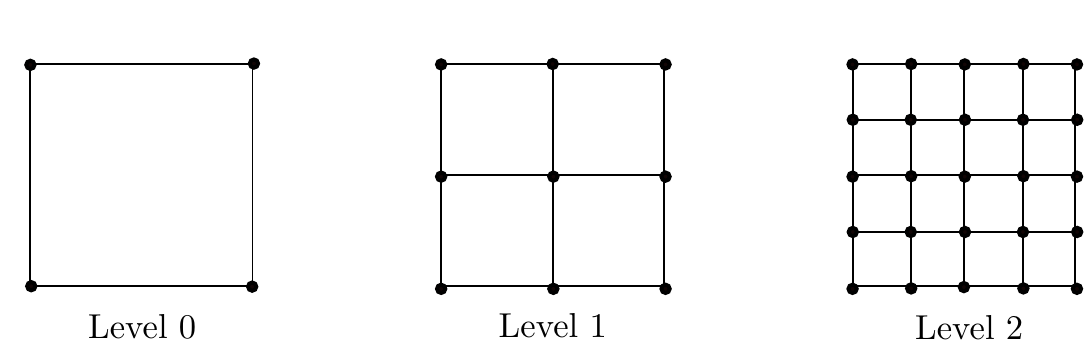}
\caption{Illustrative example of a hierarchy used in the MLMC method.}
\label{fig:gridrefinement}
\end{figure}

The Multilevel Monte Carlo method (MLMC) is an extension of the standard Monte Carlo (MC) method, see, e.g.,~\cite{Giles, Giles2}. The method relies on a clever combination of many computationally cheap low resolution samples and a relatively small number of higher resolution, but computationally more expensive samples. The application of MLMC to our problem will be based on a hierarchy of nested finite element meshes. These meshes will be indexed from $0$ to $L$, with $0$ indicating the coarsest mesh and $L$ the finest mesh. An example of such a hierarchy is shown in Fig.\,\ref{fig:gridrefinement}. It is common in the PDE setting to use a geometric relation for the number of degrees of freedom between the different levels. We set the number of finite elements for a mesh at level $\ell$  proportional to $2^{d\ell}$, where $d$ is the dimension of the problem.

Let $\E[P_L(\omega)]$, or $\E[P_L]$ for short, be the expected value of a particular quantity of interest $P$ depending on a random variable $\omega$, discretized on mesh $L$. The standard MC estimator for $\E[P_L]$ using $N_L$ samples on mesh $L$, denoted as $Q^{\textrm{MC}}_L$, can be written as
\begin{linenomath*}
\begin{equation}
Q^{\textrm{MC}}_L={\frac{1}{N_L}}\sum_{n=1}^{N_L} P_L(\omega^n)\,.
\label{eq:SimpleMC}
\end{equation}
\end{linenomath*}
Multilevel Monte Carlo, on the other hand, starts from a reformulation of $\E[P_L]$ as a telescoping sum. The expected value of the quantity of interest on the finest mesh is expressed as the expected value of the quantity of interest on the coarsest mesh, plus a series of correction terms (or \emph{differences}):
\begin{linenomath*}
\begin{equation}
\E[P_L]=\E[P_0]+\sum_{\ell=1}^L \E[P_\ell -P_{\ell-1}]\,.
\label{Eq:TelescopingSum}
\end{equation}
\end{linenomath*}
Each term in the right-hand side is then estimated separately by a standard Monte Carlo estimator with $N_\ell$ samples, i.e.,
\begin{linenomath*}
\begin{equation}
Q^{\textrm{MLMC}}_L=\frac{1}{N_0}\sum_{n=1}^{N_0} P_0(\omega^{n}) + \sum_{\ell=1}^L \left \{ \frac{1}{N_\ell} \sum_{n=1}^{N_\ell} \left( P_\ell(\omega^{n})-P_{\ell-1}(\omega^{n})\right) \right \},
\label{eq:MLMC}
\end{equation}
\end{linenomath*}

where $Q^{{\textrm{MLMC}}}_L$ is the Multilevel Monte Carlo estimator for the expected value $\E[P_L]$, which is a discrete approximation of the continuous solution $\E[P]$. The mean square error (MSE) is defined as 
\begin{linenomath*}
\begin{equation}\label{eq:MSE}
\begin{split}
\textrm{MSE}(Q^{\textrm{MLMC}}_L) & := \E\left[\left(Q^{\textrm{MLMC}}_L - \E\left[P\right]\right)^2\right] \\
& := \V\left[Q^{\textrm{MLMC}}_L\right] + \left(\E\left[Q^{\textrm{MLMC}}_L\right] - \E\left[P\right]\right)^2,
\end{split}
\end{equation}
\end{linenomath*}

with $\V\left[\mathrm{x}\right]$ denoting the variance of a random variable $\mathrm{x}$.
The MLMC estimator in \eqref{eq:MLMC} can be written as a sum of $L+1$ estimators for the expected value of the difference on each level, i.e.,
\begin{linenomath*}
\begin{equation}
Q^{\textrm{MLMC}}_L = \sum_{\ell = 0}^L Y_{\ell}, \quad \text{where} \quad Y_{\ell} = \frac{1}{N_\ell} \sum_{n=1}^{N_\ell} \left( P_\ell(\omega^{n})-P_{\ell-1}(\omega^{n})\right).
\end{equation}
\end{linenomath*}

where we defined $P_{-1}\coloneqq0$.

Because of the telescoping sum, the MLMC estimator is an unbiased estimator for the quantity of interest on the finest mesh, i.e.,
\begin{linenomath*}
\begin{equation}
\E[P_L] = \E[Q^{\textrm{MLMC}}_L].
\end{equation}
\end{linenomath*}

Denoting by $V_{\ell}$ the variance of the difference $P_{\ell} - P_{\ell-1}$, the variance of the estimator can be written as
\begin{linenomath*}
\begin{equation}
\V[Q^{\textrm{MLMC}}_L] = \sum_{\ell=0}^L \frac{V_{\ell}}{N_{\ell}}.
\end{equation}
\end{linenomath*}

In order to ensure that the MSE is below a given tolerance $\epsilon^2$, it is sufficient to enforce that the variance $\V[Q^{\textrm{MLMC}}_L]$ and the squared bias $(\E[P_L-P])^2$ are both less than $\epsilon^2/2$.
The condition on the variance of the estimator can be used to determine the number of samples needed on each level $\ell$. Following the classic argument by Giles in~\cite{Giles}, we minimize the total cost of the MLMC estimator
\begin{linenomath*}
\begin{equation}
\text{cost}(Q^{\textrm{MLMC}}) = \sum_{\ell=0}^{L} N_\ell C_\ell, 
\end{equation}
\end{linenomath*}
where $C_\ell$ denotes the cost to compute a single realization of the difference $P_\ell-P_{\ell-1}$, subject to the constraint
\begin{linenomath*}
\begin{equation}
\sum_{\ell=0}^L \frac{V_{\ell}}{N_{\ell}} \leq \frac{\epsilon^2}{2}.
\end{equation}
\end{linenomath*}
Treating the $N_\ell$ as continuous variables, we find
\begin{linenomath*}
\begin{equation}\label{eq:nopt}
 N_\ell = \frac{2}{\epsilon^2} \sqrt{\frac{V_\ell}{C_\ell}} \sum_{\ell=0}^L \sqrt{V_\ell C_\ell} .
\end{equation}
\end{linenomath*}
Note that if $\E[P_\ell]\rightarrow\E[P]$, then $V_\ell\rightarrow0$ as $\ell$ increases. Hence, the number of samples $N_\ell$ will be a decreasing function of $\ell$. This means that most samples will be taken on the coarse mesh, where samples are cheap, and increasingly fewer samples are required on the finer, but more expensive meshes. In practice, the number of samples must be truncated to $\lceil N_\ell \rceil$, the least integer larger than or equal to $N_\ell$. 

Using~\eqref{eq:nopt}, the total cost of the MLMC estimator can be written as
\begin{linenomath*}
\begin{equation}\label{eq:cost}
\text{cost}(Q^{\textrm{MLMC}}) = \frac{2}{\epsilon^2}\left(\sum_{\ell=0}^L \sqrt{V_\ell C_\ell}\right)^{2}.
\end{equation}
\end{linenomath*}
This can be interpreted as follows. When the variance $V_\ell$ decreases faster with increasing level $\ell$ than the cost increases, %\pj{i.e., $\beta>\gamma$,} 
the dominant computational cost is located on the coarsest level. The computational cost is then proportional to $V_0 C_0$, which is small because $C_0$ is small. Conversely, if the variance decreases slower with increasing level $\ell$ than the cost increases, %\pj{i.e., $\beta<\gamma$,} 
the dominant computational cost will be located on the finest level $L$, and proportional to $V_L C_L$. This quantity is small because $V_L$ is small. For comparison, the computational cost of a Monte Carlo simulation that reaches the same accuracy is proportional to $V_0 C_L$.

In our numerical results presented next, we will compare the cost of the MLMC estimator to the cost of a standard MC simulation, both in actual runtime (seconds) and in some normalized cost measure. This cost is chosen such that the time needed to obtain a sample on the coarsest mesh (level zero) is equal to a cost of one unit. When using the standard geometric mesh hierarchy, the cost for a single sample on level $\ell$ is proportional to $2^{\gamma \ell}$. The factor $\gamma$ is determined by the efficiency of the solver. The advantage of using this normalized cost is that it decouples the cost of the simulation from the specific computer hardware. 

The second term in \eqref{eq:MSE} is used to determine the maximum number of levels $L$. A typical MLMC implementation is level-adaptive, i.e., starting from a coarse finite element mesh, finer meshes are only added if required to reach a certain accuracy. Assume that the convergence $\E[P_\ell]\rightarrow\E[P]$ is bounded as $|\E[P_{\ell} - P]| = \mathcal{O}(2^{-\alpha \ell})$. Then, we can use the heuristic
\begin{linenomath*}
\begin{equation}\label{eq:bias_constraint}
\abs{\E[P_L-P]} = \abs{\sum_{\ell=L+1}^{\infty} \E[P_\ell-P_{\ell-1}]} \approx \frac{\abs{\E[P_L-P_{L-1}]}}{2^\alpha-1}
\end{equation}
\end{linenomath*}
and check for convergence using $|\E[P_L-P_{L-1}]|/(2^\alpha-1)\leq\epsilon/\sqrt{2}$, see~\cite{Giles} for details.

For completion, we now mention the central MLMC complexity theorem. We refer to~\cite{Giles2} for a proof.
\begin{theorem}
Given the  positive constants $\alpha, \beta, \gamma, c_1, c_2, c_3$ such that $\alpha \geq \dfrac{1}{2} \mathrm{min}\left(\beta,\gamma\right)$ and assume that the following conditions hold: 
\begin{enumerate}
\item $\lvert \E[P_{\ell} - P] \rvert \leq c_1 2^{-\alpha \ell}$,
\item $V_{\ell} \leq  c_2 2^{-\beta \ell}$ \;and
\item $C_{\ell} \leq  c_3 2^{\gamma \ell}$.
\end{enumerate}

Then, there exists a positive constant $c_4$ such that for any $\epsilon < \exp(-1)$ there exists an  $L$ and  a sequence $\{N_{\ell}\}_{\ell=0}^L$ for which the multilevel estimator, $Q^{\mathrm{MLMC}}_L$ has an $\mathrm{MSE}\leq\epsilon^2$, and
\begin{linenomath*}
\begin{equation}
\mathrm{cost}(Q^{\mathrm{MLMC}}) \leq    \left\{
  \begin{aligned}
& c_4 \epsilon^{-2} && \mathrm{if} \quad  \beta > \gamma, \\
& c_4 \epsilon^{-2}\left(\log \;\epsilon \right)^2 && \mathrm{if} \quad  \beta = \gamma, \\
& c_4 \epsilon^{-2-\left(\gamma-\beta\right)/\alpha} && \mathrm{if} \quad  \beta < \gamma. \\
  \end{aligned}
  \right.
  \label{eq:Algo_regime}
\end{equation}
\end{linenomath*}
\label{Theorem_1}
\end{theorem}

\subsection{Implementation aspects}\label{Implementation_aspects}

The MLMC method is non-intrusive. Only an interface between the Finite Element solver routine and the multilevel routine is necessary. All modules used in the simulation are written in \textsc{Matlab}. 

The elastic computations are run in sequential mode, without making use of parallelization. The elastoplastic computations are run in parallel mode because of their significantly larger computational cost. In particular, the computation of  the individual samples is parallelized in order to get a speedup. This is possible because of the \emph{embarrassingly parallel} nature of the MLMC and the MC methods. For the aforementioned configuration, a number of $12$ samples can be computed concurrently. For more details on load balancing of MLMC samplers, we refer to \cite{Scheichl}

Computation of the optimal number of samples per level according to \eqref{eq:nopt} is based on the variances of one finite element node on these levels. The node chosen  for this task, is the node with the biggest variance of all the nodes which make up the finite element mesh. This way, the variance constraint is guaranteed to be satisfied for all nodes that constitute the finite element mesh. The first estimation of these variances is done by computing a trial sample set on levels 0, 1 and 2.
The size of this sample set is $200$ for the elastic responses and $24$ for the elastoplastic responses. Variances on additional levels are estimated according to the second condition from Theorem \ref{Theorem_1}, following \cite{Giles, Giles2}. These trial samples are not included in the results presenting the number of samples per level. They are, however, included in the total runtime in seconds  and the normalized cost of the algorithm. For MC, the number of trial samples taken for the elastic and elastoplastic response is identical as for  level 0 in MLMC. As in the MLMC case, these samples will not be included in the tables.

\section{Numerical Results}

In this section, we discuss our numerical experiments with the MLMC method. We consider the static and dynamic response, using both a homogeneous and a heterogeneous uncertain Young's modulus. First a comparison will be made between the MC method and the MLMC method. Secondly, the rates from Theorem 1 will be estimated. For these first two parts, only the static responses (elastic and elastoplastic) are considered. This is because solutions for the static responses  require only one MLMC simulation and one MC simulation for comparison. In contrast, solutions for the dynamic response require multiple individual MLMC simulations and multiple MC simulations: one for each individual frequency of the frequency response function. The third and last part visually  presents solutions and their uncertainty. For both static responses, the solutions consist of displacements of the beam in the spatial domain. For the dynamic response, the solution is a frequency response function. 

All the results have been computed  on a  workstation equipped with 12 physical (24 logical) cores, Intel Xeon E5645 CPU's, clocked at 2.40 GHz,  and a total of 49 GB RAM.  

\subsection{Comparison between MC and MLMC}
\label{par:comparison}
First, we compare the efficiency of MLMC and MC for the static elastic and elastoplastic response of the beam clamped at both sides and loaded in the middle as shown in Fig.\,\ref{fig:bean_configurations} (right). For both the elastic and the elastoplastic case, the load is modeled as a distributed load acting on each of the vertical middle nodes of the beam. The sum of all these individual loads is independent of the refinement of the mesh.  For the elastic case, the total load equals $10000\,\mathrm{kN}$. For the elastoplastic case, the beam is loaded in steps of $135\,\mathrm{N}$ from $0\,\mathrm{N}$ to $13.5\,\mathrm{kN}$. The quantity of interest is the node with the maximum transversal deflection, which coincidently also corresponds to the node with the largest variance. The coarsest finite element mesh (level 0) consists of 410 degrees of freedom with a square element size of $0.0625\,\mathrm{m}$, while the finest finite element mesh considered (level 4) consists of 83330 degrees of freedom with a square element size of $0.0039\,\mathrm{m}$. The coarsest mesh is chosen so as to discretize the beam in the height by at least four elements.

\subsubsection{Elastic Response}

\begin{figure}[H]
\begin{subfigure}[b]{0.54\textwidth}
\centering
\scalebox{0.45}{
	\makebox[\textwidth]{
	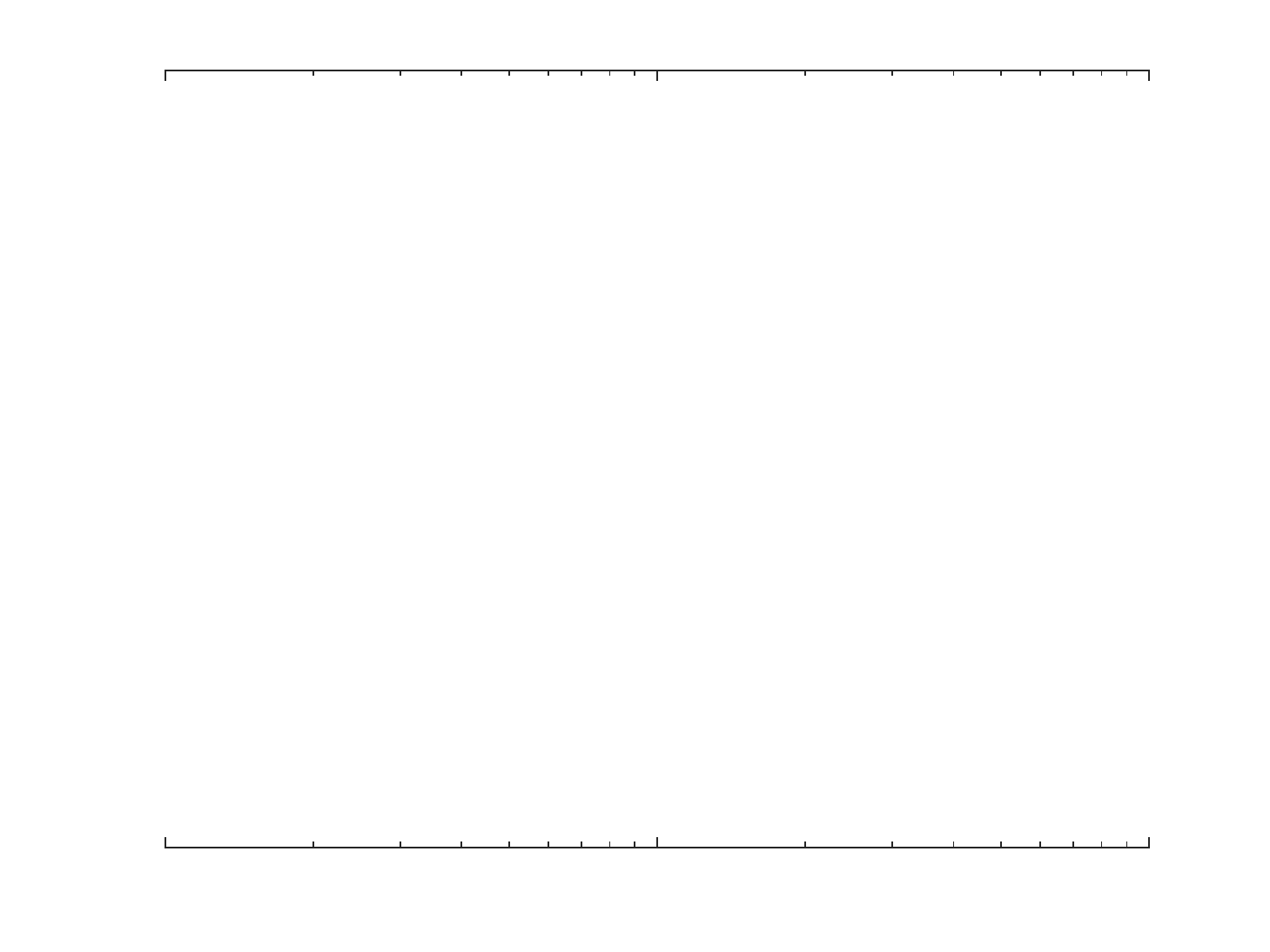}}
\end{subfigure}
%\centering
\begin{subfigure}[b]{0.45\linewidth}
\centering
	\scalebox{0.45}{
		\makebox[\textwidth]{
	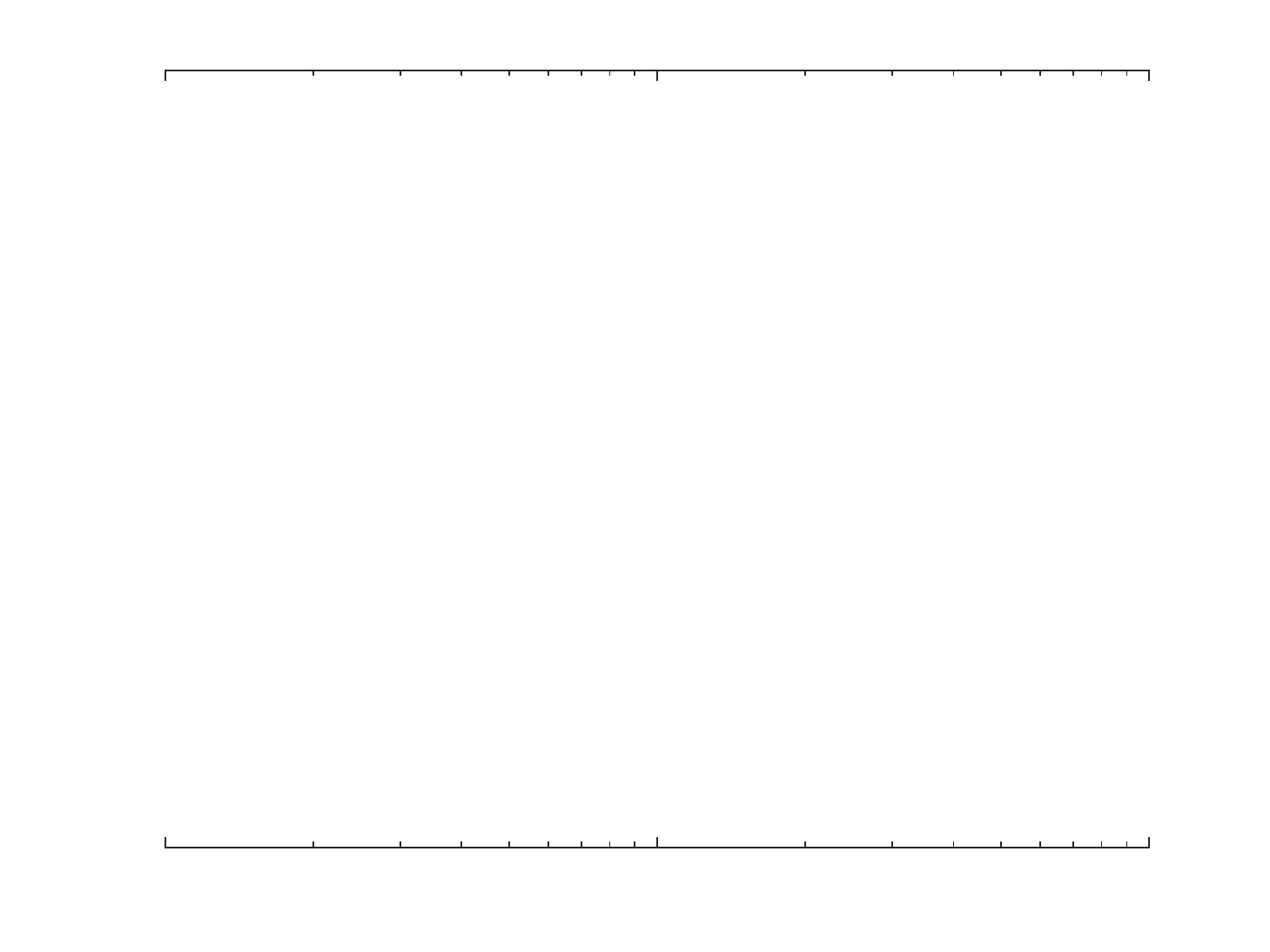}}
\end{subfigure}
\caption{Actual simulation time in seconds as a function of the desired tolerance on the RMSE applied to the concrete beam clamped at both sides in Fig.\,\ref{fig:bean_configurations} (right) for a homogeneous (left) and heterogeneous (right) Young's modulus, in the elastic domain.}
\label{fig:Times}
\end{figure}

Fig.\,\ref{fig:Times} compares the actual simulation time needed to reach a certain tolerance $\epsilon$ on the root mean square error (RMSE) for both standard MC and MLMC, for a homogeneous (left) and heterogeneous uncertain Young's modulus (right). Our MLMC method consistently outperforms the standard MC method, with speedups up to a factor ten or more. In Tab.\,\ref{table:times}, we summarize the results expressed in actual simulation time and in standard cost. For MC, some values of the RMSE have been omitted. This is due to the long computation time, which is of the order of days.
Also note that the MC simulation is run at the finest level $L$ of the corresponding MLMC simulation, where $L$ is chosen according to~\eqref{eq:bias_constraint}. This explains the sudden jump in the MC simulation time in Fig.\,\ref{fig:Times} (right). As stated in $\S$\ref{Implementation_aspects}, $200$ samples are taken on the first three levels. The average time needed for this is $35$ seconds in case of a homogeneous Young's modulus and $48$ seconds in case of a heterogeneous Young's modulus. For the lowest listed tolerance of 2.5E-4, the percentage of time used to compute these trial samples amounts to $55$ percent of the total time. For finer tolerances, however, the relative cost of this preparatory work rapidly drops and becomes negligible compared to the cost of the remaining part of the algorithm. For the finest tolerance listed of 2.5E-5, this percentage is below $1$ percent. 

\begin{table}[H]
  \scalebox{1}{
%  \hspace{-1cm}
 \begin{tabular}{lllllllll}
 \toprule
 \multirow{3}{*}{RMSE [/]} & \multicolumn{4}{c}{Homogeneous Young's modulus}   & \multicolumn{4}{c}{Heterogeneous Young's modulus}  \\ [0.5ex]
 \cmidrule(rl{4pt}){2-5} \cmidrule(rl{4pt}){6-9} 
  & \multicolumn{2}{c}{Time [sec]} & \multicolumn{2}{c}{Norm. Cost} & \multicolumn{2}{c}{Time [sec]} & \multicolumn{2}{c}{Norm. Cost}\\
  \cmidrule(rl{4pt}){2-3} \cmidrule(rl{4pt}){4-5} \cmidrule(rl{4pt}){6-7} \cmidrule(rl{4pt}){8-9}
   & MLMC & MC & MLMC & MC & MLMC & MC & MLMC & MC \\
\cmidrule{1-9}
  2.5E-4 & 84& 934 & 9.43E3 & 8.79E4 & 86& 227 & 6.91E3 & 1.69E4\\
  7.5E-5 & 514& 9671 & 5.89E4& 9.88E5 & 334& 1578 & 3.49E4 & 1.33E5\\
  5.0E-5 & 1287 & -& 1.34E5& - & 991& 17588 & 1.15E5 & 2.01E6\\
  2.5E-5 & 4650& - & 5.19E5&-& 5244& - & 5.68E5 & - \\
  \bottomrule
\end{tabular}}
\caption{Actual simulation time in seconds and normalized cost for MLMC and MC applied to the concrete beam clamped at both sides in Fig.\,\ref{fig:bean_configurations} (right) for a homogeneous and heterogeneous Young's modulus, in the elastic domain.}
\label{table:times}
\end{table}

\begin{figure}[H]
\begin{subfigure}[b]{0.54\textwidth}
\centering
\scalebox{0.45}{
	\makebox[\textwidth]{
	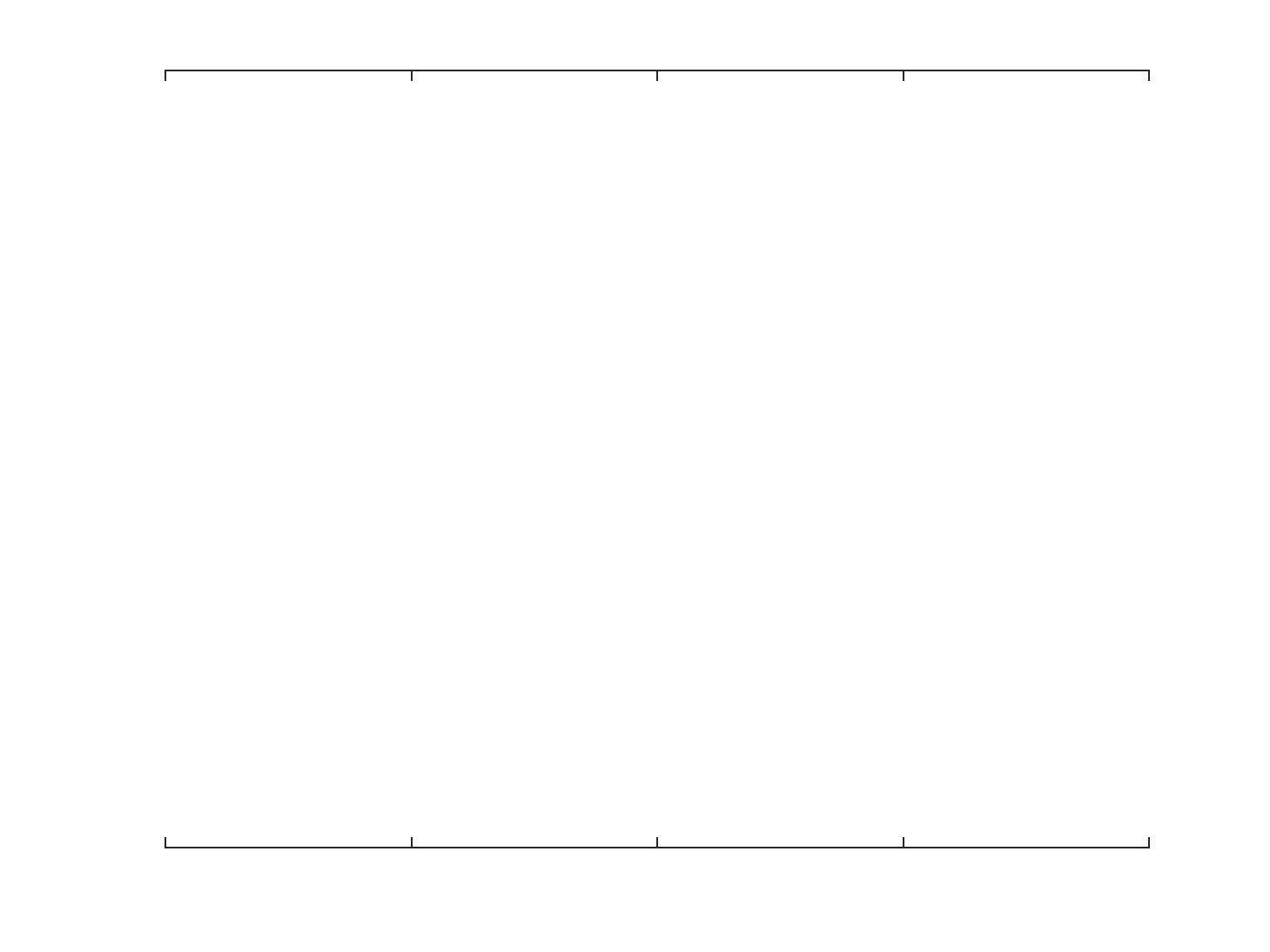}}
\end{subfigure}
%\centering
\begin{subfigure}[b]{0.45\linewidth}
\centering
	\scalebox{0.45}{
		\makebox[\textwidth]{
	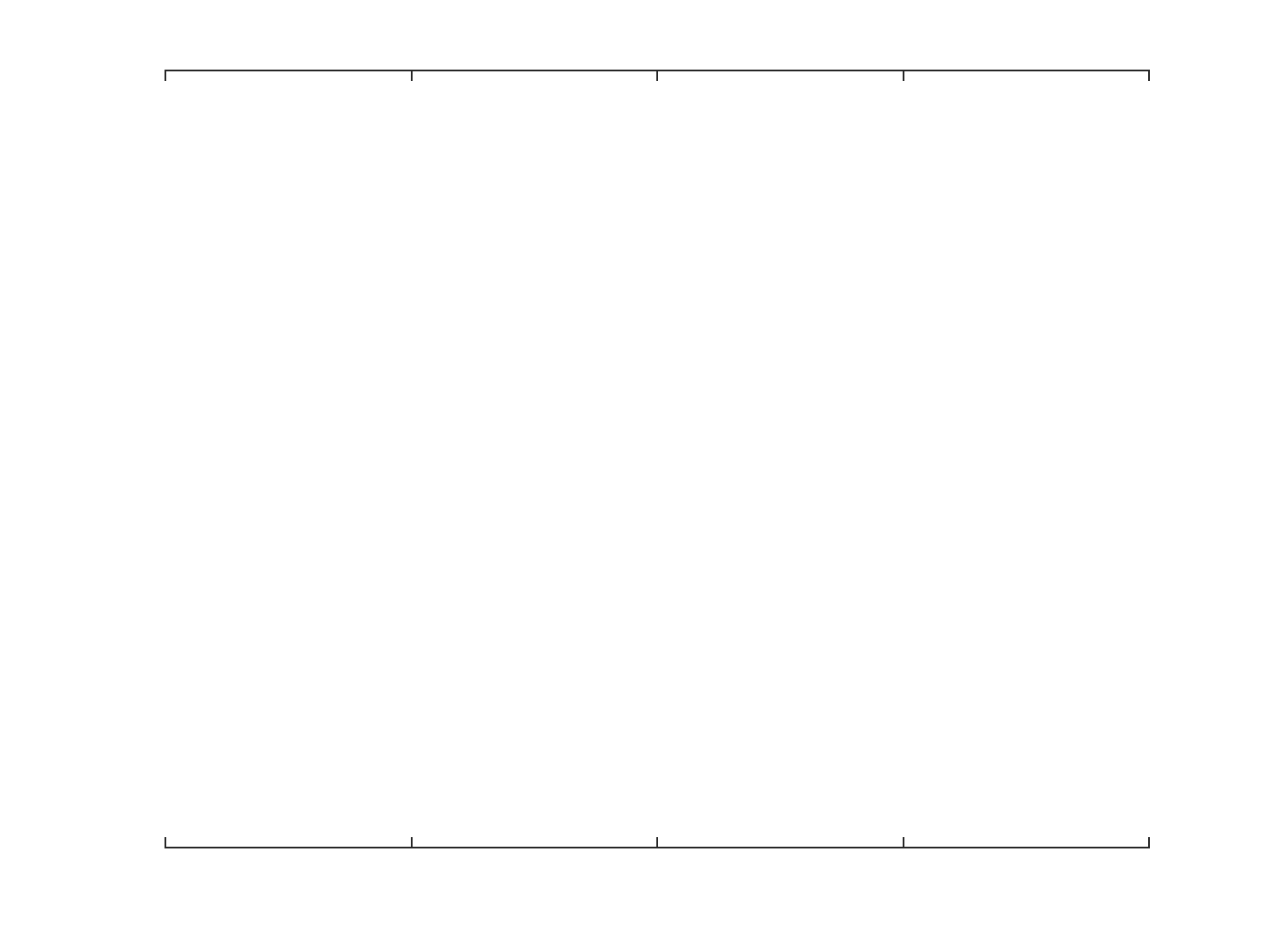}}
\end{subfigure}
\caption{Total number of samples on each level for different tolerances $\epsilon$ on the RMSE using a homogeneous (left) and heterogeneous (right) Young's modulus for the elastic response.}
\label{fig:sample_sizes}
\end{figure}

 \begin{table}[H]
 \setlength{\tabcolsep}{.16667em}
%\centering
\scalebox{1}{
%\hspace{-2cm}
 \begin{tabular}{ccccccccccccccc}
\toprule
 \multirow{5}{*}{RMSE [/]} & \multicolumn{7}{c}{Homogeneous Young's modulus} & \multicolumn{7}{c}{Heterogeneous Young's modulus}\\
\cmidrule(rl{4pt}){2-8}  \cmidrule(rl{4pt}){9-15} 
 \multicolumn{1}{c}{} & \multicolumn{6}{c}{MLMC} & \multicolumn{1}{c}{MC} &  \multicolumn{6}{c}{MLMC} & \multicolumn{1}{c}{MC}\\
\cmidrule(rl{4pt}){2-7} \cmidrule(rl{4pt}){8-8} \cmidrule(rl{4pt}){9-14} \cmidrule(rl{4pt}){15-15}
&\multicolumn{5}{c}{level}&\multirow{1}{*}{equivalent}&\multirow{1}{*}{level}&\multicolumn{5}{c}{level}&\multirow{1}{*}{equivalent}&\multirow{1}{*}{level}\\
%\cmidrule(l{8pt}){2-6} \cmidrule(l{8pt}){9-13}
  &{0} & {1} & {2} & {3} &  4& Max &Max &{0} & {1} &{2} & {3} & {4}&Max & Max \\ [0.5ex]
\cmidrule{1-15}
  2.5E-4 & 4408 & 56 & 6 & / &/&262 &4651 & 1183 & 202 & 33 & / & /&153& 771\\
   7.5E-5 & 49816 & 635 & 61 & / &/&2697& 49308 & 13477 & 2169 & 370 & / & /&1568& 6750 \\
  5.0E-5 & 116934 & 1446 & 172 & 49 &/&1651& - & 38322 & 5728 & 893 & 180 & /&918& 16700 \\
   2.5E-5 & 464180 & 5393 & 669 & 71 & 7&1091& -&173742 & 27492 & 4412 & 958 & 218&1374& - \\
\bottomrule
\end{tabular}}
\caption{Number of samples for MLMC and MC for the elastic response.}
\label{tab:Samples}
\end{table}

Fig.\,\ref{fig:sample_sizes} shows the number of samples over the different levels in function of the desired tolerance. Note that the number of samples is decreasing as the level $\ell$ increases, as required. Numerical values for $N_\ell$ are repeated in Tab.\,\ref{tab:Samples}. In order to make a better comparison between the cost of MC and MLMC, we also list, under the column label ``equivalent Max", the equivalent number of samples on the finest level $L$, that is, the number of samples on level $L$ that would yield a cost equal to the MLMC cost without including the cost of the trial samples. These results show a considerably lower sample size in favor of MLMC.

\begin{figure}[H]
\begin{subfigure}[b]{0.32\textwidth}
%\hspace{-0.1cm}
\scalebox{0.32}{
	\makebox[\textwidth]{
	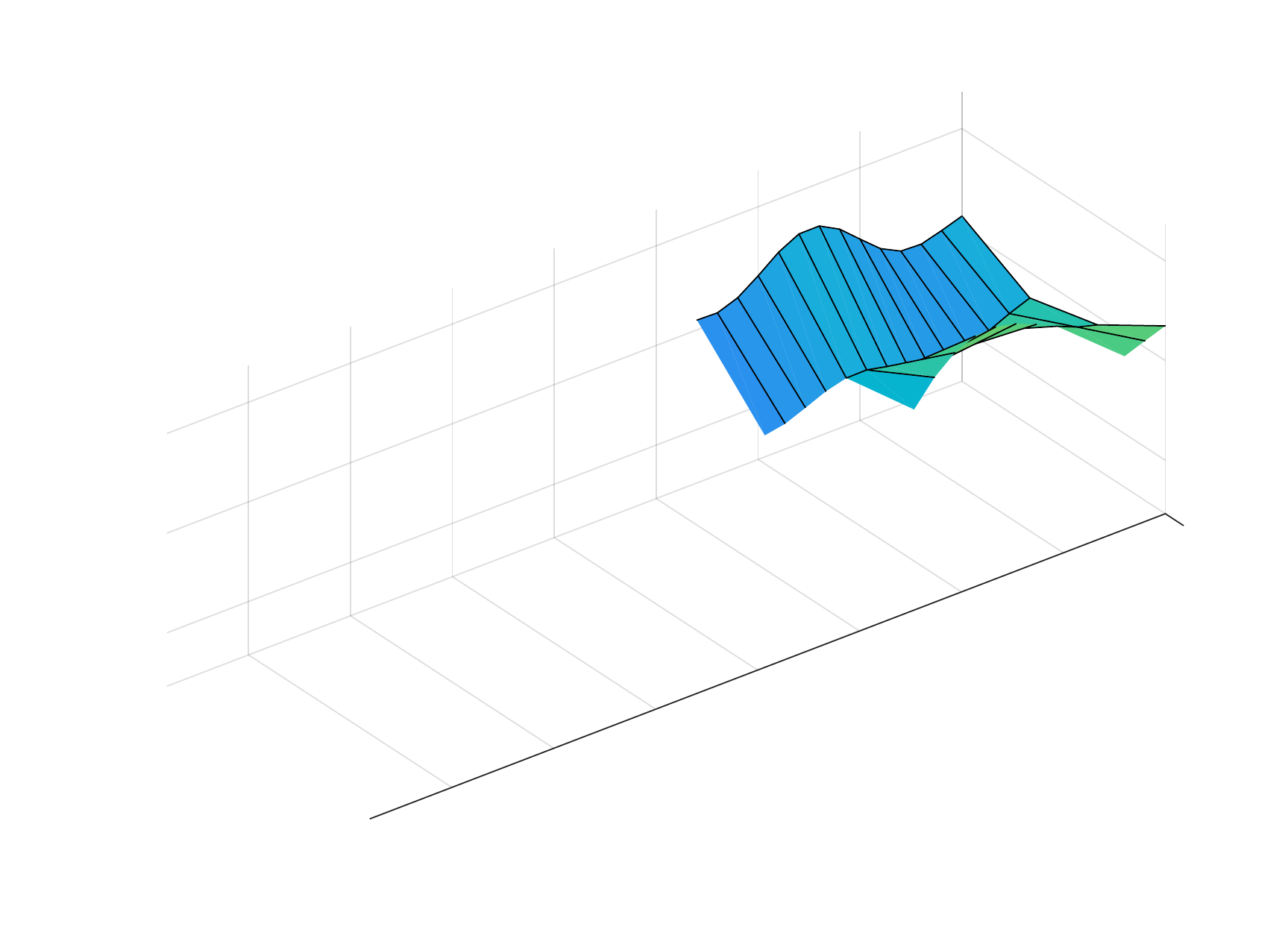}}
\end{subfigure}
%\centering
\begin{subfigure}[b]{0.32\linewidth}
\centering
	\scalebox{0.32}{
		\makebox[\textwidth]{
	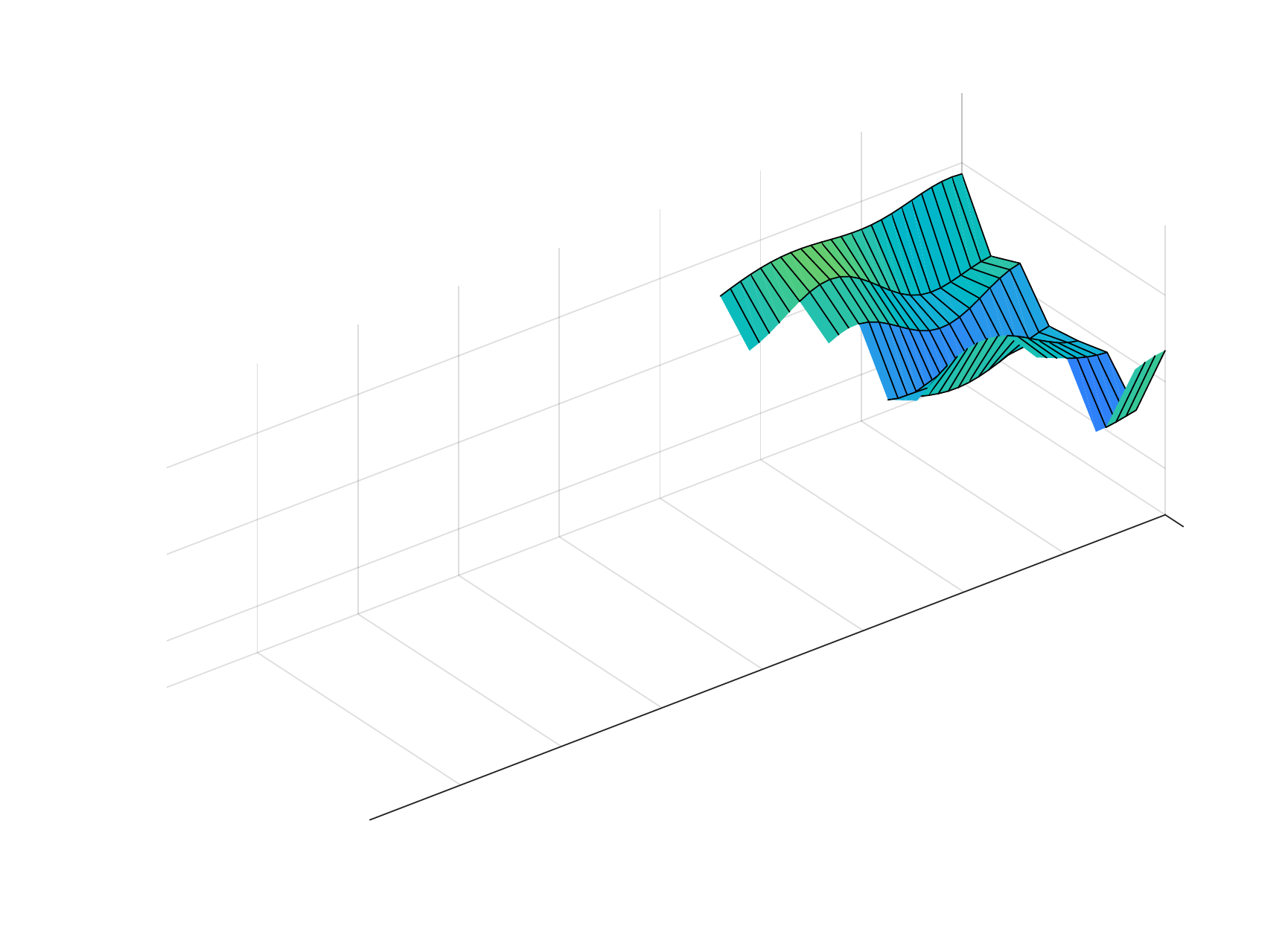}}
\end{subfigure}
\begin{subfigure}[b]{0.32\linewidth}
\hspace{2.5cm}
	\scalebox{0.32}{
			\makebox[\textwidth]{
	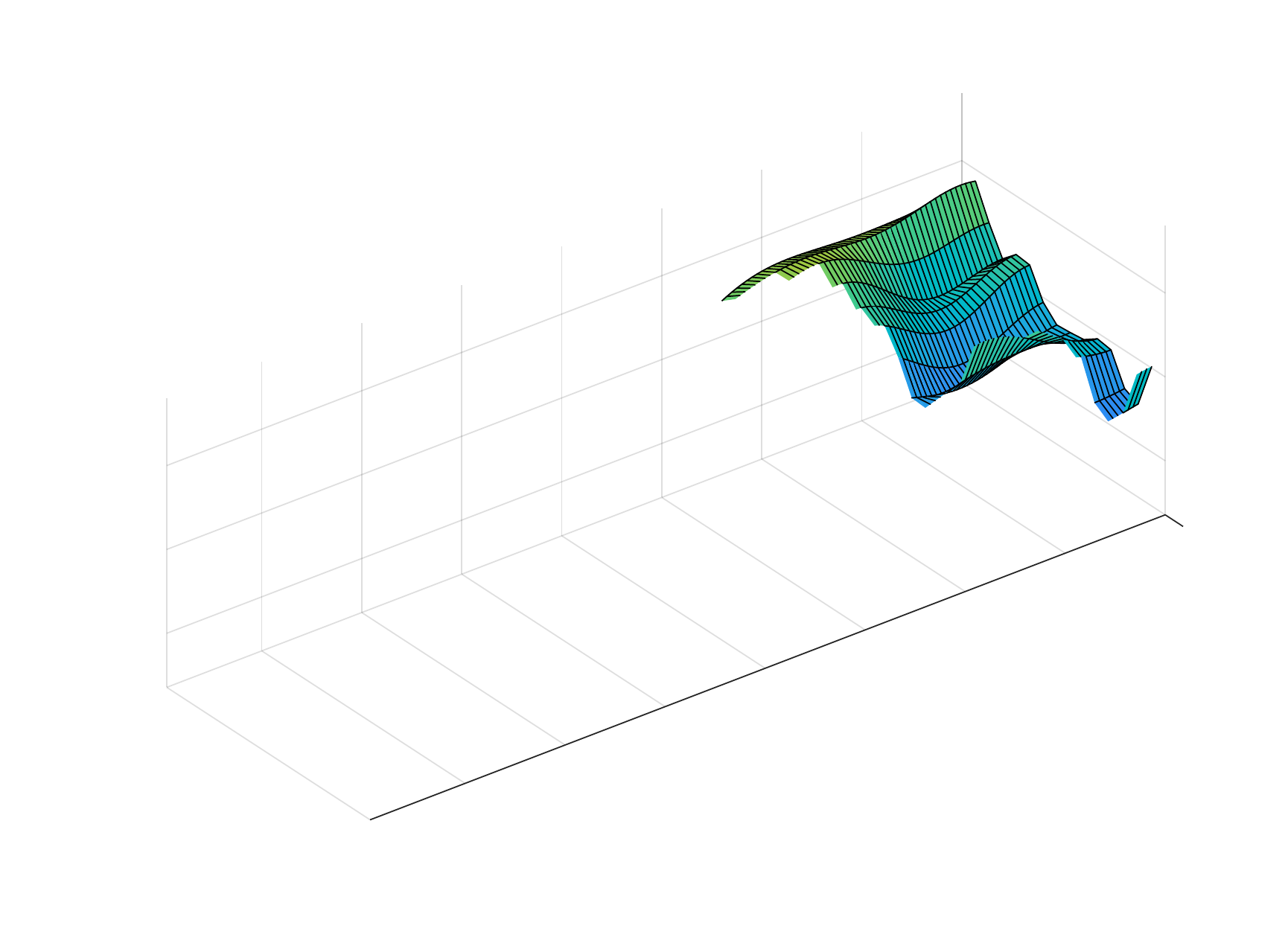}}
\end{subfigure}
	\caption{Realizations of a Gaussian random field on level 0 (left), level 1 (middle), level 2 (right). }
	\label{fig:level_overview}
\end{figure}

Comparing the actual number of samples of MLMC with MC, we notice that when the Young's modulus is homogeneous, the number of samples on the coarsest level in the MLMC simulation is of the same order of magnitude as the number of samples on the finest level $L$ in the MC simulation. This would allow  the estimation of  the number of samples needed for a MC simulation given the results of a MLMC simulation if required. However, when the Young's modulus is heterogeneous, this number of samples differs. This stems from the fact that on this coarsest level a very rough approximation of the random field is used for the MLMC simulation, while the MC simulation uses a random field corresponding to the  finest level of the MLMC simulation.  It is thus probable that on the coarsest level, the variability of the random field is not fully and adequately captured. In order to better illustrate this point, the same Gaussian random field  is shown in Fig.\,\ref{fig:level_overview} at different resolution levels. As can be seen, a sufficiently high number of points is required to appropriately capture the variability. 

In order to validate this hypothesis,  MLMC simulations with a heterogeneous Young's modulus have been rerun with a finer coarsest level. The results are presented in Tab.\,\ref{tab:Samples_GRF_refinement}. As can be seen from this table, the number of samples of MC is of similar magnitude as the number of samples on the coarsest level of MLMC.

  \begin{table}[H]
  \scalebox{1}{
 \begin{tabular}{cc c c c c  c}
\toprule
 \multirow{3}{*}{RMSE [/]}& \multicolumn{5}{c}{MLMC} & \multicolumn{1}{c}{MC}\\
\cmidrule(rl{4pt}){2-6} \cmidrule(rl{4pt}){7-7}
 & \multicolumn{5}{c}{level}& level  \\ [0.5ex]
  & 0 & 1 & 2 & 3 &4 & Max\\ [0.5ex]
\cmidrule{1-7}
  2.5E-4 & / & 738 & 38 & 10 & /& 649 \\
   7.5E-5 & / & 8628 & 468 & 97 & /& 7178  \\
  5.0E-5 & / & 19477 & 969 & 234 & /& 15947  \\
\bottomrule
\end{tabular}}
\caption{Number of samples for MLMC and MC for a heterogeneous Young's modulus with a finer coarsest level.}
\label{tab:Samples_GRF_refinement}
\end{table}

\subsubsection{Elastoplastic Response}

For the elastic response, the MLMC simulations are level adaptive. The number of extra levels to be added depends on the condition of the bias, \eqref{eq:bias_constraint}. 

However, for the elastoplastic response we chose to manually set the maximum level. This level is chosen based on a mesh convergence analysis. This is done because for the  tolerances considered in Tab.\,\ref{table:times_Plast}, samples would have to be taken on levels up to 5 or more. This would constitute a considerable time cost, which results from the iterative nature of the Finite Element implementation.

\begin{figure}[H]

\begin{subfigure}[b]{0.54\textwidth}
\centering
\scalebox{0.45}{
\hspace{-1cm}
	\makebox[\textwidth]{
	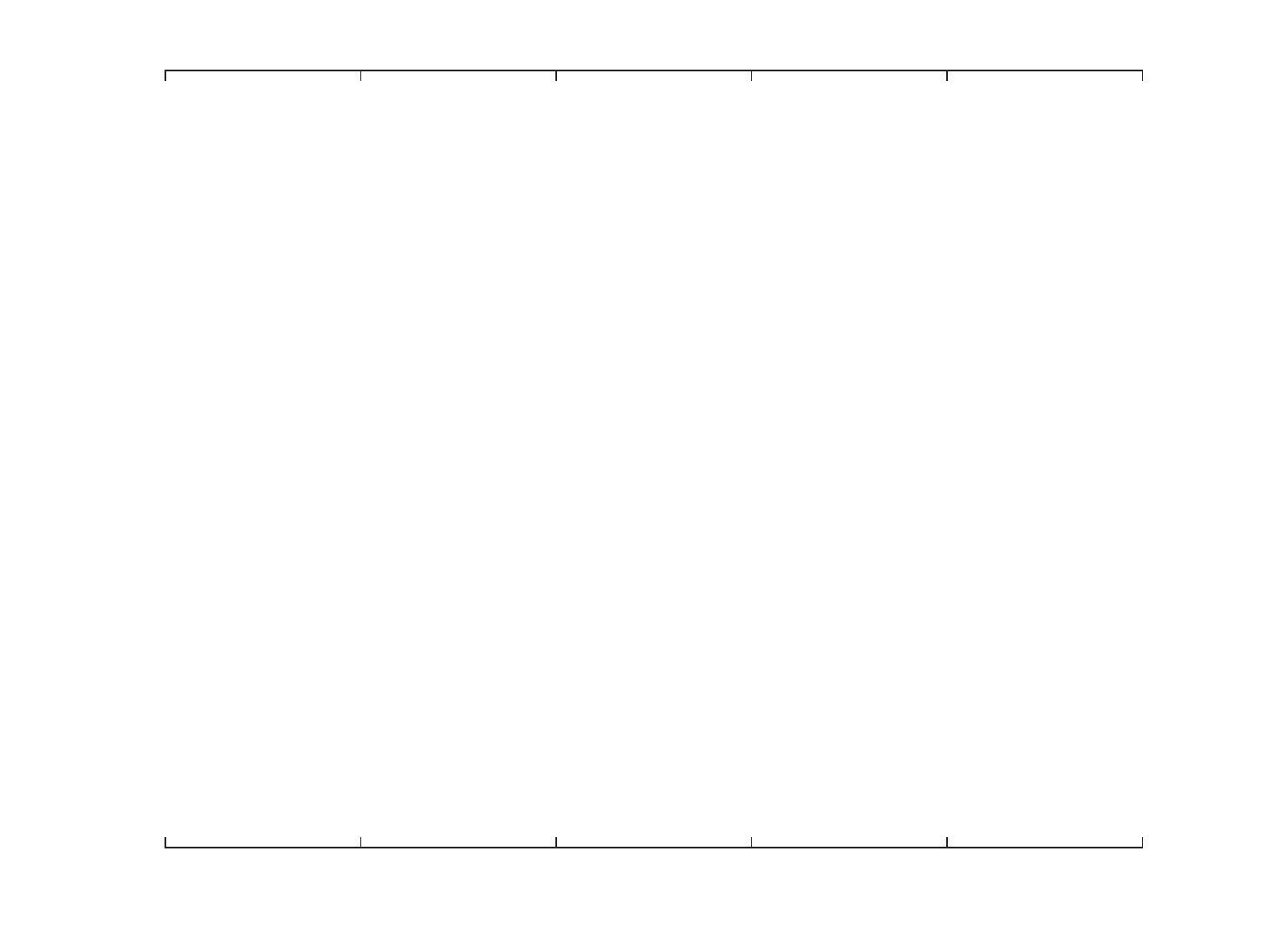}}
\end{subfigure}
\centering
\begin{subfigure}[b]{0.45\linewidth}
\centering
	\scalebox{0.45}{
		\makebox[\textwidth]{
	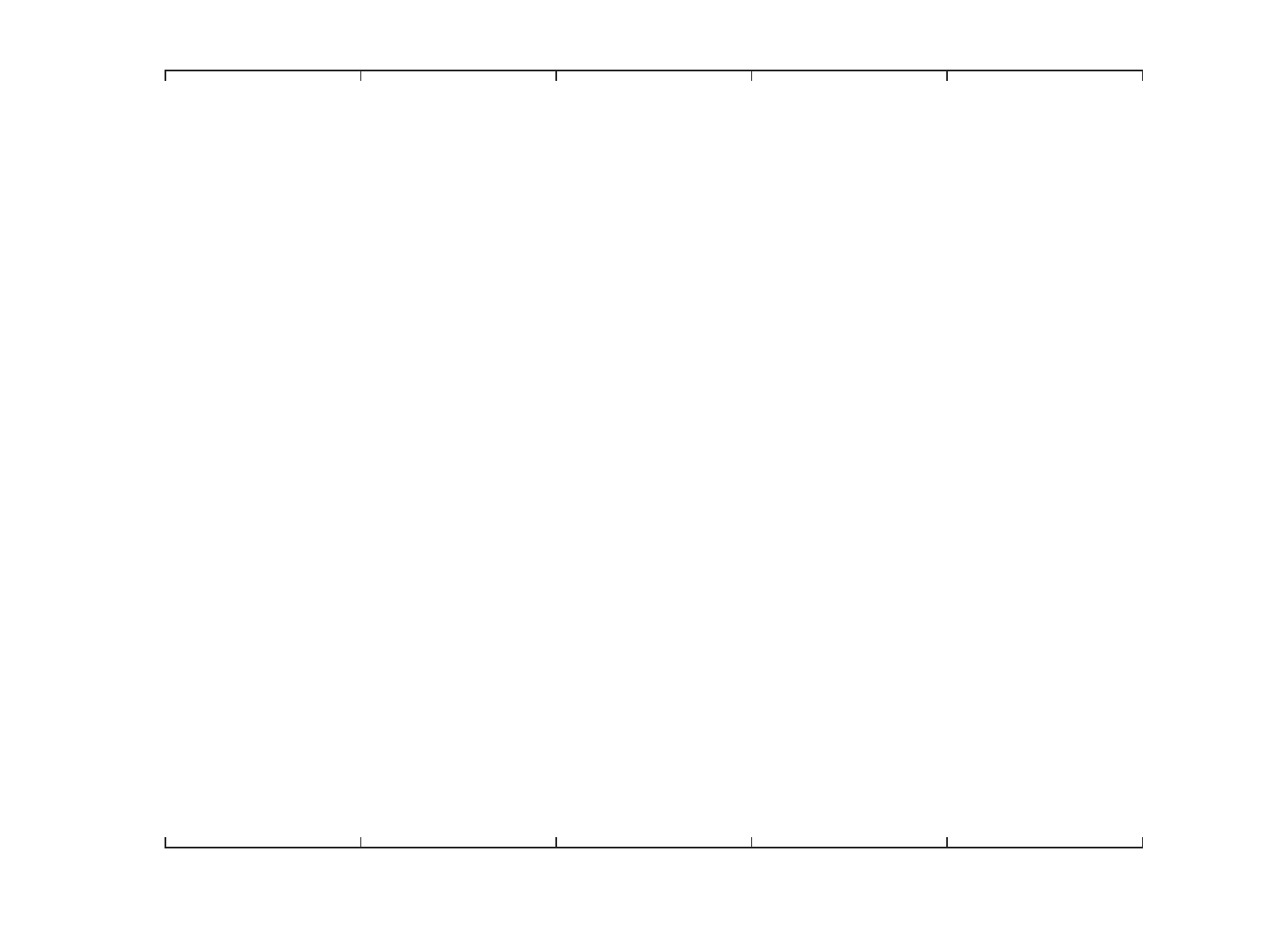}}
\end{subfigure}
\caption{Deflection and difference of the deflection of the beam's top layer middle node for the elastoplastic response (left) and elastic response (right).}
\label{fig:mesh_convergence}
\end{figure}

The results of a mesh convergence study are shown in Fig.\,\ref{fig:mesh_convergence}, respectively for the elastoplastic response (left) and the elastic response (right), with a beam configuration that is clamped at both sides and loaded in the middle. The figures show the transverse deflection of the middle node located on the beam's top layer per level, represented as a full line, and the absolute value of its difference over the levels, represented as a  dashed  line. Comparing the results for both responses, it shows that for the elastic response the absolute value of the deflection's difference on the finest mesh is much smaller  than  the one for the elastoplastic response. For the elastoplastic response, the deflection starts stagnating at around level three. Following these results, we thus state that the bias condition for the elastoplastic response is fulfilled at level three; no more than four MLMC levels are used. 

\begin{figure}[H]
\begin{subfigure}[b]{0.54\textwidth}
\centering
\scalebox{0.45}{
	\makebox[\textwidth]{
	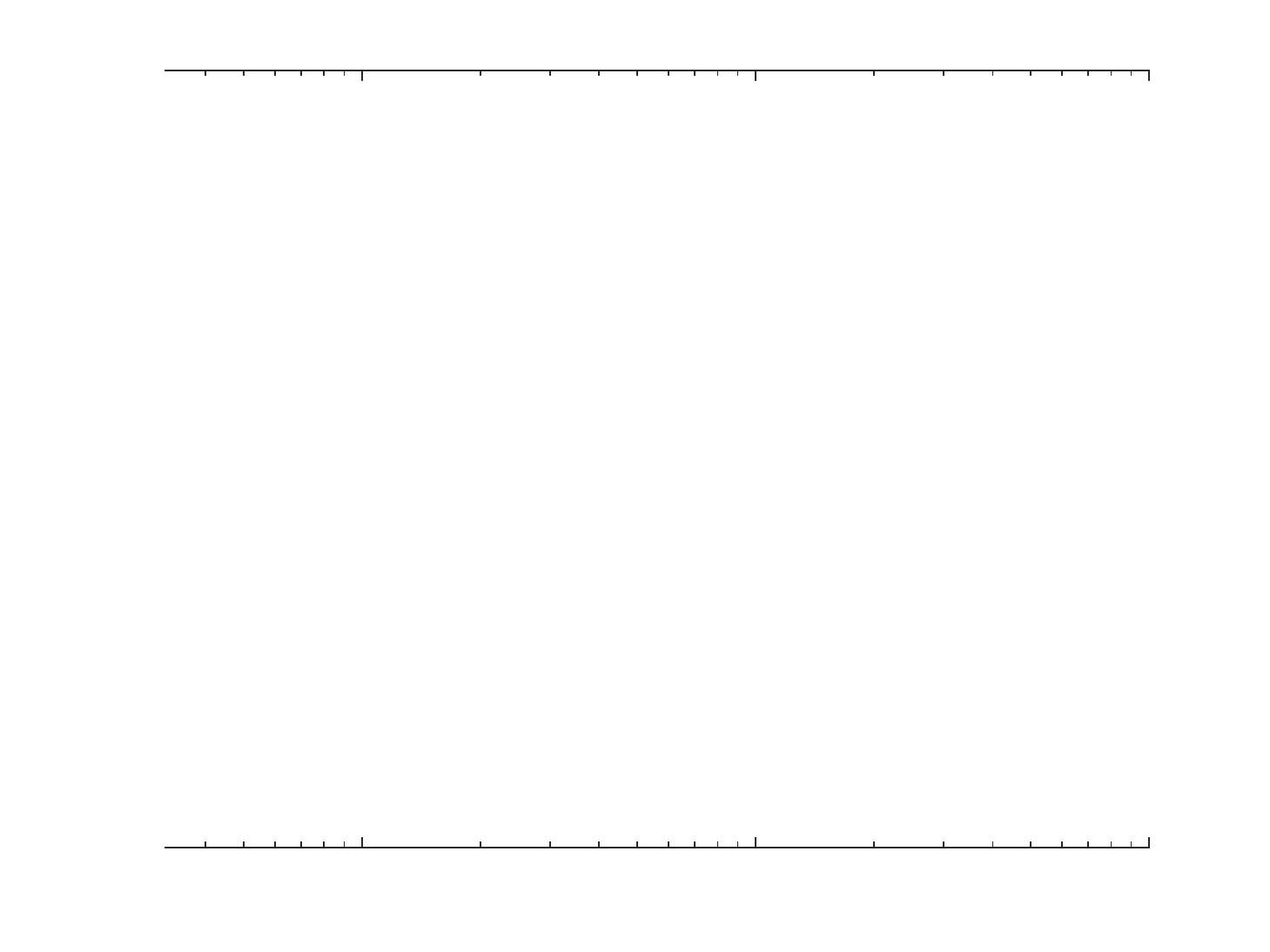}}
\end{subfigure}
%\centering
\begin{subfigure}[b]{0.45\linewidth}
\centering
	\scalebox{0.45}{
		\makebox[\textwidth]{
	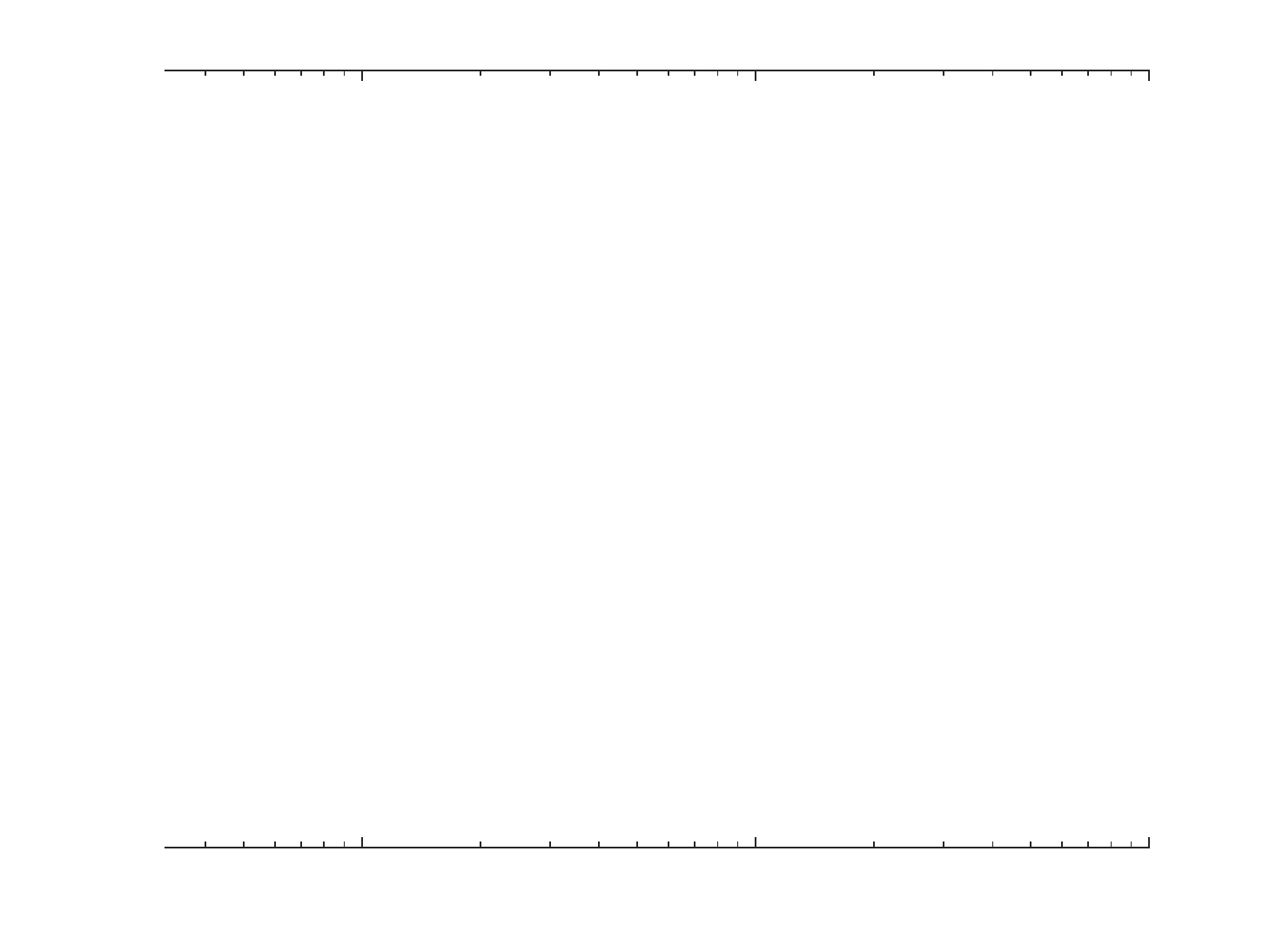}}
\end{subfigure}
\caption{Actual simulation time in seconds as a function of the desired tolerance on the $\text{RMSE}$ applied to the steel beam clamped at both sides in Fig.\,\ref{fig:bean_configurations} (right) for a homogeneous (left) and heterogeneous (right) Young's modulus in the elastoplastic domain.}
\label{fig:Times_plast}
\end{figure}

Fig.\,\ref{fig:Times_plast} lists the actual simulation time needed to reach a user specified tolerance $\epsilon$ on the root mean square error (RMSE) for MLMC for a homogeneous and a heterogeneous Young's modulus for the elastoplatic response, i.e., the steel beam clamped at both sides. These values are presented in Tab.\,\ref{table:times_Plast}. As can be observed, the MLMC simulation outperforms the MC simulation in terms of computational speed and cost. A speedup of up to a factor ten in favor of MLMC is observed. As can be seen in Tab.\,\ref{table:times_Plast},  the costs in case of lower tolerances for MLMC and MC are close together, while for finer tolerances, MLMC is significantly cheaper than MC. As elaborated in $\S$\ref{Implementation_aspects}, $24$ samples are taken on the first three levels in order to obtain an initial estimate for the variances on the coarsest levels. The average time needed for this is $3728$ seconds in case of a homogeneous Young's modulus and $3927$ seconds in case of a heterogeneous Young's modulus. It can be observed that for the lowest tolerance of 2.5E-5 listed here, the percentage of time taken  to compute these trial samples is about $33$ percent. For the finest tolerance of 2.5E-6 listed here, this percentage drops below $6$ percent.

\begin{table}[H]
  \scalebox{1}{
%  \hspace{-1cm}
 \begin{tabular}{ccccccccc}
\toprule
 \multirow{3}{*}{$\text{RMSE}$ [/]} & \multicolumn{4}{c}{Homogeneous Young's modulus}   & \multicolumn{4}{c}{Heterogeneous Young's modulus}  \\ [0.5ex]
   \cmidrule(rl{4pt}){2-5} \cmidrule(rl{4pt}){6-9} 
  & \multicolumn{2}{c}{Time [sec]} & \multicolumn{2}{c}{Norm. Cost} & \multicolumn{2}{c}{Time [sec]} & \multicolumn{2}{c}{Norm. Cost}\\
  \cmidrule(rl{4pt}){2-3} \cmidrule(rl{4pt}){4-5} \cmidrule(rl{4pt}){6-7} \cmidrule(rl{4pt}){8-9}
   & MLMC & MC & MLMC & MC & MLMC & MC & MLMC & MC \\
\cmidrule{1-9}
  2.5E-5 & 11870& 72513 & 1247& 10437 & 12000& 31427 & 1220 & 5732\\
  7.5E-6 & 18415& 868410 & 2810& 183180 & 16869& 192384 & 2014 & 34515\\
  5.0E-6 & 27249 & -& 4915& - & 20854& - & 3182 & -\\
  2.5E-6 & 68047& - & 14089&-& 69641& - & 11241 & - \\
\bottomrule
\end{tabular}}
\caption{Actual simulation time in seconds and normalized cost for MLMC and MC applied to the steel beam clamped at both sides in Fig.\,\ref{fig:bean_configurations} (right) for a homogeneous and heterogeneous Young's modulus in the elastoplastic domain.}
\label{table:times_Plast}
\end{table}

\begin{figure}[H]
\begin{subfigure}[b]{0.54\textwidth}
\centering
\scalebox{0.45}{
	\makebox[\textwidth]{
	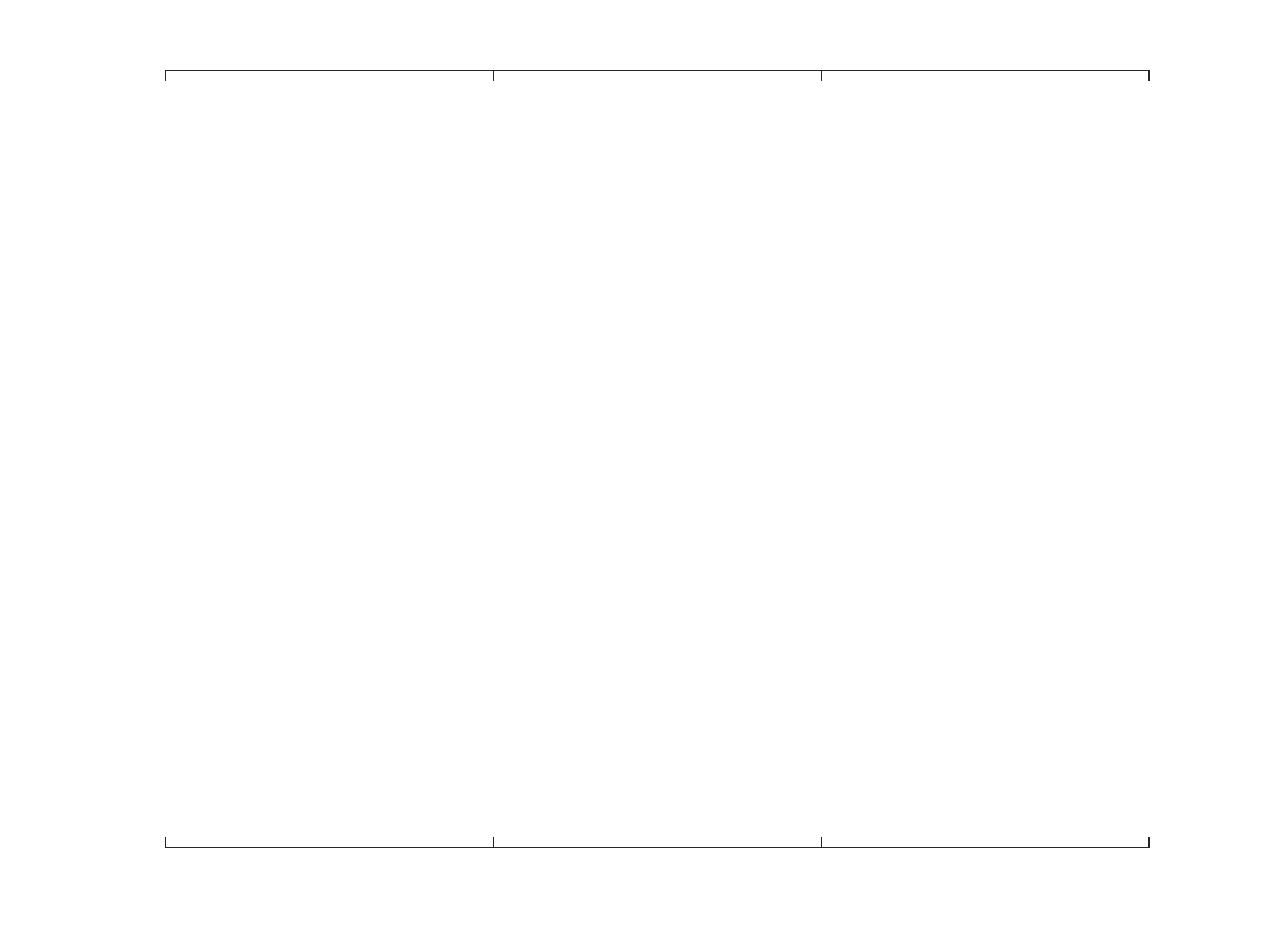}}
\end{subfigure}
%\centering
\begin{subfigure}[b]{0.45\linewidth}
\centering
	\scalebox{0.45}{
		\makebox[\textwidth]{
	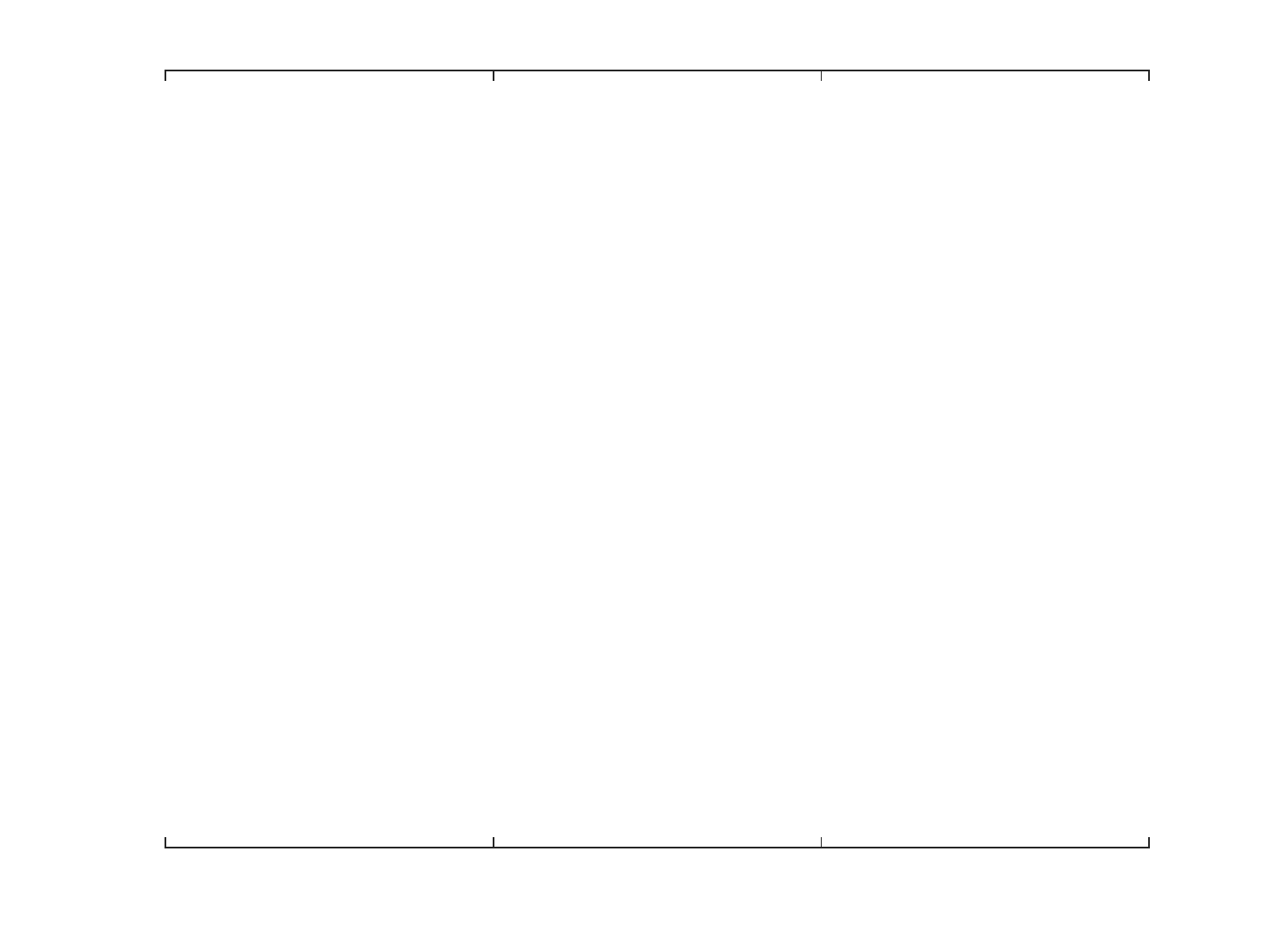}}
\end{subfigure}
\caption{Total number of samples on each level for different tolerances $\epsilon$ on the RMSE using a homogeneous (left) and heterogeneous (right) Young's modulus for the elastoplastic response.}
\label{fig:sample_sizes_plast}
\end{figure}

Fig.\,\ref{fig:sample_sizes_plast} shows again the number of samples over the different levels in function of the desired tolerance. Numerical values are listed in Tab.\,\ref{tab:Samples_plast}. As is the case for the elastic response, the sample size decreases with increasing level. Note that on the finest level, only a few samples are required. For completeness, we need to mention a technical implementation detail. When \eqref{eq:nopt} suggests a value of $N_{\ell}$ equal to or less than 2 on an extra level, i.e., a level greater than 2,  our algorithm still enforces a minimal number of 3 samples. This explains the lower limit of 3 samples  in  Tab.\,\ref{tab:Samples_plast}.

As stated before, the parallelization enables the algorithm to compute $12$ samples concurrently. This is highly efficient for levels with a large number of samples, as is the case for coarser levels, and less efficient for finer levels. This is due to the load imbalance in the latter case.

 \begin{table}[H]
 \setlength{\tabcolsep}{.36667em}
%\centering
\scalebox{1}{
%\hspace{-0.5cm}
 \begin{tabular}{c c c  c c c cc c  c c c  c}
\toprule
 \multirow{5}{*}{$\text{RMSE}$ [/]} & \multicolumn{6}{c}{Homogeneous Young's modulus} & \multicolumn{6}{c}{Heterogeneous Young's modulus}\\
\cmidrule(rl{4pt}){2-7} \cmidrule(rl{4pt}){8-13}
 \multicolumn{1}{c}{} & \multicolumn{5}{c}{MLMC} & \multicolumn{1}{c}{MC} &  \multicolumn{5}{c}{MLMC} & \multicolumn{1}{c}{MC}\\
\cmidrule(rl{4pt}){2-6} \cmidrule(rl{4pt}){7-7} \cmidrule(rl{4pt}){8-12} \cmidrule(rl{4pt}){13-13}
&\multicolumn{4}{c}{level}&\multirow{1}{*}{equivalent}&\multirow{1}{*}{level}&\multicolumn{4}{c}{level}&\multirow{1}{*}{equivalent}&\multirow{1}{*}{level}\\
  &{0} & {1} & {2} & {3} &   3 &3 &{0} & {1} &{2} & {3} & 3 & 3 \\ [0.5ex]
\cmidrule{1-13}
   2.5E-5 & 172 & 4 & 3 & 3 &6& 75 & 41 & 5 & 3 & 3 &5& 24 \\
   7.5E-6 & 1371 & 30 & 3 & 3 &15& 1302 & 477 & 58 & 7 & 3 &11& 265 \\
   5.0E-6 & 3293 & 73 & 4 & 3 &33& - & 940 & 123 & 14 & 3 &17& - \\
   2.5E-6 & 11328 & 271 & 7 & 3 &99& -&4111 & 524 & 73 & 18 &90& - \\
\bottomrule
\end{tabular}}
\caption{Number of samples for MLMC and MC for the elastoplastic response.}
\label{tab:Samples_plast}
\end{table}
\subsection{Rate verification}

In this section, we estimate the rates $\alpha$, $\beta$ and $\gamma$ from Theorem 1, first for the static elastic response, then for the static elastoplastic response.

\subsubsection{Elastic Response}

We list the online estimated rates, i.e., the rates estimated during the run of the algorithm,  for various tolerances $\epsilon$ on the RMSE, see Tab.\,\ref{tab:Parameters for the MLMC algorithm},  corresponding to the tolerances considered in Fig.\,\ref{fig:Times}. Following the results in this table, it is possible to estimate the asymptotic cost of the MLMC estimator, using the different regimes from Eq.\,\eqref{eq:Algo_regime}. Since $\beta>\gamma$ in all cases, we expect an optimal cost proportional to $\epsilon^{-2}$. This is indeed what we observed in the numerical experiments in Fig.\,\ref{fig:Times}.

\begin{table}[H]
\centering
  \scalebox{1}{
 \begin{tabular}{c c c cc c c}
\toprule
\multirow{3}{*}{RMSE [/]} & \multicolumn{3}{c}{Homogeneous } & \multicolumn{3}{c}{Heterogeneous }\\
 & \multicolumn{3}{c}{Young's modulus} & \multicolumn{3}{c}{Young's modulus}\\
 \cmidrule(rl{4pt}){2-4}  \cmidrule(rl{4pt}){5-7}
 & $\alpha$ & $\beta$ & $\gamma$ & $\alpha$ & $\beta$ & $\gamma$ \\ [0.5ex]
\cmidrule{1-7}
2.5E-4 &  1.97 & 4.38 & 2.09 & 2.18 & 3.22 & 2.03\\
7.5E-5 &  2.03 & 5.11 & 2.16 & 2.13 & 2.99 & 2.13 \\
5.0E-5 &  1.88 & 4.03 & 2.10 & 1.86 & 3.08 & 2.30 \\
 2.5E-5 & 1.87 & 3.80 & 2.22 & 1.61 & 3.12 & 2.17 \\
\bottomrule
\end{tabular}}
\caption{Parameters for the MLMC algorithm.}
\label{tab:Parameters for the MLMC algorithm}
\end{table}

\begin{figure}[H]
\begin{subfigure}[b]{0.54\textwidth}
\centering
\scalebox{0.45}{
	\makebox[\textwidth]{
	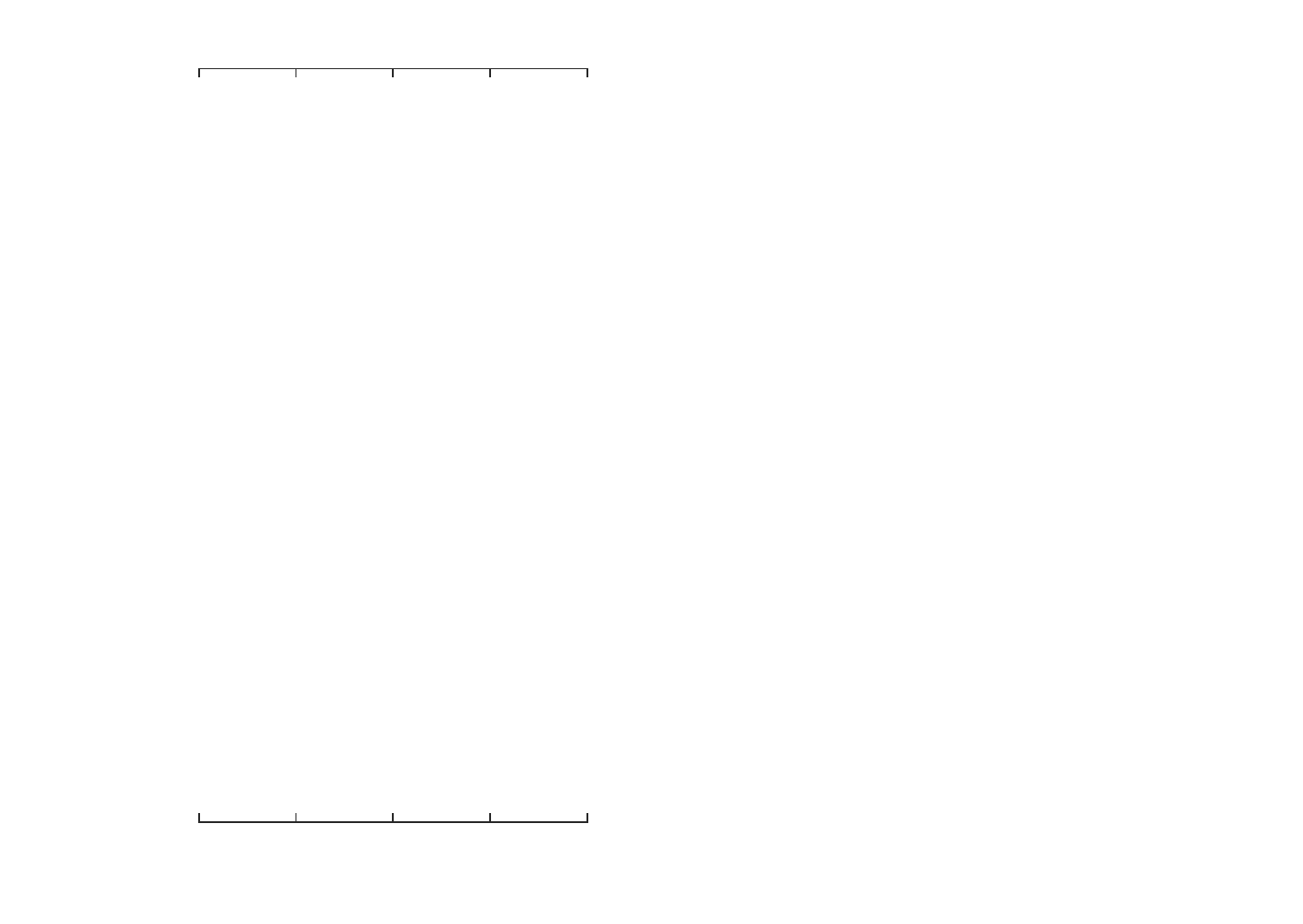}}
\end{subfigure}
%\centering
\begin{subfigure}[b]{0.45\linewidth}
\centering
	\scalebox{0.45}{
		\makebox[\textwidth]{
	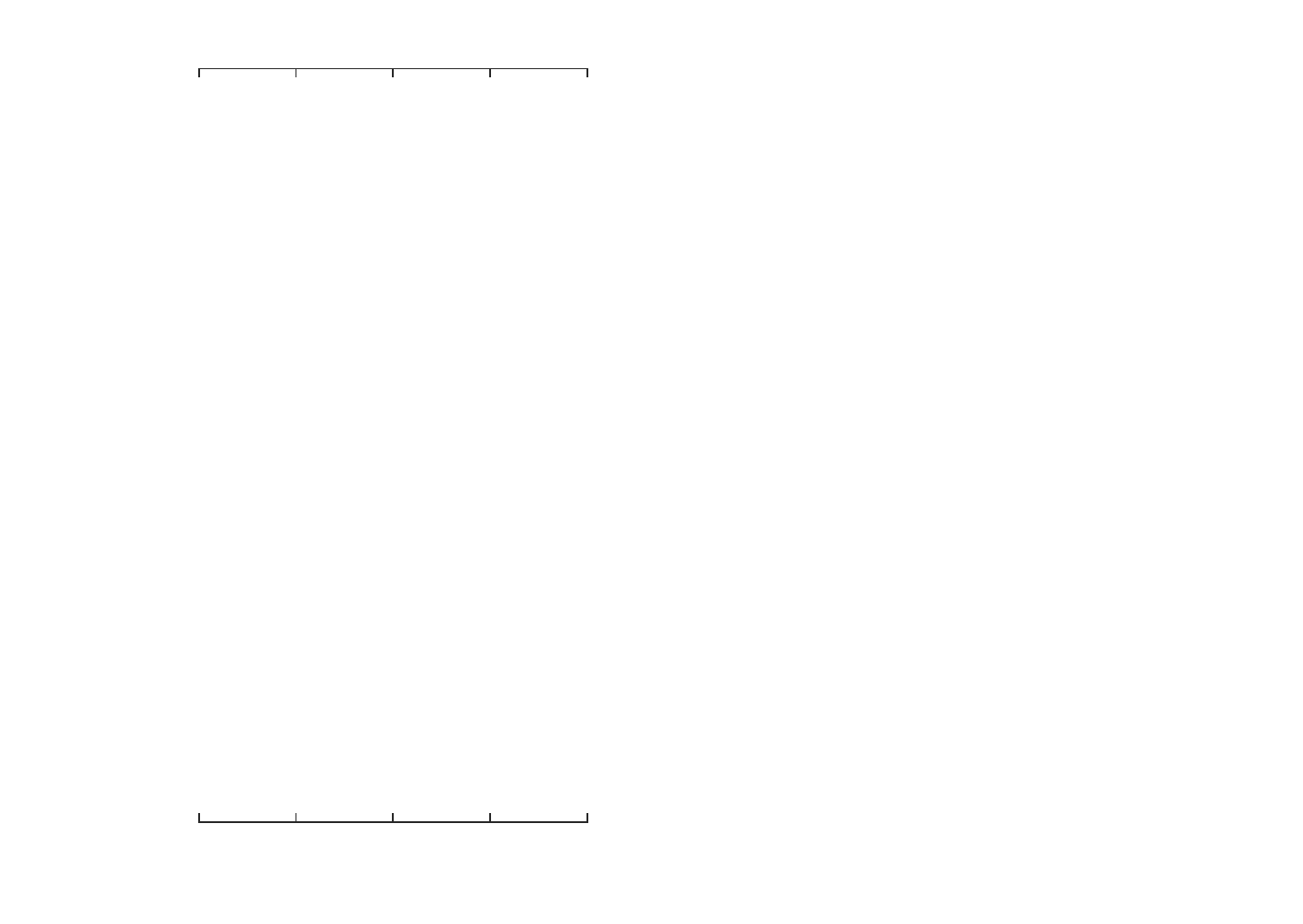}}
\end{subfigure}
\caption{Elastic response rates for a homogeneous (left) and a heterogeneous Young's modulus (right), in case of a tolerance $\epsilon$ equal to 2.5E-5.}
\label{fig:rates}
\end{figure}

Fig.\,\ref{fig:rates} shows the behavior of the expected value and the variance of the quantity of interest $P_\ell$, and of the difference $P_\ell-P_{\ell-1}$, in case of a tolerance $\epsilon$ equal to 2.5E-5. Note that the mean and the variance of $P_\ell$ over the different levels remains constant while the variance and the mean of the differences between two successive levels continuously decreases. The rates $\alpha$ and $\beta$ can be read off as the slope of the lines labeled with $\triangledown$. 

In case where the Young's modulus is homogeneous, Fig.\,\ref{fig:rates} (left), the variance corresponding to the finest level is lower than the variance on the previous levels. This phenomenon can be explained by the number of samples on that level, which is used to estimate the variance. Only 7 samples were taken. Taking a higher number of samples on that level would bring the variance to equal the variance of the other levels.

\subsubsection{Elastoplastic Response}

As for the elastic response, we list the online rates for various tolerances $\epsilon$ on the $\text{RMSE}$ in Tab.\,\ref{tab:Parameters for the MLMC algorithm plast}. For all tolerances, $\beta>\gamma$.

\begin{table}[H]
\centering
  \scalebox{1}{
 \begin{tabular}{c c c cc c c}
\toprule
\multirow{3}{*}{$\text{RMSE}$ [/]} & \multicolumn{3}{c}{Homogeneous } & \multicolumn{3}{c}{Heterogeneous }\\
 & \multicolumn{3}{c}{Young's modulus} & \multicolumn{3}{c}{Young's modulus}\\
 \cmidrule(rl{4pt}){2-4}  \cmidrule(rl{4pt}){5-7}
 & $\alpha$ & $\beta$ & $\gamma$ & $\alpha$ & $\beta$ & $\gamma$ \\ [0.5ex]
\cmidrule{1-7}
 2.5E-5 & 1.50  & 3.59  & 2.24  & 1.53 & 5.28 & 2.22 \\
7.5E-6 &  1.50 & 4.81 & 2.26 & 1.51 & 4.21 & 2.17 \\
5.0E-6 &  1.51 & 9.62 & 2.25 & 1.49 & 4.51 & 2.11 \\
 2.5E-6 & 1.50 & 8.30 & 2.21 & 1.50 & 3.43 & 2.17 \\
\bottomrule
\end{tabular}}
\caption{Parameters for the MLMC algorithm.}
\label{tab:Parameters for the MLMC algorithm plast}
\end{table}

\begin{figure}[H]
\begin{subfigure}[b]{0.54\textwidth}
\centering
\scalebox{0.45}{
	\makebox[\textwidth]{
	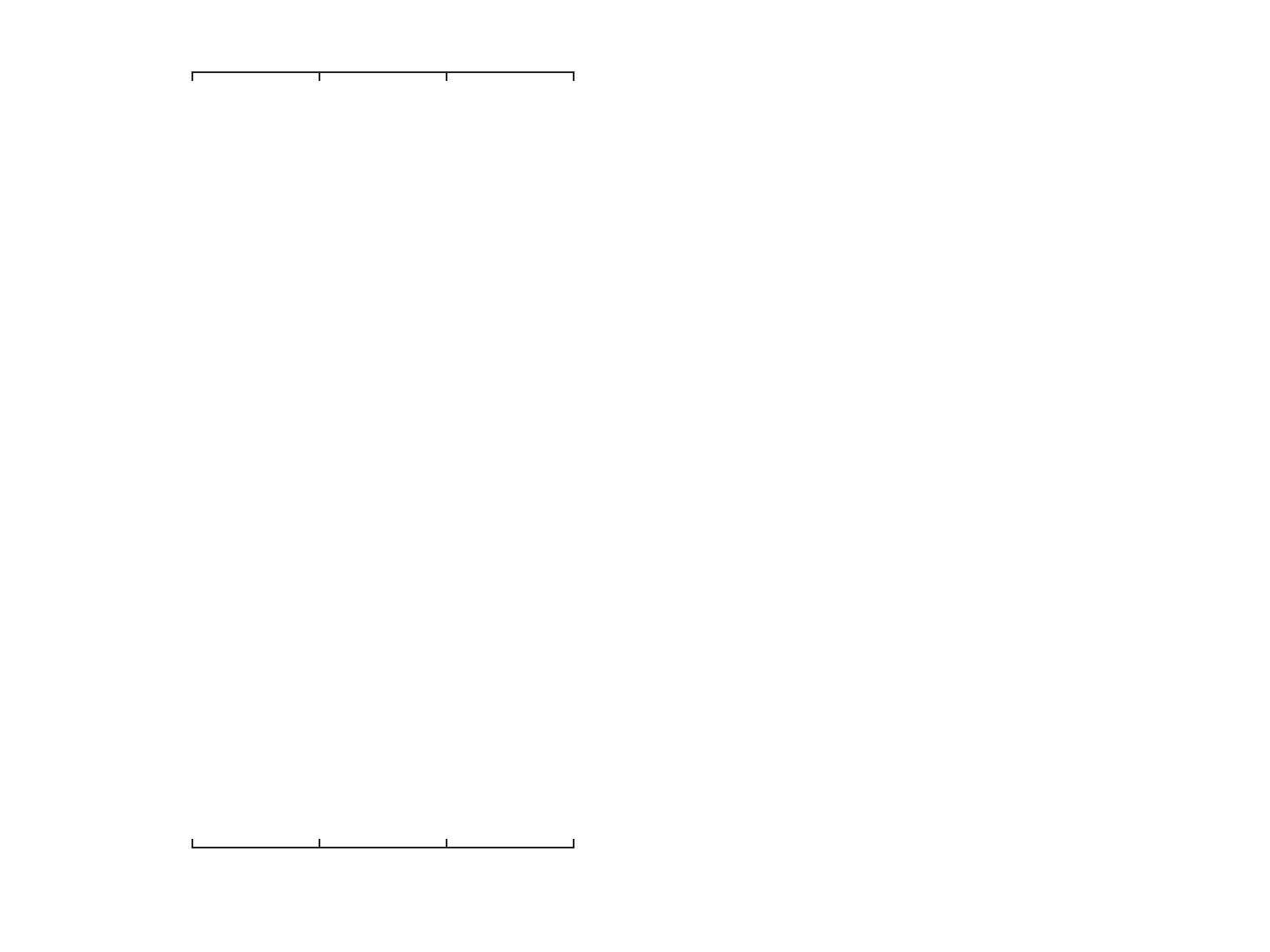}}
\end{subfigure}
%\centering
\begin{subfigure}[b]{0.45\linewidth}
\centering
	\scalebox{0.45}{
		\makebox[\textwidth]{
	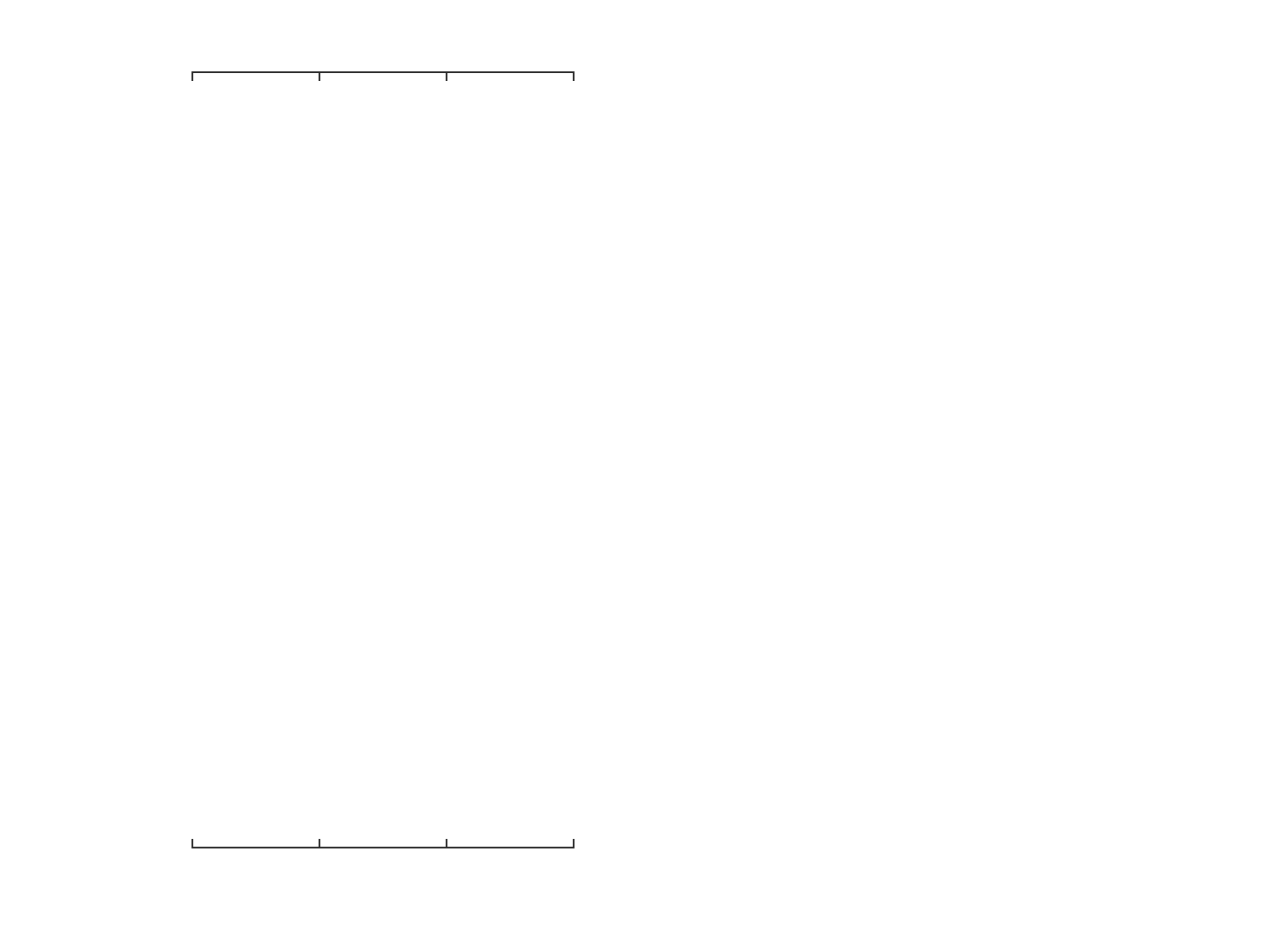}}
\end{subfigure}
\caption{Elastoplastic response rates for a homogeneous (left) and a heterogeneous Young's modulus (right), in case of a tolerance $\epsilon$ equal to 2.5E-6.}
\label{fig:rates_plast}
\end{figure}

Fig.\,\ref{fig:rates_plast} shows the behavior of the expected value and the variance of the quantity of interest $P_\ell$, and of the difference $P_\ell-P_{\ell-1}$, in case of a RMSE equal to 2.5E-6. Comparing Fig.\,\ref{fig:rates} with Fig.\,\ref{fig:rates_plast}, one notices that for the elastoplastic response the mean is not perfectly constant over the levels. This slight increase stems from the number of elements used for the discretization; the displacement is indeed dependent upon the discretization of the mesh, Fig.\,\ref{fig:mesh_convergence}, for the mesh sizes considered here. 

\subsection{Visualization of the solution}

In this part, the static response, i.e., the displacement of the beam in the spatial domain, and the dynamic response, i.e., the frequency response functions (FRF), are shown. First, the static elastic response is presented. This consists of a visualization of the transverse displacement of the nodes along the top side of the beam.  Secondly, results for the static elastoplastic response will be shown. The results consists of a visualization of a force deflection curve of the middle top side node of the beam. Thirdly and lastly,  solutions for the dynamic response are  presented. These consist of frequency responses functions for a single node of the finite element mesh. This node is chosen as the one that has the largest variance of all the nodes that make up the mesh.

\subsubsection{Static Elastic Response}\label{Sec:Viz_Sta_El}
In this part we show the displacement of the concrete beam under a static load including uncertainty bounds in the elastic domain. The  beam configuration is a beam clamped at both ends,  loaded at mid span, as shown in Fig.\,\ref{fig:bean_configurations} (right). The load is modeled as a distributed load acting on each of the vertical middle nodes of the beam. The force on each node is such that the sum of these forces equals $10000\,\mathrm{kN}$, regardless of the refinement of the mesh.

\begin{figure}[H]
\begin{subfigure}[b]{0.54\textwidth}
\centering
\scalebox{0.45}{
	\makebox[\textwidth]{
	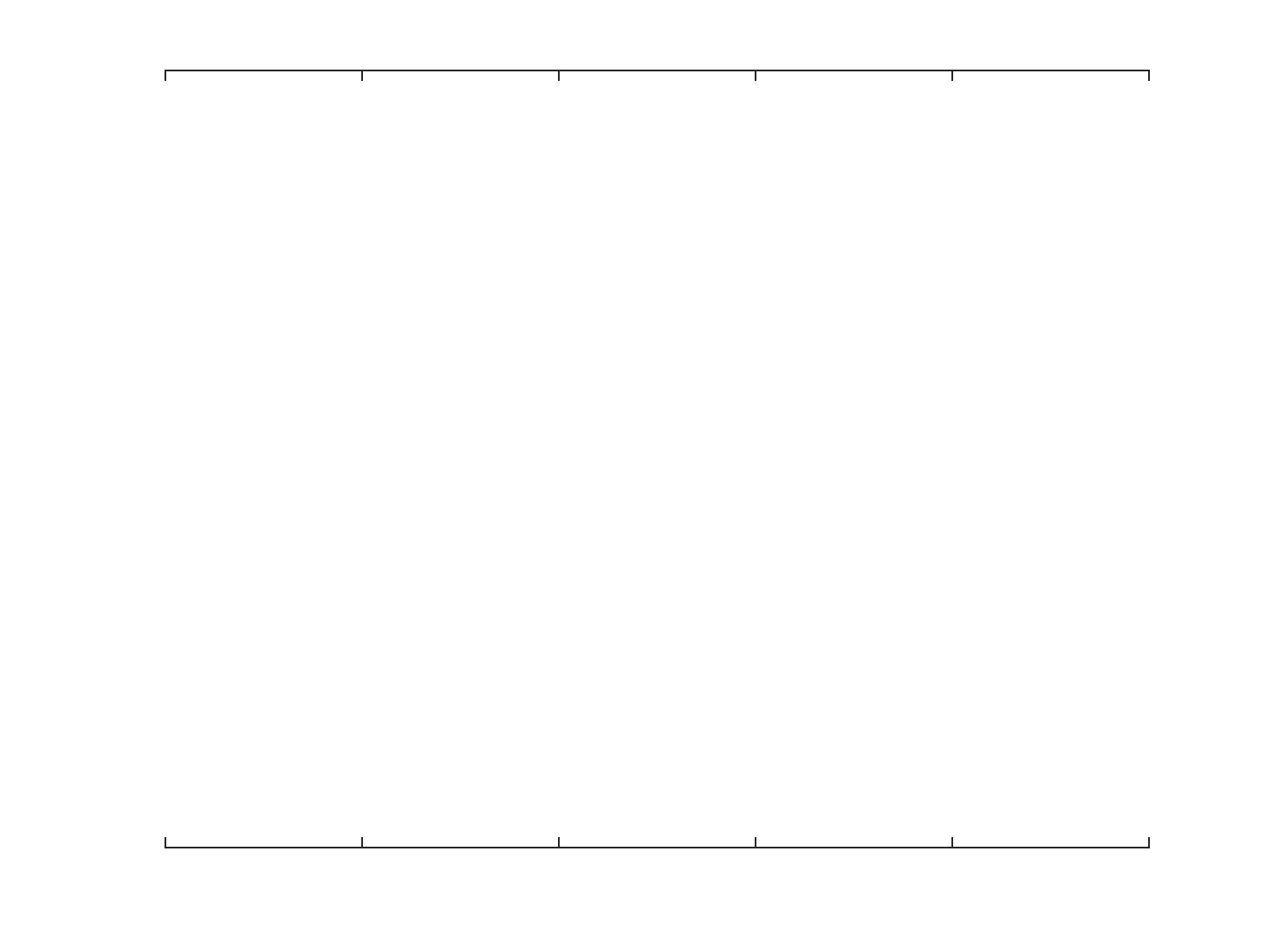}}
%	\caption{Gamma Random Field}
\end{subfigure}
%\centering
\begin{subfigure}[b]{0.45\linewidth}
\centering
	\scalebox{0.45}{
		\makebox[\textwidth]{
	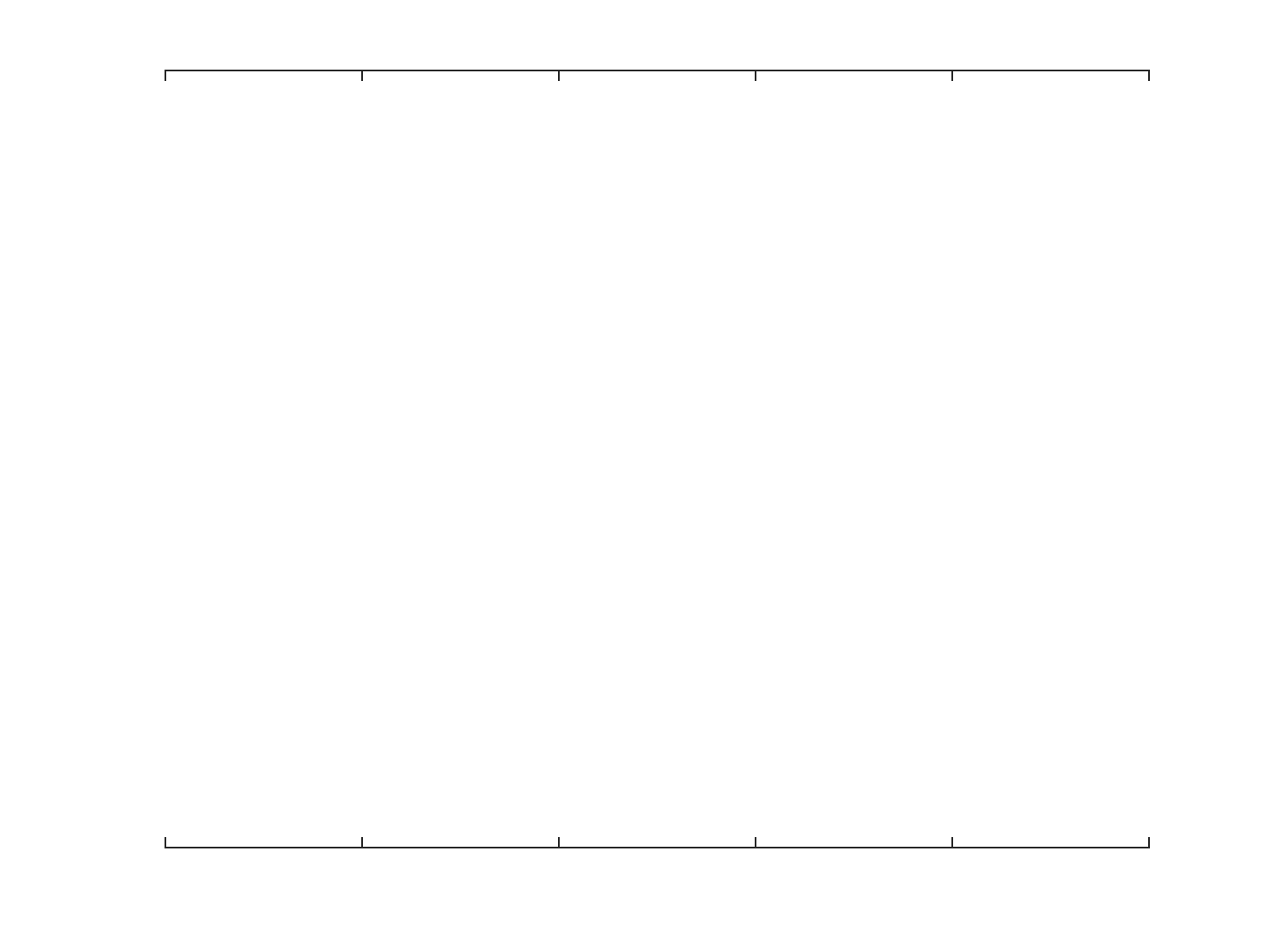}}
\end{subfigure}
\caption{Deflection of the beam for when the Young's modulus is homogeneous (left) and  heterogeneous (right).}
\label{fig:deflection_total}
\end{figure}

\begin{figure}[H]
\begin{subfigure}[b]{0.54\textwidth}
\centering
\scalebox{0.44}{
	\makebox[\textwidth]{
	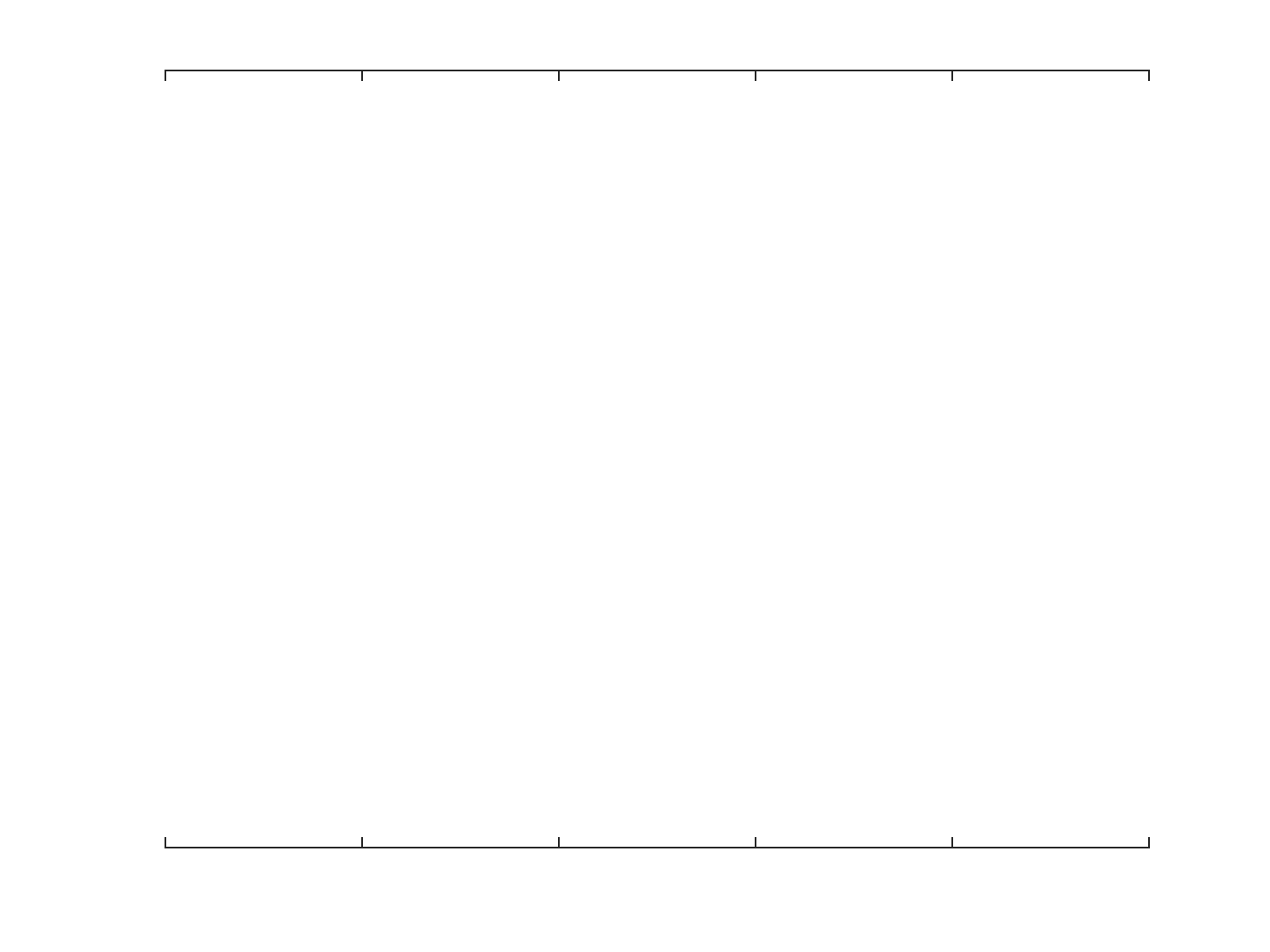}}
\end{subfigure}
%\centering
\begin{subfigure}[b]{0.45\linewidth}
\centering
	\scalebox{0.44}{
		\makebox[\textwidth]{
	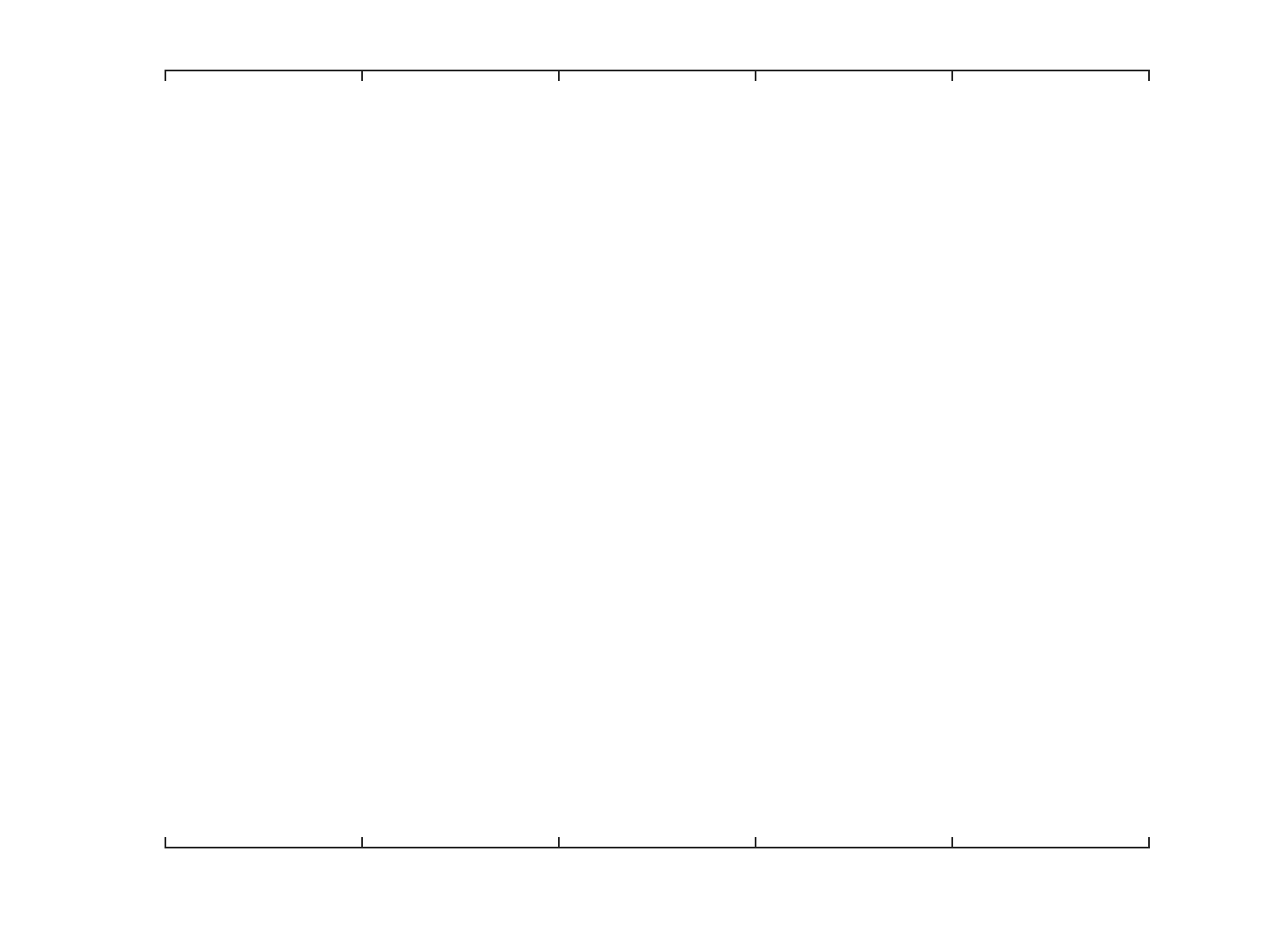}}
\end{subfigure}
\caption{Ten different deflection samples for when the Young's modulus is homogeneous (left) and heterogeneous (right).}
\label{fig:indiv_samples_elast}
\end{figure}

Fig.\,\ref{fig:deflection_total} shows the deflection of the beam with a homogeneous Young's modulus (left) and  with a heterogeneous Young's modulus (right). The full orange line represents the average of the displacement, the dashed orange lines are the 1$\sigma$ bounds equidistant around the average, which are only relevant in case of a Normal distribution. The shades of blue represent the PDF, with the dark blue line corresponding to the most probable value.
As can be observed from  Fig.\,\ref{fig:deflection_total} (right), the average value and the most probable value of the  PDF tend to coincide for the heterogeneous case. Here, the PDF of the displacement closely resembles that of a Normal distribution. This is however not the case when the Young's modulus is homogeneous. Then the distribution of the solution has a non-negligible skewness. 
The homogeneous Young's modulus case exhibits a larger uncertainty on its displacement due to the fact that the Young's modulus is uncertain for each individual computed sample but uniform in each point for that sample. Averaging all these individual samples gives rise to wider uncertainty bounds. While for the heterogeneous Young's modulus case, each individual computed sample is also uncertain but non-uniform; in each individual point the value for the Young's modulus is different. This means that in different locations of the beam the Young's modulus will be different. These locations will tend to compensate each other so that beam does not become overly stiff or weak. Fig.\,\ref{fig:indiv_samples_elast} shows ten samples in case of a homogeneous Young's modulus (left) and a heterogeneous Young's modulus (right).

\begin{figure}[H]
\begin{subfigure}[b]{0.54\linewidth}
\centering
	\scalebox{0.35}{
		\makebox[\textwidth]{
	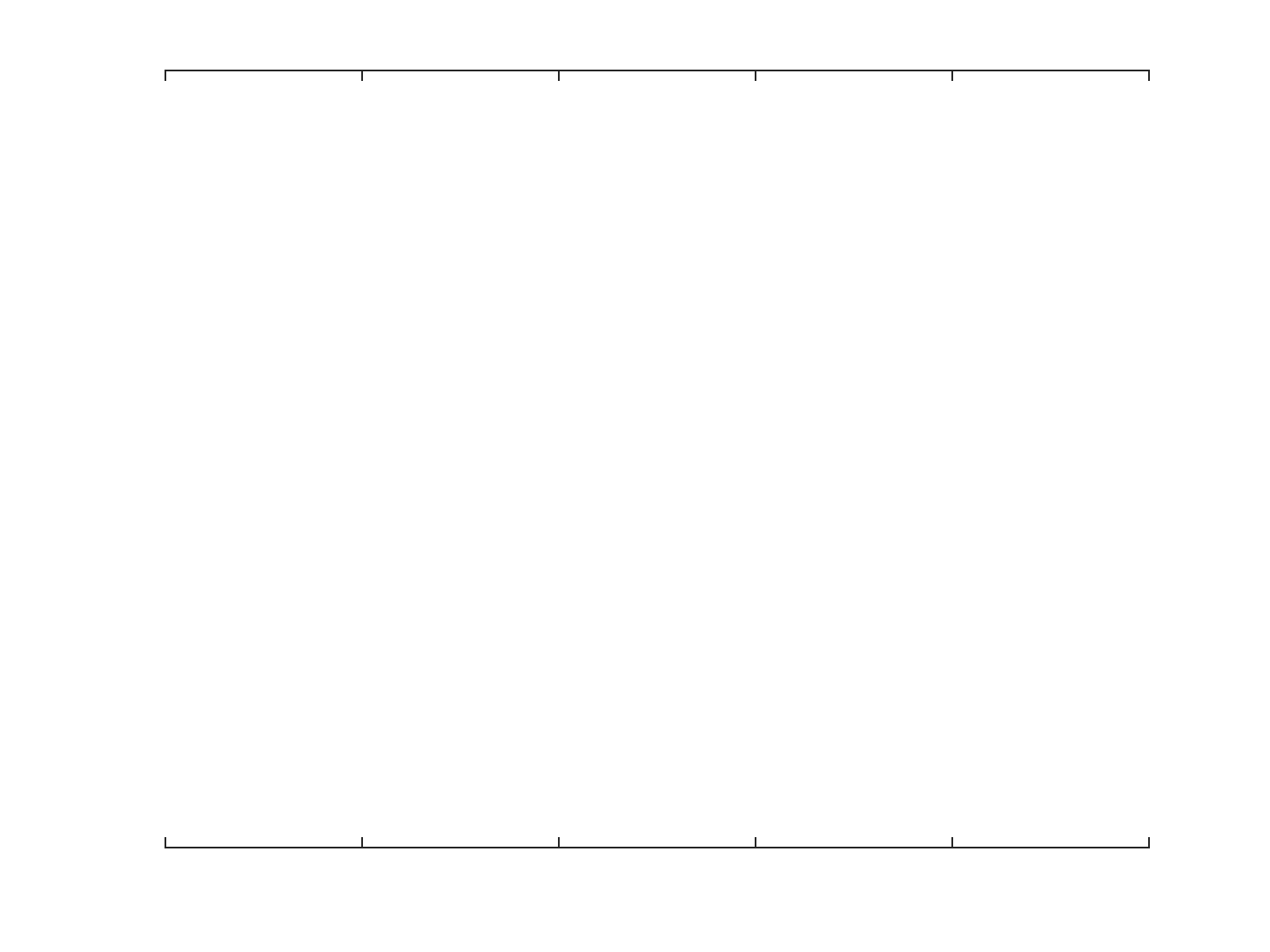}}
\end{subfigure}
\begin{subfigure}[b]{0.45\textwidth}
\centering
\scalebox{0.35}{
	\makebox[\textwidth]{
	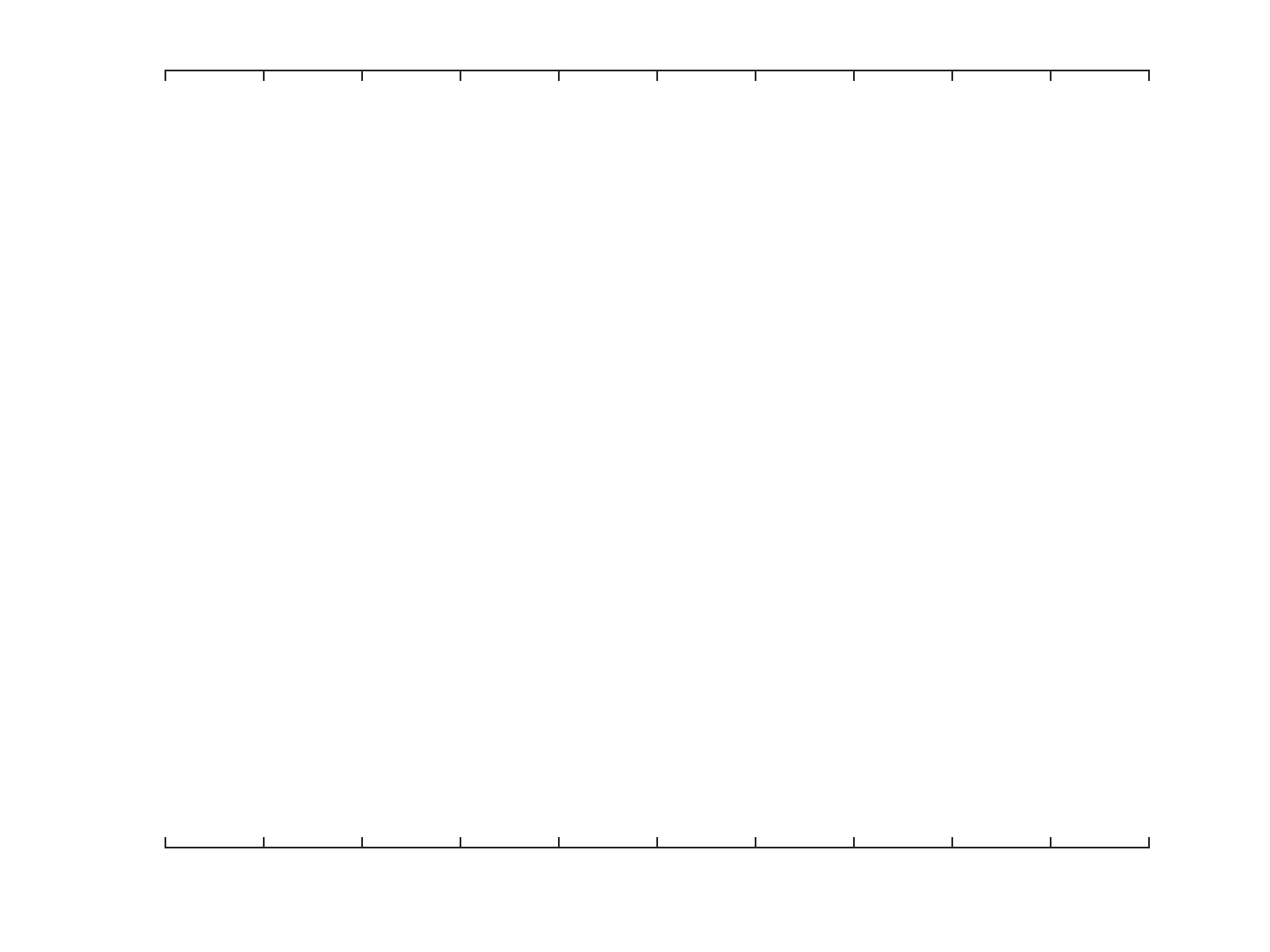}}
\end{subfigure}
%\centering
\caption{Visualization of the PDF  when the Young's modulus is homogeneous, beam displacement (left), AB cut-through (right).}
\label{fig:Construction of the PDF Gam Dist}
\end{figure}

\begin{figure}[H]
\begin{subfigure}[b]{0.54\linewidth}
\centering
	\scalebox{0.35}{
		\makebox[\textwidth]{
	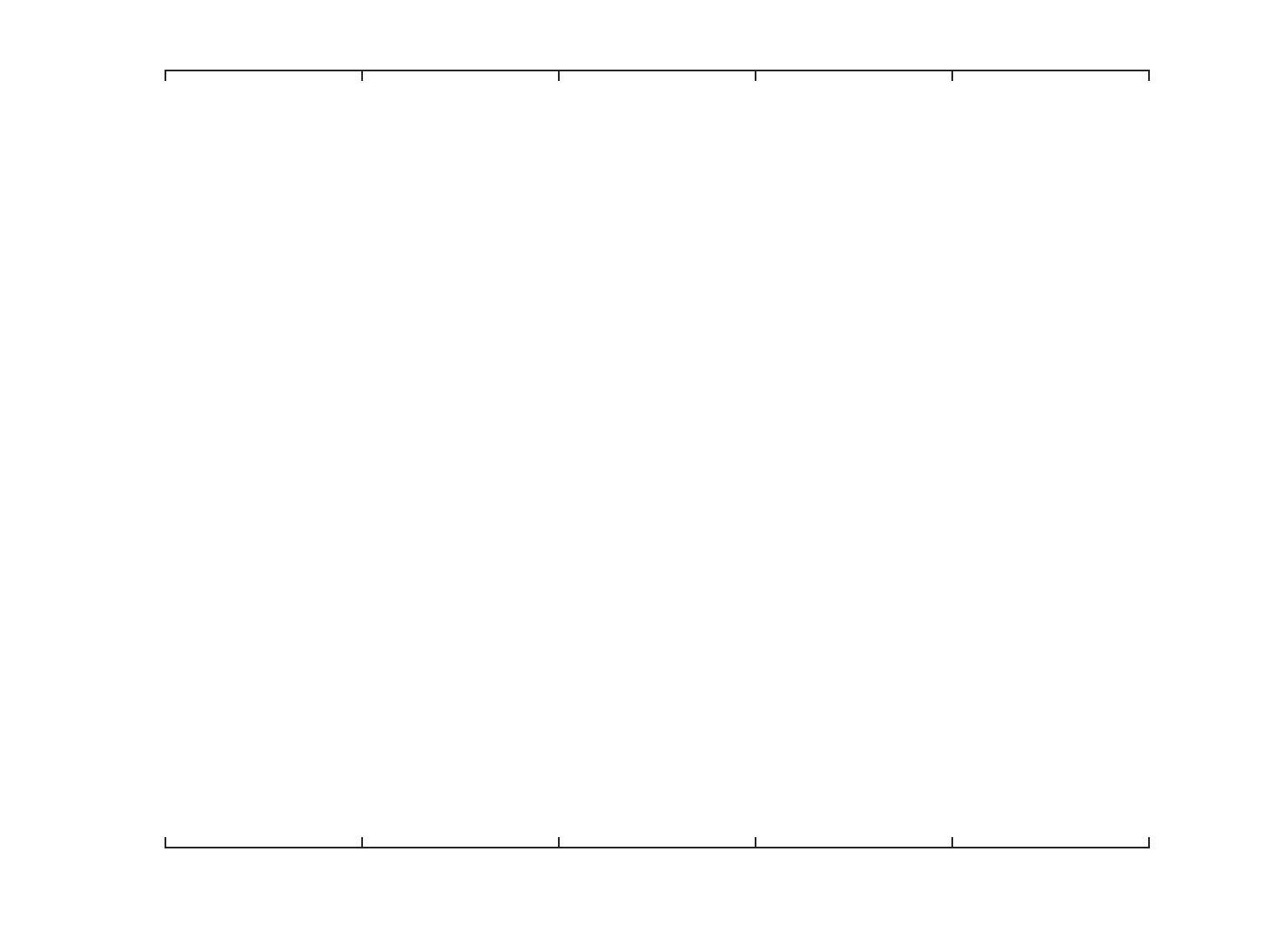}}
\end{subfigure}
\begin{subfigure}[b]{0.45\textwidth}
\centering
\scalebox{0.35}{
	\makebox[\textwidth]{
	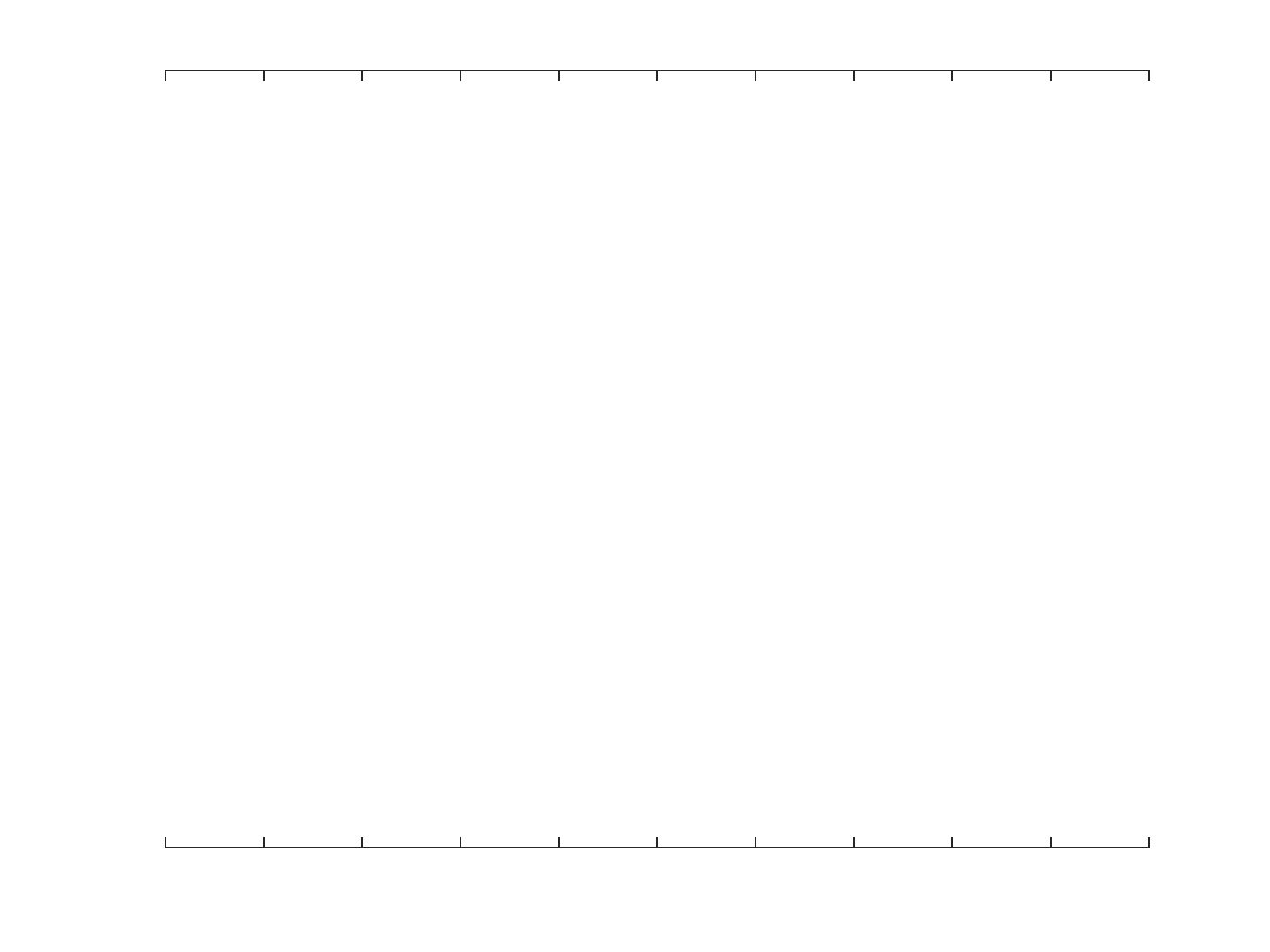}}
\end{subfigure}
%\centering
\caption{Visualization of the PDF  when the Young's modulus is heterogeneous, beam displacement (left), AB cut-through (right).}
\label{fig:Construction of the PDF Gam Field}
\end{figure}

Fig.\,\ref{fig:Construction of the PDF Gam Dist} and Fig.\,\ref{fig:Construction of the PDF Gam Field} show a cut-through in order to better illustrate how the shades of blue represent the PDF.

\subsubsection{Static Elastoplastic Response}

\begin{figure}[H]
\begin{subfigure}[b]{0.54\textwidth}
\centering
\scalebox{0.45}{
	\makebox[\textwidth]{
	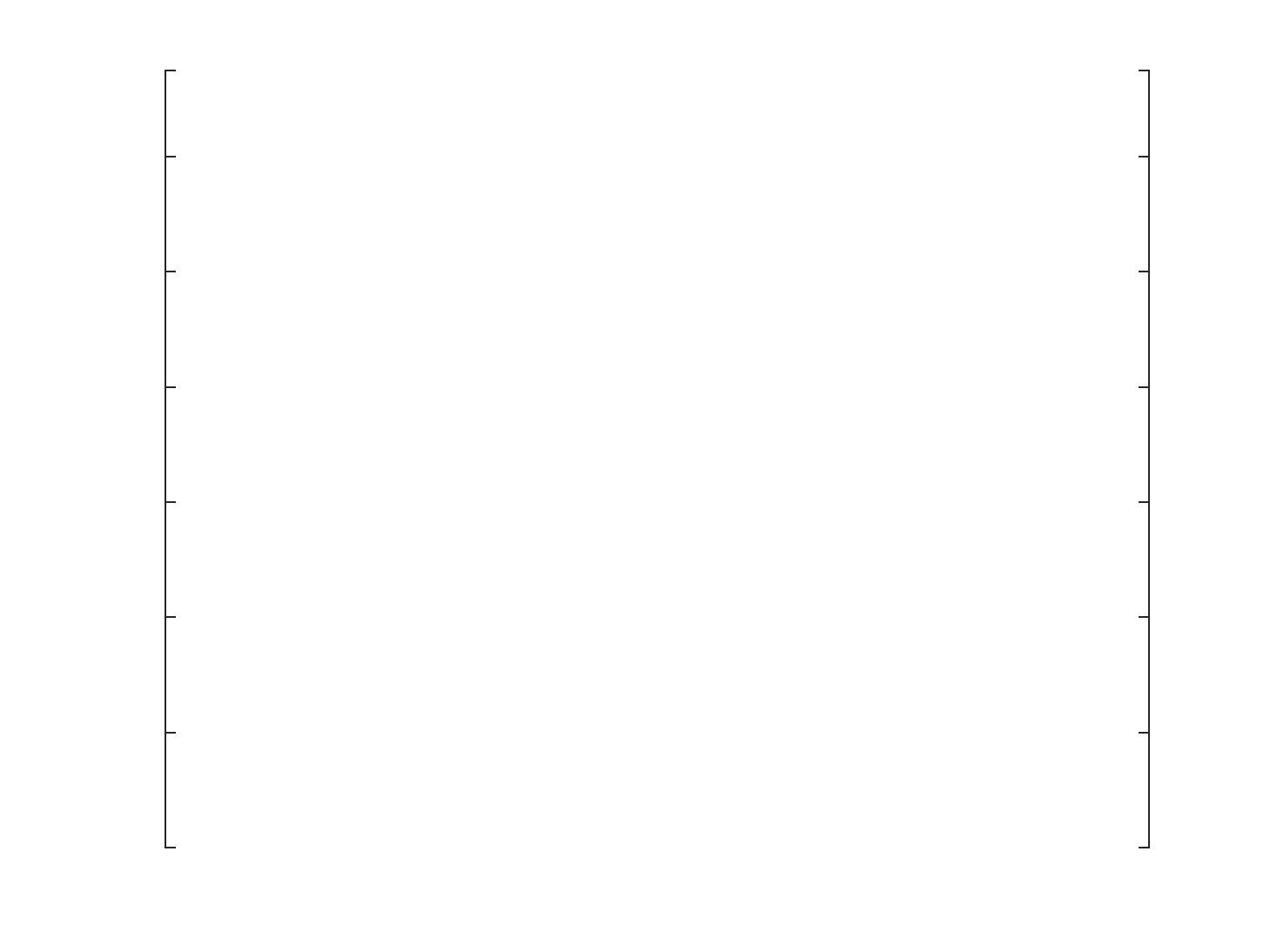}}
%	\caption{Gamma Random Field}
\end{subfigure}
%\centering
\begin{subfigure}[b]{0.45\linewidth}
\centering
	\scalebox{0.45}{
		\makebox[\textwidth]{
	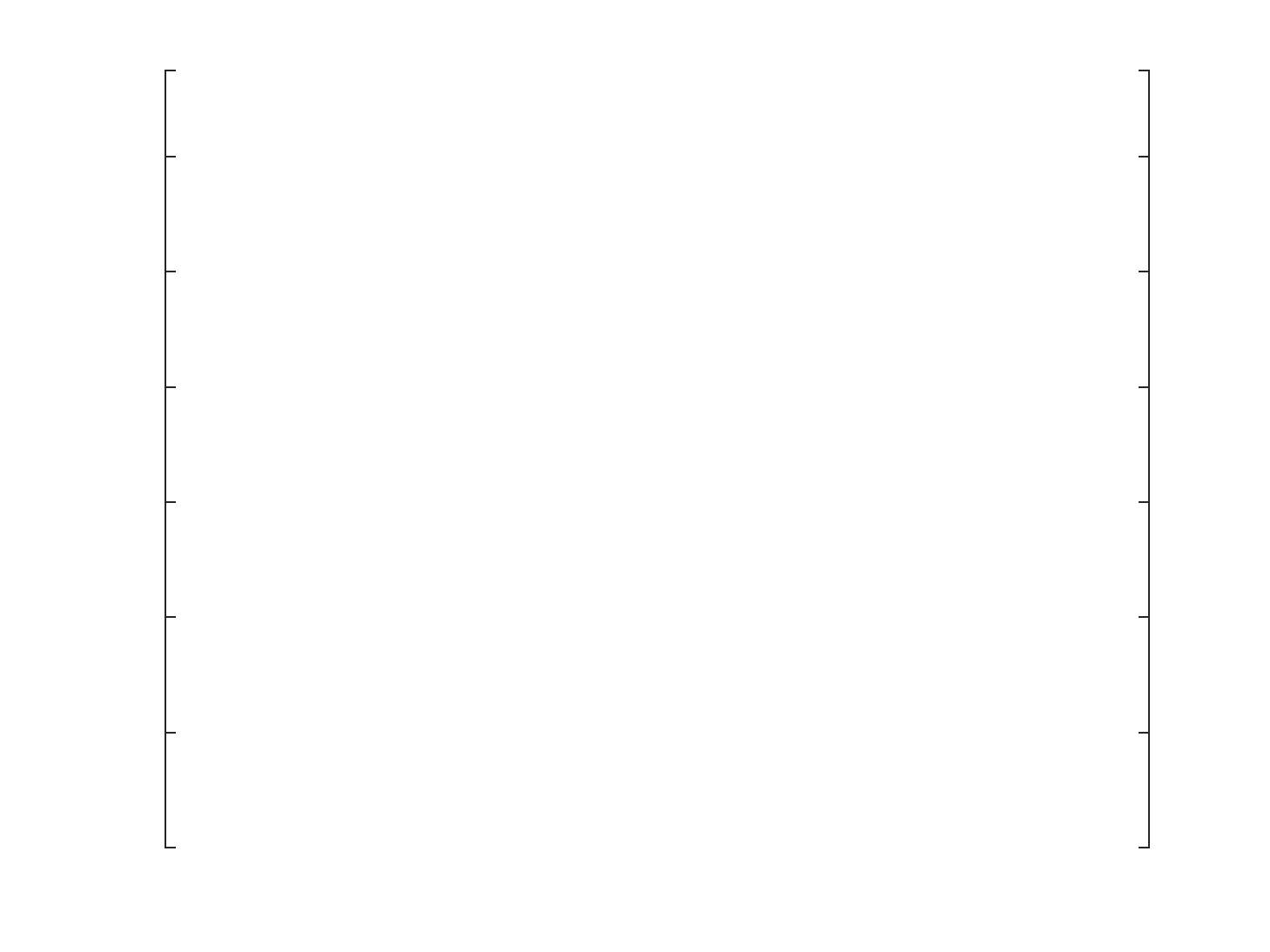}}
\end{subfigure}
\caption{Force deflection curve of the steel beam for when the Young's modulus is homogeneous (left) and  heterogeneous (right).}
\label{fig:deflection_total_plast}
\end{figure}

The solution is presented as a force deflection curve, Fig.\,\ref{fig:deflection_total_plast}. The beam configuration is the same as for the static elastic response but with a different material; a steel beam clamped at both sides loaded at mid span, Fig.\,\ref{fig:bean_configurations} (right). The load is modeled in the same way as for the static elastic response. The beam is loaded from $0\,\mathrm{N}$ to $13.5\,\mathrm{kN}$ in steps of $135\,\mathrm{N}$.

\begin{figure}[H]
\begin{subfigure}[b]{0.54\textwidth}
\centering
\scalebox{0.44}{
	\makebox[\textwidth]{
	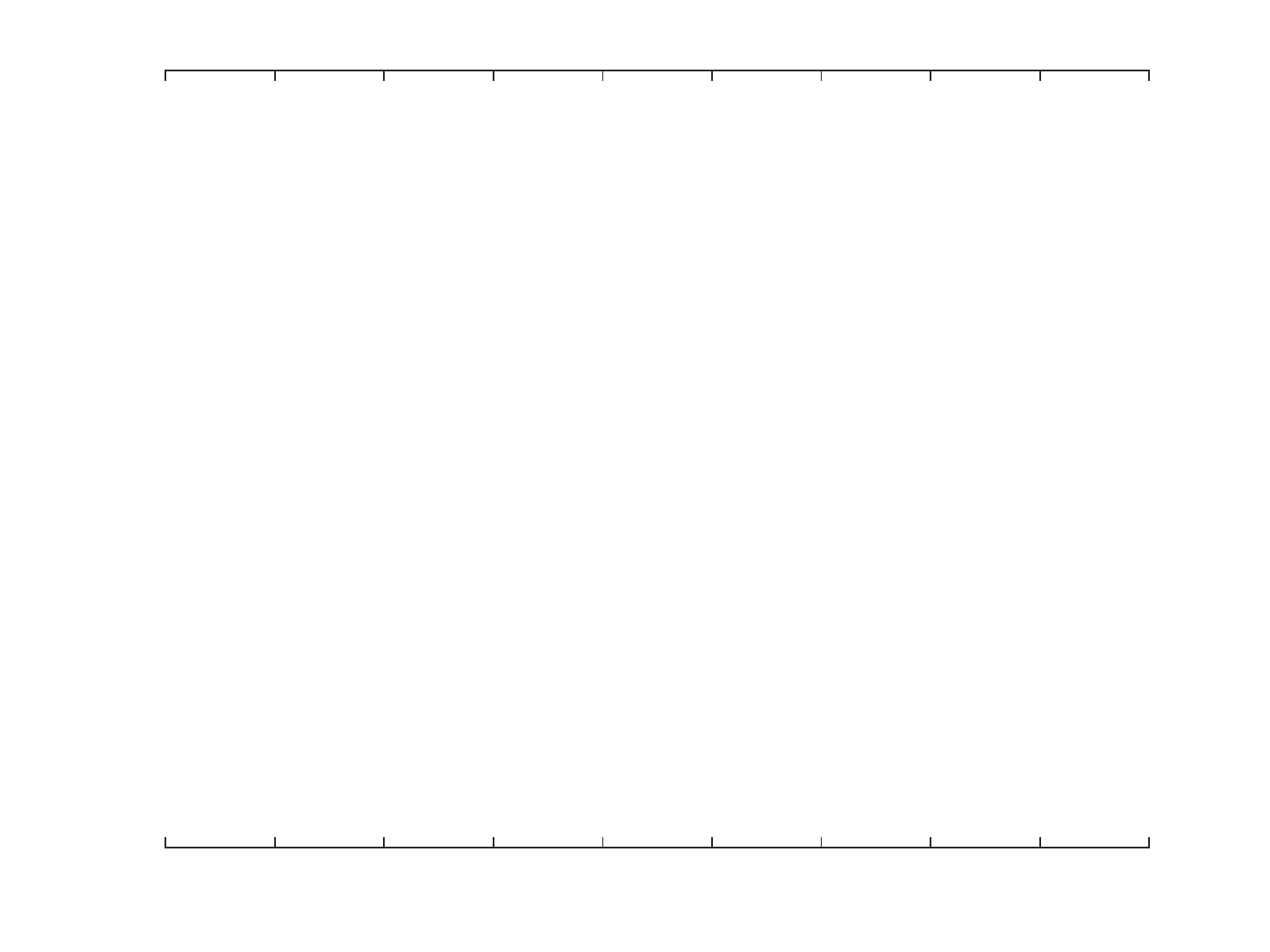}}
\end{subfigure}
%\centering
\begin{subfigure}[b]{0.45\linewidth}
\centering
	\scalebox{0.44}{
		\makebox[\textwidth]{
	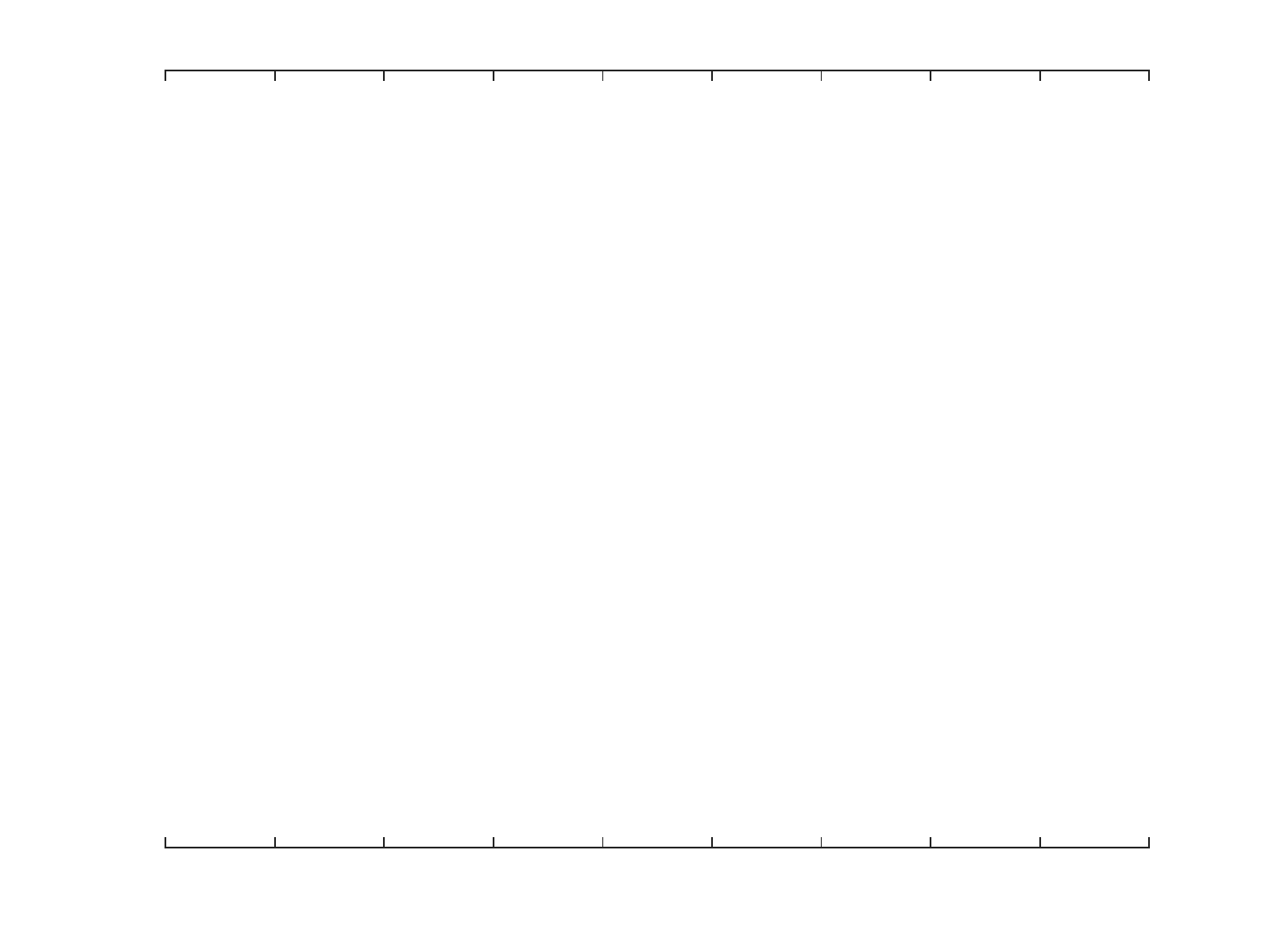}}
\end{subfigure}
\caption{Ten different force deflection curve samples for when the Young's modulus is homogeneous (left) and heterogeneous (right).}
\label{fig:indiv_samples_plast}
\end{figure}

The line style and color convention is the same as for the static elastic response. As can be observed, the uncertainty bounds in case of a homogeneous Young's modulus, Fig.\,\ref{fig:deflection_total_plast} (left) are wider and more spread out than in case of a heterogeneous modulus (right). This behavior corroborates with the ones from the  static elastic response. In Fig.\,\ref{fig:indiv_samples_plast}, ten individual samples are shown for a homogeneous Young's modulus (left) and a heterogeneous Young's modulus (right). Comparing these two sets of figures, it shows that, the deflection corresponding to the highest applied force in Fig.\,\ref{fig:indiv_samples_plast} (left and right) is less than the deflection shown in Fig.\,\ref{fig:deflection_total_plast} (left and right). This stems from the fact that these ten samples per Young's modulus are generated on the coarsest mesh. These results illustrate even better the power of the MLMC method; even whilst taking most of the samples on coarser meshes, one still gets the result as if the simulation was run on the finest mesh.

\subsubsection{Dynamic Elastic Response}
For the dynamic elastic response, the solution is a frequency response function (FRF). The beam configuration we consider is a  concrete cantilever beam clamped on the left side and loaded on the right side, as shown in Fig.\,\ref{fig:bean_configurations} (left), with a dynamic load. The load is modeled in the same way as for the static elastic response with the same magnitude. In order to ensure the correct representation of the dynamic response, and to determine the minimum number of finite elements required, the bending wavelength, $\lambda_{\mathrm{min}}$,  must be evaluated as
\begin{linenomath*}
\begin{equation}
\lambda_{\mathrm{min}} = \sqrt{\dfrac{2 \pi}{f_{\mathrm{max}}}} \sqrt[4]{\dfrac{E I}{\rho A}},
\label{eq:MinElements}
\end{equation}
\end{linenomath*}
with $E$ the mean Young's modulus, $I$ the moment of inertia, $A$ the area, $\rho$ the density, $f_{\mathrm{max}}$ the highest simulated frequency, and $\lambda_{\mathrm{min}}$ the smallest obtained wavelength for the highest input frequency.

For the considered beam configuration, it has been checked that at least six elements are used to represent the wavelength on the coarsest grid for the highest simulated frequency, which in this case is 400 Hz.

\begin{figure}[H]
\begin{subfigure}[b]{0.54\textwidth}
\centering
\scalebox{0.46}{
	\makebox[\textwidth]{
	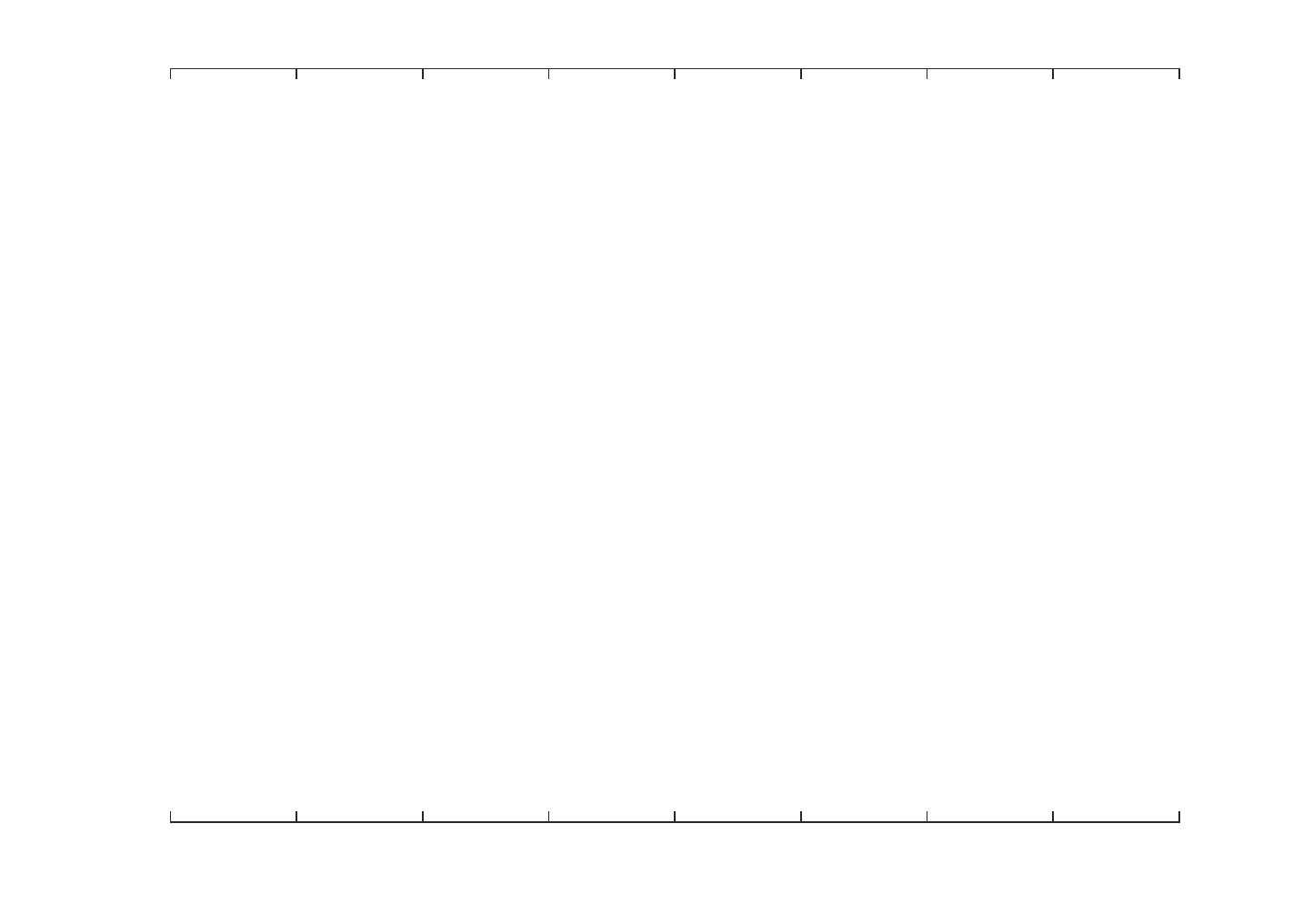}}
\end{subfigure}
%\centering
\begin{subfigure}[b]{0.45\linewidth}
\centering
	\scalebox{0.43}{
		\makebox[\textwidth]{
	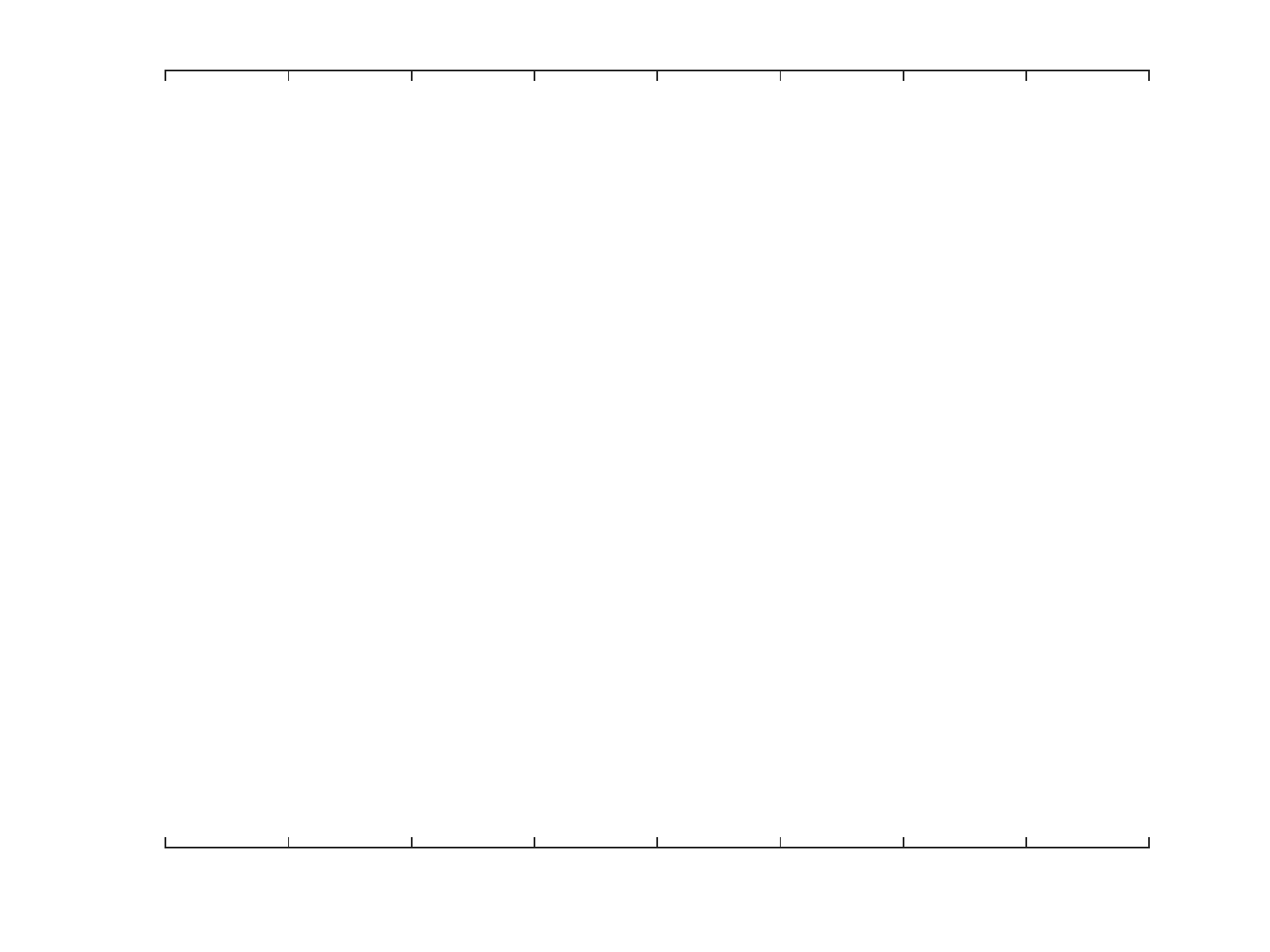}}
\end{subfigure}
\caption{Dynamic Responses of the beam for when the Young's modulus is  homogeneous (left) and  heterogeneous (right).}
\label{fig:dynamic_total}
\end{figure}
\begin{figure}[H]
\begin{subfigure}[b]{0.54\textwidth}
\centering
\scalebox{0.44}{
	\makebox[\textwidth]{
	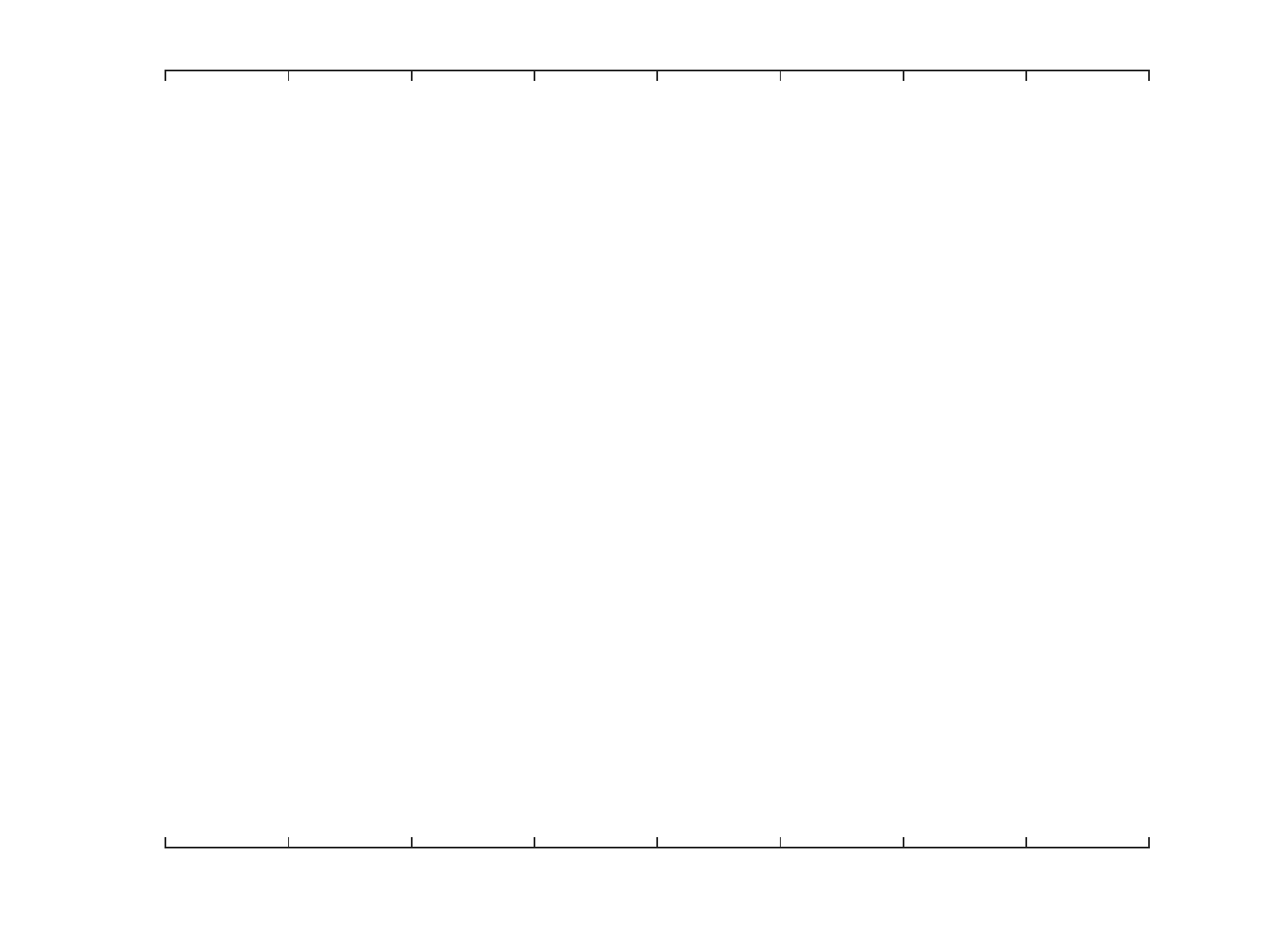}}
\end{subfigure}
%\centering
\begin{subfigure}[b]{0.45\linewidth}
\centering
	\scalebox{0.44}{
		\makebox[\textwidth]{
	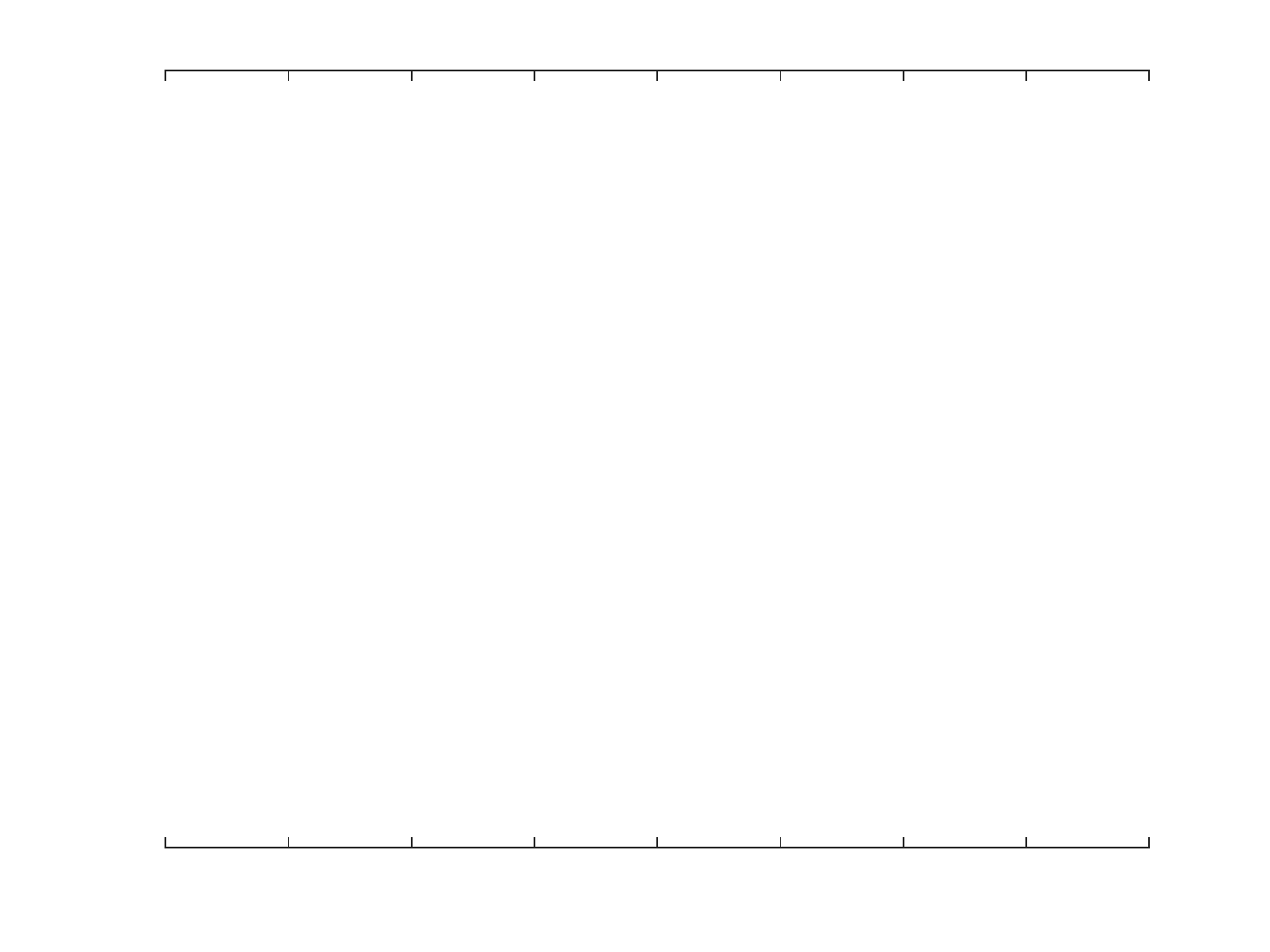}}
\end{subfigure}
\caption{Ten different FRF samples for when the Young's modulus is homogeneous (left) and heterogeneous (right).}
\label{fig:dynamic_samples}
\end{figure}

The FRF results are presented in Fig.\,\ref{fig:dynamic_total}. As was the case for the static elastic and elastoplastic response, the shades of blue represent the PDF, with the blue line being the most probable value, and the orange line the average value. As can be observed, the uncertainty bounds for the FRF  are wider and more spread out when the Young's modulus is homogeneous, Fig.\,\ref{fig:dynamic_total} (left), as opposed to a heterogeneous Young's modulus,  Fig.\,\ref{fig:dynamic_total} (right). This discrepancy is due to the fact that in case of a homogeneous Young's modulus, the resonance frequency will be shifted for each different sample. Averaging all these samples gives rise to a broad and wide uncertainty bound.  In case of a heterogeneous Young's modulus, the different samples compensate each other, in analogy with the explanation given in $\S$\ref{Sec:Viz_Sta_El}. This gives rise to much smaller uncertainty bounds. Fig.\,\ref{fig:dynamic_samples} shows the resulting FRF for ten  realizations.

An important inequality that must hold for MLMC to work well is
\begin{linenomath*}
\begin{equation}
\V[P_1-P_0] \ll \V[P_1].
\label{eq:Variance_ineq}
\end{equation}
\end{linenomath*}
It has been observed empirically that \eqref{eq:Variance_ineq} is not necessarily fulfilled for resonance frequencies and frequencies close by those resonance frequencies for the dynamic elastic response in case of a heterogeneous Young's modulus. The reason for this phenomenon appears to be closely linked to the resolution of random fields on the different levels, as discussed in \S \ref{par:comparison}.
In order to remedy to this, we use the following strategy. Using a small number of samples, the magnitude of $\V[P_1] - \V[P_1-P_0]$ is estimated. When the estimation is below a certain threshold  $T$, the coarsest level is discarded  and the algorithm is started on a finer mesh. We write this condition in a $\log_2$ base, in order  not to lose consistency with Fig.\,\ref{fig:rates}, which depict the rates in a $\log_2$ base. The estimation after simplification is written as
\begin{linenomath*}
\begin{equation}
\log_2\left(\dfrac{\V[P_1]}{\V[P_1-P_0]}\right) > \log_2{T}.
\label{eq:tresh}
\end{equation}  
\end{linenomath*} 
For the experiments reported here, we selected the right-hand side of  \eqref{eq:tresh} heuristically to be equal to $2.3$.

\section{Conclusion}
In this work, it has been demonstrated that the Multilevel Monte Carlo method provides a significant computational cost reduction and speedup compared to the standard Monte Carlo method for computing the stochastics of the response in a structural engineering problem. This has been shown by means of actual computing times and normalized computational costs. Speedups are typically of the order of a factor ten for moderate accuracies and can be even much higher for higher accuracies. In addition, figures depicting the static elastic, the static elastoplastic  and the dynamic elastic response have been presented with uncertainty bounds. To model the uncertainty in the Young's modulus, we considered both a homogeneous model, represented by a single random variable, and a heterogeneous model, modeled by means of a random field. We observed that the nature of the uncertainty in the Young's modulus has a big impact on the uncertainty characteristics of the simulation results. For the homogeneous Young's modulus, the variance in the solution is larger than for the heterogeneous Young's modulus. This is apparent from the wider and  more spread out uncertainty bounds in $\S$\ref{Sec:Viz_Sta_El}. Hence, for realistic applications, the appropriate choice of the uncertainty model is of great importance.

Further paths of research will focus on ways to exploit the similarities between the responses for neighboring frequencies and to couple the Finite Element solvers with an existing Multilevel/Multi-index framework \cite{PJ,PJ2}.

\section*{Acknowledgements}

This research was funded by project IWT/SBO EUFORIA: ``Efficient Uncertainty quantification For Optimization in Robust desgn of Industrial Applications" (IWT-140068) of the Agency for Innovation by Science and Technology, Flanders, Belgium. The authors gratefully acknowledge the support from the research council of KU Leuven though the funding of project C16/17/008 ``Efficient methods for large-scale PDE-constrained optimization in the presence of uncertainty and complex technological constraints". The authors also would like to thank the Structural Mechanics Section of the KU Leuven and Jef Wambacq for supplying their StaBIL code and providing support.

\section*{References}

\bibliography{Bib_ref.bib}

\newpage
\appendix

\end{document}

%% file: PDF_Gam_Dis.pdf_tex
%% Accompanies image file 'PDF_Gam_Dis.pdf' (pdf, eps, ps)
\begingroup%
  \makeatletter%
  \providecommand\color[2][]{%
    \errmessage{(Inkscape) Color is used for the text in Inkscape, but the package 'color.sty' is not loaded}%
    \renewcommand\color[2][]{}%
  }%
  \providecommand\transparent[1]{%
    \errmessage{(Inkscape) Transparency is used (non-zero) for the text in Inkscape, but the package 'transparent.sty' is not loaded}%
    \renewcommand\transparent[1]{}%
  }%
  \providecommand\rotatebox[2]{#2}%
  \newcommand*\fsize{\dimexpr\f@size pt\relax}%
  \newcommand*\lineheight[1]{\fontsize{\fsize}{#1\fsize}\selectfont}%
  \ifx\svgwidth\undefined%
    \setlength{\unitlength}{420bp}%
    \ifx\svgscale\undefined%
      \relax%
    \else%
      \setlength{\unitlength}{\unitlength * \real{\svgscale}}%
    \fi%
  \else%
    \setlength{\unitlength}{\svgwidth}%
  \fi%
  \global\let\svgwidth\undefined%
  \global\let\svgscale\undefined%
  \makeatother%
  \begin{picture}(1,0.75)%
    \lineheight{1}%
    \setlength\tabcolsep{0pt}%
    \put(0,0){\includegraphics[width=\unitlength,page=1]{PDF_Gam_Dis.pdf}}%
    \put(0.13035714,0.05404768){\makebox(0,0)[t]{\lineheight{1.25}\smash{\begin{tabular}[t]{c}0\end{tabular}}}}%
    \put(0.20785714,0.05404768){\makebox(0,0)[t]{\lineheight{1.25}\smash{\begin{tabular}[t]{c}10\end{tabular}}}}%
    \put(0.28535714,0.05404768){\makebox(0,0)[t]{\lineheight{1.25}\smash{\begin{tabular}[t]{c}20\end{tabular}}}}%
    \put(0.36285714,0.05404768){\makebox(0,0)[t]{\lineheight{1.25}\smash{\begin{tabular}[t]{c}30\end{tabular}}}}%
    \put(0.44035714,0.05404768){\makebox(0,0)[t]{\lineheight{1.25}\smash{\begin{tabular}[t]{c}40\end{tabular}}}}%
    \put(0.51785714,0.05404768){\makebox(0,0)[t]{\lineheight{1.25}\smash{\begin{tabular}[t]{c}50\end{tabular}}}}%
    \put(0.59535714,0.05404768){\makebox(0,0)[t]{\lineheight{1.25}\smash{\begin{tabular}[t]{c}60\end{tabular}}}}%
    \put(0.67285714,0.05404768){\makebox(0,0)[t]{\lineheight{1.25}\smash{\begin{tabular}[t]{c}70\end{tabular}}}}%
    \put(0.75035714,0.05404768){\makebox(0,0)[t]{\lineheight{1.25}\smash{\begin{tabular}[t]{c}80\end{tabular}}}}%
    \put(0.82785714,0.05404768){\makebox(0,0)[t]{\lineheight{1.25}\smash{\begin{tabular}[t]{c}90\end{tabular}}}}%
    \put(0.90535714,0.05404768){\makebox(0,0)[t]{\lineheight{1.25}\smash{\begin{tabular}[t]{c}100\end{tabular}}}}%
    \put(0.5178575,0.01988089){\makebox(0,0)[t]{\lineheight{1.25}\smash{\begin{tabular}[t]{c}\scalebox{1.5}{Young's modulus [GPa]}\end{tabular}}}}%
    \put(0,0){\includegraphics[width=\unitlength,page=2]{PDF_Gam_Dis.pdf}}%
    \put(0.12083339,0.07571429){\makebox(0,0)[rt]{\lineheight{1.25}\smash{\begin{tabular}[t]{r}0\end{tabular}}}}%
    \put(0.12083339,0.15227679){\makebox(0,0)[rt]{\lineheight{1.25}\smash{\begin{tabular}[t]{r}0.005\end{tabular}}}}%
    \put(0.12083339,0.22883929){\makebox(0,0)[rt]{\lineheight{1.25}\smash{\begin{tabular}[t]{r}0.01\end{tabular}}}}%
    \put(0.12083339,0.30540179){\makebox(0,0)[rt]{\lineheight{1.25}\smash{\begin{tabular}[t]{r}0.015\end{tabular}}}}%
    \put(0.12083339,0.38196429){\makebox(0,0)[rt]{\lineheight{1.25}\smash{\begin{tabular}[t]{r}0.02\end{tabular}}}}%
    \put(0.12083339,0.45852679){\makebox(0,0)[rt]{\lineheight{1.25}\smash{\begin{tabular}[t]{r}0.025\end{tabular}}}}%
    \put(0.12083339,0.53508929){\makebox(0,0)[rt]{\lineheight{1.25}\smash{\begin{tabular}[t]{r}0.03\end{tabular}}}}%
    \put(0.12083339,0.61165179){\makebox(0,0)[rt]{\lineheight{1.25}\smash{\begin{tabular}[t]{r}0.035\end{tabular}}}}%
    \put(0.12083339,0.68821429){\makebox(0,0)[rt]{\lineheight{1.25}\smash{\begin{tabular}[t]{r}0.04\end{tabular}}}}%
    \put(0.04726196,0.38839321){\rotatebox{90}{\makebox(0,0)[t]{\lineheight{1.25}\smash{\begin{tabular}[t]{c}\scalebox{1.5}{PDF [1/GPa]}\end{tabular}}}}}%
    \put(0,0){\includegraphics[width=\unitlength,page=3]{PDF_Gam_Dis.pdf}}%
    \put(0.37060714,0.22741071){\makebox(0,0)[rt]{\lineheight{1.25}\smash{\scalebox{2.}{\begin{tabular}[t]{r}$\mu$\end{tabular}}}}}%
    \put(0,0){\includegraphics[width=\unitlength,page=4]{PDF_Gam_Dis.pdf}}%
    \put(0.40629196,0.29631696){\makebox(0,0)[t]{\lineheight{1.25}\smash{\scalebox{2.}{\begin{tabular}[t]{c}$+\sigma$\end{tabular}}}}}%
    \put(0.31942232,0.29631696){\makebox(0,0)[t]{\lineheight{1.25}\smash{\scalebox{2.}{\begin{tabular}[t]{c}$-\sigma$\end{tabular}}}}}%
  \end{picture}%
\endgroup%

%% file: PDF_steel.pdf_tex
%% Accompanies image file 'PDF_steel.pdf' (pdf, eps, ps)
\begingroup%
  \makeatletter%
  \providecommand\color[2][]{%
    \errmessage{(Inkscape) Color is used for the text in Inkscape, but the package 'color.sty' is not loaded}%
    \renewcommand\color[2][]{}%
  }%
  \providecommand\transparent[1]{%
    \errmessage{(Inkscape) Transparency is used (non-zero) for the text in Inkscape, but the package 'transparent.sty' is not loaded}%
    \renewcommand\transparent[1]{}%
  }%
  \providecommand\rotatebox[2]{#2}%
  \newcommand*\fsize{\dimexpr\f@size pt\relax}%
  \newcommand*\lineheight[1]{\fontsize{\fsize}{#1\fsize}\selectfont}%
  \ifx\svgwidth\undefined%
    \setlength{\unitlength}{420bp}%
    \ifx\svgscale\undefined%
      \relax%
    \else%
      \setlength{\unitlength}{\unitlength * \real{\svgscale}}%
    \fi%
  \else%
    \setlength{\unitlength}{\svgwidth}%
  \fi%
  \global\let\svgwidth\undefined%
  \global\let\svgscale\undefined%
  \makeatother%
  \begin{picture}(1,0.75)%
    \lineheight{1}%
    \setlength\tabcolsep{0pt}%
    \put(0,0){\includegraphics[width=\unitlength,page=1]{PDF_steel.pdf}}%
    \put(0.13035714,0.05414679){\makebox(0,0)[t]{\lineheight{1.25}\smash{\begin{tabular}[t]{c}170\end{tabular}}}}%
    \put(0.25952375,0.05414679){\makebox(0,0)[t]{\lineheight{1.25}\smash{\begin{tabular}[t]{c}180\end{tabular}}}}%
    \put(0.38869054,0.05414679){\makebox(0,0)[t]{\lineheight{1.25}\smash{\begin{tabular}[t]{c}190\end{tabular}}}}%
    \put(0.51785714,0.05414679){\makebox(0,0)[t]{\lineheight{1.25}\smash{\begin{tabular}[t]{c}200\end{tabular}}}}%
    \put(0.64702375,0.05414679){\makebox(0,0)[t]{\lineheight{1.25}\smash{\begin{tabular}[t]{c}210\end{tabular}}}}%
    \put(0.77619054,0.05414679){\makebox(0,0)[t]{\lineheight{1.25}\smash{\begin{tabular}[t]{c}220\end{tabular}}}}%
    \put(0.90535714,0.05414679){\makebox(0,0)[t]{\lineheight{1.25}\smash{\begin{tabular}[t]{c}230\end{tabular}}}}%
    \put(0.5178575,0.020005){\makebox(0,0)[t]{\lineheight{1.25}\smash{\begin{tabular}[t]{c}\scalebox{1.5}{Young's modulus [GPa]}\end{tabular}}}}%
    \put(0,0){\includegraphics[width=\unitlength,page=2]{PDF_steel.pdf}}%
    \put(0.1209325,0.07571429){\makebox(0,0)[rt]{\lineheight{1.25}\smash{\begin{tabular}[t]{r}0\end{tabular}}}}%
    \put(0.1209325,0.16321429){\makebox(0,0)[rt]{\lineheight{1.25}\smash{\begin{tabular}[t]{r}0.01\end{tabular}}}}%
    \put(0.1209325,0.25071429){\makebox(0,0)[rt]{\lineheight{1.25}\smash{\begin{tabular}[t]{r}0.02\end{tabular}}}}%
    \put(0.1209325,0.33821429){\makebox(0,0)[rt]{\lineheight{1.25}\smash{\begin{tabular}[t]{r}0.03\end{tabular}}}}%
    \put(0.1209325,0.42571429){\makebox(0,0)[rt]{\lineheight{1.25}\smash{\begin{tabular}[t]{r}0.04\end{tabular}}}}%
    \put(0.1209325,0.51321429){\makebox(0,0)[rt]{\lineheight{1.25}\smash{\begin{tabular}[t]{r}0.05\end{tabular}}}}%
    \put(0.1209325,0.60071429){\makebox(0,0)[rt]{\lineheight{1.25}\smash{\begin{tabular}[t]{r}0.06\end{tabular}}}}%
    \put(0.1209325,0.68821429){\makebox(0,0)[rt]{\lineheight{1.25}\smash{\begin{tabular}[t]{r}0.07\end{tabular}}}}%
    \put(0.04743554,0.38839321){\rotatebox{90}{\makebox(0,0)[t]{\lineheight{1.25}\smash{\begin{tabular}[t]{c}\scalebox{1.5}{PDF [1/GPa]}\end{tabular}}}}}%
    \put(0,0){\includegraphics[width=\unitlength,page=3]{PDF_steel.pdf}}%
    \put(0.53077375,0.16178571){\makebox(0,0)[rt]{\lineheight{1.25}\smash{\scalebox{2.}{\begin{tabular}[t]{r}$\mu$\end{tabular}}}}}%
    \put(0,0){\includegraphics[width=\unitlength,page=4]{PDF_steel.pdf}}%
    \put(0.56011714,0.20116071){\makebox(0,0)[t]{\lineheight{1.25}\smash{\scalebox{2.}{\begin{tabular}[t]{c}$+\sigma$\end{tabular}}}}}%
    \put(0.47559714,0.20116071){\makebox(0,0)[t]{\lineheight{1.25}\smash{\scalebox{2.}{\begin{tabular}[t]{c}$-\sigma$\end{tabular}}}}}%
  \end{picture}%
\endgroup%

%% file: EigenValues.pdf_tex
%% Accompanies image file 'EigenValues.pdf' (pdf, eps, ps)
\begingroup%
  \makeatletter%
  \providecommand\color[2][]{%
    \errmessage{(Inkscape) Color is used for the text in Inkscape, but the package 'color.sty' is not loaded}%
    \renewcommand\color[2][]{}%
  }%
  \providecommand\transparent[1]{%
    \errmessage{(Inkscape) Transparency is used (non-zero) for the text in Inkscape, but the package 'transparent.sty' is not loaded}%
    \renewcommand\transparent[1]{}%
  }%
  \providecommand\rotatebox[2]{#2}%
  \newcommand*\fsize{\dimexpr\f@size pt\relax}%
  \newcommand*\lineheight[1]{\fontsize{\fsize}{#1\fsize}\selectfont}%
  \ifx\svgwidth\undefined%
    \setlength{\unitlength}{420bp}%
    \ifx\svgscale\undefined%
      \relax%
    \else%
      \setlength{\unitlength}{\unitlength * \real{\svgscale}}%
    \fi%
  \else%
    \setlength{\unitlength}{\svgwidth}%
  \fi%
  \global\let\svgwidth\undefined%
  \global\let\svgscale\undefined%
  \makeatother%
  \begin{picture}(1,0.75)%
    \lineheight{1}%
    \setlength\tabcolsep{0pt}%
    \put(0,0){\includegraphics[width=\unitlength,page=1]{EigenValues.pdf}}%
    \put(0.13035714,0.05404768){\makebox(0,0)[t]{\lineheight{1.25}\smash{\begin{tabular}[t]{c}$10^{0}$\end{tabular}}}}%
    \put(0.38333339,0.05404768){\makebox(0,0)[t]{\lineheight{1.25}\smash{\begin{tabular}[t]{c}$10^{1}$\end{tabular}}}}%
    \put(0.63630946,0.05404768){\makebox(0,0)[t]{\lineheight{1.25}\smash{\begin{tabular}[t]{c}$10^{2}$\end{tabular}}}}%
    \put(0.88928571,0.05404768){\makebox(0,0)[t]{\lineheight{1.25}\smash{\begin{tabular}[t]{c}$10^{3}$\end{tabular}}}}%
    \put(0.50982179,0.01988089){\makebox(0,0)[t]{\lineheight{1.25}\smash{\begin{tabular}[t]{c}\scalebox{1.5}{$\theta^{2\mathrm{D}} {n^{th}}$ eigenvalue}\end{tabular}}}}%
    \put(0,0){\includegraphics[width=\unitlength,page=2]{EigenValues.pdf}}%
    \put(0.08333339,0.09292768){\makebox(0,0)[lt]{\lineheight{1.25}\smash{\begin{tabular}[t]{l}0.2\end{tabular}}}}%
    \put(0.08333339,0.24663125){\makebox(0,0)[lt]{\lineheight{1.25}\smash{\begin{tabular}[t]{l}0.3\end{tabular}}}}%
    \put(0.08333339,0.35568554){\makebox(0,0)[lt]{\lineheight{1.25}\smash{\begin{tabular}[t]{l}0.4\end{tabular}}}}%
    \put(0.08333339,0.44027464){\makebox(0,0)[lt]{\lineheight{1.25}\smash{\begin{tabular}[t]{l}0.5\end{tabular}}}}%
    \put(0.08333339,0.50938893){\makebox(0,0)[lt]{\lineheight{1.25}\smash{\begin{tabular}[t]{l}0.6\end{tabular}}}}%
    \put(0.08333339,0.56782446){\makebox(0,0)[lt]{\lineheight{1.25}\smash{\begin{tabular}[t]{l}0.7\end{tabular}}}}%
    \put(0.08333339,0.61844339){\makebox(0,0)[lt]{\lineheight{1.25}\smash{\begin{tabular}[t]{l}0.8\end{tabular}}}}%
    \put(0.08333339,0.6630925){\makebox(0,0)[lt]{\lineheight{1.25}\smash{\begin{tabular}[t]{l}0.9\end{tabular}}}}%
    \put(0.07047625,0.38839321){\rotatebox{90}{\makebox(0,0)[t]{\lineheight{1.25}\smash{\begin{tabular}[t]{c}\scalebox{1.5}{Cumulative value of $\theta^{2\mathrm{D}} {n}$ eigenvalues}\end{tabular}}}}}%
    \put(0,0){\includegraphics[width=\unitlength,page=3]{EigenValues.pdf}}%
    \put(0.89880946,0.07571429){\makebox(0,0)[lt]{\lineheight{1.25}\smash{\begin{tabular}[t]{l}$10^{-7}$\end{tabular}}}}%
    \put(0.89880946,0.16321429){\makebox(0,0)[lt]{\lineheight{1.25}\smash{\begin{tabular}[t]{l}$10^{-6}$\end{tabular}}}}%
    \put(0.89880946,0.25071429){\makebox(0,0)[lt]{\lineheight{1.25}\smash{\begin{tabular}[t]{l}$10^{-5}$\end{tabular}}}}%
    \put(0.89880946,0.33821429){\makebox(0,0)[lt]{\lineheight{1.25}\smash{\begin{tabular}[t]{l}$10^{-4}$\end{tabular}}}}%
    \put(0.89880946,0.42571429){\makebox(0,0)[lt]{\lineheight{1.25}\smash{\begin{tabular}[t]{l}$10^{-3}$\end{tabular}}}}%
    \put(0.89880946,0.51321429){\makebox(0,0)[lt]{\lineheight{1.25}\smash{\begin{tabular}[t]{l}$10^{-2}$\end{tabular}}}}%
    \put(0.89880946,0.60071429){\makebox(0,0)[lt]{\lineheight{1.25}\smash{\begin{tabular}[t]{l}$10^{-1}$\end{tabular}}}}%
    \put(0.89880946,0.68821429){\makebox(0,0)[lt]{\lineheight{1.25}\smash{\begin{tabular}[t]{l}$10^{0}$\end{tabular}}}}%
    \put(0.98845232,0.38839321){\rotatebox{90}{\makebox(0,0)[t]{\lineheight{1.25}\smash{\begin{tabular}[t]{c}\scalebox{1.5}{Magnitude of $\theta^{2\mathrm{D}} {n^{th}}$ eigenvalue}\end{tabular}}}}}%
    \put(0,0){\includegraphics[width=\unitlength,page=4]{EigenValues.pdf}}%
    \put(0.3375,0.16971018){\makebox(0,0)[lt]{\lineheight{1.25}\smash{\begin{tabular}[t]{l}Cumulative value of $\theta^{2\mathrm{D}} {n}$ eigenvalues $\quad$\end{tabular}}}}%
    \put(0,0){\includegraphics[width=\unitlength,page=5]{EigenValues.pdf}}%
    \put(0.3375,0.14243268){\makebox(0,0)[lt]{\lineheight{1.25}\smash{\begin{tabular}[t]{l}Magnitude of $\theta^{2\mathrm{D}} {n^{th}}$ eigenvalue\end{tabular}}}}%
    \put(0,0){\includegraphics[width=\unitlength,page=6]{EigenValues.pdf}}%
  \end{picture}%
\endgroup%

%% file: Mem_Trans.pdf_tex
%% Creator: Inkscape inkscape 0.92.2, www.inkscape.org
%% Accompanies image file 'Mem_Trans.pdf' (pdf, eps, ps)
\begingroup%
  \makeatletter%
  \providecommand\color[2][]{%
    \errmessage{(Inkscape) Color is used for the text in Inkscape, but the package 'color.sty' is not loaded}%
    \renewcommand\color[2][]{}%
  }%
  \providecommand\transparent[1]{%
    \errmessage{(Inkscape) Transparency is used (non-zero) for the text in Inkscape, but the package 'transparent.sty' is not loaded}%
    \renewcommand\transparent[1]{}%
  }%
  \providecommand\rotatebox[2]{#2}%
  \ifx\svgwidth\undefined%
    \setlength{\unitlength}{420bp}%
    \ifx\svgscale\undefined%
      \relax%
    \else%
      \setlength{\unitlength}{\unitlength * \real{\svgscale}}%
    \fi%
  \else%
    \setlength{\unitlength}{\svgwidth}%
  \fi%
  \global\let\svgwidth\undefined%
  \global\let\svgscale\undefined%
  \makeatother%
  \begin{picture}(1,0.75)%
    \put(0,0){\includegraphics[width=\unitlength,page=1]{Mem_Trans.pdf}}%
    \put(0.13035714,0.05404768){\makebox(0,0)[b]{\smash{-4}}}%
    \put(0.22723214,0.05404768){\makebox(0,0)[b]{\smash{-3}}}%
    \put(0.32410714,0.05404768){\makebox(0,0)[b]{\smash{-2}}}%
    \put(0.42098214,0.05404768){\makebox(0,0)[b]{\smash{-1}}}%
    \put(0.51785714,0.05404768){\makebox(0,0)[b]{\smash{0}}}%
    \put(0.61473214,0.05404768){\makebox(0,0)[b]{\smash{1}}}%
    \put(0.71160714,0.05404768){\makebox(0,0)[b]{\smash{2}}}%
    \put(0.80848214,0.05404768){\makebox(0,0)[b]{\smash{3}}}%
    \put(0.90535714,0.05404768){\makebox(0,0)[b]{\smash{4}}}%
    \put(0.5178575,0.01988089){\makebox(0,0)[b]{\smash{\scalebox{1.5}{Value}}}}%
    \put(0,0){\includegraphics[width=\unitlength,page=2]{Mem_Trans.pdf}}%
    \put(0.12083339,0.07571429){\makebox(0,0)[rb]{\smash{0}}}%
    \put(0.12083339,0.13696429){\makebox(0,0)[rb]{\smash{0.1}}}%
    \put(0.12083339,0.19821429){\makebox(0,0)[rb]{\smash{0.2}}}%
    \put(0.12083339,0.25946429){\makebox(0,0)[rb]{\smash{0.3}}}%
    \put(0.12083339,0.32071429){\makebox(0,0)[rb]{\smash{0.4}}}%
    \put(0.12083339,0.38196429){\makebox(0,0)[rb]{\smash{0.5}}}%
    \put(0.12083339,0.44321429){\makebox(0,0)[rb]{\smash{0.6}}}%
    \put(0.12083339,0.50446429){\makebox(0,0)[rb]{\smash{0.7}}}%
    \put(0.12083339,0.56571429){\makebox(0,0)[rb]{\smash{0.8}}}%
    \put(0.12083339,0.62696429){\makebox(0,0)[rb]{\smash{0.9}}}%
    \put(0.12083339,0.68821429){\makebox(0,0)[rb]{\smash{1}}}%
    \put(0.04726196,0.38839321){\rotatebox{90}{\makebox(0,0)[b]{\smash{\scalebox{1.5}{Normalized PDF value}}}}}%
    \put(0,0){\includegraphics[width=\unitlength,page=3]{Mem_Trans.pdf}}%
    \put(0.64379464,0.57703679){\makebox(0,0)[lb]{\smash{$\leftarrow F$}}}%
    \put(0.61473214,0.58960946){\makebox(0,0)[rb]{\smash{$\rightarrow$}}}%
    \put(0.58566964,0.58960946){\makebox(0,0)[rb]{\smash{$\Phi$}}}%
    \put(0.13035714,0.49780643){\makebox(0,0)[b]{\smash{$\bullet$}}}%
    \put(0.12066964,0.47943143){\makebox(0,0)[rb]{\smash{ $u$}}}%
    \put(0.56629464,0.07428571){\makebox(0,0)[b]{\smash{$\bullet$}}}%
    \put(0.56629464,0.05928571){\makebox(0,0)[b]{\smash{  $y$}}}%
    \put(0.62787036,0.07428571){\makebox(0,0)[b]{\smash{$\bullet$}}}%
    \put(0.65208911,0.05928571){\makebox(0,0)[b]{\smash{  $g(y)$}}}%
    \put(0.63658911,0.28604607){\makebox(0,0)[lb]{\smash{$F^{-1}(u)$}}}%
    \put(0.56532589,0.28604607){\makebox(0,0)[rb]{\smash{$\Phi(y)$}}}%
  \end{picture}%
\endgroup%

%% file: Contour_Gaus_Field.pdf_tex
%% Accompanies image file 'Contour_Gaus_Field.pdf' (pdf, eps, ps)
\begingroup%
  \makeatletter%
  \providecommand\color[2][]{%
    \errmessage{(Inkscape) Color is used for the text in Inkscape, but the package 'color.sty' is not loaded}%
    \renewcommand\color[2][]{}%
  }%
  \providecommand\transparent[1]{%
    \errmessage{(Inkscape) Transparency is used (non-zero) for the text in Inkscape, but the package 'transparent.sty' is not loaded}%
    \renewcommand\transparent[1]{}%
  }%
  \providecommand\rotatebox[2]{#2}%
  \newcommand*\fsize{\dimexpr\f@size pt\relax}%
  \newcommand*\lineheight[1]{\fontsize{\fsize}{#1\fsize}\selectfont}%
  \ifx\svgwidth\undefined%
    \setlength{\unitlength}{420bp}%
    \ifx\svgscale\undefined%
      \relax%
    \else%
      \setlength{\unitlength}{\unitlength * \real{\svgscale}}%
    \fi%
  \else%
    \setlength{\unitlength}{\svgwidth}%
  \fi%
  \global\let\svgwidth\undefined%
  \global\let\svgscale\undefined%
  \makeatother%
  \begin{picture}(1,0.75)%
    \lineheight{1}%
    \setlength\tabcolsep{0pt}%
    \put(0,0){\includegraphics[width=\unitlength,page=1]{Contour_Gaus_Field.pdf}}%
  \end{picture}%
\endgroup%

%% file: Contour_Gam_Field.pdf_tex
%% Accompanies image file 'Contour_Gam_Field.pdf' (pdf, eps, ps)
\begingroup%
  \makeatletter%
  \providecommand\color[2][]{%
    \errmessage{(Inkscape) Color is used for the text in Inkscape, but the package 'color.sty' is not loaded}%
    \renewcommand\color[2][]{}%
  }%
  \providecommand\transparent[1]{%
    \errmessage{(Inkscape) Transparency is used (non-zero) for the text in Inkscape, but the package 'transparent.sty' is not loaded}%
    \renewcommand\transparent[1]{}%
  }%
  \providecommand\rotatebox[2]{#2}%
  \newcommand*\fsize{\dimexpr\f@size pt\relax}%
  \newcommand*\lineheight[1]{\fontsize{\fsize}{#1\fsize}\selectfont}%
  \ifx\svgwidth\undefined%
    \setlength{\unitlength}{420bp}%
    \ifx\svgscale\undefined%
      \relax%
    \else%
      \setlength{\unitlength}{\unitlength * \real{\svgscale}}%
    \fi%
  \else%
    \setlength{\unitlength}{\svgwidth}%
  \fi%
  \global\let\svgwidth\undefined%
  \global\let\svgscale\undefined%
  \makeatother%
  \begin{picture}(1,0.75)%
    \lineheight{1}%
    \setlength\tabcolsep{0pt}%
    \put(0,0){\includegraphics[width=\unitlength,page=1]{Contour_Gam_Field.pdf}}%
  \end{picture}%
\endgroup%

%% file: Times_Gam_Dist.pdf_tex
%% Accompanies image file 'Times_Gam_Dist.pdf' (pdf, eps, ps)
\begingroup%
  \makeatletter%
  \providecommand\color[2][]{%
    \errmessage{(Inkscape) Color is used for the text in Inkscape, but the package 'color.sty' is not loaded}%
    \renewcommand\color[2][]{}%
  }%
  \providecommand\transparent[1]{%
    \errmessage{(Inkscape) Transparency is used (non-zero) for the text in Inkscape, but the package 'transparent.sty' is not loaded}%
    \renewcommand\transparent[1]{}%
  }%
  \providecommand\rotatebox[2]{#2}%
  \newcommand*\fsize{\dimexpr\f@size pt\relax}%
  \newcommand*\lineheight[1]{\fontsize{\fsize}{#1\fsize}\selectfont}%
  \ifx\svgwidth\undefined%
    \setlength{\unitlength}{420bp}%
    \ifx\svgscale\undefined%
      \relax%
    \else%
      \setlength{\unitlength}{\unitlength * \real{\svgscale}}%
    \fi%
  \else%
    \setlength{\unitlength}{\svgwidth}%
  \fi%
  \global\let\svgwidth\undefined%
  \global\let\svgscale\undefined%
  \makeatother%
  \begin{picture}(1,0.75)%
    \lineheight{1}%
    \setlength\tabcolsep{0pt}%
    \put(0,0){\includegraphics[width=\unitlength,page=1]{Times_Gam_Dist.pdf}}%
    \put(0.13035714,0.05404768){\makebox(0,0)[t]{\lineheight{1.25}\smash{\begin{tabular}[t]{c}$10^{-5}$\end{tabular}}}}%
    \put(0.51785714,0.05404768){\makebox(0,0)[t]{\lineheight{1.25}\smash{\begin{tabular}[t]{c}$10^{-4}$\end{tabular}}}}%
    \put(0.90535714,0.05404768){\makebox(0,0)[t]{\lineheight{1.25}\smash{\begin{tabular}[t]{c}$10^{-3}$\end{tabular}}}}%
    \put(0.5178575,0.01988089){\makebox(0,0)[t]{\lineheight{1.25}\smash{\begin{tabular}[t]{c}\scalebox{1.5}{RMSE [/]}\end{tabular}}}}%
    \put(0,0){\includegraphics[width=\unitlength,page=2]{Times_Gam_Dist.pdf}}%
    \put(0.12083339,0.07571429){\makebox(0,0)[rt]{\lineheight{1.25}\smash{\begin{tabular}[t]{r}$10^{1}$\end{tabular}}}}%
    \put(0.12083339,0.22883929){\makebox(0,0)[rt]{\lineheight{1.25}\smash{\begin{tabular}[t]{r}$10^{2}$\end{tabular}}}}%
    \put(0.12083339,0.38196429){\makebox(0,0)[rt]{\lineheight{1.25}\smash{\begin{tabular}[t]{r}$10^{3}$\end{tabular}}}}%
    \put(0.12083339,0.53508929){\makebox(0,0)[rt]{\lineheight{1.25}\smash{\begin{tabular}[t]{r}$10^{4}$\end{tabular}}}}%
    \put(0.12083339,0.68821429){\makebox(0,0)[rt]{\lineheight{1.25}\smash{\begin{tabular}[t]{r}$10^{5}$\end{tabular}}}}%
    \put(0.04726196,0.38839321){\rotatebox{90}{\makebox(0,0)[t]{\lineheight{1.25}\smash{\begin{tabular}[t]{c}\scalebox{1.5}{Time [sec]}\end{tabular}}}}}%
    \put(0,0){\includegraphics[width=\unitlength,page=3]{Times_Gam_Dist.pdf}}%
    \put(0.61607143,0.64803571){\makebox(0,0)[lt]{\lineheight{1.25}\smash{\begin{tabular}[t]{l}MC Homogeneous\end{tabular}}}}%
    \put(0,0){\includegraphics[width=\unitlength,page=4]{Times_Gam_Dist.pdf}}%
    \put(0.61607143,0.62035714){\makebox(0,0)[lt]{\lineheight{1.25}\smash{\begin{tabular}[t]{l}MLMC Homogeneous\end{tabular}}}}%
    \put(0,0){\includegraphics[width=\unitlength,page=5]{Times_Gam_Dist.pdf}}%
    \put(0.61607143,0.59267857){\makebox(0,0)[lt]{\lineheight{1.25}\smash{\begin{tabular}[t]{l}$\epsilon^{-2}$\end{tabular}}}}%
    \put(0,0){\includegraphics[width=\unitlength,page=6]{Times_Gam_Dist.pdf}}%
  \end{picture}%
\endgroup%

%% file: Times_Gam_Field.pdf_tex
%% Accompanies image file 'Times_Gam_Field.pdf' (pdf, eps, ps)
\begingroup%
  \makeatletter%
  \providecommand\color[2][]{%
    \errmessage{(Inkscape) Color is used for the text in Inkscape, but the package 'color.sty' is not loaded}%
    \renewcommand\color[2][]{}%
  }%
  \providecommand\transparent[1]{%
    \errmessage{(Inkscape) Transparency is used (non-zero) for the text in Inkscape, but the package 'transparent.sty' is not loaded}%
    \renewcommand\transparent[1]{}%
  }%
  \providecommand\rotatebox[2]{#2}%
  \newcommand*\fsize{\dimexpr\f@size pt\relax}%
  \newcommand*\lineheight[1]{\fontsize{\fsize}{#1\fsize}\selectfont}%
  \ifx\svgwidth\undefined%
    \setlength{\unitlength}{420bp}%
    \ifx\svgscale\undefined%
      \relax%
    \else%
      \setlength{\unitlength}{\unitlength * \real{\svgscale}}%
    \fi%
  \else%
    \setlength{\unitlength}{\svgwidth}%
  \fi%
  \global\let\svgwidth\undefined%
  \global\let\svgscale\undefined%
  \makeatother%
  \begin{picture}(1,0.75)%
    \lineheight{1}%
    \setlength\tabcolsep{0pt}%
    \put(0,0){\includegraphics[width=\unitlength,page=1]{Times_Gam_Field.pdf}}%
    \put(0.13035714,0.05404768){\makebox(0,0)[t]{\lineheight{1.25}\smash{\begin{tabular}[t]{c}$10^{-5}$\end{tabular}}}}%
    \put(0.51785714,0.05404768){\makebox(0,0)[t]{\lineheight{1.25}\smash{\begin{tabular}[t]{c}$10^{-4}$\end{tabular}}}}%
    \put(0.90535714,0.05404768){\makebox(0,0)[t]{\lineheight{1.25}\smash{\begin{tabular}[t]{c}$10^{-3}$\end{tabular}}}}%
    \put(0.5178575,0.01988089){\makebox(0,0)[t]{\lineheight{1.25}\smash{\begin{tabular}[t]{c}\scalebox{1.5}{RMSE [/]}\end{tabular}}}}%
    \put(0,0){\includegraphics[width=\unitlength,page=2]{Times_Gam_Field.pdf}}%
    \put(0.12083339,0.07571429){\makebox(0,0)[rt]{\lineheight{1.25}\smash{\begin{tabular}[t]{r}$10^{1}$\end{tabular}}}}%
    \put(0.12083339,0.22883929){\makebox(0,0)[rt]{\lineheight{1.25}\smash{\begin{tabular}[t]{r}$10^{2}$\end{tabular}}}}%
    \put(0.12083339,0.38196429){\makebox(0,0)[rt]{\lineheight{1.25}\smash{\begin{tabular}[t]{r}$10^{3}$\end{tabular}}}}%
    \put(0.12083339,0.53508929){\makebox(0,0)[rt]{\lineheight{1.25}\smash{\begin{tabular}[t]{r}$10^{4}$\end{tabular}}}}%
    \put(0.12083339,0.68821429){\makebox(0,0)[rt]{\lineheight{1.25}\smash{\begin{tabular}[t]{r}$10^{5}$\end{tabular}}}}%
    \put(0.04726196,0.38839321){\rotatebox{90}{\makebox(0,0)[t]{\lineheight{1.25}\smash{\begin{tabular}[t]{c}\scalebox{1.5}{Time [sec]}\end{tabular}}}}}%
    \put(0,0){\includegraphics[width=\unitlength,page=3]{Times_Gam_Field.pdf}}%
    \put(0.60892857,0.64803571){\makebox(0,0)[lt]{\lineheight{1.25}\smash{\begin{tabular}[t]{l}MC Heterogeneous\end{tabular}}}}%
    \put(0,0){\includegraphics[width=\unitlength,page=4]{Times_Gam_Field.pdf}}%
    \put(0.60892857,0.62035714){\makebox(0,0)[lt]{\lineheight{1.25}\smash{\begin{tabular}[t]{l}MLMC Heterogeneous\end{tabular}}}}%
    \put(0,0){\includegraphics[width=\unitlength,page=5]{Times_Gam_Field.pdf}}%
    \put(0.60892857,0.59267857){\makebox(0,0)[lt]{\lineheight{1.25}\smash{\begin{tabular}[t]{l}$\epsilon^{-2}$\end{tabular}}}}%
    \put(0,0){\includegraphics[width=\unitlength,page=6]{Times_Gam_Field.pdf}}%
  \end{picture}%
\endgroup%

%% file: Samples_Gam_Dist.pdf_tex
%% Accompanies image file 'Samples_Gam_Dist.pdf' (pdf, eps, ps)
\begingroup%
  \makeatletter%
  \providecommand\color[2][]{%
    \errmessage{(Inkscape) Color is used for the text in Inkscape, but the package 'color.sty' is not loaded}%
    \renewcommand\color[2][]{}%
  }%
  \providecommand\transparent[1]{%
    \errmessage{(Inkscape) Transparency is used (non-zero) for the text in Inkscape, but the package 'transparent.sty' is not loaded}%
    \renewcommand\transparent[1]{}%
  }%
  \providecommand\rotatebox[2]{#2}%
  \newcommand*\fsize{\dimexpr\f@size pt\relax}%
  \newcommand*\lineheight[1]{\fontsize{\fsize}{#1\fsize}\selectfont}%
  \ifx\svgwidth\undefined%
    \setlength{\unitlength}{420bp}%
    \ifx\svgscale\undefined%
      \relax%
    \else%
      \setlength{\unitlength}{\unitlength * \real{\svgscale}}%
    \fi%
  \else%
    \setlength{\unitlength}{\svgwidth}%
  \fi%
  \global\let\svgwidth\undefined%
  \global\let\svgscale\undefined%
  \makeatother%
  \begin{picture}(1,0.75)%
    \lineheight{1}%
    \setlength\tabcolsep{0pt}%
    \put(0,0){\includegraphics[width=\unitlength,page=1]{Samples_Gam_Dist.pdf}}%
    \put(0.13035714,0.05404768){\makebox(0,0)[t]{\lineheight{1.25}\smash{\begin{tabular}[t]{c}0\end{tabular}}}}%
    \put(0.32410714,0.05404768){\makebox(0,0)[t]{\lineheight{1.25}\smash{\begin{tabular}[t]{c}1\end{tabular}}}}%
    \put(0.51785714,0.05404768){\makebox(0,0)[t]{\lineheight{1.25}\smash{\begin{tabular}[t]{c}2\end{tabular}}}}%
    \put(0.71160714,0.05404768){\makebox(0,0)[t]{\lineheight{1.25}\smash{\begin{tabular}[t]{c}3\end{tabular}}}}%
    \put(0.90535714,0.05404768){\makebox(0,0)[t]{\lineheight{1.25}\smash{\begin{tabular}[t]{c}4\end{tabular}}}}%
    \put(0.5178575,0.01988089){\makebox(0,0)[t]{\lineheight{1.25}\smash{\begin{tabular}[t]{c}\scalebox{1.5}{Level}\end{tabular}}}}%
    \put(0,0){\includegraphics[width=\unitlength,page=2]{Samples_Gam_Dist.pdf}}%
    \put(0.12083339,0.07571429){\makebox(0,0)[rt]{\lineheight{1.25}\smash{\begin{tabular}[t]{r}$10^{0}$\end{tabular}}}}%
    \put(0.12083339,0.17779768){\makebox(0,0)[rt]{\lineheight{1.25}\smash{\begin{tabular}[t]{r}$10^{1}$\end{tabular}}}}%
    \put(0.12083339,0.27988089){\makebox(0,0)[rt]{\lineheight{1.25}\smash{\begin{tabular}[t]{r}$10^{2}$\end{tabular}}}}%
    \put(0.12083339,0.38196429){\makebox(0,0)[rt]{\lineheight{1.25}\smash{\begin{tabular}[t]{r}$10^{3}$\end{tabular}}}}%
    \put(0.12083339,0.48404768){\makebox(0,0)[rt]{\lineheight{1.25}\smash{\begin{tabular}[t]{r}$10^{4}$\end{tabular}}}}%
    \put(0.12083339,0.58613089){\makebox(0,0)[rt]{\lineheight{1.25}\smash{\begin{tabular}[t]{r}$10^{5}$\end{tabular}}}}%
    \put(0.12083339,0.68821429){\makebox(0,0)[rt]{\lineheight{1.25}\smash{\begin{tabular}[t]{r}$10^{6}$\end{tabular}}}}%
    \put(0.04726196,0.38839321){\rotatebox{90}{\makebox(0,0)[t]{\lineheight{1.25}\smash{\begin{tabular}[t]{c}\scalebox{1.5}{Number of samples}\end{tabular}}}}}%
    \put(0,0){\includegraphics[width=\unitlength,page=3]{Samples_Gam_Dist.pdf}}%
    \put(0.69107143,0.64790625){\makebox(0,0)[lt]{\lineheight{1.25}\smash{\begin{tabular}[t]{l}RMSE = 2.5E-4\end{tabular}}}}%
    \put(0,0){\includegraphics[width=\unitlength,page=4]{Samples_Gam_Dist.pdf}}%
    \put(0.69107143,0.62001643){\makebox(0,0)[lt]{\lineheight{1.25}\smash{\begin{tabular}[t]{l}RMSE = 7.5E-5\end{tabular}}}}%
    \put(0,0){\includegraphics[width=\unitlength,page=5]{Samples_Gam_Dist.pdf}}%
    \put(0.69107143,0.59212643){\makebox(0,0)[lt]{\lineheight{1.25}\smash{\begin{tabular}[t]{l}RMSE = 5.0E-5\end{tabular}}}}%
    \put(0,0){\includegraphics[width=\unitlength,page=6]{Samples_Gam_Dist.pdf}}%
    \put(0.69107143,0.56423661){\makebox(0,0)[lt]{\lineheight{1.25}\smash{\begin{tabular}[t]{l}RMSE = 2.5E-5\end{tabular}}}}%
    \put(0,0){\includegraphics[width=\unitlength,page=7]{Samples_Gam_Dist.pdf}}%
  \end{picture}%
\endgroup%

%% file: Samples_Gam_Field.pdf_tex
%% Accompanies image file 'Samples_Gam_Field.pdf' (pdf, eps, ps)
\begingroup%
  \makeatletter%
  \providecommand\color[2][]{%
    \errmessage{(Inkscape) Color is used for the text in Inkscape, but the package 'color.sty' is not loaded}%
    \renewcommand\color[2][]{}%
  }%
  \providecommand\transparent[1]{%
    \errmessage{(Inkscape) Transparency is used (non-zero) for the text in Inkscape, but the package 'transparent.sty' is not loaded}%
    \renewcommand\transparent[1]{}%
  }%
  \providecommand\rotatebox[2]{#2}%
  \newcommand*\fsize{\dimexpr\f@size pt\relax}%
  \newcommand*\lineheight[1]{\fontsize{\fsize}{#1\fsize}\selectfont}%
  \ifx\svgwidth\undefined%
    \setlength{\unitlength}{420bp}%
    \ifx\svgscale\undefined%
      \relax%
    \else%
      \setlength{\unitlength}{\unitlength * \real{\svgscale}}%
    \fi%
  \else%
    \setlength{\unitlength}{\svgwidth}%
  \fi%
  \global\let\svgwidth\undefined%
  \global\let\svgscale\undefined%
  \makeatother%
  \begin{picture}(1,0.75)%
    \lineheight{1}%
    \setlength\tabcolsep{0pt}%
    \put(0,0){\includegraphics[width=\unitlength,page=1]{Samples_Gam_Field.pdf}}%
    \put(0.13035714,0.05404768){\makebox(0,0)[t]{\lineheight{1.25}\smash{\begin{tabular}[t]{c}0\end{tabular}}}}%
    \put(0.32410714,0.05404768){\makebox(0,0)[t]{\lineheight{1.25}\smash{\begin{tabular}[t]{c}1\end{tabular}}}}%
    \put(0.51785714,0.05404768){\makebox(0,0)[t]{\lineheight{1.25}\smash{\begin{tabular}[t]{c}2\end{tabular}}}}%
    \put(0.71160714,0.05404768){\makebox(0,0)[t]{\lineheight{1.25}\smash{\begin{tabular}[t]{c}3\end{tabular}}}}%
    \put(0.90535714,0.05404768){\makebox(0,0)[t]{\lineheight{1.25}\smash{\begin{tabular}[t]{c}4\end{tabular}}}}%
    \put(0.5178575,0.01988089){\makebox(0,0)[t]{\lineheight{1.25}\smash{\begin{tabular}[t]{c}\scalebox{1.5}{Level}\end{tabular}}}}%
    \put(0,0){\includegraphics[width=\unitlength,page=2]{Samples_Gam_Field.pdf}}%
    \put(0.12083339,0.07571429){\makebox(0,0)[rt]{\lineheight{1.25}\smash{\begin{tabular}[t]{r}$10^{0}$\end{tabular}}}}%
    \put(0.12083339,0.17779768){\makebox(0,0)[rt]{\lineheight{1.25}\smash{\begin{tabular}[t]{r}$10^{1}$\end{tabular}}}}%
    \put(0.12083339,0.27988089){\makebox(0,0)[rt]{\lineheight{1.25}\smash{\begin{tabular}[t]{r}$10^{2}$\end{tabular}}}}%
    \put(0.12083339,0.38196429){\makebox(0,0)[rt]{\lineheight{1.25}\smash{\begin{tabular}[t]{r}$10^{3}$\end{tabular}}}}%
    \put(0.12083339,0.48404768){\makebox(0,0)[rt]{\lineheight{1.25}\smash{\begin{tabular}[t]{r}$10^{4}$\end{tabular}}}}%
    \put(0.12083339,0.58613089){\makebox(0,0)[rt]{\lineheight{1.25}\smash{\begin{tabular}[t]{r}$10^{5}$\end{tabular}}}}%
    \put(0.12083339,0.68821429){\makebox(0,0)[rt]{\lineheight{1.25}\smash{\begin{tabular}[t]{r}$10^{6}$\end{tabular}}}}%
    \put(0.04726196,0.38839321){\rotatebox{90}{\makebox(0,0)[t]{\lineheight{1.25}\smash{\begin{tabular}[t]{c}\scalebox{1.5}{Number of samples}\end{tabular}}}}}%
    \put(0,0){\includegraphics[width=\unitlength,page=3]{Samples_Gam_Field.pdf}}%
    \put(0.69107143,0.64790625){\makebox(0,0)[lt]{\lineheight{1.25}\smash{\begin{tabular}[t]{l}RMSE = 2.5E-4\end{tabular}}}}%
    \put(0,0){\includegraphics[width=\unitlength,page=4]{Samples_Gam_Field.pdf}}%
    \put(0.69107143,0.62001643){\makebox(0,0)[lt]{\lineheight{1.25}\smash{\begin{tabular}[t]{l}RMSE = 7.5E-5\end{tabular}}}}%
    \put(0,0){\includegraphics[width=\unitlength,page=5]{Samples_Gam_Field.pdf}}%
    \put(0.69107143,0.59212643){\makebox(0,0)[lt]{\lineheight{1.25}\smash{\begin{tabular}[t]{l}RMSE = 5.0E-5\end{tabular}}}}%
    \put(0,0){\includegraphics[width=\unitlength,page=6]{Samples_Gam_Field.pdf}}%
    \put(0.69107143,0.56423661){\makebox(0,0)[lt]{\lineheight{1.25}\smash{\begin{tabular}[t]{l}RMSE = 2.5E-5\end{tabular}}}}%
    \put(0,0){\includegraphics[width=\unitlength,page=7]{Samples_Gam_Field.pdf}}%
  \end{picture}%
\endgroup%

%% file: Level_0.pdf_tex
%% Accompanies image file 'Level_0.pdf' (pdf, eps, ps)
\begingroup%
  \makeatletter%
  \providecommand\color[2][]{%
    \errmessage{(Inkscape) Color is used for the text in Inkscape, but the package 'color.sty' is not loaded}%
    \renewcommand\color[2][]{}%
  }%
  \providecommand\transparent[1]{%
    \errmessage{(Inkscape) Transparency is used (non-zero) for the text in Inkscape, but the package 'transparent.sty' is not loaded}%
    \renewcommand\transparent[1]{}%
  }%
  \providecommand\rotatebox[2]{#2}%
  \newcommand*\fsize{\dimexpr\f@size pt\relax}%
  \newcommand*\lineheight[1]{\fontsize{\fsize}{#1\fsize}\selectfont}%
  \ifx\svgwidth\undefined%
    \setlength{\unitlength}{497.25bp}%
    \ifx\svgscale\undefined%
      \relax%
    \else%
      \setlength{\unitlength}{\unitlength * \real{\svgscale}}%
    \fi%
  \else%
    \setlength{\unitlength}{\svgwidth}%
  \fi%
  \global\let\svgwidth\undefined%
  \global\let\svgscale\undefined%
  \makeatother%
  \begin{picture}(1,0.73303167)%
    \lineheight{1}%
    \setlength\tabcolsep{0pt}%
    \put(0,0){\includegraphics[width=\unitlength,page=1]{Level_0.pdf}}%
    \put(0.92556697,0.30489774){\makebox(0,0)[t]{\lineheight{1.25}\smash{\begin{tabular}[t]{c}40\end{tabular}}}}%
    \put(0,0){\includegraphics[width=\unitlength,page=2]{Level_0.pdf}}%
    \put(0.84641855,0.27453137){\makebox(0,0)[t]{\lineheight{1.25}\smash{\begin{tabular}[t]{c}35\end{tabular}}}}%
    \put(0,0){\includegraphics[width=\unitlength,page=3]{Level_0.pdf}}%
    \put(0.76726998,0.24416501){\makebox(0,0)[t]{\lineheight{1.25}\smash{\begin{tabular}[t]{c}30\end{tabular}}}}%
    \put(0,0){\includegraphics[width=\unitlength,page=4]{Level_0.pdf}}%
    \put(0.68812142,0.21379849){\makebox(0,0)[t]{\lineheight{1.25}\smash{\begin{tabular}[t]{c}25\end{tabular}}}}%
    \put(0,0){\includegraphics[width=\unitlength,page=5]{Level_0.pdf}}%
    \put(0.10912368,0.2495083){\makebox(0,0)[rt]{\lineheight{1.25}\smash{\begin{tabular}[t]{r}-1\end{tabular}}}}%
    \put(0.60897285,0.18343213){\makebox(0,0)[t]{\lineheight{1.25}\smash{\begin{tabular}[t]{c}20\end{tabular}}}}%
    \put(0,0){\includegraphics[width=\unitlength,page=6]{Level_0.pdf}}%
    \put(0.10676893,0.17554646){\makebox(0,0)[rt]{\lineheight{1.25}\smash{\begin{tabular}[t]{r}4\end{tabular}}}}%
    \put(0,0){\includegraphics[width=\unitlength,page=7]{Level_0.pdf}}%
    \put(0.10912368,0.32691599){\makebox(0,0)[rt]{\lineheight{1.25}\smash{\begin{tabular}[t]{r}0\end{tabular}}}}%
    \put(0,0){\includegraphics[width=\unitlength,page=8]{Level_0.pdf}}%
    \put(0.52982428,0.15306576){\makebox(0,0)[t]{\lineheight{1.25}\smash{\begin{tabular}[t]{c}15\end{tabular}}}}%
    \put(0,0){\includegraphics[width=\unitlength,page=9]{Level_0.pdf}}%
    \put(0.10912368,0.40432368){\makebox(0,0)[rt]{\lineheight{1.25}\smash{\begin{tabular}[t]{r}1\end{tabular}}}}%
    \put(0,0){\includegraphics[width=\unitlength,page=10]{Level_0.pdf}}%
    \put(0.45067587,0.1226994){\makebox(0,0)[t]{\lineheight{1.25}\smash{\begin{tabular}[t]{c}10\end{tabular}}}}%
    \put(0,0){\includegraphics[width=\unitlength,page=11]{Level_0.pdf}}%
    \put(0.3715273,0.09233288){\makebox(0,0)[t]{\lineheight{1.25}\smash{\begin{tabular}[t]{c}5\end{tabular}}}}%
    \put(0,0){\includegraphics[width=\unitlength,page=12]{Level_0.pdf}}%
    \put(0.26467436,0.07265339){\makebox(0,0)[rt]{\lineheight{1.25}\smash{\begin{tabular}[t]{r}1\end{tabular}}}}%
    \put(0,0){\includegraphics[width=\unitlength,page=13]{Level_0.pdf}}%
    \put(0.62780714,0.12251709){\rotatebox{20.9902}{\makebox(0,0)[t]{\lineheight{1.25}\smash{\begin{tabular}[t]{c}\scalebox{1.5}{x coord. element number [/]}\end{tabular}}}}}%
    \put(0.17562978,0.09136923){\rotatebox{-33.0889}{\makebox(0,0)[t]{\lineheight{1.25}\smash{\begin{tabular}[t]{c}\scalebox{1.5}{y coord. element number [/]}\end{tabular}}}}}%
    \put(0.05381599,0.41235385){\rotatebox{90}{\makebox(0,0)[t]{\lineheight{1.25}\smash{\begin{tabular}[t]{c}\scalebox{1.3}{Gaussian random field value [/]}\end{tabular}}}}}%
  \end{picture}%
\endgroup%

%% file: Level_1.pdf_tex
%% Accompanies image file 'Level_1.pdf' (pdf, eps, ps)
\begingroup%
  \makeatletter%
  \providecommand\color[2][]{%
    \errmessage{(Inkscape) Color is used for the text in Inkscape, but the package 'color.sty' is not loaded}%
    \renewcommand\color[2][]{}%
  }%
  \providecommand\transparent[1]{%
    \errmessage{(Inkscape) Transparency is used (non-zero) for the text in Inkscape, but the package 'transparent.sty' is not loaded}%
    \renewcommand\transparent[1]{}%
  }%
  \providecommand\rotatebox[2]{#2}%
  \newcommand*\fsize{\dimexpr\f@size pt\relax}%
  \newcommand*\lineheight[1]{\fontsize{\fsize}{#1\fsize}\selectfont}%
  \ifx\svgwidth\undefined%
    \setlength{\unitlength}{498bp}%
    \ifx\svgscale\undefined%
      \relax%
    \else%
      \setlength{\unitlength}{\unitlength * \real{\svgscale}}%
    \fi%
  \else%
    \setlength{\unitlength}{\svgwidth}%
  \fi%
  \global\let\svgwidth\undefined%
  \global\let\svgscale\undefined%
  \makeatother%
  \begin{picture}(1,0.73493976)%
    \lineheight{1}%
    \setlength\tabcolsep{0pt}%
    \put(0,0){\includegraphics[width=\unitlength,page=1]{Level_1.pdf}}%
    \put(0.92570602,0.30583901){\makebox(0,0)[t]{\lineheight{1.25}\smash{\begin{tabular}[t]{c}80\end{tabular}}}}%
    \put(0,0){\includegraphics[width=\unitlength,page=2]{Level_1.pdf}}%
    \put(0.8475253,0.27584398){\makebox(0,0)[t]{\lineheight{1.25}\smash{\begin{tabular}[t]{c}70\end{tabular}}}}%
    \put(0,0){\includegraphics[width=\unitlength,page=3]{Level_1.pdf}}%
    \put(0.76934458,0.2458488){\makebox(0,0)[t]{\lineheight{1.25}\smash{\begin{tabular}[t]{c}60\end{tabular}}}}%
    \put(0,0){\includegraphics[width=\unitlength,page=4]{Level_1.pdf}}%
    \put(0.6911637,0.21585362){\makebox(0,0)[t]{\lineheight{1.25}\smash{\begin{tabular}[t]{c}50\end{tabular}}}}%
    \put(0,0){\includegraphics[width=\unitlength,page=5]{Level_1.pdf}}%
    \put(0.10893253,0.24494066){\makebox(0,0)[rt]{\lineheight{1.25}\smash{\begin{tabular}[t]{r}-1\end{tabular}}}}%
    \put(0,0){\includegraphics[width=\unitlength,page=6]{Level_1.pdf}}%
    \put(0.61298298,0.18585858){\makebox(0,0)[t]{\lineheight{1.25}\smash{\begin{tabular}[t]{c}40\end{tabular}}}}%
    \put(0,0){\includegraphics[width=\unitlength,page=7]{Level_1.pdf}}%
    \put(0.10657816,0.17642831){\makebox(0,0)[rt]{\lineheight{1.25}\smash{\begin{tabular}[t]{r}8\end{tabular}}}}%
    \put(0,0){\includegraphics[width=\unitlength,page=8]{Level_1.pdf}}%
    \put(0.10893253,0.31221747){\makebox(0,0)[rt]{\lineheight{1.25}\smash{\begin{tabular}[t]{r}0\end{tabular}}}}%
    \put(0,0){\includegraphics[width=\unitlength,page=9]{Level_1.pdf}}%
    \put(0.53480211,0.1558634){\makebox(0,0)[t]{\lineheight{1.25}\smash{\begin{tabular}[t]{c}30\end{tabular}}}}%
    \put(0,0){\includegraphics[width=\unitlength,page=10]{Level_1.pdf}}%
    \put(0.10893253,0.37949443){\makebox(0,0)[rt]{\lineheight{1.25}\smash{\begin{tabular}[t]{r}1\end{tabular}}}}%
    \put(0,0){\includegraphics[width=\unitlength,page=11]{Level_1.pdf}}%
    \put(0.45662139,0.12586837){\makebox(0,0)[t]{\lineheight{1.25}\smash{\begin{tabular}[t]{c}20\end{tabular}}}}%
    \put(0,0){\includegraphics[width=\unitlength,page=12]{Level_1.pdf}}%
    \put(0.37844066,0.09587319){\makebox(0,0)[t]{\lineheight{1.25}\smash{\begin{tabular}[t]{c}10\end{tabular}}}}%
    \put(0,0){\includegraphics[width=\unitlength,page=13]{Level_1.pdf}}%
    \put(0.26455241,0.07349021){\makebox(0,0)[rt]{\lineheight{1.25}\smash{\begin{tabular}[t]{r}1\end{tabular}}}}%
    \put(0,0){\includegraphics[width=\unitlength,page=14]{Level_1.pdf}}%
    \put(0.62776797,0.12352038){\rotatebox{20.9902}{\makebox(0,0)[t]{\lineheight{1.25}\smash{\begin{tabular}[t]{c}\scalebox{1.5}{x coord. element number [/]}\end{tabular}}}}}%
    \put(0.17551874,0.09228946){\rotatebox{-33.0889}{\makebox(0,0)[t]{\lineheight{1.25}\smash{\begin{tabular}[t]{c}\scalebox{1.5}{y coord. element number [/]}\end{tabular}}}}}%
    \put(0.05373494,0.41325934){\rotatebox{90}{\makebox(0,0)[t]{\lineheight{1.25}\smash{\begin{tabular}[t]{c}\scalebox{1.3}{Gaussian random field value [/]}\end{tabular}}}}}%
  \end{picture}%
\endgroup%

%% file: Level_2.pdf_tex
%% Accompanies image file 'Level_2.pdf' (pdf, eps, ps)
\begingroup%
  \makeatletter%
  \providecommand\color[2][]{%
    \errmessage{(Inkscape) Color is used for the text in Inkscape, but the package 'color.sty' is not loaded}%
    \renewcommand\color[2][]{}%
  }%
  \providecommand\transparent[1]{%
    \errmessage{(Inkscape) Transparency is used (non-zero) for the text in Inkscape, but the package 'transparent.sty' is not loaded}%
    \renewcommand\transparent[1]{}%
  }%
  \providecommand\rotatebox[2]{#2}%
  \newcommand*\fsize{\dimexpr\f@size pt\relax}%
  \newcommand*\lineheight[1]{\fontsize{\fsize}{#1\fsize}\selectfont}%
  \ifx\svgwidth\undefined%
    \setlength{\unitlength}{498bp}%
    \ifx\svgscale\undefined%
      \relax%
    \else%
      \setlength{\unitlength}{\unitlength * \real{\svgscale}}%
    \fi%
  \else%
    \setlength{\unitlength}{\svgwidth}%
  \fi%
  \global\let\svgwidth\undefined%
  \global\let\svgscale\undefined%
  \makeatother%
  \begin{picture}(1,0.73493976)%
    \lineheight{1}%
    \setlength\tabcolsep{0pt}%
    \put(0,0){\includegraphics[width=\unitlength,page=1]{Level_2.pdf}}%
    \put(0.92570602,0.30583901){\makebox(0,0)[t]{\lineheight{1.25}\smash{\begin{tabular}[t]{c}160\end{tabular}}}}%
    \put(0,0){\includegraphics[width=\unitlength,page=2]{Level_2.pdf}}%
    \put(0.84801702,0.27603253){\makebox(0,0)[t]{\lineheight{1.25}\smash{\begin{tabular}[t]{c}140\end{tabular}}}}%
    \put(0,0){\includegraphics[width=\unitlength,page=3]{Level_2.pdf}}%
    \put(0.77032786,0.24622605){\makebox(0,0)[t]{\lineheight{1.25}\smash{\begin{tabular}[t]{c}120\end{tabular}}}}%
    \put(0,0){\includegraphics[width=\unitlength,page=4]{Level_2.pdf}}%
    \put(0.69263886,0.21641958){\makebox(0,0)[t]{\lineheight{1.25}\smash{\begin{tabular}[t]{c}100\end{tabular}}}}%
    \put(0,0){\includegraphics[width=\unitlength,page=5]{Level_2.pdf}}%
    \put(0.6149497,0.1866131){\makebox(0,0)[t]{\lineheight{1.25}\smash{\begin{tabular}[t]{c}80\end{tabular}}}}%
    \put(0,0){\includegraphics[width=\unitlength,page=6]{Level_2.pdf}}%
    \put(0.10893253,0.25086446){\makebox(0,0)[rt]{\lineheight{1.25}\smash{\begin{tabular}[t]{r}-1\end{tabular}}}}%
    \put(0,0){\includegraphics[width=\unitlength,page=7]{Level_2.pdf}}%
    \put(0.10657816,0.17642831){\makebox(0,0)[rt]{\lineheight{1.25}\smash{\begin{tabular}[t]{r}16\end{tabular}}}}%
    \put(0,0){\includegraphics[width=\unitlength,page=8]{Level_2.pdf}}%
    \put(0.10893253,0.31602907){\makebox(0,0)[rt]{\lineheight{1.25}\smash{\begin{tabular}[t]{r}0\end{tabular}}}}%
    \put(0,0){\includegraphics[width=\unitlength,page=9]{Level_2.pdf}}%
    \put(0.53726069,0.15680678){\makebox(0,0)[t]{\lineheight{1.25}\smash{\begin{tabular}[t]{c}60\end{tabular}}}}%
    \put(0,0){\includegraphics[width=\unitlength,page=10]{Level_2.pdf}}%
    \put(0.10893253,0.38119367){\makebox(0,0)[rt]{\lineheight{1.25}\smash{\begin{tabular}[t]{r}1\end{tabular}}}}%
    \put(0,0){\includegraphics[width=\unitlength,page=11]{Level_2.pdf}}%
    \put(0.45957154,0.12700015){\makebox(0,0)[t]{\lineheight{1.25}\smash{\begin{tabular}[t]{c}40\end{tabular}}}}%
    \put(0,0){\includegraphics[width=\unitlength,page=12]{Level_2.pdf}}%
    \put(0.38188253,0.09719383){\makebox(0,0)[t]{\lineheight{1.25}\smash{\begin{tabular}[t]{c}20\end{tabular}}}}%
    \put(0,0){\includegraphics[width=\unitlength,page=13]{Level_2.pdf}}%
    \put(0.26455241,0.07349021){\makebox(0,0)[rt]{\lineheight{1.25}\smash{\begin{tabular}[t]{r}1\end{tabular}}}}%
    \put(0,0){\includegraphics[width=\unitlength,page=14]{Level_2.pdf}}%
    \put(0.30807786,0.06887771){\makebox(0,0)[t]{\lineheight{1.25}\smash{\begin{tabular}[t]{c}1\end{tabular}}}}%
    \put(0,0){\includegraphics[width=\unitlength,page=15]{Level_2.pdf}}%
    \put(0.63305472,0.10777219){\rotatebox{20.9902}{\makebox(0,0)[t]{\lineheight{1.25}\smash{\begin{tabular}[t]{c}\scalebox{1.5}{x coord. element number [/]}\end{tabular}}}}}%
    \put(0.1651591,0.07411973){\rotatebox{-33.0889}{\makebox(0,0)[t]{\lineheight{1.25}\smash{\begin{tabular}[t]{c}\scalebox{1.5}{y coord. element number [/]}\end{tabular}}}}}%
    \put(0.06216867,0.41325934){\rotatebox{90}{\makebox(0,0)[t]{\lineheight{1.25}\smash{\begin{tabular}[t]{c}\scalebox{1.3}{Gaussian random field value [/]}\end{tabular}}}}}%
  \end{picture}%
\endgroup%

%% file: Mesh_Convergence_plast.pdf_tex
%% Accompanies image file 'Mesh_Convergence_plast.pdf' (pdf, eps, ps)
\begingroup%
  \makeatletter%
  \providecommand\color[2][]{%
    \errmessage{(Inkscape) Color is used for the text in Inkscape, but the package 'color.sty' is not loaded}%
    \renewcommand\color[2][]{}%
  }%
  \providecommand\transparent[1]{%
    \errmessage{(Inkscape) Transparency is used (non-zero) for the text in Inkscape, but the package 'transparent.sty' is not loaded}%
    \renewcommand\transparent[1]{}%
  }%
  \providecommand\rotatebox[2]{#2}%
  \newcommand*\fsize{\dimexpr\f@size pt\relax}%
  \newcommand*\lineheight[1]{\fontsize{\fsize}{#1\fsize}\selectfont}%
  \ifx\svgwidth\undefined%
    \setlength{\unitlength}{420bp}%
    \ifx\svgscale\undefined%
      \relax%
    \else%
      \setlength{\unitlength}{\unitlength * \real{\svgscale}}%
    \fi%
  \else%
    \setlength{\unitlength}{\svgwidth}%
  \fi%
  \global\let\svgwidth\undefined%
  \global\let\svgscale\undefined%
  \makeatother%
  \begin{picture}(1,0.75)%
    \lineheight{1}%
    \setlength\tabcolsep{0pt}%
    \put(0,0){\includegraphics[width=\unitlength,page=1]{Mesh_Convergence_plast.pdf}}%
    \put(0.13035714,0.0540475){\makebox(0,0)[t]{\lineheight{1.25}\smash{\begin{tabular}[t]{c}0\end{tabular}}}}%
    \put(0.28428571,0.0540475){\makebox(0,0)[t]{\lineheight{1.25}\smash{\begin{tabular}[t]{c}1\end{tabular}}}}%
    \put(0.43821429,0.0540475){\makebox(0,0)[t]{\lineheight{1.25}\smash{\begin{tabular}[t]{c}2\end{tabular}}}}%
    \put(0.59214286,0.0540475){\makebox(0,0)[t]{\lineheight{1.25}\smash{\begin{tabular}[t]{c}3\end{tabular}}}}%
    \put(0.74607143,0.0540475){\makebox(0,0)[t]{\lineheight{1.25}\smash{\begin{tabular}[t]{c}4\end{tabular}}}}%
    \put(0.9,0.0540475){\makebox(0,0)[t]{\lineheight{1.25}\smash{\begin{tabular}[t]{c}5\end{tabular}}}}%
    \put(0.51517893,0.01988089){\makebox(0,0)[t]{\lineheight{1.25}\smash{\begin{tabular}[t]{c}\scalebox{1.5}{Level} \end{tabular}}}}%
    \put(0,0){\includegraphics[width=\unitlength,page=2]{Mesh_Convergence_plast.pdf}}%
    \put(0.12083339,0.07571411){\makebox(0,0)[rt]{\lineheight{1.25}\smash{\begin{tabular}[t]{r}-7\end{tabular}}}}%
    \put(0.12083339,0.1777975){\makebox(0,0)[rt]{\lineheight{1.25}\smash{\begin{tabular}[t]{r}-6.8\end{tabular}}}}%
    \put(0.12083339,0.27988089){\makebox(0,0)[rt]{\lineheight{1.25}\smash{\begin{tabular}[t]{r}-6.6\end{tabular}}}}%
    \put(0.12083339,0.38196429){\makebox(0,0)[rt]{\lineheight{1.25}\smash{\begin{tabular}[t]{r}-6.4\end{tabular}}}}%
    \put(0.12083339,0.4840475){\makebox(0,0)[rt]{\lineheight{1.25}\smash{\begin{tabular}[t]{r}-6.2\end{tabular}}}}%
    \put(0.12083339,0.58613089){\makebox(0,0)[rt]{\lineheight{1.25}\smash{\begin{tabular}[t]{r}-6\end{tabular}}}}%
    \put(0.12083339,0.68821429){\makebox(0,0)[rt]{\lineheight{1.25}\smash{\begin{tabular}[t]{r}-5.8\end{tabular}}}}%
    \put(0.04726196,0.38839321){\rotatebox{90}{\makebox(0,0)[t]{\lineheight{1.25}\smash{\begin{tabular}[t]{c}\scalebox{1.5}{Deflection [m]}\end{tabular}}}}}%
    \put(0,0){\includegraphics[width=\unitlength,page=3]{Mesh_Convergence_plast.pdf}}%
    \put(0.90892857,0.06964286){\makebox(0,0)[lt]{\lineheight{1.25}\smash{\begin{tabular}[t]{l}10\end{tabular}}}}%
    \put(0.93928571,0.08035714){\makebox(0,0)[lt]{\lineheight{1.25}\smash{\begin{tabular}[t]{l}-7\end{tabular}}}}%
    \put(0.90892857,0.19285714){\makebox(0,0)[lt]{\lineheight{1.25}\smash{\begin{tabular}[t]{l}10\end{tabular}}}}%
    \put(0.93928571,0.20357143){\makebox(0,0)[lt]{\lineheight{1.25}\smash{\begin{tabular}[t]{l}-6\end{tabular}}}}%
    \put(0.90892857,0.31428571){\makebox(0,0)[lt]{\lineheight{1.25}\smash{\begin{tabular}[t]{l}10\end{tabular}}}}%
    \put(0.93928571,0.325){\makebox(0,0)[lt]{\lineheight{1.25}\smash{\begin{tabular}[t]{l}-5\end{tabular}}}}%
    \put(0.90892857,0.4375){\makebox(0,0)[lt]{\lineheight{1.25}\smash{\begin{tabular}[t]{l}10\end{tabular}}}}%
    \put(0.93928571,0.44821429){\makebox(0,0)[lt]{\lineheight{1.25}\smash{\begin{tabular}[t]{l}-4\end{tabular}}}}%
    \put(0.90892857,0.55892857){\makebox(0,0)[lt]{\lineheight{1.25}\smash{\begin{tabular}[t]{l}10\end{tabular}}}}%
    \put(0.93928571,0.56964286){\makebox(0,0)[lt]{\lineheight{1.25}\smash{\begin{tabular}[t]{l}-3\end{tabular}}}}%
    \put(0.90892857,0.68214286){\makebox(0,0)[lt]{\lineheight{1.25}\smash{\begin{tabular}[t]{l}10\end{tabular}}}}%
    \put(0.93928571,0.69285714){\makebox(0,0)[lt]{\lineheight{1.25}\smash{\begin{tabular}[t]{l}-2\end{tabular}}}}%
    \put(0.9884525,0.38839321){\rotatebox{90}{\makebox(0,0)[t]{\lineheight{1.25}\smash{\begin{tabular}[t]{c}\scalebox{1.3}{Absolute value of deflection's differences [m]}\end{tabular}}}}}%
    \put(0,0){\includegraphics[width=\unitlength,page=4]{Mesh_Convergence_plast.pdf}}%
    \put(0.13035714,0.70035714){\makebox(0,0)[lt]{\lineheight{1.25}\smash{\begin{tabular}[t]{l}$\times 10^{\:{-3}}$\end{tabular}}}}%
    \put(0,0){\includegraphics[width=\unitlength,page=5]{Mesh_Convergence_plast.pdf}}%
    \put(0.33035714,0.64600411){\makebox(0,0)[lt]{\lineheight{1.25}\smash{\begin{tabular}[t]{l}Deflection\end{tabular}}}}%
    \put(0,0){\includegraphics[width=\unitlength,page=6]{Mesh_Convergence_plast.pdf}}%
    \put(0.33035714,0.61792446){\makebox(0,0)[lt]{\lineheight{1.25}\smash{\begin{tabular}[t]{l}Abs. val. of deflection's differences\end{tabular}}}}%
    \put(0,0){\includegraphics[width=\unitlength,page=7]{Mesh_Convergence_plast.pdf}}%
  \end{picture}%
\endgroup%

%% file: Mesh_Convergence_elast.pdf_tex
%% Accompanies image file 'Mesh_Convergence_elast.pdf' (pdf, eps, ps)
\begingroup%
  \makeatletter%
  \providecommand\color[2][]{%
    \errmessage{(Inkscape) Color is used for the text in Inkscape, but the package 'color.sty' is not loaded}%
    \renewcommand\color[2][]{}%
  }%
  \providecommand\transparent[1]{%
    \errmessage{(Inkscape) Transparency is used (non-zero) for the text in Inkscape, but the package 'transparent.sty' is not loaded}%
    \renewcommand\transparent[1]{}%
  }%
  \providecommand\rotatebox[2]{#2}%
  \newcommand*\fsize{\dimexpr\f@size pt\relax}%
  \newcommand*\lineheight[1]{\fontsize{\fsize}{#1\fsize}\selectfont}%
  \ifx\svgwidth\undefined%
    \setlength{\unitlength}{420bp}%
    \ifx\svgscale\undefined%
      \relax%
    \else%
      \setlength{\unitlength}{\unitlength * \real{\svgscale}}%
    \fi%
  \else%
    \setlength{\unitlength}{\svgwidth}%
  \fi%
  \global\let\svgwidth\undefined%
  \global\let\svgscale\undefined%
  \makeatother%
  \begin{picture}(1,0.75)%
    \lineheight{1}%
    \setlength\tabcolsep{0pt}%
    \put(0,0){\includegraphics[width=\unitlength,page=1]{Mesh_Convergence_elast.pdf}}%
    \put(0.13035714,0.0540325){\makebox(0,0)[t]{\lineheight{1.25}\smash{\begin{tabular}[t]{c}0\end{tabular}}}}%
    \put(0.28428571,0.0540325){\makebox(0,0)[t]{\lineheight{1.25}\smash{\begin{tabular}[t]{c}1\end{tabular}}}}%
    \put(0.43821429,0.0540325){\makebox(0,0)[t]{\lineheight{1.25}\smash{\begin{tabular}[t]{c}2\end{tabular}}}}%
    \put(0.59214286,0.0540325){\makebox(0,0)[t]{\lineheight{1.25}\smash{\begin{tabular}[t]{c}3\end{tabular}}}}%
    \put(0.74607143,0.0540325){\makebox(0,0)[t]{\lineheight{1.25}\smash{\begin{tabular}[t]{c}4\end{tabular}}}}%
    \put(0.9,0.0540325){\makebox(0,0)[t]{\lineheight{1.25}\smash{\begin{tabular}[t]{c}5\end{tabular}}}}%
    \put(0.51517893,0.01986107){\makebox(0,0)[t]{\lineheight{1.25}\smash{\begin{tabular}[t]{c}\scalebox{1.5}{Level} \end{tabular}}}}%
    \put(0,0){\includegraphics[width=\unitlength,page=2]{Mesh_Convergence_elast.pdf}}%
    \put(0.12083339,0.1522825){\makebox(0,0)[rt]{\lineheight{1.25}\smash{\begin{tabular}[t]{r}-0.022975\end{tabular}}}}%
    \put(0.12083339,0.30541589){\makebox(0,0)[rt]{\lineheight{1.25}\smash{\begin{tabular}[t]{r}-0.022965\end{tabular}}}}%
    \put(0.12083339,0.45852071){\makebox(0,0)[rt]{\lineheight{1.25}\smash{\begin{tabular}[t]{r}-0.022955\end{tabular}}}}%
    \put(0.12083339,0.61165393){\makebox(0,0)[rt]{\lineheight{1.25}\smash{\begin{tabular}[t]{r}-0.022945\end{tabular}}}}%
    \put(0.01526196,0.38839679){\rotatebox{90}{\makebox(0,0)[t]{\lineheight{1.25}\smash{\begin{tabular}[t]{c}\scalebox{1.5}{Deflection [m]}\end{tabular}}}}}%
    \put(0,0){\includegraphics[width=\unitlength,page=3]{Mesh_Convergence_elast.pdf}}%
    \put(0.90892857,0.06964286){\makebox(0,0)[lt]{\lineheight{1.25}\smash{\begin{tabular}[t]{l}10\end{tabular}}}}%
    \put(0.93928571,0.08035714){\makebox(0,0)[lt]{\lineheight{1.25}\smash{\begin{tabular}[t]{l}-7\end{tabular}}}}%
    \put(0.90892857,0.19285714){\makebox(0,0)[lt]{\lineheight{1.25}\smash{\begin{tabular}[t]{l}10\end{tabular}}}}%
    \put(0.93928571,0.20357143){\makebox(0,0)[lt]{\lineheight{1.25}\smash{\begin{tabular}[t]{l}-6\end{tabular}}}}%
    \put(0.90892857,0.31428571){\makebox(0,0)[lt]{\lineheight{1.25}\smash{\begin{tabular}[t]{l}10\end{tabular}}}}%
    \put(0.93928571,0.325){\makebox(0,0)[lt]{\lineheight{1.25}\smash{\begin{tabular}[t]{l}-5\end{tabular}}}}%
    \put(0.90892857,0.4375){\makebox(0,0)[lt]{\lineheight{1.25}\smash{\begin{tabular}[t]{l}10\end{tabular}}}}%
    \put(0.93928571,0.44821429){\makebox(0,0)[lt]{\lineheight{1.25}\smash{\begin{tabular}[t]{l}-4\end{tabular}}}}%
    \put(0.90892857,0.55892857){\makebox(0,0)[lt]{\lineheight{1.25}\smash{\begin{tabular}[t]{l}10\end{tabular}}}}%
    \put(0.93928571,0.56964286){\makebox(0,0)[lt]{\lineheight{1.25}\smash{\begin{tabular}[t]{l}-3\end{tabular}}}}%
    \put(0.90892857,0.68214286){\makebox(0,0)[lt]{\lineheight{1.25}\smash{\begin{tabular}[t]{l}10\end{tabular}}}}%
    \put(0.93928571,0.69285714){\makebox(0,0)[lt]{\lineheight{1.25}\smash{\begin{tabular}[t]{l}-2\end{tabular}}}}%
    \put(0.9884525,0.38839321){\rotatebox{90}{\makebox(0,0)[t]{\lineheight{1.25}\smash{\begin{tabular}[t]{c}\scalebox{1.3}{Absolute value of deflection's differences [m]}\end{tabular}}}}}%
    \put(0,0){\includegraphics[width=\unitlength,page=4]{Mesh_Convergence_elast.pdf}}%
    \put(0.34285714,0.64243268){\makebox(0,0)[lt]{\lineheight{1.25}\smash{\begin{tabular}[t]{l}Deflection\end{tabular}}}}%
    \put(0,0){\includegraphics[width=\unitlength,page=5]{Mesh_Convergence_elast.pdf}}%
    \put(0.34285714,0.61435304){\makebox(0,0)[lt]{\lineheight{1.25}\smash{\begin{tabular}[t]{l}Abs. val. of deflection's differences\end{tabular}}}}%
    \put(0,0){\includegraphics[width=\unitlength,page=6]{Mesh_Convergence_elast.pdf}}%
  \end{picture}%
\endgroup%

%% file: Times_levl_Dist_plast.pdf_tex
%% Accompanies image file 'Times_levl_Dist_plast.pdf' (pdf, eps, ps)
\begingroup%
  \makeatletter%
  \providecommand\color[2][]{%
    \errmessage{(Inkscape) Color is used for the text in Inkscape, but the package 'color.sty' is not loaded}%
    \renewcommand\color[2][]{}%
  }%
  \providecommand\transparent[1]{%
    \errmessage{(Inkscape) Transparency is used (non-zero) for the text in Inkscape, but the package 'transparent.sty' is not loaded}%
    \renewcommand\transparent[1]{}%
  }%
  \providecommand\rotatebox[2]{#2}%
  \newcommand*\fsize{\dimexpr\f@size pt\relax}%
  \newcommand*\lineheight[1]{\fontsize{\fsize}{#1\fsize}\selectfont}%
  \ifx\svgwidth\undefined%
    \setlength{\unitlength}{420bp}%
    \ifx\svgscale\undefined%
      \relax%
    \else%
      \setlength{\unitlength}{\unitlength * \real{\svgscale}}%
    \fi%
  \else%
    \setlength{\unitlength}{\svgwidth}%
  \fi%
  \global\let\svgwidth\undefined%
  \global\let\svgscale\undefined%
  \makeatother%
  \begin{picture}(1,0.75)%
    \lineheight{1}%
    \setlength\tabcolsep{0pt}%
    \put(0,0){\includegraphics[width=\unitlength,page=1]{Times_levl_Dist_plast.pdf}}%
    \put(0.28535714,0.05404768){\makebox(0,0)[t]{\lineheight{1.25}\smash{\begin{tabular}[t]{c}$10^{-6}$\end{tabular}}}}%
    \put(0.59535714,0.05404768){\makebox(0,0)[t]{\lineheight{1.25}\smash{\begin{tabular}[t]{c}$10^{-5}$\end{tabular}}}}%
    \put(0.90535714,0.05404768){\makebox(0,0)[t]{\lineheight{1.25}\smash{\begin{tabular}[t]{c}$10^{-4}$\end{tabular}}}}%
    \put(0.5178575,0.01988089){\makebox(0,0)[t]{\lineheight{1.25}\smash{\begin{tabular}[t]{c}\scalebox{1.5}{$\text{RMSE}$ [/]}\end{tabular}}}}%
    \put(0,0){\includegraphics[width=\unitlength,page=2]{Times_levl_Dist_plast.pdf}}%
    \put(0.12083339,0.07571429){\makebox(0,0)[rt]{\lineheight{1.25}\smash{\begin{tabular}[t]{r}$10^{2}$\end{tabular}}}}%
    \put(0.12083339,0.19821429){\makebox(0,0)[rt]{\lineheight{1.25}\smash{\begin{tabular}[t]{r}$10^{3}$\end{tabular}}}}%
    \put(0.12083339,0.32071429){\makebox(0,0)[rt]{\lineheight{1.25}\smash{\begin{tabular}[t]{r}$10^{4}$\end{tabular}}}}%
    \put(0.12083339,0.44321429){\makebox(0,0)[rt]{\lineheight{1.25}\smash{\begin{tabular}[t]{r}$10^{5}$\end{tabular}}}}%
    \put(0.12083339,0.56571429){\makebox(0,0)[rt]{\lineheight{1.25}\smash{\begin{tabular}[t]{r}$10^{6}$\end{tabular}}}}%
    \put(0.12083339,0.68821429){\makebox(0,0)[rt]{\lineheight{1.25}\smash{\begin{tabular}[t]{r}$10^{7}$\end{tabular}}}}%
    \put(0.04726196,0.38839321){\rotatebox{90}{\makebox(0,0)[t]{\lineheight{1.25}\smash{\begin{tabular}[t]{c}\scalebox{1.5}{Time [sec]}\end{tabular}}}}}%
    \put(0,0){\includegraphics[width=\unitlength,page=3]{Times_levl_Dist_plast.pdf}}%
    \put(0.59642857,0.65339286){\makebox(0,0)[lt]{\lineheight{1.25}\smash{\begin{tabular}[t]{l}MC Homogeneous\end{tabular}}}}%
    \put(0,0){\includegraphics[width=\unitlength,page=4]{Times_levl_Dist_plast.pdf}}%
    \put(0.59642857,0.62571429){\makebox(0,0)[lt]{\lineheight{1.25}\smash{\begin{tabular}[t]{l}MLMC Homogeneous\end{tabular}}}}%
    \put(0,0){\includegraphics[width=\unitlength,page=5]{Times_levl_Dist_plast.pdf}}%
    \put(0.59642857,0.59803571){\makebox(0,0)[lt]{\lineheight{1.25}\smash{\begin{tabular}[t]{l}$\epsilon^{-2}$\end{tabular}}}}%
    \put(0,0){\includegraphics[width=\unitlength,page=6]{Times_levl_Dist_plast.pdf}}%
  \end{picture}%
\endgroup%

%% file: Times_levl_Field_plast.pdf_tex
%% Accompanies image file 'Times_levl_Field_plast.pdf' (pdf, eps, ps)
\begingroup%
  \makeatletter%
  \providecommand\color[2][]{%
    \errmessage{(Inkscape) Color is used for the text in Inkscape, but the package 'color.sty' is not loaded}%
    \renewcommand\color[2][]{}%
  }%
  \providecommand\transparent[1]{%
    \errmessage{(Inkscape) Transparency is used (non-zero) for the text in Inkscape, but the package 'transparent.sty' is not loaded}%
    \renewcommand\transparent[1]{}%
  }%
  \providecommand\rotatebox[2]{#2}%
  \newcommand*\fsize{\dimexpr\f@size pt\relax}%
  \newcommand*\lineheight[1]{\fontsize{\fsize}{#1\fsize}\selectfont}%
  \ifx\svgwidth\undefined%
    \setlength{\unitlength}{420bp}%
    \ifx\svgscale\undefined%
      \relax%
    \else%
      \setlength{\unitlength}{\unitlength * \real{\svgscale}}%
    \fi%
  \else%
    \setlength{\unitlength}{\svgwidth}%
  \fi%
  \global\let\svgwidth\undefined%
  \global\let\svgscale\undefined%
  \makeatother%
  \begin{picture}(1,0.75)%
    \lineheight{1}%
    \setlength\tabcolsep{0pt}%
    \put(0,0){\includegraphics[width=\unitlength,page=1]{Times_levl_Field_plast.pdf}}%
    \put(0.28535714,0.05404768){\makebox(0,0)[t]{\lineheight{1.25}\smash{\begin{tabular}[t]{c}$10^{-6}$\end{tabular}}}}%
    \put(0.59535714,0.05404768){\makebox(0,0)[t]{\lineheight{1.25}\smash{\begin{tabular}[t]{c}$10^{-5}$\end{tabular}}}}%
    \put(0.90535714,0.05404768){\makebox(0,0)[t]{\lineheight{1.25}\smash{\begin{tabular}[t]{c}$10^{-4}$\end{tabular}}}}%
    \put(0.5178575,0.01988089){\makebox(0,0)[t]{\lineheight{1.25}\smash{\begin{tabular}[t]{c}\scalebox{1.5}{$\text{RMSE}$ [/]}\end{tabular}}}}%
    \put(0,0){\includegraphics[width=\unitlength,page=2]{Times_levl_Field_plast.pdf}}%
    \put(0.12083339,0.07571429){\makebox(0,0)[rt]{\lineheight{1.25}\smash{\begin{tabular}[t]{r}$10^{2}$\end{tabular}}}}%
    \put(0.12083339,0.19821429){\makebox(0,0)[rt]{\lineheight{1.25}\smash{\begin{tabular}[t]{r}$10^{3}$\end{tabular}}}}%
    \put(0.12083339,0.32071429){\makebox(0,0)[rt]{\lineheight{1.25}\smash{\begin{tabular}[t]{r}$10^{4}$\end{tabular}}}}%
    \put(0.12083339,0.44321429){\makebox(0,0)[rt]{\lineheight{1.25}\smash{\begin{tabular}[t]{r}$10^{5}$\end{tabular}}}}%
    \put(0.12083339,0.56571429){\makebox(0,0)[rt]{\lineheight{1.25}\smash{\begin{tabular}[t]{r}$10^{6}$\end{tabular}}}}%
    \put(0.12083339,0.68821429){\makebox(0,0)[rt]{\lineheight{1.25}\smash{\begin{tabular}[t]{r}$10^{7}$\end{tabular}}}}%
    \put(0.04726196,0.38839321){\rotatebox{90}{\makebox(0,0)[t]{\lineheight{1.25}\smash{\begin{tabular}[t]{c}\scalebox{1.5}{Time [sec]}\end{tabular}}}}}%
    \put(0,0){\includegraphics[width=\unitlength,page=3]{Times_levl_Field_plast.pdf}}%
    \put(0.54464286,0.63553571){\makebox(0,0)[lt]{\lineheight{1.25}\smash{\begin{tabular}[t]{l}MC Heterogeneous\end{tabular}}}}%
    \put(0,0){\includegraphics[width=\unitlength,page=4]{Times_levl_Field_plast.pdf}}%
    \put(0.54464286,0.60785714){\makebox(0,0)[lt]{\lineheight{1.25}\smash{\begin{tabular}[t]{l}MLMC Heterogeneous\end{tabular}}}}%
    \put(0,0){\includegraphics[width=\unitlength,page=5]{Times_levl_Field_plast.pdf}}%
    \put(0.54464286,0.58017857){\makebox(0,0)[lt]{\lineheight{1.25}\smash{\begin{tabular}[t]{l}$\epsilon^{-2}$\end{tabular}}}}%
    \put(0,0){\includegraphics[width=\unitlength,page=6]{Times_levl_Field_plast.pdf}}%
  \end{picture}%
\endgroup%

%% file: Samples_Gam_Dist_Plast.pdf_tex
%% Accompanies image file 'Samples_Gam_Dist_Plast.pdf' (pdf, eps, ps)
\begingroup%
  \makeatletter%
  \providecommand\color[2][]{%
    \errmessage{(Inkscape) Color is used for the text in Inkscape, but the package 'color.sty' is not loaded}%
    \renewcommand\color[2][]{}%
  }%
  \providecommand\transparent[1]{%
    \errmessage{(Inkscape) Transparency is used (non-zero) for the text in Inkscape, but the package 'transparent.sty' is not loaded}%
    \renewcommand\transparent[1]{}%
  }%
  \providecommand\rotatebox[2]{#2}%
  \newcommand*\fsize{\dimexpr\f@size pt\relax}%
  \newcommand*\lineheight[1]{\fontsize{\fsize}{#1\fsize}\selectfont}%
  \ifx\svgwidth\undefined%
    \setlength{\unitlength}{420bp}%
    \ifx\svgscale\undefined%
      \relax%
    \else%
      \setlength{\unitlength}{\unitlength * \real{\svgscale}}%
    \fi%
  \else%
    \setlength{\unitlength}{\svgwidth}%
  \fi%
  \global\let\svgwidth\undefined%
  \global\let\svgscale\undefined%
  \makeatother%
  \begin{picture}(1,0.75)%
    \lineheight{1}%
    \setlength\tabcolsep{0pt}%
    \put(0,0){\includegraphics[width=\unitlength,page=1]{Samples_Gam_Dist_Plast.pdf}}%
    \put(0.13035714,0.05414679){\makebox(0,0)[t]{\lineheight{1.25}\smash{\begin{tabular}[t]{c}0\end{tabular}}}}%
    \put(0.38869054,0.05414679){\makebox(0,0)[t]{\lineheight{1.25}\smash{\begin{tabular}[t]{c}1\end{tabular}}}}%
    \put(0.64702375,0.05414679){\makebox(0,0)[t]{\lineheight{1.25}\smash{\begin{tabular}[t]{c}2\end{tabular}}}}%
    \put(0.90535714,0.05414679){\makebox(0,0)[t]{\lineheight{1.25}\smash{\begin{tabular}[t]{c}3\end{tabular}}}}%
    \put(0.5178575,0.020005){\makebox(0,0)[t]{\lineheight{1.25}\smash{\begin{tabular}[t]{c}\scalebox{1.5}{Level}\end{tabular}}}}%
    \put(0,0){\includegraphics[width=\unitlength,page=2]{Samples_Gam_Dist_Plast.pdf}}%
    \put(0.1209325,0.07571429){\makebox(0,0)[rt]{\lineheight{1.25}\smash{\begin{tabular}[t]{r}$10^{0}$\end{tabular}}}}%
    \put(0.1209325,0.19821429){\makebox(0,0)[rt]{\lineheight{1.25}\smash{\begin{tabular}[t]{r}$10^{1}$\end{tabular}}}}%
    \put(0.1209325,0.32071429){\makebox(0,0)[rt]{\lineheight{1.25}\smash{\begin{tabular}[t]{r}$10^{2}$\end{tabular}}}}%
    \put(0.1209325,0.44321429){\makebox(0,0)[rt]{\lineheight{1.25}\smash{\begin{tabular}[t]{r}$10^{3}$\end{tabular}}}}%
    \put(0.1209325,0.56571429){\makebox(0,0)[rt]{\lineheight{1.25}\smash{\begin{tabular}[t]{r}$10^{4}$\end{tabular}}}}%
    \put(0.1209325,0.68821429){\makebox(0,0)[rt]{\lineheight{1.25}\smash{\begin{tabular}[t]{r}$10^{5}$\end{tabular}}}}%
    \put(0.04743554,0.38839321){\rotatebox{90}{\makebox(0,0)[t]{\lineheight{1.25}\smash{\begin{tabular}[t]{c}\scalebox{1.5}{Number of samples}\end{tabular}}}}}%
    \put(0,0){\includegraphics[width=\unitlength,page=3]{Samples_Gam_Dist_Plast.pdf}}%
    \put(0.70306393,0.64790625){\makebox(0,0)[lt]{\lineheight{1.25}\smash{\begin{tabular}[t]{l}RMSE=2.5E-5\end{tabular}}}}%
    \put(0,0){\includegraphics[width=\unitlength,page=4]{Samples_Gam_Dist_Plast.pdf}}%
    \put(0.70306393,0.62001643){\makebox(0,0)[lt]{\lineheight{1.25}\smash{\begin{tabular}[t]{l}RMSE=7.5E-6\end{tabular}}}}%
    \put(0,0){\includegraphics[width=\unitlength,page=5]{Samples_Gam_Dist_Plast.pdf}}%
    \put(0.70306393,0.59212643){\makebox(0,0)[lt]{\lineheight{1.25}\smash{\begin{tabular}[t]{l}RMSE=5.0E-6\end{tabular}}}}%
    \put(0,0){\includegraphics[width=\unitlength,page=6]{Samples_Gam_Dist_Plast.pdf}}%
    \put(0.70306393,0.56423661){\makebox(0,0)[lt]{\lineheight{1.25}\smash{\begin{tabular}[t]{l}RMSE=2.5E-6\end{tabular}}}}%
    \put(0,0){\includegraphics[width=\unitlength,page=7]{Samples_Gam_Dist_Plast.pdf}}%
  \end{picture}%
\endgroup%

%% file: Samples_Gam_Field_Plast.pdf_tex
%% Accompanies image file 'Samples_Gam_Field_Plast.pdf' (pdf, eps, ps)
\begingroup%
  \makeatletter%
  \providecommand\color[2][]{%
    \errmessage{(Inkscape) Color is used for the text in Inkscape, but the package 'color.sty' is not loaded}%
    \renewcommand\color[2][]{}%
  }%
  \providecommand\transparent[1]{%
    \errmessage{(Inkscape) Transparency is used (non-zero) for the text in Inkscape, but the package 'transparent.sty' is not loaded}%
    \renewcommand\transparent[1]{}%
  }%
  \providecommand\rotatebox[2]{#2}%
  \newcommand*\fsize{\dimexpr\f@size pt\relax}%
  \newcommand*\lineheight[1]{\fontsize{\fsize}{#1\fsize}\selectfont}%
  \ifx\svgwidth\undefined%
    \setlength{\unitlength}{420bp}%
    \ifx\svgscale\undefined%
      \relax%
    \else%
      \setlength{\unitlength}{\unitlength * \real{\svgscale}}%
    \fi%
  \else%
    \setlength{\unitlength}{\svgwidth}%
  \fi%
  \global\let\svgwidth\undefined%
  \global\let\svgscale\undefined%
  \makeatother%
  \begin{picture}(1,0.75)%
    \lineheight{1}%
    \setlength\tabcolsep{0pt}%
    \put(0,0){\includegraphics[width=\unitlength,page=1]{Samples_Gam_Field_Plast.pdf}}%
    \put(0.13035714,0.05414679){\makebox(0,0)[t]{\lineheight{1.25}\smash{\begin{tabular}[t]{c}0\end{tabular}}}}%
    \put(0.38869054,0.05414679){\makebox(0,0)[t]{\lineheight{1.25}\smash{\begin{tabular}[t]{c}1\end{tabular}}}}%
    \put(0.64702375,0.05414679){\makebox(0,0)[t]{\lineheight{1.25}\smash{\begin{tabular}[t]{c}2\end{tabular}}}}%
    \put(0.90535714,0.05414679){\makebox(0,0)[t]{\lineheight{1.25}\smash{\begin{tabular}[t]{c}3\end{tabular}}}}%
    \put(0.5178575,0.020005){\makebox(0,0)[t]{\lineheight{1.25}\smash{\begin{tabular}[t]{c}\scalebox{1.5}{Level}\end{tabular}}}}%
    \put(0,0){\includegraphics[width=\unitlength,page=2]{Samples_Gam_Field_Plast.pdf}}%
    \put(0.1209325,0.07571429){\makebox(0,0)[rt]{\lineheight{1.25}\smash{\begin{tabular}[t]{r}$10^{0}$\end{tabular}}}}%
    \put(0.1209325,0.19821429){\makebox(0,0)[rt]{\lineheight{1.25}\smash{\begin{tabular}[t]{r}$10^{1}$\end{tabular}}}}%
    \put(0.1209325,0.32071429){\makebox(0,0)[rt]{\lineheight{1.25}\smash{\begin{tabular}[t]{r}$10^{2}$\end{tabular}}}}%
    \put(0.1209325,0.44321429){\makebox(0,0)[rt]{\lineheight{1.25}\smash{\begin{tabular}[t]{r}$10^{3}$\end{tabular}}}}%
    \put(0.1209325,0.56571429){\makebox(0,0)[rt]{\lineheight{1.25}\smash{\begin{tabular}[t]{r}$10^{4}$\end{tabular}}}}%
    \put(0.1209325,0.68821429){\makebox(0,0)[rt]{\lineheight{1.25}\smash{\begin{tabular}[t]{r}$10^{5}$\end{tabular}}}}%
    \put(0.04743554,0.38839321){\rotatebox{90}{\makebox(0,0)[t]{\lineheight{1.25}\smash{\begin{tabular}[t]{c}\scalebox{1.5}{Number of samples}\end{tabular}}}}}%
    \put(0,0){\includegraphics[width=\unitlength,page=3]{Samples_Gam_Field_Plast.pdf}}%
    \put(0.70306393,0.64790625){\makebox(0,0)[lt]{\lineheight{1.25}\smash{\begin{tabular}[t]{l}RMSE=2.5E-5\end{tabular}}}}%
    \put(0,0){\includegraphics[width=\unitlength,page=4]{Samples_Gam_Field_Plast.pdf}}%
    \put(0.70306393,0.62001643){\makebox(0,0)[lt]{\lineheight{1.25}\smash{\begin{tabular}[t]{l}RMSE=7.5E-6\end{tabular}}}}%
    \put(0,0){\includegraphics[width=\unitlength,page=5]{Samples_Gam_Field_Plast.pdf}}%
    \put(0.70306393,0.59212643){\makebox(0,0)[lt]{\lineheight{1.25}\smash{\begin{tabular}[t]{l}RMSE=5.0E-6\end{tabular}}}}%
    \put(0,0){\includegraphics[width=\unitlength,page=6]{Samples_Gam_Field_Plast.pdf}}%
    \put(0.70306393,0.56423661){\makebox(0,0)[lt]{\lineheight{1.25}\smash{\begin{tabular}[t]{l}RMSE=2.5E-6\end{tabular}}}}%
    \put(0,0){\includegraphics[width=\unitlength,page=7]{Samples_Gam_Field_Plast.pdf}}%
  \end{picture}%
\endgroup%

%% file: Rates_stat_gam_dist.pdf_tex
%% Accompanies image file 'Rates_stat_gam_dist.pdf' (pdf, eps, ps)
\begingroup%
  \makeatletter%
  \providecommand\color[2][]{%
    \errmessage{(Inkscape) Color is used for the text in Inkscape, but the package 'color.sty' is not loaded}%
    \renewcommand\color[2][]{}%
  }%
  \providecommand\transparent[1]{%
    \errmessage{(Inkscape) Transparency is used (non-zero) for the text in Inkscape, but the package 'transparent.sty' is not loaded}%
    \renewcommand\transparent[1]{}%
  }%
  \providecommand\rotatebox[2]{#2}%
  \newcommand*\fsize{\dimexpr\f@size pt\relax}%
  \newcommand*\lineheight[1]{\fontsize{\fsize}{#1\fsize}\selectfont}%
  \ifx\svgwidth\undefined%
    \setlength{\unitlength}{412.5bp}%
    \ifx\svgscale\undefined%
      \relax%
    \else%
      \setlength{\unitlength}{\unitlength * \real{\svgscale}}%
    \fi%
  \else%
    \setlength{\unitlength}{\svgwidth}%
  \fi%
  \global\let\svgwidth\undefined%
  \global\let\svgscale\undefined%
  \makeatother%
  \begin{picture}(1,0.70909091)%
    \lineheight{1}%
    \setlength\tabcolsep{0pt}%
    \put(0,0){\includegraphics[width=\unitlength,page=1]{Rates_stat_gam_dist.pdf}}%
    \put(0.15272727,0.04957582){\makebox(0,0)[t]{\lineheight{1.25}\smash{\begin{tabular}[t]{c}0\end{tabular}}}}%
    \put(0.22727273,0.04957582){\makebox(0,0)[t]{\lineheight{1.25}\smash{\begin{tabular}[t]{c}1\end{tabular}}}}%
    \put(0.30181818,0.04957582){\makebox(0,0)[t]{\lineheight{1.25}\smash{\begin{tabular}[t]{c}2\end{tabular}}}}%
    \put(0.37636364,0.04957582){\makebox(0,0)[t]{\lineheight{1.25}\smash{\begin{tabular}[t]{c}3\end{tabular}}}}%
    \put(0.45090909,0.04957582){\makebox(0,0)[t]{\lineheight{1.25}\smash{\begin{tabular}[t]{c}4\end{tabular}}}}%
    \put(0.30181836,0.01478782){\makebox(0,0)[t]{\lineheight{1.25}\smash{\begin{tabular}[t]{c}\scalebox{1.5}{Level}\end{tabular}}}}%
    \put(0,0){\includegraphics[width=\unitlength,page=2]{Rates_stat_gam_dist.pdf}}%
    \put(0.14303036,0.07163636){\makebox(0,0)[rt]{\lineheight{1.25}\smash{\begin{tabular}[t]{r}-40\end{tabular}}}}%
    \put(0.14303036,0.168){\makebox(0,0)[rt]{\lineheight{1.25}\smash{\begin{tabular}[t]{r}-35\end{tabular}}}}%
    \put(0.14303036,0.26436364){\makebox(0,0)[rt]{\lineheight{1.25}\smash{\begin{tabular}[t]{r}-30\end{tabular}}}}%
    \put(0.14303036,0.36072727){\makebox(0,0)[rt]{\lineheight{1.25}\smash{\begin{tabular}[t]{r}-25\end{tabular}}}}%
    \put(0.14303036,0.45709091){\makebox(0,0)[rt]{\lineheight{1.25}\smash{\begin{tabular}[t]{r}-20\end{tabular}}}}%
    \put(0.14303036,0.55345455){\makebox(0,0)[rt]{\lineheight{1.25}\smash{\begin{tabular}[t]{r}-15\end{tabular}}}}%
    \put(0.14303036,0.64981818){\makebox(0,0)[rt]{\lineheight{1.25}\smash{\begin{tabular}[t]{r}-10\end{tabular}}}}%
    \put(0.06812127,0.36727291){\rotatebox{90}{\makebox(0,0)[t]{\lineheight{1.25}\smash{\begin{tabular}[t]{c}\scalebox{1.5}{$\log_2$ variance}\end{tabular}}}}}%
    \put(0,0){\includegraphics[width=\unitlength,page=3]{Rates_stat_gam_dist.pdf}}%
    \put(0.24909091,0.54915945){\makebox(0,0)[lt]{\lineheight{1.25}\smash{\begin{tabular}[t]{l}$\mathbf{V}[P_\ell]$\end{tabular}}}}%
    \put(0,0){\includegraphics[width=\unitlength,page=4]{Rates_stat_gam_dist.pdf}}%
    \put(0.24909091,0.521386){\makebox(0,0)[lt]{\lineheight{1.25}\smash{\begin{tabular}[t]{l}$\mathbf{V}[P_\ell-P_{\ell-1}]$\end{tabular}}}}%
    \put(0,0){\includegraphics[width=\unitlength,page=5]{Rates_stat_gam_dist.pdf}}%
    \put(0.59454545,0.04957582){\makebox(0,0)[t]{\lineheight{1.25}\smash{\begin{tabular}[t]{c}0\end{tabular}}}}%
    \put(0.66863636,0.04957582){\makebox(0,0)[t]{\lineheight{1.25}\smash{\begin{tabular}[t]{c}1\end{tabular}}}}%
    \put(0.74272727,0.04957582){\makebox(0,0)[t]{\lineheight{1.25}\smash{\begin{tabular}[t]{c}2\end{tabular}}}}%
    \put(0.81681818,0.04957582){\makebox(0,0)[t]{\lineheight{1.25}\smash{\begin{tabular}[t]{c}3\end{tabular}}}}%
    \put(0.89090909,0.04957582){\makebox(0,0)[t]{\lineheight{1.25}\smash{\begin{tabular}[t]{c}4\end{tabular}}}}%
    \put(0.74272745,0.01478782){\makebox(0,0)[t]{\lineheight{1.25}\smash{\begin{tabular}[t]{c}\scalebox{1.5}{Level}\end{tabular}}}}%
    \put(0,0){\includegraphics[width=\unitlength,page=6]{Rates_stat_gam_dist.pdf}}%
    \put(0.58484855,0.07163636){\makebox(0,0)[rt]{\lineheight{1.25}\smash{\begin{tabular}[t]{r}-18\end{tabular}}}}%
    \put(0.58484855,0.15423382){\makebox(0,0)[rt]{\lineheight{1.25}\smash{\begin{tabular}[t]{r}-16\end{tabular}}}}%
    \put(0.58484855,0.23683109){\makebox(0,0)[rt]{\lineheight{1.25}\smash{\begin{tabular}[t]{r}-14\end{tabular}}}}%
    \put(0.58484855,0.31942855){\makebox(0,0)[rt]{\lineheight{1.25}\smash{\begin{tabular}[t]{r}-12\end{tabular}}}}%
    \put(0.58484855,0.402026){\makebox(0,0)[rt]{\lineheight{1.25}\smash{\begin{tabular}[t]{r}-10\end{tabular}}}}%
    \put(0.58484855,0.48462345){\makebox(0,0)[rt]{\lineheight{1.25}\smash{\begin{tabular}[t]{r}-8\end{tabular}}}}%
    \put(0.58484855,0.56722073){\makebox(0,0)[rt]{\lineheight{1.25}\smash{\begin{tabular}[t]{r}-6\end{tabular}}}}%
    \put(0.58484855,0.64981818){\makebox(0,0)[rt]{\lineheight{1.25}\smash{\begin{tabular}[t]{r}-4\end{tabular}}}}%
    \put(0.50993945,0.36727309){\rotatebox{90}{\makebox(0,0)[t]{\lineheight{1.25}\smash{\begin{tabular}[t]{c}\scalebox{1.5}{$\log_2$ mean}\end{tabular}}}}}%
    \put(0,0){\includegraphics[width=\unitlength,page=7]{Rates_stat_gam_dist.pdf}}%
    \put(0.69272727,0.55825036){\makebox(0,0)[lt]{\lineheight{1.25}\smash{\begin{tabular}[t]{l}$\mathbf{E}[P_\ell]$\end{tabular}}}}%
    \put(0,0){\includegraphics[width=\unitlength,page=8]{Rates_stat_gam_dist.pdf}}%
    \put(0.69272727,0.53047691){\makebox(0,0)[lt]{\lineheight{1.25}\smash{\begin{tabular}[t]{l}$\mathbf{E}[P_\ell-P_{\ell-1}]$\end{tabular}}}}%
    \put(0,0){\includegraphics[width=\unitlength,page=9]{Rates_stat_gam_dist.pdf}}%
  \end{picture}%
\endgroup%

%% file: Rates_stat_gam_field.pdf_tex
%% Accompanies image file 'Rates_stat_gam_field.pdf' (pdf, eps, ps)
\begingroup%
  \makeatletter%
  \providecommand\color[2][]{%
    \errmessage{(Inkscape) Color is used for the text in Inkscape, but the package 'color.sty' is not loaded}%
    \renewcommand\color[2][]{}%
  }%
  \providecommand\transparent[1]{%
    \errmessage{(Inkscape) Transparency is used (non-zero) for the text in Inkscape, but the package 'transparent.sty' is not loaded}%
    \renewcommand\transparent[1]{}%
  }%
  \providecommand\rotatebox[2]{#2}%
  \newcommand*\fsize{\dimexpr\f@size pt\relax}%
  \newcommand*\lineheight[1]{\fontsize{\fsize}{#1\fsize}\selectfont}%
  \ifx\svgwidth\undefined%
    \setlength{\unitlength}{412.5bp}%
    \ifx\svgscale\undefined%
      \relax%
    \else%
      \setlength{\unitlength}{\unitlength * \real{\svgscale}}%
    \fi%
  \else%
    \setlength{\unitlength}{\svgwidth}%
  \fi%
  \global\let\svgwidth\undefined%
  \global\let\svgscale\undefined%
  \makeatother%
  \begin{picture}(1,0.70909091)%
    \lineheight{1}%
    \setlength\tabcolsep{0pt}%
    \put(0,0){\includegraphics[width=\unitlength,page=1]{Rates_stat_gam_field.pdf}}%
    \put(0.15272727,0.04957582){\makebox(0,0)[t]{\lineheight{1.25}\smash{\begin{tabular}[t]{c}0\end{tabular}}}}%
    \put(0.22727273,0.04957582){\makebox(0,0)[t]{\lineheight{1.25}\smash{\begin{tabular}[t]{c}1\end{tabular}}}}%
    \put(0.30181818,0.04957582){\makebox(0,0)[t]{\lineheight{1.25}\smash{\begin{tabular}[t]{c}2\end{tabular}}}}%
    \put(0.37636364,0.04957582){\makebox(0,0)[t]{\lineheight{1.25}\smash{\begin{tabular}[t]{c}3\end{tabular}}}}%
    \put(0.45090909,0.04957582){\makebox(0,0)[t]{\lineheight{1.25}\smash{\begin{tabular}[t]{c}4\end{tabular}}}}%
    \put(0.30181836,0.01478782){\makebox(0,0)[t]{\lineheight{1.25}\smash{\begin{tabular}[t]{c}\scalebox{1.5}{Level}\end{tabular}}}}%
    \put(0,0){\includegraphics[width=\unitlength,page=2]{Rates_stat_gam_field.pdf}}%
    \put(0.14303036,0.07163636){\makebox(0,0)[rt]{\lineheight{1.25}\smash{\begin{tabular}[t]{r}-28\end{tabular}}}}%
    \put(0.14303036,0.15423382){\makebox(0,0)[rt]{\lineheight{1.25}\smash{\begin{tabular}[t]{r}-26\end{tabular}}}}%
    \put(0.14303036,0.23683109){\makebox(0,0)[rt]{\lineheight{1.25}\smash{\begin{tabular}[t]{r}-24\end{tabular}}}}%
    \put(0.14303036,0.31942855){\makebox(0,0)[rt]{\lineheight{1.25}\smash{\begin{tabular}[t]{r}-22\end{tabular}}}}%
    \put(0.14303036,0.402026){\makebox(0,0)[rt]{\lineheight{1.25}\smash{\begin{tabular}[t]{r}-20\end{tabular}}}}%
    \put(0.14303036,0.48462345){\makebox(0,0)[rt]{\lineheight{1.25}\smash{\begin{tabular}[t]{r}-18\end{tabular}}}}%
    \put(0.14303036,0.56722073){\makebox(0,0)[rt]{\lineheight{1.25}\smash{\begin{tabular}[t]{r}-16\end{tabular}}}}%
    \put(0.14303036,0.64981818){\makebox(0,0)[rt]{\lineheight{1.25}\smash{\begin{tabular}[t]{r}-14\end{tabular}}}}%
    \put(0.06812127,0.36727309){\rotatebox{90}{\makebox(0,0)[t]{\lineheight{1.25}\smash{\begin{tabular}[t]{c}\scalebox{1.5}{$\log_2$ variance}\end{tabular}}}}}%
    \put(0,0){\includegraphics[width=\unitlength,page=3]{Rates_stat_gam_field.pdf}}%
    \put(0.25454545,0.52370491){\makebox(0,0)[lt]{\lineheight{1.25}\smash{\begin{tabular}[t]{l}$\mathbf{V}[P_\ell]$\end{tabular}}}}%
    \put(0,0){\includegraphics[width=\unitlength,page=4]{Rates_stat_gam_field.pdf}}%
    \put(0.25454545,0.49593145){\makebox(0,0)[lt]{\lineheight{1.25}\smash{\begin{tabular}[t]{l}$\mathbf{V}[P_\ell-P_{\ell-1}]$\end{tabular}}}}%
    \put(0,0){\includegraphics[width=\unitlength,page=5]{Rates_stat_gam_field.pdf}}%
    \put(0.59454545,0.04957582){\makebox(0,0)[t]{\lineheight{1.25}\smash{\begin{tabular}[t]{c}0\end{tabular}}}}%
    \put(0.66863636,0.04957582){\makebox(0,0)[t]{\lineheight{1.25}\smash{\begin{tabular}[t]{c}1\end{tabular}}}}%
    \put(0.74272727,0.04957582){\makebox(0,0)[t]{\lineheight{1.25}\smash{\begin{tabular}[t]{c}2\end{tabular}}}}%
    \put(0.81681818,0.04957582){\makebox(0,0)[t]{\lineheight{1.25}\smash{\begin{tabular}[t]{c}3\end{tabular}}}}%
    \put(0.89090909,0.04957582){\makebox(0,0)[t]{\lineheight{1.25}\smash{\begin{tabular}[t]{c}4\end{tabular}}}}%
    \put(0.74272745,0.01478782){\makebox(0,0)[t]{\lineheight{1.25}\smash{\begin{tabular}[t]{c}\scalebox{1.5}{Level}\end{tabular}}}}%
    \put(0,0){\includegraphics[width=\unitlength,page=6]{Rates_stat_gam_field.pdf}}%
    \put(0.58484855,0.07163636){\makebox(0,0)[rt]{\lineheight{1.25}\smash{\begin{tabular}[t]{r}-18\end{tabular}}}}%
    \put(0.58484855,0.15423382){\makebox(0,0)[rt]{\lineheight{1.25}\smash{\begin{tabular}[t]{r}-16\end{tabular}}}}%
    \put(0.58484855,0.23683109){\makebox(0,0)[rt]{\lineheight{1.25}\smash{\begin{tabular}[t]{r}-14\end{tabular}}}}%
    \put(0.58484855,0.31942855){\makebox(0,0)[rt]{\lineheight{1.25}\smash{\begin{tabular}[t]{r}-12\end{tabular}}}}%
    \put(0.58484855,0.402026){\makebox(0,0)[rt]{\lineheight{1.25}\smash{\begin{tabular}[t]{r}-10\end{tabular}}}}%
    \put(0.58484855,0.48462345){\makebox(0,0)[rt]{\lineheight{1.25}\smash{\begin{tabular}[t]{r}-8\end{tabular}}}}%
    \put(0.58484855,0.56722073){\makebox(0,0)[rt]{\lineheight{1.25}\smash{\begin{tabular}[t]{r}-6\end{tabular}}}}%
    \put(0.58484855,0.64981818){\makebox(0,0)[rt]{\lineheight{1.25}\smash{\begin{tabular}[t]{r}-4\end{tabular}}}}%
    \put(0.50993945,0.36727309){\rotatebox{90}{\makebox(0,0)[t]{\lineheight{1.25}\smash{\begin{tabular}[t]{c}\scalebox{1.5}{$\log_2$ mean}\end{tabular}}}}}%
    \put(0,0){\includegraphics[width=\unitlength,page=7]{Rates_stat_gam_field.pdf}}%
    \put(0.70545455,0.52370491){\makebox(0,0)[lt]{\lineheight{1.25}\smash{\begin{tabular}[t]{l}$\mathbf{E}[P_\ell]$\end{tabular}}}}%
    \put(0,0){\includegraphics[width=\unitlength,page=8]{Rates_stat_gam_field.pdf}}%
    \put(0.70545455,0.49593145){\makebox(0,0)[lt]{\lineheight{1.25}\smash{\begin{tabular}[t]{l}$\mathbf{E}[P_\ell-P_{\ell-1}]$\end{tabular}}}}%
    \put(0,0){\includegraphics[width=\unitlength,page=9]{Rates_stat_gam_field.pdf}}%
  \end{picture}%
\endgroup%

%% file: Rates_stat_gam_dist_plast.pdf_tex
%% Accompanies image file 'Rates_stat_gam_dist_plast.pdf' (pdf, eps, ps)
\begingroup%
  \makeatletter%
  \providecommand\color[2][]{%
    \errmessage{(Inkscape) Color is used for the text in Inkscape, but the package 'color.sty' is not loaded}%
    \renewcommand\color[2][]{}%
  }%
  \providecommand\transparent[1]{%
    \errmessage{(Inkscape) Transparency is used (non-zero) for the text in Inkscape, but the package 'transparent.sty' is not loaded}%
    \renewcommand\transparent[1]{}%
  }%
  \providecommand\rotatebox[2]{#2}%
  \newcommand*\fsize{\dimexpr\f@size pt\relax}%
  \newcommand*\lineheight[1]{\fontsize{\fsize}{#1\fsize}\selectfont}%
  \ifx\svgwidth\undefined%
    \setlength{\unitlength}{420bp}%
    \ifx\svgscale\undefined%
      \relax%
    \else%
      \setlength{\unitlength}{\unitlength * \real{\svgscale}}%
    \fi%
  \else%
    \setlength{\unitlength}{\svgwidth}%
  \fi%
  \global\let\svgwidth\undefined%
  \global\let\svgscale\undefined%
  \makeatother%
  \begin{picture}(1,0.75)%
    \lineheight{1}%
    \setlength\tabcolsep{0pt}%
    \put(0,0){\includegraphics[width=\unitlength,page=1]{Rates_stat_gam_dist_plast.pdf}}%
    \put(0.15178571,0.05404768){\makebox(0,0)[t]{\lineheight{1.25}\smash{\begin{tabular}[t]{c}0\end{tabular}}}}%
    \put(0.25178571,0.05404768){\makebox(0,0)[t]{\lineheight{1.25}\smash{\begin{tabular}[t]{c}1\end{tabular}}}}%
    \put(0.35178571,0.05404768){\makebox(0,0)[t]{\lineheight{1.25}\smash{\begin{tabular}[t]{c}2\end{tabular}}}}%
    \put(0.45178571,0.05404768){\makebox(0,0)[t]{\lineheight{1.25}\smash{\begin{tabular}[t]{c}3\end{tabular}}}}%
    \put(0.30178589,0.01988089){\makebox(0,0)[t]{\lineheight{1.25}\smash{\begin{tabular}[t]{c}\scalebox{1.5}{Level }\end{tabular}}}}%
    \put(0,0){\includegraphics[width=\unitlength,page=2]{Rates_stat_gam_dist_plast.pdf}}%
    \put(0.14226196,0.07571429){\makebox(0,0)[rt]{\lineheight{1.25}\smash{\begin{tabular}[t]{r}-45\end{tabular}}}}%
    \put(0.14226196,0.19785714){\makebox(0,0)[rt]{\lineheight{1.25}\smash{\begin{tabular}[t]{r}-40\end{tabular}}}}%
    \put(0.14226196,0.32){\makebox(0,0)[rt]{\lineheight{1.25}\smash{\begin{tabular}[t]{r}-35\end{tabular}}}}%
    \put(0.14226196,0.44214286){\makebox(0,0)[rt]{\lineheight{1.25}\smash{\begin{tabular}[t]{r}-30\end{tabular}}}}%
    \put(0.14226196,0.56428571){\makebox(0,0)[rt]{\lineheight{1.25}\smash{\begin{tabular}[t]{r}-25\end{tabular}}}}%
    \put(0.14226196,0.68642857){\makebox(0,0)[rt]{\lineheight{1.25}\smash{\begin{tabular}[t]{r}-20\end{tabular}}}}%
    \put(0.06869054,0.38750036){\rotatebox{90}{\makebox(0,0)[t]{\lineheight{1.25}\smash{\begin{tabular}[t]{c}\scalebox{1.5}{$\log_2$ variance}\end{tabular}}}}}%
    \put(0,0){\includegraphics[width=\unitlength,page=3]{Rates_stat_gam_dist_plast.pdf}}%
    \put(0.24107143,0.49471018){\makebox(0,0)[lt]{\lineheight{1.25}\smash{\begin{tabular}[t]{l}$\mathbf{V}[P_\ell]$\end{tabular}}}}%
    \put(0,0){\includegraphics[width=\unitlength,page=4]{Rates_stat_gam_dist_plast.pdf}}%
    \put(0.24107143,0.46743268){\makebox(0,0)[lt]{\lineheight{1.25}\smash{\begin{tabular}[t]{l}$\mathbf{V}[P_\ell-P_{\ell-1}]$\end{tabular}}}}%
    \put(0,0){\includegraphics[width=\unitlength,page=5]{Rates_stat_gam_dist_plast.pdf}}%
    \put(0.59107143,0.05404768){\makebox(0,0)[t]{\lineheight{1.25}\smash{\begin{tabular}[t]{c}0\end{tabular}}}}%
    \put(0.69166661,0.05404768){\makebox(0,0)[t]{\lineheight{1.25}\smash{\begin{tabular}[t]{c}1\end{tabular}}}}%
    \put(0.79226196,0.05404768){\makebox(0,0)[t]{\lineheight{1.25}\smash{\begin{tabular}[t]{c}2\end{tabular}}}}%
    \put(0.89285714,0.05404768){\makebox(0,0)[t]{\lineheight{1.25}\smash{\begin{tabular}[t]{c}3\end{tabular}}}}%
    \put(0.74196446,0.01988089){\makebox(0,0)[t]{\lineheight{1.25}\smash{\begin{tabular}[t]{c}\scalebox{1.5}{Level }\end{tabular}}}}%
    \put(0,0){\includegraphics[width=\unitlength,page=6]{Rates_stat_gam_dist_plast.pdf}}%
    \put(0.58154768,0.07571429){\makebox(0,0)[rt]{\lineheight{1.25}\smash{\begin{tabular}[t]{r}-14\end{tabular}}}}%
    \put(0.58154768,0.16295911){\makebox(0,0)[rt]{\lineheight{1.25}\smash{\begin{tabular}[t]{r}-13\end{tabular}}}}%
    \put(0.58154768,0.25020411){\makebox(0,0)[rt]{\lineheight{1.25}\smash{\begin{tabular}[t]{r}-12\end{tabular}}}}%
    \put(0.58154768,0.33744893){\makebox(0,0)[rt]{\lineheight{1.25}\smash{\begin{tabular}[t]{r}-11\end{tabular}}}}%
    \put(0.58154768,0.42469393){\makebox(0,0)[rt]{\lineheight{1.25}\smash{\begin{tabular}[t]{r}-10\end{tabular}}}}%
    \put(0.58154768,0.51193875){\makebox(0,0)[rt]{\lineheight{1.25}\smash{\begin{tabular}[t]{r}-9\end{tabular}}}}%
    \put(0.58154768,0.59918375){\makebox(0,0)[rt]{\lineheight{1.25}\smash{\begin{tabular}[t]{r}-8\end{tabular}}}}%
    \put(0.58154768,0.68642857){\makebox(0,0)[rt]{\lineheight{1.25}\smash{\begin{tabular}[t]{r}-7\end{tabular}}}}%
    \put(0.50797625,0.38750036){\rotatebox{90}{\makebox(0,0)[t]{\lineheight{1.25}\smash{\begin{tabular}[t]{c}\scalebox{1.5}{$\log_2$ mean}\end{tabular}}}}}%
    \put(0,0){\includegraphics[width=\unitlength,page=7]{Rates_stat_gam_dist_plast.pdf}}%
    \put(0.70357143,0.58756732){\makebox(0,0)[lt]{\lineheight{1.25}\smash{\begin{tabular}[t]{l}$\mathbf{E}[P_\ell]$\end{tabular}}}}%
    \put(0,0){\includegraphics[width=\unitlength,page=8]{Rates_stat_gam_dist_plast.pdf}}%
    \put(0.70357143,0.56028982){\makebox(0,0)[lt]{\lineheight{1.25}\smash{\begin{tabular}[t]{l}$\mathbf{E}[P_\ell-P_{\ell-1}]$\end{tabular}}}}%
    \put(0,0){\includegraphics[width=\unitlength,page=9]{Rates_stat_gam_dist_plast.pdf}}%
  \end{picture}%
\endgroup%

%% file: Rates_stat_gam_field_plast.pdf_tex
%% Accompanies image file 'Rates_stat_gam_field_plast.pdf' (pdf, eps, ps)
\begingroup%
  \makeatletter%
  \providecommand\color[2][]{%
    \errmessage{(Inkscape) Color is used for the text in Inkscape, but the package 'color.sty' is not loaded}%
    \renewcommand\color[2][]{}%
  }%
  \providecommand\transparent[1]{%
    \errmessage{(Inkscape) Transparency is used (non-zero) for the text in Inkscape, but the package 'transparent.sty' is not loaded}%
    \renewcommand\transparent[1]{}%
  }%
  \providecommand\rotatebox[2]{#2}%
  \newcommand*\fsize{\dimexpr\f@size pt\relax}%
  \newcommand*\lineheight[1]{\fontsize{\fsize}{#1\fsize}\selectfont}%
  \ifx\svgwidth\undefined%
    \setlength{\unitlength}{420bp}%
    \ifx\svgscale\undefined%
      \relax%
    \else%
      \setlength{\unitlength}{\unitlength * \real{\svgscale}}%
    \fi%
  \else%
    \setlength{\unitlength}{\svgwidth}%
  \fi%
  \global\let\svgwidth\undefined%
  \global\let\svgscale\undefined%
  \makeatother%
  \begin{picture}(1,0.75)%
    \lineheight{1}%
    \setlength\tabcolsep{0pt}%
    \put(0,0){\includegraphics[width=\unitlength,page=1]{Rates_stat_gam_field_plast.pdf}}%
    \put(0.15178571,0.05404768){\makebox(0,0)[t]{\lineheight{1.25}\smash{\begin{tabular}[t]{c}0\end{tabular}}}}%
    \put(0.25178571,0.05404768){\makebox(0,0)[t]{\lineheight{1.25}\smash{\begin{tabular}[t]{c}1\end{tabular}}}}%
    \put(0.35178571,0.05404768){\makebox(0,0)[t]{\lineheight{1.25}\smash{\begin{tabular}[t]{c}2\end{tabular}}}}%
    \put(0.45178571,0.05404768){\makebox(0,0)[t]{\lineheight{1.25}\smash{\begin{tabular}[t]{c}3\end{tabular}}}}%
    \put(0.30178589,0.01988089){\makebox(0,0)[t]{\lineheight{1.25}\smash{\begin{tabular}[t]{c}\scalebox{1.5}{Level }\end{tabular}}}}%
    \put(0,0){\includegraphics[width=\unitlength,page=2]{Rates_stat_gam_field_plast.pdf}}%
    \put(0.14226196,0.07571429){\makebox(0,0)[rt]{\lineheight{1.25}\smash{\begin{tabular}[t]{r}-37\end{tabular}}}}%
    \put(0.14226196,0.13678571){\makebox(0,0)[rt]{\lineheight{1.25}\smash{\begin{tabular}[t]{r}-36\end{tabular}}}}%
    \put(0.14226196,0.19785714){\makebox(0,0)[rt]{\lineheight{1.25}\smash{\begin{tabular}[t]{r}-35\end{tabular}}}}%
    \put(0.14226196,0.25892857){\makebox(0,0)[rt]{\lineheight{1.25}\smash{\begin{tabular}[t]{r}-34\end{tabular}}}}%
    \put(0.14226196,0.32){\makebox(0,0)[rt]{\lineheight{1.25}\smash{\begin{tabular}[t]{r}-33\end{tabular}}}}%
    \put(0.14226196,0.38107143){\makebox(0,0)[rt]{\lineheight{1.25}\smash{\begin{tabular}[t]{r}-32\end{tabular}}}}%
    \put(0.14226196,0.44214286){\makebox(0,0)[rt]{\lineheight{1.25}\smash{\begin{tabular}[t]{r}-31\end{tabular}}}}%
    \put(0.14226196,0.50321429){\makebox(0,0)[rt]{\lineheight{1.25}\smash{\begin{tabular}[t]{r}-30\end{tabular}}}}%
    \put(0.14226196,0.56428571){\makebox(0,0)[rt]{\lineheight{1.25}\smash{\begin{tabular}[t]{r}-29\end{tabular}}}}%
    \put(0.14226196,0.62535714){\makebox(0,0)[rt]{\lineheight{1.25}\smash{\begin{tabular}[t]{r}-28\end{tabular}}}}%
    \put(0.14226196,0.68642857){\makebox(0,0)[rt]{\lineheight{1.25}\smash{\begin{tabular}[t]{r}-27\end{tabular}}}}%
    \put(0.06869054,0.38750036){\rotatebox{90}{\makebox(0,0)[t]{\lineheight{1.25}\smash{\begin{tabular}[t]{c}\scalebox{1.5}{$\log_2$ variance}\end{tabular}}}}}%
    \put(0,0){\includegraphics[width=\unitlength,page=3]{Rates_stat_gam_field_plast.pdf}}%
    \put(0.2625,0.61971018){\makebox(0,0)[lt]{\lineheight{1.25}\smash{\begin{tabular}[t]{l}$\mathbf{V}[P_\ell]$\end{tabular}}}}%
    \put(0,0){\includegraphics[width=\unitlength,page=4]{Rates_stat_gam_field_plast.pdf}}%
    \put(0.2625,0.59243268){\makebox(0,0)[lt]{\lineheight{1.25}\smash{\begin{tabular}[t]{l}$\mathbf{V}[P_\ell-P_{\ell-1}]$\end{tabular}}}}%
    \put(0,0){\includegraphics[width=\unitlength,page=5]{Rates_stat_gam_field_plast.pdf}}%
    \put(0.59107143,0.05404768){\makebox(0,0)[t]{\lineheight{1.25}\smash{\begin{tabular}[t]{c}0\end{tabular}}}}%
    \put(0.69166661,0.05404768){\makebox(0,0)[t]{\lineheight{1.25}\smash{\begin{tabular}[t]{c}1\end{tabular}}}}%
    \put(0.79226196,0.05404768){\makebox(0,0)[t]{\lineheight{1.25}\smash{\begin{tabular}[t]{c}2\end{tabular}}}}%
    \put(0.89285714,0.05404768){\makebox(0,0)[t]{\lineheight{1.25}\smash{\begin{tabular}[t]{c}3\end{tabular}}}}%
    \put(0.74196446,0.01988089){\makebox(0,0)[t]{\lineheight{1.25}\smash{\begin{tabular}[t]{c}\scalebox{1.5}{Level }\end{tabular}}}}%
    \put(0,0){\includegraphics[width=\unitlength,page=6]{Rates_stat_gam_field_plast.pdf}}%
    \put(0.58154768,0.07571429){\makebox(0,0)[rt]{\lineheight{1.25}\smash{\begin{tabular}[t]{r}-14\end{tabular}}}}%
    \put(0.58154768,0.16295911){\makebox(0,0)[rt]{\lineheight{1.25}\smash{\begin{tabular}[t]{r}-13\end{tabular}}}}%
    \put(0.58154768,0.25020411){\makebox(0,0)[rt]{\lineheight{1.25}\smash{\begin{tabular}[t]{r}-12\end{tabular}}}}%
    \put(0.58154768,0.33744893){\makebox(0,0)[rt]{\lineheight{1.25}\smash{\begin{tabular}[t]{r}-11\end{tabular}}}}%
    \put(0.58154768,0.42469393){\makebox(0,0)[rt]{\lineheight{1.25}\smash{\begin{tabular}[t]{r}-10\end{tabular}}}}%
    \put(0.58154768,0.51193875){\makebox(0,0)[rt]{\lineheight{1.25}\smash{\begin{tabular}[t]{r}-9\end{tabular}}}}%
    \put(0.58154768,0.59918375){\makebox(0,0)[rt]{\lineheight{1.25}\smash{\begin{tabular}[t]{r}-8\end{tabular}}}}%
    \put(0.58154768,0.68642857){\makebox(0,0)[rt]{\lineheight{1.25}\smash{\begin{tabular}[t]{r}-7\end{tabular}}}}%
    \put(0.50797625,0.38750036){\rotatebox{90}{\makebox(0,0)[t]{\lineheight{1.25}\smash{\begin{tabular}[t]{c}\scalebox{1.5}{$\log_2$ mean}\end{tabular}}}}}%
    \put(0,0){\includegraphics[width=\unitlength,page=7]{Rates_stat_gam_field_plast.pdf}}%
    \put(0.69821429,0.60185304){\makebox(0,0)[lt]{\lineheight{1.25}\smash{\begin{tabular}[t]{l}$\mathbf{E}[P_\ell]$\end{tabular}}}}%
    \put(0,0){\includegraphics[width=\unitlength,page=8]{Rates_stat_gam_field_plast.pdf}}%
    \put(0.69821429,0.57457554){\makebox(0,0)[lt]{\lineheight{1.25}\smash{\begin{tabular}[t]{l}$\mathbf{E}[P_\ell-P_{\ell-1}]$\end{tabular}}}}%
    \put(0,0){\includegraphics[width=\unitlength,page=9]{Rates_stat_gam_field_plast.pdf}}%
  \end{picture}%
\endgroup%

%% file: Deflection_stat_gam_dist.pdf_tex
%% Accompanies image file 'Deflection_stat_gam_dist.pdf' (pdf, eps, ps)
\begingroup%
  \makeatletter%
  \providecommand\color[2][]{%
    \errmessage{(Inkscape) Color is used for the text in Inkscape, but the package 'color.sty' is not loaded}%
    \renewcommand\color[2][]{}%
  }%
  \providecommand\transparent[1]{%
    \errmessage{(Inkscape) Transparency is used (non-zero) for the text in Inkscape, but the package 'transparent.sty' is not loaded}%
    \renewcommand\transparent[1]{}%
  }%
  \providecommand\rotatebox[2]{#2}%
  \newcommand*\fsize{\dimexpr\f@size pt\relax}%
  \newcommand*\lineheight[1]{\fontsize{\fsize}{#1\fsize}\selectfont}%
  \ifx\svgwidth\undefined%
    \setlength{\unitlength}{420bp}%
    \ifx\svgscale\undefined%
      \relax%
    \else%
      \setlength{\unitlength}{\unitlength * \real{\svgscale}}%
    \fi%
  \else%
    \setlength{\unitlength}{\svgwidth}%
  \fi%
  \global\let\svgwidth\undefined%
  \global\let\svgscale\undefined%
  \makeatother%
  \begin{picture}(1,0.75)%
    \lineheight{1}%
    \setlength\tabcolsep{0pt}%
    \put(0,0){\includegraphics[width=\unitlength,page=1]{Deflection_stat_gam_dist.pdf}}%
    \put(0.13035714,0.05404768){\makebox(0,0)[t]{\lineheight{1.25}\smash{\begin{tabular}[t]{c}0\end{tabular}}}}%
    \put(0.28535714,0.05404768){\makebox(0,0)[t]{\lineheight{1.25}\smash{\begin{tabular}[t]{c}0.5\end{tabular}}}}%
    \put(0.44035714,0.05404768){\makebox(0,0)[t]{\lineheight{1.25}\smash{\begin{tabular}[t]{c}1\end{tabular}}}}%
    \put(0.59535714,0.05404768){\makebox(0,0)[t]{\lineheight{1.25}\smash{\begin{tabular}[t]{c}1.5\end{tabular}}}}%
    \put(0.75035714,0.05404768){\makebox(0,0)[t]{\lineheight{1.25}\smash{\begin{tabular}[t]{c}2\end{tabular}}}}%
    \put(0.90535714,0.05404768){\makebox(0,0)[t]{\lineheight{1.25}\smash{\begin{tabular}[t]{c}2.5\end{tabular}}}}%
    \put(0.5178575,0.01988089){\makebox(0,0)[t]{\lineheight{1.25}\smash{\begin{tabular}[t]{c}\scalebox{1.5}{x-coordinate [m]}\end{tabular}}}}%
    \put(0,0){\includegraphics[width=\unitlength,page=2]{Deflection_stat_gam_dist.pdf}}%
    \put(0.12083339,0.07571429){\makebox(0,0)[rt]{\lineheight{1.25}\smash{\begin{tabular}[t]{r}-0.05\end{tabular}}}}%
    \put(0.12083339,0.13696429){\makebox(0,0)[rt]{\lineheight{1.25}\smash{\begin{tabular}[t]{r}-0.045\end{tabular}}}}%
    \put(0.12083339,0.19821429){\makebox(0,0)[rt]{\lineheight{1.25}\smash{\begin{tabular}[t]{r}-0.04\end{tabular}}}}%
    \put(0.12083339,0.25946429){\makebox(0,0)[rt]{\lineheight{1.25}\smash{\begin{tabular}[t]{r}-0.035\end{tabular}}}}%
    \put(0.12083339,0.32071429){\makebox(0,0)[rt]{\lineheight{1.25}\smash{\begin{tabular}[t]{r}-0.03\end{tabular}}}}%
    \put(0.12083339,0.38196429){\makebox(0,0)[rt]{\lineheight{1.25}\smash{\begin{tabular}[t]{r}-0.025\end{tabular}}}}%
    \put(0.12083339,0.44321429){\makebox(0,0)[rt]{\lineheight{1.25}\smash{\begin{tabular}[t]{r}-0.02\end{tabular}}}}%
    \put(0.12083339,0.50446429){\makebox(0,0)[rt]{\lineheight{1.25}\smash{\begin{tabular}[t]{r}-0.015\end{tabular}}}}%
    \put(0.12083339,0.56571429){\makebox(0,0)[rt]{\lineheight{1.25}\smash{\begin{tabular}[t]{r}-0.01\end{tabular}}}}%
    \put(0.12083339,0.62696429){\makebox(0,0)[rt]{\lineheight{1.25}\smash{\begin{tabular}[t]{r}-0.005\end{tabular}}}}%
    \put(0.12083339,0.68821429){\makebox(0,0)[rt]{\lineheight{1.25}\smash{\begin{tabular}[t]{r}0\end{tabular}}}}%
    \put(0.04726196,0.38839321){\rotatebox{90}{\makebox(0,0)[t]{\lineheight{1.25}\smash{\begin{tabular}[t]{c}\scalebox{1.5}{Deflection [m]}\end{tabular}}}}}%
    \put(0,0){\includegraphics[width=\unitlength,page=3]{Deflection_stat_gam_dist.pdf}}%
  \end{picture}%
\endgroup%

%% file: Deflection_stat_gam_field.pdf_tex
%% Accompanies image file 'Deflection_stat_gam_field.pdf' (pdf, eps, ps)
\begingroup%
  \makeatletter%
  \providecommand\color[2][]{%
    \errmessage{(Inkscape) Color is used for the text in Inkscape, but the package 'color.sty' is not loaded}%
    \renewcommand\color[2][]{}%
  }%
  \providecommand\transparent[1]{%
    \errmessage{(Inkscape) Transparency is used (non-zero) for the text in Inkscape, but the package 'transparent.sty' is not loaded}%
    \renewcommand\transparent[1]{}%
  }%
  \providecommand\rotatebox[2]{#2}%
  \newcommand*\fsize{\dimexpr\f@size pt\relax}%
  \newcommand*\lineheight[1]{\fontsize{\fsize}{#1\fsize}\selectfont}%
  \ifx\svgwidth\undefined%
    \setlength{\unitlength}{420bp}%
    \ifx\svgscale\undefined%
      \relax%
    \else%
      \setlength{\unitlength}{\unitlength * \real{\svgscale}}%
    \fi%
  \else%
    \setlength{\unitlength}{\svgwidth}%
  \fi%
  \global\let\svgwidth\undefined%
  \global\let\svgscale\undefined%
  \makeatother%
  \begin{picture}(1,0.75)%
    \lineheight{1}%
    \setlength\tabcolsep{0pt}%
    \put(0,0){\includegraphics[width=\unitlength,page=1]{Deflection_stat_gam_field.pdf}}%
    \put(0.13035714,0.05404768){\makebox(0,0)[t]{\lineheight{1.25}\smash{\begin{tabular}[t]{c}0\end{tabular}}}}%
    \put(0.28535714,0.05404768){\makebox(0,0)[t]{\lineheight{1.25}\smash{\begin{tabular}[t]{c}0.5\end{tabular}}}}%
    \put(0.44035714,0.05404768){\makebox(0,0)[t]{\lineheight{1.25}\smash{\begin{tabular}[t]{c}1\end{tabular}}}}%
    \put(0.59535714,0.05404768){\makebox(0,0)[t]{\lineheight{1.25}\smash{\begin{tabular}[t]{c}1.5\end{tabular}}}}%
    \put(0.75035714,0.05404768){\makebox(0,0)[t]{\lineheight{1.25}\smash{\begin{tabular}[t]{c}2\end{tabular}}}}%
    \put(0.90535714,0.05404768){\makebox(0,0)[t]{\lineheight{1.25}\smash{\begin{tabular}[t]{c}2.5\end{tabular}}}}%
    \put(0.5178575,0.01988089){\makebox(0,0)[t]{\lineheight{1.25}\smash{\begin{tabular}[t]{c}\scalebox{1.5}{x-coordinate [m]}\end{tabular}}}}%
    \put(0,0){\includegraphics[width=\unitlength,page=2]{Deflection_stat_gam_field.pdf}}%
    \put(0.12083339,0.07571429){\makebox(0,0)[rt]{\lineheight{1.25}\smash{\begin{tabular}[t]{r}-0.05\end{tabular}}}}%
    \put(0.12083339,0.13696429){\makebox(0,0)[rt]{\lineheight{1.25}\smash{\begin{tabular}[t]{r}-0.045\end{tabular}}}}%
    \put(0.12083339,0.19821429){\makebox(0,0)[rt]{\lineheight{1.25}\smash{\begin{tabular}[t]{r}-0.04\end{tabular}}}}%
    \put(0.12083339,0.25946429){\makebox(0,0)[rt]{\lineheight{1.25}\smash{\begin{tabular}[t]{r}-0.035\end{tabular}}}}%
    \put(0.12083339,0.32071429){\makebox(0,0)[rt]{\lineheight{1.25}\smash{\begin{tabular}[t]{r}-0.03\end{tabular}}}}%
    \put(0.12083339,0.38196429){\makebox(0,0)[rt]{\lineheight{1.25}\smash{\begin{tabular}[t]{r}-0.025\end{tabular}}}}%
    \put(0.12083339,0.44321429){\makebox(0,0)[rt]{\lineheight{1.25}\smash{\begin{tabular}[t]{r}-0.02\end{tabular}}}}%
    \put(0.12083339,0.50446429){\makebox(0,0)[rt]{\lineheight{1.25}\smash{\begin{tabular}[t]{r}-0.015\end{tabular}}}}%
    \put(0.12083339,0.56571429){\makebox(0,0)[rt]{\lineheight{1.25}\smash{\begin{tabular}[t]{r}-0.01\end{tabular}}}}%
    \put(0.12083339,0.62696429){\makebox(0,0)[rt]{\lineheight{1.25}\smash{\begin{tabular}[t]{r}-0.005\end{tabular}}}}%
    \put(0.12083339,0.68821429){\makebox(0,0)[rt]{\lineheight{1.25}\smash{\begin{tabular}[t]{r}0\end{tabular}}}}%
    \put(0.04726196,0.38839321){\rotatebox{90}{\makebox(0,0)[t]{\lineheight{1.25}\smash{\begin{tabular}[t]{c}\scalebox{1.5}{Deflection [m]}\end{tabular}}}}}%
    \put(0,0){\includegraphics[width=\unitlength,page=3]{Deflection_stat_gam_field.pdf}}%
  \end{picture}%
\endgroup%

%% file: Real_Gam_Dist_stat_el.pdf_tex
%% Accompanies image file 'Real_Gam_Dist_stat_el.pdf' (pdf, eps, ps)
\begingroup%
  \makeatletter%
  \providecommand\color[2][]{%
    \errmessage{(Inkscape) Color is used for the text in Inkscape, but the package 'color.sty' is not loaded}%
    \renewcommand\color[2][]{}%
  }%
  \providecommand\transparent[1]{%
    \errmessage{(Inkscape) Transparency is used (non-zero) for the text in Inkscape, but the package 'transparent.sty' is not loaded}%
    \renewcommand\transparent[1]{}%
  }%
  \providecommand\rotatebox[2]{#2}%
  \newcommand*\fsize{\dimexpr\f@size pt\relax}%
  \newcommand*\lineheight[1]{\fontsize{\fsize}{#1\fsize}\selectfont}%
  \ifx\svgwidth\undefined%
    \setlength{\unitlength}{420bp}%
    \ifx\svgscale\undefined%
      \relax%
    \else%
      \setlength{\unitlength}{\unitlength * \real{\svgscale}}%
    \fi%
  \else%
    \setlength{\unitlength}{\svgwidth}%
  \fi%
  \global\let\svgwidth\undefined%
  \global\let\svgscale\undefined%
  \makeatother%
  \begin{picture}(1,0.75)%
    \lineheight{1}%
    \setlength\tabcolsep{0pt}%
    \put(0,0){\includegraphics[width=\unitlength,page=1]{Real_Gam_Dist_stat_el.pdf}}%
    \put(0.13035714,0.05404768){\makebox(0,0)[t]{\lineheight{1.25}\smash{\begin{tabular}[t]{c}0\end{tabular}}}}%
    \put(0.28535714,0.05404768){\makebox(0,0)[t]{\lineheight{1.25}\smash{\begin{tabular}[t]{c}0.5\end{tabular}}}}%
    \put(0.44035714,0.05404768){\makebox(0,0)[t]{\lineheight{1.25}\smash{\begin{tabular}[t]{c}1\end{tabular}}}}%
    \put(0.59535714,0.05404768){\makebox(0,0)[t]{\lineheight{1.25}\smash{\begin{tabular}[t]{c}1.5\end{tabular}}}}%
    \put(0.75035714,0.05404768){\makebox(0,0)[t]{\lineheight{1.25}\smash{\begin{tabular}[t]{c}2\end{tabular}}}}%
    \put(0.90535714,0.05404768){\makebox(0,0)[t]{\lineheight{1.25}\smash{\begin{tabular}[t]{c}2.5\end{tabular}}}}%
    \put(0.5178575,0.01988089){\makebox(0,0)[t]{\lineheight{1.25}\smash{\begin{tabular}[t]{c}\scalebox{1.5}{x-coordinate [m]}\end{tabular}}}}%
    \put(0,0){\includegraphics[width=\unitlength,page=2]{Real_Gam_Dist_stat_el.pdf}}%
    \put(0.12083339,0.07571429){\makebox(0,0)[rt]{\lineheight{1.25}\smash{\begin{tabular}[t]{r}-0.05\end{tabular}}}}%
    \put(0.12083339,0.13696429){\makebox(0,0)[rt]{\lineheight{1.25}\smash{\begin{tabular}[t]{r}-0.045\end{tabular}}}}%
    \put(0.12083339,0.19821429){\makebox(0,0)[rt]{\lineheight{1.25}\smash{\begin{tabular}[t]{r}-0.04\end{tabular}}}}%
    \put(0.12083339,0.25946429){\makebox(0,0)[rt]{\lineheight{1.25}\smash{\begin{tabular}[t]{r}-0.035\end{tabular}}}}%
    \put(0.12083339,0.32071429){\makebox(0,0)[rt]{\lineheight{1.25}\smash{\begin{tabular}[t]{r}-0.03\end{tabular}}}}%
    \put(0.12083339,0.38196429){\makebox(0,0)[rt]{\lineheight{1.25}\smash{\begin{tabular}[t]{r}-0.025\end{tabular}}}}%
    \put(0.12083339,0.44321429){\makebox(0,0)[rt]{\lineheight{1.25}\smash{\begin{tabular}[t]{r}-0.02\end{tabular}}}}%
    \put(0.12083339,0.50446429){\makebox(0,0)[rt]{\lineheight{1.25}\smash{\begin{tabular}[t]{r}-0.015\end{tabular}}}}%
    \put(0.12083339,0.56571429){\makebox(0,0)[rt]{\lineheight{1.25}\smash{\begin{tabular}[t]{r}-0.01\end{tabular}}}}%
    \put(0.12083339,0.62696429){\makebox(0,0)[rt]{\lineheight{1.25}\smash{\begin{tabular}[t]{r}-0.005\end{tabular}}}}%
    \put(0.12083339,0.68821429){\makebox(0,0)[rt]{\lineheight{1.25}\smash{\begin{tabular}[t]{r}0\end{tabular}}}}%
    \put(0.04726196,0.38839321){\rotatebox{90}{\makebox(0,0)[t]{\lineheight{1.25}\smash{\begin{tabular}[t]{c}\scalebox{1.5}{Deflection [m]} \end{tabular}}}}}%
    \put(0,0){\includegraphics[width=\unitlength,page=3]{Real_Gam_Dist_stat_el.pdf}}%
  \end{picture}%
\endgroup%

%% file: Real_Gam_Field_stat_el.pdf_tex
%% Accompanies image file 'Real_Gam_Field_stat_el.pdf' (pdf, eps, ps)
\begingroup%
  \makeatletter%
  \providecommand\color[2][]{%
    \errmessage{(Inkscape) Color is used for the text in Inkscape, but the package 'color.sty' is not loaded}%
    \renewcommand\color[2][]{}%
  }%
  \providecommand\transparent[1]{%
    \errmessage{(Inkscape) Transparency is used (non-zero) for the text in Inkscape, but the package 'transparent.sty' is not loaded}%
    \renewcommand\transparent[1]{}%
  }%
  \providecommand\rotatebox[2]{#2}%
  \newcommand*\fsize{\dimexpr\f@size pt\relax}%
  \newcommand*\lineheight[1]{\fontsize{\fsize}{#1\fsize}\selectfont}%
  \ifx\svgwidth\undefined%
    \setlength{\unitlength}{420bp}%
    \ifx\svgscale\undefined%
      \relax%
    \else%
      \setlength{\unitlength}{\unitlength * \real{\svgscale}}%
    \fi%
  \else%
    \setlength{\unitlength}{\svgwidth}%
  \fi%
  \global\let\svgwidth\undefined%
  \global\let\svgscale\undefined%
  \makeatother%
  \begin{picture}(1,0.75)%
    \lineheight{1}%
    \setlength\tabcolsep{0pt}%
    \put(0,0){\includegraphics[width=\unitlength,page=1]{Real_Gam_Field_stat_el.pdf}}%
    \put(0.13035714,0.05404768){\makebox(0,0)[t]{\lineheight{1.25}\smash{\begin{tabular}[t]{c}0\end{tabular}}}}%
    \put(0.28535714,0.05404768){\makebox(0,0)[t]{\lineheight{1.25}\smash{\begin{tabular}[t]{c}0.5\end{tabular}}}}%
    \put(0.44035714,0.05404768){\makebox(0,0)[t]{\lineheight{1.25}\smash{\begin{tabular}[t]{c}1\end{tabular}}}}%
    \put(0.59535714,0.05404768){\makebox(0,0)[t]{\lineheight{1.25}\smash{\begin{tabular}[t]{c}1.5\end{tabular}}}}%
    \put(0.75035714,0.05404768){\makebox(0,0)[t]{\lineheight{1.25}\smash{\begin{tabular}[t]{c}2\end{tabular}}}}%
    \put(0.90535714,0.05404768){\makebox(0,0)[t]{\lineheight{1.25}\smash{\begin{tabular}[t]{c}2.5\end{tabular}}}}%
    \put(0.5178575,0.01988089){\makebox(0,0)[t]{\lineheight{1.25}\smash{\begin{tabular}[t]{c}\scalebox{1.5}{x-coordinate [m]}\end{tabular}}}}%
    \put(0,0){\includegraphics[width=\unitlength,page=2]{Real_Gam_Field_stat_el.pdf}}%
    \put(0.12083339,0.07571429){\makebox(0,0)[rt]{\lineheight{1.25}\smash{\begin{tabular}[t]{r}-0.05\end{tabular}}}}%
    \put(0.12083339,0.13696429){\makebox(0,0)[rt]{\lineheight{1.25}\smash{\begin{tabular}[t]{r}-0.045\end{tabular}}}}%
    \put(0.12083339,0.19821429){\makebox(0,0)[rt]{\lineheight{1.25}\smash{\begin{tabular}[t]{r}-0.04\end{tabular}}}}%
    \put(0.12083339,0.25946429){\makebox(0,0)[rt]{\lineheight{1.25}\smash{\begin{tabular}[t]{r}-0.035\end{tabular}}}}%
    \put(0.12083339,0.32071429){\makebox(0,0)[rt]{\lineheight{1.25}\smash{\begin{tabular}[t]{r}-0.03\end{tabular}}}}%
    \put(0.12083339,0.38196429){\makebox(0,0)[rt]{\lineheight{1.25}\smash{\begin{tabular}[t]{r}-0.025\end{tabular}}}}%
    \put(0.12083339,0.44321429){\makebox(0,0)[rt]{\lineheight{1.25}\smash{\begin{tabular}[t]{r}-0.02\end{tabular}}}}%
    \put(0.12083339,0.50446429){\makebox(0,0)[rt]{\lineheight{1.25}\smash{\begin{tabular}[t]{r}-0.015\end{tabular}}}}%
    \put(0.12083339,0.56571429){\makebox(0,0)[rt]{\lineheight{1.25}\smash{\begin{tabular}[t]{r}-0.01\end{tabular}}}}%
    \put(0.12083339,0.62696429){\makebox(0,0)[rt]{\lineheight{1.25}\smash{\begin{tabular}[t]{r}-0.005\end{tabular}}}}%
    \put(0.12083339,0.68821429){\makebox(0,0)[rt]{\lineheight{1.25}\smash{\begin{tabular}[t]{r}0\end{tabular}}}}%
    \put(0.04726196,0.38839321){\rotatebox{90}{\makebox(0,0)[t]{\lineheight{1.25}\smash{\begin{tabular}[t]{c}\scalebox{1.5}{Deflection [m]} \end{tabular}}}}}%
    \put(0,0){\includegraphics[width=\unitlength,page=3]{Real_Gam_Field_stat_el.pdf}}%
  \end{picture}%
\endgroup%

%% file: Interm_Deflection_stat_gam_dist.pdf_tex
%% Accompanies image file 'Interm_Deflection_stat_gam_dist.pdf' (pdf, eps, ps)
\begingroup%
  \makeatletter%
  \providecommand\color[2][]{%
    \errmessage{(Inkscape) Color is used for the text in Inkscape, but the package 'color.sty' is not loaded}%
    \renewcommand\color[2][]{}%
  }%
  \providecommand\transparent[1]{%
    \errmessage{(Inkscape) Transparency is used (non-zero) for the text in Inkscape, but the package 'transparent.sty' is not loaded}%
    \renewcommand\transparent[1]{}%
  }%
  \providecommand\rotatebox[2]{#2}%
  \newcommand*\fsize{\dimexpr\f@size pt\relax}%
  \newcommand*\lineheight[1]{\fontsize{\fsize}{#1\fsize}\selectfont}%
  \ifx\svgwidth\undefined%
    \setlength{\unitlength}{420bp}%
    \ifx\svgscale\undefined%
      \relax%
    \else%
      \setlength{\unitlength}{\unitlength * \real{\svgscale}}%
    \fi%
  \else%
    \setlength{\unitlength}{\svgwidth}%
  \fi%
  \global\let\svgwidth\undefined%
  \global\let\svgscale\undefined%
  \makeatother%
  \begin{picture}(1,0.75)%
    \lineheight{1}%
    \setlength\tabcolsep{0pt}%
    \put(0,0){\includegraphics[width=\unitlength,page=1]{Interm_Deflection_stat_gam_dist.pdf}}%
    \put(0.13035714,0.05404768){\makebox(0,0)[t]{\lineheight{1.25}\smash{\begin{tabular}[t]{c}0\end{tabular}}}}%
    \put(0.28535714,0.05404768){\makebox(0,0)[t]{\lineheight{1.25}\smash{\begin{tabular}[t]{c}0.5\end{tabular}}}}%
    \put(0.44035714,0.05404768){\makebox(0,0)[t]{\lineheight{1.25}\smash{\begin{tabular}[t]{c}1\end{tabular}}}}%
    \put(0.59535714,0.05404768){\makebox(0,0)[t]{\lineheight{1.25}\smash{\begin{tabular}[t]{c}1.5\end{tabular}}}}%
    \put(0.75035714,0.05404768){\makebox(0,0)[t]{\lineheight{1.25}\smash{\begin{tabular}[t]{c}2\end{tabular}}}}%
    \put(0.90535714,0.05404768){\makebox(0,0)[t]{\lineheight{1.25}\smash{\begin{tabular}[t]{c}2.5\end{tabular}}}}%
    \put(0.5178575,0.01988089){\makebox(0,0)[t]{\lineheight{1.25}\smash{\begin{tabular}[t]{c}\scalebox{1.5}{x-coordinate [m]}\end{tabular}}}}%
    \put(0,0){\includegraphics[width=\unitlength,page=2]{Interm_Deflection_stat_gam_dist.pdf}}%
    \put(0.12083339,0.07571429){\makebox(0,0)[rt]{\lineheight{1.25}\smash{\begin{tabular}[t]{r}-0.05\end{tabular}}}}%
    \put(0.12083339,0.13696429){\makebox(0,0)[rt]{\lineheight{1.25}\smash{\begin{tabular}[t]{r}-0.045\end{tabular}}}}%
    \put(0.12083339,0.19821429){\makebox(0,0)[rt]{\lineheight{1.25}\smash{\begin{tabular}[t]{r}-0.04\end{tabular}}}}%
    \put(0.12083339,0.25946429){\makebox(0,0)[rt]{\lineheight{1.25}\smash{\begin{tabular}[t]{r}-0.035\end{tabular}}}}%
    \put(0.12083339,0.32071429){\makebox(0,0)[rt]{\lineheight{1.25}\smash{\begin{tabular}[t]{r}-0.03\end{tabular}}}}%
    \put(0.12083339,0.38196429){\makebox(0,0)[rt]{\lineheight{1.25}\smash{\begin{tabular}[t]{r}-0.025\end{tabular}}}}%
    \put(0.12083339,0.44321429){\makebox(0,0)[rt]{\lineheight{1.25}\smash{\begin{tabular}[t]{r}-0.02\end{tabular}}}}%
    \put(0.12083339,0.50446429){\makebox(0,0)[rt]{\lineheight{1.25}\smash{\begin{tabular}[t]{r}-0.015\end{tabular}}}}%
    \put(0.12083339,0.56571429){\makebox(0,0)[rt]{\lineheight{1.25}\smash{\begin{tabular}[t]{r}-0.01\end{tabular}}}}%
    \put(0.12083339,0.62696429){\makebox(0,0)[rt]{\lineheight{1.25}\smash{\begin{tabular}[t]{r}-0.005\end{tabular}}}}%
    \put(0.12083339,0.68821429){\makebox(0,0)[rt]{\lineheight{1.25}\smash{\begin{tabular}[t]{r}0\end{tabular}}}}%
    \put(0.04726196,0.38839321){\rotatebox{90}{\makebox(0,0)[t]{\lineheight{1.25}\smash{\begin{tabular}[t]{c}\scalebox{1.5}{Deflection [m]}\end{tabular}}}}}%
    \put(0,0){\includegraphics[width=\unitlength,page=3]{Interm_Deflection_stat_gam_dist.pdf}}%
    \put(0.52043661,0.09439286){\makebox(0,0)[lt]{\lineheight{1.25}\smash{\begin{tabular}[t]{l}\scalebox{1.5}{$\color{red}{\leftarrow} A$}\end{tabular}}}}%
    \put(0.52043661,0.66239286){\makebox(0,0)[lt]{\lineheight{1.25}\smash{\begin{tabular}[t]{l}\scalebox{1.5}{$\color{red}{\leftarrow} B$}\end{tabular}}}}%
  \end{picture}%
\endgroup%

%% file: Interm_Deflection_stat_gam_dist_cut_1.pdf_tex
%% Accompanies image file 'Interm_Deflection_stat_gam_dist_cut_1.pdf' (pdf, eps, ps)
\begingroup%
  \makeatletter%
  \providecommand\color[2][]{%
    \errmessage{(Inkscape) Color is used for the text in Inkscape, but the package 'color.sty' is not loaded}%
    \renewcommand\color[2][]{}%
  }%
  \providecommand\transparent[1]{%
    \errmessage{(Inkscape) Transparency is used (non-zero) for the text in Inkscape, but the package 'transparent.sty' is not loaded}%
    \renewcommand\transparent[1]{}%
  }%
  \providecommand\rotatebox[2]{#2}%
  \newcommand*\fsize{\dimexpr\f@size pt\relax}%
  \newcommand*\lineheight[1]{\fontsize{\fsize}{#1\fsize}\selectfont}%
  \ifx\svgwidth\undefined%
    \setlength{\unitlength}{420bp}%
    \ifx\svgscale\undefined%
      \relax%
    \else%
      \setlength{\unitlength}{\unitlength * \real{\svgscale}}%
    \fi%
  \else%
    \setlength{\unitlength}{\svgwidth}%
  \fi%
  \global\let\svgwidth\undefined%
  \global\let\svgscale\undefined%
  \makeatother%
  \begin{picture}(1,0.75)%
    \lineheight{1}%
    \setlength\tabcolsep{0pt}%
    \put(0,0){\includegraphics[width=\unitlength,page=1]{Interm_Deflection_stat_gam_dist_cut_1.pdf}}%
    \put(0.13035714,0.05404768){\makebox(0,0)[t]{\lineheight{1.25}\smash{\begin{tabular}[t]{c}-0.1\end{tabular}}}}%
    \put(0.20785714,0.05404768){\makebox(0,0)[t]{\lineheight{1.25}\smash{\begin{tabular}[t]{c}-0.09\end{tabular}}}}%
    \put(0.28535714,0.05404768){\makebox(0,0)[t]{\lineheight{1.25}\smash{\begin{tabular}[t]{c}-0.08\end{tabular}}}}%
    \put(0.36285714,0.05404768){\makebox(0,0)[t]{\lineheight{1.25}\smash{\begin{tabular}[t]{c}-0.07\end{tabular}}}}%
    \put(0.44035714,0.05404768){\makebox(0,0)[t]{\lineheight{1.25}\smash{\begin{tabular}[t]{c}-0.06\end{tabular}}}}%
    \put(0.51785714,0.05404768){\makebox(0,0)[t]{\lineheight{1.25}\smash{\begin{tabular}[t]{c}-0.05\end{tabular}}}}%
    \put(0.59535714,0.05404768){\makebox(0,0)[t]{\lineheight{1.25}\smash{\begin{tabular}[t]{c}-0.04\end{tabular}}}}%
    \put(0.67285714,0.05404768){\makebox(0,0)[t]{\lineheight{1.25}\smash{\begin{tabular}[t]{c}-0.03\end{tabular}}}}%
    \put(0.75035714,0.05404768){\makebox(0,0)[t]{\lineheight{1.25}\smash{\begin{tabular}[t]{c}-0.02\end{tabular}}}}%
    \put(0.82785714,0.05404768){\makebox(0,0)[t]{\lineheight{1.25}\smash{\begin{tabular}[t]{c}-0.01\end{tabular}}}}%
    \put(0.90535714,0.05404768){\makebox(0,0)[t]{\lineheight{1.25}\smash{\begin{tabular}[t]{c}0\end{tabular}}}}%
    \put(0.5178575,0.01988089){\makebox(0,0)[t]{\lineheight{1.25}\smash{\begin{tabular}[t]{c}\scalebox{1.5}{Deflection [m]}\end{tabular}}}}%
    \put(0,0){\includegraphics[width=\unitlength,page=2]{Interm_Deflection_stat_gam_dist_cut_1.pdf}}%
    \put(0.12083339,0.07571429){\makebox(0,0)[rt]{\lineheight{1.25}\smash{\begin{tabular}[t]{r}0\end{tabular}}}}%
    \put(0.12083339,0.13349732){\makebox(0,0)[rt]{\lineheight{1.25}\smash{\begin{tabular}[t]{r}5\end{tabular}}}}%
    \put(0.12083339,0.19128036){\makebox(0,0)[rt]{\lineheight{1.25}\smash{\begin{tabular}[t]{r}10\end{tabular}}}}%
    \put(0.12083339,0.24906339){\makebox(0,0)[rt]{\lineheight{1.25}\smash{\begin{tabular}[t]{r}15\end{tabular}}}}%
    \put(0.12083339,0.30684643){\makebox(0,0)[rt]{\lineheight{1.25}\smash{\begin{tabular}[t]{r}20\end{tabular}}}}%
    \put(0.12083339,0.36462929){\makebox(0,0)[rt]{\lineheight{1.25}\smash{\begin{tabular}[t]{r}25\end{tabular}}}}%
    \put(0.12083339,0.42241232){\makebox(0,0)[rt]{\lineheight{1.25}\smash{\begin{tabular}[t]{r}30\end{tabular}}}}%
    \put(0.12083339,0.48019536){\makebox(0,0)[rt]{\lineheight{1.25}\smash{\begin{tabular}[t]{r}35\end{tabular}}}}%
    \put(0.12083339,0.53797839){\makebox(0,0)[rt]{\lineheight{1.25}\smash{\begin{tabular}[t]{r}40\end{tabular}}}}%
    \put(0.12083339,0.59576143){\makebox(0,0)[rt]{\lineheight{1.25}\smash{\begin{tabular}[t]{r}45\end{tabular}}}}%
    \put(0.12083339,0.65354446){\makebox(0,0)[rt]{\lineheight{1.25}\smash{\begin{tabular}[t]{r}50\end{tabular}}}}%
    \put(0.04726196,0.38839321){\rotatebox{90}{\makebox(0,0)[t]{\lineheight{1.25}\smash{\begin{tabular}[t]{c}\scalebox{1.5}{PDF [1/m]}\end{tabular}}}}}%
    \put(0,0){\includegraphics[width=\unitlength,page=3]{Interm_Deflection_stat_gam_dist_cut_1.pdf}}%
    \put(0.13423214,0.64419482){\makebox(0,0)[lt]{\lineheight{1.25}\smash{\begin{tabular}[t]{l}\scalebox{1.5}{$\color{red}{\uparrow} A$}\end{tabular}}}}%
    \put(0.84335714,0.64419482){\makebox(0,0)[lt]{\lineheight{1.25}\smash{\begin{tabular}[t]{l}\scalebox{1.5}{$\color{red}{\uparrow} B$}\end{tabular}}}}%
  \end{picture}%
\endgroup%

%% file: Interm_Deflection_stat_gam_field.pdf_tex
%% Accompanies image file 'Interm_Deflection_stat_gam_field.pdf' (pdf, eps, ps)
\begingroup%
  \makeatletter%
  \providecommand\color[2][]{%
    \errmessage{(Inkscape) Color is used for the text in Inkscape, but the package 'color.sty' is not loaded}%
    \renewcommand\color[2][]{}%
  }%
  \providecommand\transparent[1]{%
    \errmessage{(Inkscape) Transparency is used (non-zero) for the text in Inkscape, but the package 'transparent.sty' is not loaded}%
    \renewcommand\transparent[1]{}%
  }%
  \providecommand\rotatebox[2]{#2}%
  \newcommand*\fsize{\dimexpr\f@size pt\relax}%
  \newcommand*\lineheight[1]{\fontsize{\fsize}{#1\fsize}\selectfont}%
  \ifx\svgwidth\undefined%
    \setlength{\unitlength}{420bp}%
    \ifx\svgscale\undefined%
      \relax%
    \else%
      \setlength{\unitlength}{\unitlength * \real{\svgscale}}%
    \fi%
  \else%
    \setlength{\unitlength}{\svgwidth}%
  \fi%
  \global\let\svgwidth\undefined%
  \global\let\svgscale\undefined%
  \makeatother%
  \begin{picture}(1,0.75)%
    \lineheight{1}%
    \setlength\tabcolsep{0pt}%
    \put(0,0){\includegraphics[width=\unitlength,page=1]{Interm_Deflection_stat_gam_field.pdf}}%
    \put(0.13035714,0.05404768){\makebox(0,0)[t]{\lineheight{1.25}\smash{\begin{tabular}[t]{c}0\end{tabular}}}}%
    \put(0.28535714,0.05404768){\makebox(0,0)[t]{\lineheight{1.25}\smash{\begin{tabular}[t]{c}0.5\end{tabular}}}}%
    \put(0.44035714,0.05404768){\makebox(0,0)[t]{\lineheight{1.25}\smash{\begin{tabular}[t]{c}1\end{tabular}}}}%
    \put(0.59535714,0.05404768){\makebox(0,0)[t]{\lineheight{1.25}\smash{\begin{tabular}[t]{c}1.5\end{tabular}}}}%
    \put(0.75035714,0.05404768){\makebox(0,0)[t]{\lineheight{1.25}\smash{\begin{tabular}[t]{c}2\end{tabular}}}}%
    \put(0.90535714,0.05404768){\makebox(0,0)[t]{\lineheight{1.25}\smash{\begin{tabular}[t]{c}2.5\end{tabular}}}}%
    \put(0.5178575,0.01988089){\makebox(0,0)[t]{\lineheight{1.25}\smash{\begin{tabular}[t]{c}\scalebox{1.5}{x-coordinate [m]}\end{tabular}}}}%
    \put(0,0){\includegraphics[width=\unitlength,page=2]{Interm_Deflection_stat_gam_field.pdf}}%
    \put(0.12083339,0.07571429){\makebox(0,0)[rt]{\lineheight{1.25}\smash{\begin{tabular}[t]{r}-0.05\end{tabular}}}}%
    \put(0.12083339,0.13696429){\makebox(0,0)[rt]{\lineheight{1.25}\smash{\begin{tabular}[t]{r}-0.045\end{tabular}}}}%
    \put(0.12083339,0.19821429){\makebox(0,0)[rt]{\lineheight{1.25}\smash{\begin{tabular}[t]{r}-0.04\end{tabular}}}}%
    \put(0.12083339,0.25946429){\makebox(0,0)[rt]{\lineheight{1.25}\smash{\begin{tabular}[t]{r}-0.035\end{tabular}}}}%
    \put(0.12083339,0.32071429){\makebox(0,0)[rt]{\lineheight{1.25}\smash{\begin{tabular}[t]{r}-0.03\end{tabular}}}}%
    \put(0.12083339,0.38196429){\makebox(0,0)[rt]{\lineheight{1.25}\smash{\begin{tabular}[t]{r}-0.025\end{tabular}}}}%
    \put(0.12083339,0.44321429){\makebox(0,0)[rt]{\lineheight{1.25}\smash{\begin{tabular}[t]{r}-0.02\end{tabular}}}}%
    \put(0.12083339,0.50446429){\makebox(0,0)[rt]{\lineheight{1.25}\smash{\begin{tabular}[t]{r}-0.015\end{tabular}}}}%
    \put(0.12083339,0.56571429){\makebox(0,0)[rt]{\lineheight{1.25}\smash{\begin{tabular}[t]{r}-0.01\end{tabular}}}}%
    \put(0.12083339,0.62696429){\makebox(0,0)[rt]{\lineheight{1.25}\smash{\begin{tabular}[t]{r}-0.005\end{tabular}}}}%
    \put(0.12083339,0.68821429){\makebox(0,0)[rt]{\lineheight{1.25}\smash{\begin{tabular}[t]{r}0\end{tabular}}}}%
    \put(0.04726196,0.38839321){\rotatebox{90}{\makebox(0,0)[t]{\lineheight{1.25}\smash{\begin{tabular}[t]{c}\scalebox{1.5}{Deflection [m]}\end{tabular}}}}}%
    \put(0,0){\includegraphics[width=\unitlength,page=3]{Interm_Deflection_stat_gam_field.pdf}}%
    \put(0.52043661,0.09439286){\makebox(0,0)[lt]{\lineheight{1.25}\smash{\begin{tabular}[t]{l}\scalebox{1.5}{$\color{red}{\leftarrow} A$}\end{tabular}}}}%
    \put(0.52043661,0.66239286){\makebox(0,0)[lt]{\lineheight{1.25}\smash{\begin{tabular}[t]{l}\scalebox{1.5}{$\color{red}{\leftarrow} B$}\end{tabular}}}}%
  \end{picture}%
\endgroup%

%% file: Interm_Deflection_stat_gam_field_cut_1.pdf_tex
%% Accompanies image file 'Interm_Deflection_stat_gam_field_cut_1.pdf' (pdf, eps, ps)
\begingroup%
  \makeatletter%
  \providecommand\color[2][]{%
    \errmessage{(Inkscape) Color is used for the text in Inkscape, but the package 'color.sty' is not loaded}%
    \renewcommand\color[2][]{}%
  }%
  \providecommand\transparent[1]{%
    \errmessage{(Inkscape) Transparency is used (non-zero) for the text in Inkscape, but the package 'transparent.sty' is not loaded}%
    \renewcommand\transparent[1]{}%
  }%
  \providecommand\rotatebox[2]{#2}%
  \newcommand*\fsize{\dimexpr\f@size pt\relax}%
  \newcommand*\lineheight[1]{\fontsize{\fsize}{#1\fsize}\selectfont}%
  \ifx\svgwidth\undefined%
    \setlength{\unitlength}{420bp}%
    \ifx\svgscale\undefined%
      \relax%
    \else%
      \setlength{\unitlength}{\unitlength * \real{\svgscale}}%
    \fi%
  \else%
    \setlength{\unitlength}{\svgwidth}%
  \fi%
  \global\let\svgwidth\undefined%
  \global\let\svgscale\undefined%
  \makeatother%
  \begin{picture}(1,0.75)%
    \lineheight{1}%
    \setlength\tabcolsep{0pt}%
    \put(0,0){\includegraphics[width=\unitlength,page=1]{Interm_Deflection_stat_gam_field_cut_1.pdf}}%
    \put(0.13035714,0.05404768){\makebox(0,0)[t]{\lineheight{1.25}\smash{\begin{tabular}[t]{c}-0.1\end{tabular}}}}%
    \put(0.20785714,0.05404768){\makebox(0,0)[t]{\lineheight{1.25}\smash{\begin{tabular}[t]{c}-0.09\end{tabular}}}}%
    \put(0.28535714,0.05404768){\makebox(0,0)[t]{\lineheight{1.25}\smash{\begin{tabular}[t]{c}-0.08\end{tabular}}}}%
    \put(0.36285714,0.05404768){\makebox(0,0)[t]{\lineheight{1.25}\smash{\begin{tabular}[t]{c}-0.07\end{tabular}}}}%
    \put(0.44035714,0.05404768){\makebox(0,0)[t]{\lineheight{1.25}\smash{\begin{tabular}[t]{c}-0.06\end{tabular}}}}%
    \put(0.51785714,0.05404768){\makebox(0,0)[t]{\lineheight{1.25}\smash{\begin{tabular}[t]{c}-0.05\end{tabular}}}}%
    \put(0.59535714,0.05404768){\makebox(0,0)[t]{\lineheight{1.25}\smash{\begin{tabular}[t]{c}-0.04\end{tabular}}}}%
    \put(0.67285714,0.05404768){\makebox(0,0)[t]{\lineheight{1.25}\smash{\begin{tabular}[t]{c}-0.03\end{tabular}}}}%
    \put(0.75035714,0.05404768){\makebox(0,0)[t]{\lineheight{1.25}\smash{\begin{tabular}[t]{c}-0.02\end{tabular}}}}%
    \put(0.82785714,0.05404768){\makebox(0,0)[t]{\lineheight{1.25}\smash{\begin{tabular}[t]{c}-0.01\end{tabular}}}}%
    \put(0.90535714,0.05404768){\makebox(0,0)[t]{\lineheight{1.25}\smash{\begin{tabular}[t]{c}0\end{tabular}}}}%
    \put(0.5178575,0.01988089){\makebox(0,0)[t]{\lineheight{1.25}\smash{\begin{tabular}[t]{c}\scalebox{1.5}{Deflection [m]}\end{tabular}}}}%
    \put(0,0){\includegraphics[width=\unitlength,page=2]{Interm_Deflection_stat_gam_field_cut_1.pdf}}%
    \put(0.12083339,0.07571429){\makebox(0,0)[rt]{\lineheight{1.25}\smash{\begin{tabular}[t]{r}0\end{tabular}}}}%
    \put(0.12083339,0.13764482){\makebox(0,0)[rt]{\lineheight{1.25}\smash{\begin{tabular}[t]{r}10\end{tabular}}}}%
    \put(0.12083339,0.19957536){\makebox(0,0)[rt]{\lineheight{1.25}\smash{\begin{tabular}[t]{r}20\end{tabular}}}}%
    \put(0.12083339,0.26150589){\makebox(0,0)[rt]{\lineheight{1.25}\smash{\begin{tabular}[t]{r}30\end{tabular}}}}%
    \put(0.12083339,0.32343643){\makebox(0,0)[rt]{\lineheight{1.25}\smash{\begin{tabular}[t]{r}40\end{tabular}}}}%
    \put(0.12083339,0.38536696){\makebox(0,0)[rt]{\lineheight{1.25}\smash{\begin{tabular}[t]{r}50\end{tabular}}}}%
    \put(0.12083339,0.4472975){\makebox(0,0)[rt]{\lineheight{1.25}\smash{\begin{tabular}[t]{r}60\end{tabular}}}}%
    \put(0.12083339,0.50922804){\makebox(0,0)[rt]{\lineheight{1.25}\smash{\begin{tabular}[t]{r}70\end{tabular}}}}%
    \put(0.12083339,0.57115875){\makebox(0,0)[rt]{\lineheight{1.25}\smash{\begin{tabular}[t]{r}80\end{tabular}}}}%
    \put(0.12083339,0.63308929){\makebox(0,0)[rt]{\lineheight{1.25}\smash{\begin{tabular}[t]{r}90\end{tabular}}}}%
    \put(0.04726196,0.38839321){\rotatebox{90}{\makebox(0,0)[t]{\lineheight{1.25}\smash{\begin{tabular}[t]{c}\scalebox{1.5}{PDF [1/m]}\end{tabular}}}}}%
    \put(0,0){\includegraphics[width=\unitlength,page=3]{Interm_Deflection_stat_gam_field_cut_1.pdf}}%
    \put(0.13423214,0.64748446){\makebox(0,0)[lt]{\lineheight{1.25}\smash{\begin{tabular}[t]{l}\scalebox{1.5}{$\color{red}{\uparrow} A$}\end{tabular}}}}%
    \put(0.84335714,0.64748446){\makebox(0,0)[lt]{\lineheight{1.25}\smash{\begin{tabular}[t]{l}\scalebox{1.5}{$\color{red}{\uparrow} B$}\end{tabular}}}}%
  \end{picture}%
\endgroup%

%% file: Plast_dist_sol.pdf_tex
%% Accompanies image file 'Plast_dist_sol.pdf' (pdf, eps, ps)
\begingroup%
  \makeatletter%
  \providecommand\color[2][]{%
    \errmessage{(Inkscape) Color is used for the text in Inkscape, but the package 'color.sty' is not loaded}%
    \renewcommand\color[2][]{}%
  }%
  \providecommand\transparent[1]{%
    \errmessage{(Inkscape) Transparency is used (non-zero) for the text in Inkscape, but the package 'transparent.sty' is not loaded}%
    \renewcommand\transparent[1]{}%
  }%
  \providecommand\rotatebox[2]{#2}%
  \newcommand*\fsize{\dimexpr\f@size pt\relax}%
  \newcommand*\lineheight[1]{\fontsize{\fsize}{#1\fsize}\selectfont}%
  \ifx\svgwidth\undefined%
    \setlength{\unitlength}{420bp}%
    \ifx\svgscale\undefined%
      \relax%
    \else%
      \setlength{\unitlength}{\unitlength * \real{\svgscale}}%
    \fi%
  \else%
    \setlength{\unitlength}{\svgwidth}%
  \fi%
  \global\let\svgwidth\undefined%
  \global\let\svgscale\undefined%
  \makeatother%
  \begin{picture}(1,0.75)%
    \lineheight{1}%
    \setlength\tabcolsep{0pt}%
    \put(0,0){\includegraphics[width=\unitlength,page=1]{Plast_dist_sol.pdf}}%
    \put(0.12083339,0.07571429){\makebox(0,0)[rt]{\lineheight{1.25}\smash{\begin{tabular}[t]{c}0\end{tabular}}}}%
    \put(0.12083339,0.166455){\makebox(0,0)[rt]{\lineheight{1.25}\smash{\begin{tabular}[t]{c}2\end{tabular}}}}%
    \put(0.12083339,0.25719571){\makebox(0,0)[rt]{\lineheight{1.25}\smash{\begin{tabular}[t]{c}4\end{tabular}}}}%
    \put(0.12083339,0.34793643){\makebox(0,0)[rt]{\lineheight{1.25}\smash{\begin{tabular}[t]{c}6\end{tabular}}}}%
    \put(0.12083339,0.43867732){\makebox(0,0)[rt]{\lineheight{1.25}\smash{\begin{tabular}[t]{c}8\end{tabular}}}}%
    \put(0.12083339,0.52941804){\makebox(0,0)[rt]{\lineheight{1.25}\smash{\begin{tabular}[t]{c}10\end{tabular}}}}%
    \put(0.12083339,0.62015875){\makebox(0,0)[rt]{\lineheight{1.25}\smash{\begin{tabular}[t]{c}12\end{tabular}}}}%
    \put(0.12083339,0.68821429){\makebox(0,0)[rt]{\lineheight{1.25}\smash{\begin{tabular}[t]{c}13.5\end{tabular}}}}%
    \put(0.04726196,0.38839321){\rotatebox{90}{\makebox(0,0)[t]{\lineheight{1.25}\smash{\begin{tabular}[t]{c}\scalebox{1.5}{Force [kN]}\end{tabular}}}}}%
    \put(0,0){\includegraphics[width=\unitlength,page=2]{Plast_dist_sol.pdf}}%
    \put(0.13035714,0.05404768){\makebox(0,0)[rt]{\lineheight{1.25}\smash{\begin{tabular}[t]{r}0\end{tabular}}}}%
    \put(0.21646821,0.05404768){\makebox(0,0)[rt]{\lineheight{1.25}\smash{\begin{tabular}[t]{r}1\end{tabular}}}}%
    \put(0.30257929,0.05404768){\makebox(0,0)[rt]{\lineheight{1.25}\smash{\begin{tabular}[t]{r}2\end{tabular}}}}%
    \put(0.38869054,0.05404768){\makebox(0,0)[rt]{\lineheight{1.25}\smash{\begin{tabular}[t]{r}3\end{tabular}}}}%
    \put(0.47480161,0.05404768){\makebox(0,0)[rt]{\lineheight{1.25}\smash{\begin{tabular}[t]{r}4\end{tabular}}}}%
    \put(0.56091268,0.05404768){\makebox(0,0)[rt]{\lineheight{1.25}\smash{\begin{tabular}[t]{r}5\end{tabular}}}}%
    \put(0.64702375,0.05404768){\makebox(0,0)[rt]{\lineheight{1.25}\smash{\begin{tabular}[t]{r}6\end{tabular}}}}%
    \put(0.73313482,0.05404768){\makebox(0,0)[rt]{\lineheight{1.25}\smash{\begin{tabular}[t]{r}7\end{tabular}}}}%
    \put(0.81924607,0.05404768){\makebox(0,0)[rt]{\lineheight{1.25}\smash{\begin{tabular}[t]{r}8\end{tabular}}}}%
    \put(0.90535714,0.05404768){\makebox(0,0)[rt]{\lineheight{1.25}\smash{\begin{tabular}[t]{r}9\end{tabular}}}}%
    \put(0.5178575,0.01988089){\makebox(0,0)[t]{\lineheight{1.25}\smash{\begin{tabular}[t]{c}\scalebox{1.5}{Absolute value of the Deflection [mm]}\end{tabular}}}}%
    \put(0,0){\includegraphics[width=\unitlength,page=3]{Plast_dist_sol.pdf}}%
    \put(0.39047625,0.16321429){\makebox(0,0)[rt]{\lineheight{1.25}\smash{\begin{tabular}[t]{c}9\end{tabular}}}}%
    \put(0.39047625,0.25696429){\makebox(0,0)[rt]{\lineheight{1.25}\smash{\begin{tabular}[t]{c}9.5\end{tabular}}}}%
    \put(0.39047625,0.35071429){\makebox(0,0)[rt]{\lineheight{1.25}\smash{\begin{tabular}[t]{c}10\end{tabular}}}}%
  %  \put(0.31690464,0.26339286){\rotatebox{90}{\makebox(0,0)[t]{\lineheight{1.25}\smash{\begin{tabular}[t]{c}Force [kN]\end{tabular}}}}}%
    \put(0,0){\includegraphics[width=\unitlength,page=4]{Plast_dist_sol.pdf}}%
    \put(0.4,0.14154768){\makebox(0,0)[rt]{\lineheight{1.25}\smash{\begin{tabular}[t]{r}3\end{tabular}}}}%
    \put(0.51428571,0.14154768){\makebox(0,0)[rt]{\lineheight{1.25}\smash{\begin{tabular}[t]{r}3.2\end{tabular}}}}%
    \put(0.62857143,0.14154768){\makebox(0,0)[rt]{\lineheight{1.25}\smash{\begin{tabular}[t]{r}3.4\end{tabular}}}}%
    \put(0.74285714,0.14154768){\makebox(0,0)[rt]{\lineheight{1.25}\smash{\begin{tabular}[t]{r}3.6\end{tabular}}}}%
 %   \put(0.60000018,0.10738089){\makebox(0,0)[t]{\lineheight{1.25}\smash{\begin{tabular}[t]{c}Absolute value of the Deflection [mm]\end{tabular}}}}%
    \put(0,0){\includegraphics[width=\unitlength,page=5]{Plast_dist_sol.pdf}}%
  \end{picture}%
\endgroup%

%% file: Plast_field_sol.pdf_tex
%% Accompanies image file 'Plast_field_sol.pdf' (pdf, eps, ps)
\begingroup%
  \makeatletter%
  \providecommand\color[2][]{%
    \errmessage{(Inkscape) Color is used for the text in Inkscape, but the package 'color.sty' is not loaded}%
    \renewcommand\color[2][]{}%
  }%
  \providecommand\transparent[1]{%
    \errmessage{(Inkscape) Transparency is used (non-zero) for the text in Inkscape, but the package 'transparent.sty' is not loaded}%
    \renewcommand\transparent[1]{}%
  }%
  \providecommand\rotatebox[2]{#2}%
  \newcommand*\fsize{\dimexpr\f@size pt\relax}%
  \newcommand*\lineheight[1]{\fontsize{\fsize}{#1\fsize}\selectfont}%
  \ifx\svgwidth\undefined%
    \setlength{\unitlength}{420bp}%
    \ifx\svgscale\undefined%
      \relax%
    \else%
      \setlength{\unitlength}{\unitlength * \real{\svgscale}}%
    \fi%
  \else%
    \setlength{\unitlength}{\svgwidth}%
  \fi%
  \global\let\svgwidth\undefined%
  \global\let\svgscale\undefined%
  \makeatother%
  \begin{picture}(1,0.75)%
    \lineheight{1}%
    \setlength\tabcolsep{0pt}%
    \put(0,0){\includegraphics[width=\unitlength,page=1]{Plast_field_sol.pdf}}%
    \put(0.12083339,0.07571429){\makebox(0,0)[rt]{\lineheight{1.25}\smash{\begin{tabular}[t]{c}0\end{tabular}}}}%
    \put(0.12083339,0.166455){\makebox(0,0)[rt]{\lineheight{1.25}\smash{\begin{tabular}[t]{c}2\end{tabular}}}}%
    \put(0.12083339,0.25719571){\makebox(0,0)[rt]{\lineheight{1.25}\smash{\begin{tabular}[t]{c}4\end{tabular}}}}%
    \put(0.12083339,0.34793643){\makebox(0,0)[rt]{\lineheight{1.25}\smash{\begin{tabular}[t]{c}6\end{tabular}}}}%
    \put(0.12083339,0.43867732){\makebox(0,0)[rt]{\lineheight{1.25}\smash{\begin{tabular}[t]{c}8\end{tabular}}}}%
    \put(0.12083339,0.52941804){\makebox(0,0)[rt]{\lineheight{1.25}\smash{\begin{tabular}[t]{c}10\end{tabular}}}}%
    \put(0.12083339,0.62015875){\makebox(0,0)[rt]{\lineheight{1.25}\smash{\begin{tabular}[t]{c}12\end{tabular}}}}%
    \put(0.12083339,0.68821429){\makebox(0,0)[rt]{\lineheight{1.25}\smash{\begin{tabular}[t]{c}13.5\end{tabular}}}}%
    \put(0.04726196,0.38839321){\rotatebox{90}{\makebox(0,0)[t]{\lineheight{1.25}\smash{\begin{tabular}[t]{c}\scalebox{1.5}{Force [kN]}\end{tabular}}}}}%
    \put(0,0){\includegraphics[width=\unitlength,page=2]{Plast_field_sol.pdf}}%
    \put(0.13035714,0.05404768){\makebox(0,0)[rt]{\lineheight{1.25}\smash{\begin{tabular}[t]{r}0\end{tabular}}}}%
    \put(0.21646821,0.05404768){\makebox(0,0)[rt]{\lineheight{1.25}\smash{\begin{tabular}[t]{r}1\end{tabular}}}}%
    \put(0.30257929,0.05404768){\makebox(0,0)[rt]{\lineheight{1.25}\smash{\begin{tabular}[t]{r}2\end{tabular}}}}%
    \put(0.38869054,0.05404768){\makebox(0,0)[rt]{\lineheight{1.25}\smash{\begin{tabular}[t]{r}3\end{tabular}}}}%
    \put(0.47480161,0.05404768){\makebox(0,0)[rt]{\lineheight{1.25}\smash{\begin{tabular}[t]{r}4\end{tabular}}}}%
    \put(0.56091268,0.05404768){\makebox(0,0)[rt]{\lineheight{1.25}\smash{\begin{tabular}[t]{r}5\end{tabular}}}}%
    \put(0.64702375,0.05404768){\makebox(0,0)[rt]{\lineheight{1.25}\smash{\begin{tabular}[t]{r}6\end{tabular}}}}%
    \put(0.73313482,0.05404768){\makebox(0,0)[rt]{\lineheight{1.25}\smash{\begin{tabular}[t]{r}7\end{tabular}}}}%
    \put(0.81924607,0.05404768){\makebox(0,0)[rt]{\lineheight{1.25}\smash{\begin{tabular}[t]{r}8\end{tabular}}}}%
    \put(0.90535714,0.05404768){\makebox(0,0)[rt]{\lineheight{1.25}\smash{\begin{tabular}[t]{r}9\end{tabular}}}}%
    \put(0.5178575,0.01988089){\makebox(0,0)[t]{\lineheight{1.25}\smash{\begin{tabular}[t]{c}\scalebox{1.5}{Absolute value of the Deflection [mm]}\end{tabular}}}}%
    \put(0,0){\includegraphics[width=\unitlength,page=3]{Plast_field_sol.pdf}}%
    \put(0.39047625,0.16321429){\makebox(0,0)[rt]{\lineheight{1.25}\smash{\begin{tabular}[t]{c}9\end{tabular}}}}%
    \put(0.39047625,0.25696429){\makebox(0,0)[rt]{\lineheight{1.25}\smash{\begin{tabular}[t]{c}9.5\end{tabular}}}}%
    \put(0.39047625,0.35071429){\makebox(0,0)[rt]{\lineheight{1.25}\smash{\begin{tabular}[t]{c}10\end{tabular}}}}%
  %  \put(0.31690464,0.26339286){\rotatebox{90}{\makebox(0,0)[t]{\lineheight{1.25}\smash{\begin{tabular}[t]{c}Force [kN]\end{tabular}}}}}%
    \put(0,0){\includegraphics[width=\unitlength,page=4]{Plast_field_sol.pdf}}%
    \put(0.4,0.14154768){\makebox(0,0)[rt]{\lineheight{1.25}\smash{\begin{tabular}[t]{r}3\end{tabular}}}}%
    \put(0.51428571,0.14154768){\makebox(0,0)[rt]{\lineheight{1.25}\smash{\begin{tabular}[t]{r}3.2\end{tabular}}}}%
    \put(0.62857143,0.14154768){\makebox(0,0)[rt]{\lineheight{1.25}\smash{\begin{tabular}[t]{r}3.4\end{tabular}}}}%
    \put(0.74285714,0.14154768){\makebox(0,0)[rt]{\lineheight{1.25}\smash{\begin{tabular}[t]{r}3.6\end{tabular}}}}%
 %   \put(0.60000018,0.10738089){\makebox(0,0)[t]{\lineheight{1.25}\smash{\begin{tabular}[t]{c}Absolute value of the Deflection [mm]\end{tabular}}}}%
    \put(0,0){\includegraphics[width=\unitlength,page=5]{Plast_field_sol.pdf}}%
  \end{picture}%
\endgroup%

%% file: Indiv_Samples_Dist.pdf_tex
%% Accompanies image file 'Indiv_Samples_Dist.pdf' (pdf, eps, ps)
\begingroup%
  \makeatletter%
  \providecommand\color[2][]{%
    \errmessage{(Inkscape) Color is used for the text in Inkscape, but the package 'color.sty' is not loaded}%
    \renewcommand\color[2][]{}%
  }%
  \providecommand\transparent[1]{%
    \errmessage{(Inkscape) Transparency is used (non-zero) for the text in Inkscape, but the package 'transparent.sty' is not loaded}%
    \renewcommand\transparent[1]{}%
  }%
  \providecommand\rotatebox[2]{#2}%
  \newcommand*\fsize{\dimexpr\f@size pt\relax}%
  \newcommand*\lineheight[1]{\fontsize{\fsize}{#1\fsize}\selectfont}%
  \ifx\svgwidth\undefined%
    \setlength{\unitlength}{420bp}%
    \ifx\svgscale\undefined%
      \relax%
    \else%
      \setlength{\unitlength}{\unitlength * \real{\svgscale}}%
    \fi%
  \else%
    \setlength{\unitlength}{\svgwidth}%
  \fi%
  \global\let\svgwidth\undefined%
  \global\let\svgscale\undefined%
  \makeatother%
  \begin{picture}(1,0.75)%
    \lineheight{1}%
    \setlength\tabcolsep{0pt}%
    \put(0,0){\includegraphics[width=\unitlength,page=1]{Indiv_Samples_Dist.pdf}}%
    \put(0.13035714,0.05404768){\makebox(0,0)[t]{\lineheight{1.25}\smash{\begin{tabular}[t]{c}0\end{tabular}}}}%
    \put(0.21646821,0.05404768){\makebox(0,0)[t]{\lineheight{1.25}\smash{\begin{tabular}[t]{c}1\end{tabular}}}}%
    \put(0.30257929,0.05404768){\makebox(0,0)[t]{\lineheight{1.25}\smash{\begin{tabular}[t]{c}2\end{tabular}}}}%
    \put(0.38869054,0.05404768){\makebox(0,0)[t]{\lineheight{1.25}\smash{\begin{tabular}[t]{c}3\end{tabular}}}}%
    \put(0.47480161,0.05404768){\makebox(0,0)[t]{\lineheight{1.25}\smash{\begin{tabular}[t]{c}4\end{tabular}}}}%
    \put(0.56091268,0.05404768){\makebox(0,0)[t]{\lineheight{1.25}\smash{\begin{tabular}[t]{c}5\end{tabular}}}}%
    \put(0.64702375,0.05404768){\makebox(0,0)[t]{\lineheight{1.25}\smash{\begin{tabular}[t]{c}6\end{tabular}}}}%
    \put(0.73313482,0.05404768){\makebox(0,0)[t]{\lineheight{1.25}\smash{\begin{tabular}[t]{c}7\end{tabular}}}}%
    \put(0.81924607,0.05404768){\makebox(0,0)[t]{\lineheight{1.25}\smash{\begin{tabular}[t]{c}8\end{tabular}}}}%
    \put(0.90535714,0.05404768){\makebox(0,0)[t]{\lineheight{1.25}\smash{\begin{tabular}[t]{c}9\end{tabular}}}}%
    \put(0.5178575,0.01988089){\makebox(0,0)[t]{\lineheight{1.25}\smash{\begin{tabular}[t]{c}\scalebox{1.5}{Absolute value of the Deflection [mm]}\end{tabular}}}}%
    \put(0,0){\includegraphics[width=\unitlength,page=2]{Indiv_Samples_Dist.pdf}}%
    \put(0.12083339,0.07571429){\makebox(0,0)[rt]{\lineheight{1.25}\smash{\begin{tabular}[t]{r}0\end{tabular}}}}%
    \put(0.12083339,0.166455){\makebox(0,0)[rt]{\lineheight{1.25}\smash{\begin{tabular}[t]{r}2\end{tabular}}}}%
    \put(0.12083339,0.25719571){\makebox(0,0)[rt]{\lineheight{1.25}\smash{\begin{tabular}[t]{r}4\end{tabular}}}}%
    \put(0.12083339,0.34793643){\makebox(0,0)[rt]{\lineheight{1.25}\smash{\begin{tabular}[t]{r}6\end{tabular}}}}%
    \put(0.12083339,0.43867732){\makebox(0,0)[rt]{\lineheight{1.25}\smash{\begin{tabular}[t]{r}8\end{tabular}}}}%
    \put(0.12083339,0.52941804){\makebox(0,0)[rt]{\lineheight{1.25}\smash{\begin{tabular}[t]{r}10\end{tabular}}}}%
    \put(0.12083339,0.62015875){\makebox(0,0)[rt]{\lineheight{1.25}\smash{\begin{tabular}[t]{r}12\end{tabular}}}}%
    \put(0.12083339,0.68821429){\makebox(0,0)[rt]{\lineheight{1.25}\smash{\begin{tabular}[t]{r}13.5\end{tabular}}}}%
    \put(0.04726196,0.38839321){\rotatebox{90}{\makebox(0,0)[t]{\lineheight{1.25}\smash{\begin{tabular}[t]{c}\scalebox{1.5}{Force [kN]}\end{tabular}}}}}%
    \put(0,0){\includegraphics[width=\unitlength,page=3]{Indiv_Samples_Dist.pdf}}%
    \put(0.4,0.14154768){\makebox(0,0)[t]{\lineheight{1.25}\smash{\begin{tabular}[t]{c}3\end{tabular}}}}%
    \put(0.51428571,0.14154768){\makebox(0,0)[t]{\lineheight{1.25}\smash{\begin{tabular}[t]{c}3.2\end{tabular}}}}%
    \put(0.62857143,0.14154768){\makebox(0,0)[t]{\lineheight{1.25}\smash{\begin{tabular}[t]{c}3.4\end{tabular}}}}%
    \put(0.74285714,0.14154768){\makebox(0,0)[t]{\lineheight{1.25}\smash{\begin{tabular}[t]{c}3.6\end{tabular}}}}%
 %   \put(0.60000018,0.10738089){\makebox(0,0)[t]{\lineheight{1.25}\smash{\begin{tabular}[t]{c}Absolute value of the Deflection [mm]\end{tabular}}}}%
    \put(0,0){\includegraphics[width=\unitlength,page=4]{Indiv_Samples_Dist.pdf}}%
    \put(0.39047625,0.16321429){\makebox(0,0)[rt]{\lineheight{1.25}\smash{\begin{tabular}[t]{r}9\end{tabular}}}}%
    \put(0.39047625,0.25696429){\makebox(0,0)[rt]{\lineheight{1.25}\smash{\begin{tabular}[t]{r}9.5\end{tabular}}}}%
    \put(0.39047625,0.35071429){\makebox(0,0)[rt]{\lineheight{1.25}\smash{\begin{tabular}[t]{r}10\end{tabular}}}}%
 %   \put(0.31690464,0.26339286){\rotatebox{90}{\makebox(0,0)[t]{\lineheight{1.25}\smash{\begin{tabular}[t]{c}Force [kN]\end{tabular}}}}}%
    \put(0,0){\includegraphics[width=\unitlength,page=5]{Indiv_Samples_Dist.pdf}}%
  \end{picture}%
\endgroup%

%% file: Indiv_Samples_Field.pdf_tex
%% Accompanies image file 'Indiv_Samples_Field.pdf' (pdf, eps, ps)
\begingroup%
  \makeatletter%
  \providecommand\color[2][]{%
    \errmessage{(Inkscape) Color is used for the text in Inkscape, but the package 'color.sty' is not loaded}%
    \renewcommand\color[2][]{}%
  }%
  \providecommand\transparent[1]{%
    \errmessage{(Inkscape) Transparency is used (non-zero) for the text in Inkscape, but the package 'transparent.sty' is not loaded}%
    \renewcommand\transparent[1]{}%
  }%
  \providecommand\rotatebox[2]{#2}%
  \newcommand*\fsize{\dimexpr\f@size pt\relax}%
  \newcommand*\lineheight[1]{\fontsize{\fsize}{#1\fsize}\selectfont}%
  \ifx\svgwidth\undefined%
    \setlength{\unitlength}{420bp}%
    \ifx\svgscale\undefined%
      \relax%
    \else%
      \setlength{\unitlength}{\unitlength * \real{\svgscale}}%
    \fi%
  \else%
    \setlength{\unitlength}{\svgwidth}%
  \fi%
  \global\let\svgwidth\undefined%
  \global\let\svgscale\undefined%
  \makeatother%
  \begin{picture}(1,0.75)%
    \lineheight{1}%
    \setlength\tabcolsep{0pt}%
    \put(0,0){\includegraphics[width=\unitlength,page=1]{Indiv_Samples_Field.pdf}}%
    \put(0.13035714,0.05404768){\makebox(0,0)[t]{\lineheight{1.25}\smash{\begin{tabular}[t]{c}0\end{tabular}}}}%
    \put(0.21646821,0.05404768){\makebox(0,0)[t]{\lineheight{1.25}\smash{\begin{tabular}[t]{c}1\end{tabular}}}}%
    \put(0.30257929,0.05404768){\makebox(0,0)[t]{\lineheight{1.25}\smash{\begin{tabular}[t]{c}2\end{tabular}}}}%
    \put(0.38869054,0.05404768){\makebox(0,0)[t]{\lineheight{1.25}\smash{\begin{tabular}[t]{c}3\end{tabular}}}}%
    \put(0.47480161,0.05404768){\makebox(0,0)[t]{\lineheight{1.25}\smash{\begin{tabular}[t]{c}4\end{tabular}}}}%
    \put(0.56091268,0.05404768){\makebox(0,0)[t]{\lineheight{1.25}\smash{\begin{tabular}[t]{c}5\end{tabular}}}}%
    \put(0.64702375,0.05404768){\makebox(0,0)[t]{\lineheight{1.25}\smash{\begin{tabular}[t]{c}6\end{tabular}}}}%
    \put(0.73313482,0.05404768){\makebox(0,0)[t]{\lineheight{1.25}\smash{\begin{tabular}[t]{c}7\end{tabular}}}}%
    \put(0.81924607,0.05404768){\makebox(0,0)[t]{\lineheight{1.25}\smash{\begin{tabular}[t]{c}8\end{tabular}}}}%
    \put(0.90535714,0.05404768){\makebox(0,0)[t]{\lineheight{1.25}\smash{\begin{tabular}[t]{c}9\end{tabular}}}}%
    \put(0.5178575,0.01988089){\makebox(0,0)[t]{\lineheight{1.25}\smash{\begin{tabular}[t]{c}\scalebox{1.5}{Absolute value of the Deflection [mm]}\end{tabular}}}}%
    \put(0,0){\includegraphics[width=\unitlength,page=2]{Indiv_Samples_Field.pdf}}%
    \put(0.12083339,0.07571429){\makebox(0,0)[rt]{\lineheight{1.25}\smash{\begin{tabular}[t]{r}0\end{tabular}}}}%
    \put(0.12083339,0.166455){\makebox(0,0)[rt]{\lineheight{1.25}\smash{\begin{tabular}[t]{r}2\end{tabular}}}}%
    \put(0.12083339,0.25719571){\makebox(0,0)[rt]{\lineheight{1.25}\smash{\begin{tabular}[t]{r}4\end{tabular}}}}%
    \put(0.12083339,0.34793643){\makebox(0,0)[rt]{\lineheight{1.25}\smash{\begin{tabular}[t]{r}6\end{tabular}}}}%
    \put(0.12083339,0.43867732){\makebox(0,0)[rt]{\lineheight{1.25}\smash{\begin{tabular}[t]{r}8\end{tabular}}}}%
    \put(0.12083339,0.52941804){\makebox(0,0)[rt]{\lineheight{1.25}\smash{\begin{tabular}[t]{r}10\end{tabular}}}}%
    \put(0.12083339,0.62015875){\makebox(0,0)[rt]{\lineheight{1.25}\smash{\begin{tabular}[t]{r}12\end{tabular}}}}%
    \put(0.12083339,0.68821429){\makebox(0,0)[rt]{\lineheight{1.25}\smash{\begin{tabular}[t]{r}13.5\end{tabular}}}}%
    \put(0.04726196,0.38839321){\rotatebox{90}{\makebox(0,0)[t]{\lineheight{1.25}\smash{\begin{tabular}[t]{c}\scalebox{1.5}{Force [kN]}\end{tabular}}}}}%
    \put(0,0){\includegraphics[width=\unitlength,page=3]{Indiv_Samples_Field.pdf}}%
    \put(0.4,0.14154768){\makebox(0,0)[t]{\lineheight{1.25}\smash{\begin{tabular}[t]{c}3\end{tabular}}}}%
    \put(0.51428571,0.14154768){\makebox(0,0)[t]{\lineheight{1.25}\smash{\begin{tabular}[t]{c}3.2\end{tabular}}}}%
    \put(0.62857143,0.14154768){\makebox(0,0)[t]{\lineheight{1.25}\smash{\begin{tabular}[t]{c}3.4\end{tabular}}}}%
    \put(0.74285714,0.14154768){\makebox(0,0)[t]{\lineheight{1.25}\smash{\begin{tabular}[t]{c}3.6\end{tabular}}}}%
  %  \put(0.60000018,0.10738089){\makebox(0,0)[t]{\lineheight{1.25}\smash{\begin{tabular}[t]{c}Absolute value of the Deflection [mm]\end{tabular}}}}%
    \put(0,0){\includegraphics[width=\unitlength,page=4]{Indiv_Samples_Field.pdf}}%
    \put(0.39047625,0.16321429){\makebox(0,0)[rt]{\lineheight{1.25}\smash{\begin{tabular}[t]{r}9\end{tabular}}}}%
    \put(0.39047625,0.25696429){\makebox(0,0)[rt]{\lineheight{1.25}\smash{\begin{tabular}[t]{r}9.5\end{tabular}}}}%
    \put(0.39047625,0.35071429){\makebox(0,0)[rt]{\lineheight{1.25}\smash{\begin{tabular}[t]{r}10\end{tabular}}}}%
  %  \put(0.31690464,0.26339286){\rotatebox{90}{\makebox(0,0)[t]{\lineheight{1.25}\smash{\begin{tabular}[t]{c}Force [kN]\end{tabular}}}}}%
    \put(0,0){\includegraphics[width=\unitlength,page=5]{Indiv_Samples_Field.pdf}}%
  \end{picture}%
\endgroup%

%% file: DynRes_cantilever_Gam_Dist.pdf_tex
%% Accompanies image file 'DynRes_cantilever_Gam_Dist.pdf' (pdf, eps, ps)
\begingroup%
  \makeatletter%
  \providecommand\color[2][]{%
    \errmessage{(Inkscape) Color is used for the text in Inkscape, but the package 'color.sty' is not loaded}%
    \renewcommand\color[2][]{}%
  }%
  \providecommand\transparent[1]{%
    \errmessage{(Inkscape) Transparency is used (non-zero) for the text in Inkscape, but the package 'transparent.sty' is not loaded}%
    \renewcommand\transparent[1]{}%
  }%
  \providecommand\rotatebox[2]{#2}%
  \newcommand*\fsize{\dimexpr\f@size pt\relax}%
  \newcommand*\lineheight[1]{\fontsize{\fsize}{#1\fsize}\selectfont}%
  \ifx\svgwidth\undefined%
    \setlength{\unitlength}{412.5bp}%
    \ifx\svgscale\undefined%
      \relax%
    \else%
      \setlength{\unitlength}{\unitlength * \real{\svgscale}}%
    \fi%
  \else%
    \setlength{\unitlength}{\svgwidth}%
  \fi%
  \global\let\svgwidth\undefined%
  \global\let\svgscale\undefined%
  \makeatother%
  \begin{picture}(1,0.70909091)%
    \lineheight{1}%
    \setlength\tabcolsep{0pt}%
    \put(0,0){\includegraphics[width=\unitlength,page=1]{DynRes_cantilever_Gam_Dist.pdf}}%
    \put(0.13090909,0.04957582){\makebox(0,0)[t]{\lineheight{1.25}\smash{\begin{tabular}[t]{c}0\end{tabular}}}}%
    \put(0.22772727,0.04957582){\makebox(0,0)[t]{\lineheight{1.25}\smash{\begin{tabular}[t]{c}50\end{tabular}}}}%
    \put(0.32454545,0.04957582){\makebox(0,0)[t]{\lineheight{1.25}\smash{\begin{tabular}[t]{c}100\end{tabular}}}}%
    \put(0.42136364,0.04957582){\makebox(0,0)[t]{\lineheight{1.25}\smash{\begin{tabular}[t]{c}150\end{tabular}}}}%
    \put(0.51818182,0.04957582){\makebox(0,0)[t]{\lineheight{1.25}\smash{\begin{tabular}[t]{c}200\end{tabular}}}}%
    \put(0.615,0.04957582){\makebox(0,0)[t]{\lineheight{1.25}\smash{\begin{tabular}[t]{c}250\end{tabular}}}}%
    \put(0.71181818,0.04957582){\makebox(0,0)[t]{\lineheight{1.25}\smash{\begin{tabular}[t]{c}300\end{tabular}}}}%
    \put(0.80863636,0.04957582){\makebox(0,0)[t]{\lineheight{1.25}\smash{\begin{tabular}[t]{c}350\end{tabular}}}}%
    \put(0.90545455,0.04957582){\makebox(0,0)[t]{\lineheight{1.25}\smash{\begin{tabular}[t]{c}400\end{tabular}}}}%
    \put(0.51818218,0.01478782){\makebox(0,0)[t]{\lineheight{1.25}\smash{\begin{tabular}[t]{c}\scalebox{1.5}{Frequency [Hz]}\end{tabular}}}}%
    \put(0,0){\includegraphics[width=\unitlength,page=2]{DynRes_cantilever_Gam_Dist.pdf}}%
    \put(0.12121218,0.07163636){\makebox(0,0)[rt]{\lineheight{1.25}\smash{\begin{tabular}[t]{r}-3\end{tabular}}}}%
    \put(0.12121218,0.12945455){\makebox(0,0)[rt]{\lineheight{1.25}\smash{\begin{tabular}[t]{r}-2.5\end{tabular}}}}%
    \put(0.12121218,0.18727273){\makebox(0,0)[rt]{\lineheight{1.25}\smash{\begin{tabular}[t]{r}-2\end{tabular}}}}%
    \put(0.12121218,0.24509091){\makebox(0,0)[rt]{\lineheight{1.25}\smash{\begin{tabular}[t]{r}-1.5\end{tabular}}}}%
    \put(0.12121218,0.30290909){\makebox(0,0)[rt]{\lineheight{1.25}\smash{\begin{tabular}[t]{r}-1\end{tabular}}}}%
    \put(0.12121218,0.36072727){\makebox(0,0)[rt]{\lineheight{1.25}\smash{\begin{tabular}[t]{r}-0.5\end{tabular}}}}%
    \put(0.12121218,0.41854545){\makebox(0,0)[rt]{\lineheight{1.25}\smash{\begin{tabular}[t]{r}0\end{tabular}}}}%
    \put(0.12121218,0.47636364){\makebox(0,0)[rt]{\lineheight{1.25}\smash{\begin{tabular}[t]{r}0.5\end{tabular}}}}%
    \put(0.12121218,0.53418182){\makebox(0,0)[rt]{\lineheight{1.25}\smash{\begin{tabular}[t]{r}1\end{tabular}}}}%
    \put(0.12121218,0.592){\makebox(0,0)[rt]{\lineheight{1.25}\smash{\begin{tabular}[t]{r}1.5\end{tabular}}}}%
    \put(0.12121218,0.64981818){\makebox(0,0)[rt]{\lineheight{1.25}\smash{\begin{tabular}[t]{r}2\end{tabular}}}}%
    \put(0.04630309,0.36727309){\rotatebox{90}{\makebox(0,0)[t]{\lineheight{1.25}\smash{\begin{tabular}[t]{c}\scalebox{1.5}{$\log_{10}$ Amplitude of Response }\end{tabular}}}}}%
    \put(0,0){\includegraphics[width=\unitlength,page=3]{DynRes_cantilever_Gam_Dist.pdf}}%
  \end{picture}%
\endgroup%

%% file: DynRes_cantilever_Gam_Field.pdf_tex
%% Accompanies image file 'DynRes_cantilever_Gam_Field.pdf' (pdf, eps, ps)
\begingroup%
  \makeatletter%
  \providecommand\color[2][]{%
    \errmessage{(Inkscape) Color is used for the text in Inkscape, but the package 'color.sty' is not loaded}%
    \renewcommand\color[2][]{}%
  }%
  \providecommand\transparent[1]{%
    \errmessage{(Inkscape) Transparency is used (non-zero) for the text in Inkscape, but the package 'transparent.sty' is not loaded}%
    \renewcommand\transparent[1]{}%
  }%
  \providecommand\rotatebox[2]{#2}%
  \newcommand*\fsize{\dimexpr\f@size pt\relax}%
  \newcommand*\lineheight[1]{\fontsize{\fsize}{#1\fsize}\selectfont}%
  \ifx\svgwidth\undefined%
    \setlength{\unitlength}{420bp}%
    \ifx\svgscale\undefined%
      \relax%
    \else%
      \setlength{\unitlength}{\unitlength * \real{\svgscale}}%
    \fi%
  \else%
    \setlength{\unitlength}{\svgwidth}%
  \fi%
  \global\let\svgwidth\undefined%
  \global\let\svgscale\undefined%
  \makeatother%
  \begin{picture}(1,0.75)%
    \lineheight{1}%
    \setlength\tabcolsep{0pt}%
    \put(0,0){\includegraphics[width=\unitlength,page=1]{DynRes_cantilever_Gam_Field.pdf}}%
    \put(0.13035714,0.05404768){\makebox(0,0)[t]{\lineheight{1.25}\smash{\begin{tabular}[t]{c}0\end{tabular}}}}%
    \put(0.22723214,0.05404768){\makebox(0,0)[t]{\lineheight{1.25}\smash{\begin{tabular}[t]{c}50\end{tabular}}}}%
    \put(0.32410714,0.05404768){\makebox(0,0)[t]{\lineheight{1.25}\smash{\begin{tabular}[t]{c}100\end{tabular}}}}%
    \put(0.42098214,0.05404768){\makebox(0,0)[t]{\lineheight{1.25}\smash{\begin{tabular}[t]{c}150\end{tabular}}}}%
    \put(0.51785714,0.05404768){\makebox(0,0)[t]{\lineheight{1.25}\smash{\begin{tabular}[t]{c}200\end{tabular}}}}%
    \put(0.61473214,0.05404768){\makebox(0,0)[t]{\lineheight{1.25}\smash{\begin{tabular}[t]{c}250\end{tabular}}}}%
    \put(0.71160714,0.05404768){\makebox(0,0)[t]{\lineheight{1.25}\smash{\begin{tabular}[t]{c}300\end{tabular}}}}%
    \put(0.80848214,0.05404768){\makebox(0,0)[t]{\lineheight{1.25}\smash{\begin{tabular}[t]{c}350\end{tabular}}}}%
    \put(0.90535714,0.05404768){\makebox(0,0)[t]{\lineheight{1.25}\smash{\begin{tabular}[t]{c}400\end{tabular}}}}%
    \put(0.5178575,0.01988089){\makebox(0,0)[t]{\lineheight{1.25}\smash{\begin{tabular}[t]{c}\scalebox{1.5}{Frequency [Hz]}\end{tabular}}}}%
    \put(0,0){\includegraphics[width=\unitlength,page=2]{DynRes_cantilever_Gam_Field.pdf}}%
    \put(0.12083339,0.07571429){\makebox(0,0)[rt]{\lineheight{1.25}\smash{\begin{tabular}[t]{r}-3\end{tabular}}}}%
    \put(0.12083339,0.13696429){\makebox(0,0)[rt]{\lineheight{1.25}\smash{\begin{tabular}[t]{r}-2.5\end{tabular}}}}%
    \put(0.12083339,0.19821429){\makebox(0,0)[rt]{\lineheight{1.25}\smash{\begin{tabular}[t]{r}-2\end{tabular}}}}%
    \put(0.12083339,0.25946429){\makebox(0,0)[rt]{\lineheight{1.25}\smash{\begin{tabular}[t]{r}-1.5\end{tabular}}}}%
    \put(0.12083339,0.32071429){\makebox(0,0)[rt]{\lineheight{1.25}\smash{\begin{tabular}[t]{r}-1\end{tabular}}}}%
    \put(0.12083339,0.38196429){\makebox(0,0)[rt]{\lineheight{1.25}\smash{\begin{tabular}[t]{r}-0.5\end{tabular}}}}%
    \put(0.12083339,0.44321429){\makebox(0,0)[rt]{\lineheight{1.25}\smash{\begin{tabular}[t]{r}0\end{tabular}}}}%
    \put(0.12083339,0.50446429){\makebox(0,0)[rt]{\lineheight{1.25}\smash{\begin{tabular}[t]{r}0.5\end{tabular}}}}%
    \put(0.12083339,0.56571429){\makebox(0,0)[rt]{\lineheight{1.25}\smash{\begin{tabular}[t]{r}1\end{tabular}}}}%
    \put(0.12083339,0.62696429){\makebox(0,0)[rt]{\lineheight{1.25}\smash{\begin{tabular}[t]{r}1.5\end{tabular}}}}%
    \put(0.12083339,0.68821429){\makebox(0,0)[rt]{\lineheight{1.25}\smash{\begin{tabular}[t]{r}2\end{tabular}}}}%
    \put(0.04726196,0.38839321){\rotatebox{90}{\makebox(0,0)[t]{\lineheight{1.25}\smash{\begin{tabular}[t]{c}\scalebox{1.5}{log$_{10}$ Amplitude of Response} \end{tabular}}}}}%
    \put(0,0){\includegraphics[width=\unitlength,page=3]{DynRes_cantilever_Gam_Field.pdf}}%
  \end{picture}%
\endgroup%

%% file: Real_Gam_Dist.pdf_tex
%% Accompanies image file 'Real_Gam_Dist.pdf' (pdf, eps, ps)
\begingroup%
  \makeatletter%
  \providecommand\color[2][]{%
    \errmessage{(Inkscape) Color is used for the text in Inkscape, but the package 'color.sty' is not loaded}%
    \renewcommand\color[2][]{}%
  }%
  \providecommand\transparent[1]{%
    \errmessage{(Inkscape) Transparency is used (non-zero) for the text in Inkscape, but the package 'transparent.sty' is not loaded}%
    \renewcommand\transparent[1]{}%
  }%
  \providecommand\rotatebox[2]{#2}%
  \newcommand*\fsize{\dimexpr\f@size pt\relax}%
  \newcommand*\lineheight[1]{\fontsize{\fsize}{#1\fsize}\selectfont}%
  \ifx\svgwidth\undefined%
    \setlength{\unitlength}{420bp}%
    \ifx\svgscale\undefined%
      \relax%
    \else%
      \setlength{\unitlength}{\unitlength * \real{\svgscale}}%
    \fi%
  \else%
    \setlength{\unitlength}{\svgwidth}%
  \fi%
  \global\let\svgwidth\undefined%
  \global\let\svgscale\undefined%
  \makeatother%
  \begin{picture}(1,0.75)%
    \lineheight{1}%
    \setlength\tabcolsep{0pt}%
    \put(0,0){\includegraphics[width=\unitlength,page=1]{Real_Gam_Dist.pdf}}%
    \put(0.13035714,0.05404768){\makebox(0,0)[t]{\lineheight{1.25}\smash{\begin{tabular}[t]{c}0\end{tabular}}}}%
    \put(0.22723214,0.05404768){\makebox(0,0)[t]{\lineheight{1.25}\smash{\begin{tabular}[t]{c}50\end{tabular}}}}%
    \put(0.32410714,0.05404768){\makebox(0,0)[t]{\lineheight{1.25}\smash{\begin{tabular}[t]{c}100\end{tabular}}}}%
    \put(0.42098214,0.05404768){\makebox(0,0)[t]{\lineheight{1.25}\smash{\begin{tabular}[t]{c}150\end{tabular}}}}%
    \put(0.51785714,0.05404768){\makebox(0,0)[t]{\lineheight{1.25}\smash{\begin{tabular}[t]{c}200\end{tabular}}}}%
    \put(0.61473214,0.05404768){\makebox(0,0)[t]{\lineheight{1.25}\smash{\begin{tabular}[t]{c}250\end{tabular}}}}%
    \put(0.71160714,0.05404768){\makebox(0,0)[t]{\lineheight{1.25}\smash{\begin{tabular}[t]{c}300\end{tabular}}}}%
    \put(0.80848214,0.05404768){\makebox(0,0)[t]{\lineheight{1.25}\smash{\begin{tabular}[t]{c}350\end{tabular}}}}%
    \put(0.90535714,0.05404768){\makebox(0,0)[t]{\lineheight{1.25}\smash{\begin{tabular}[t]{c}400\end{tabular}}}}%
    \put(0.5178575,0.01988089){\makebox(0,0)[t]{\lineheight{1.25}\smash{\begin{tabular}[t]{c}\scalebox{1.5}{Frequency [Hz]}\end{tabular}}}}%
    \put(0,0){\includegraphics[width=\unitlength,page=2]{Real_Gam_Dist.pdf}}%
    \put(0.12083339,0.07571429){\makebox(0,0)[rt]{\lineheight{1.25}\smash{\begin{tabular}[t]{r}-3\end{tabular}}}}%
    \put(0.12083339,0.14376982){\makebox(0,0)[rt]{\lineheight{1.25}\smash{\begin{tabular}[t]{r}-2.5\end{tabular}}}}%
    \put(0.12083339,0.21182536){\makebox(0,0)[rt]{\lineheight{1.25}\smash{\begin{tabular}[t]{r}-2\end{tabular}}}}%
    \put(0.12083339,0.27988089){\makebox(0,0)[rt]{\lineheight{1.25}\smash{\begin{tabular}[t]{r}-1.5\end{tabular}}}}%
    \put(0.12083339,0.34793643){\makebox(0,0)[rt]{\lineheight{1.25}\smash{\begin{tabular}[t]{r}-1\end{tabular}}}}%
    \put(0.12083339,0.41599214){\makebox(0,0)[rt]{\lineheight{1.25}\smash{\begin{tabular}[t]{r}-0.5\end{tabular}}}}%
    \put(0.12083339,0.48404768){\makebox(0,0)[rt]{\lineheight{1.25}\smash{\begin{tabular}[t]{r}0\end{tabular}}}}%
    \put(0.12083339,0.55210321){\makebox(0,0)[rt]{\lineheight{1.25}\smash{\begin{tabular}[t]{r}0.5\end{tabular}}}}%
    \put(0.12083339,0.62015875){\makebox(0,0)[rt]{\lineheight{1.25}\smash{\begin{tabular}[t]{r}1\end{tabular}}}}%
    \put(0.12083339,0.68821429){\makebox(0,0)[rt]{\lineheight{1.25}\smash{\begin{tabular}[t]{r}1.5\end{tabular}}}}%
    \put(0.04726196,0.38839321){\rotatebox{90}{\makebox(0,0)[t]{\lineheight{1.25}\smash{\begin{tabular}[t]{c}\scalebox{1.5}{log$_{10}$ Amplitude of Response} \end{tabular}}}}}%
    \put(0,0){\includegraphics[width=\unitlength,page=3]{Real_Gam_Dist.pdf}}%
  \end{picture}%
\endgroup%

%% file: Real_Gam_Field.pdf_tex
%% Accompanies image file 'Real_Gam_Field.pdf' (pdf, eps, ps)
\begingroup%
  \makeatletter%
  \providecommand\color[2][]{%
    \errmessage{(Inkscape) Color is used for the text in Inkscape, but the package 'color.sty' is not loaded}%
    \renewcommand\color[2][]{}%
  }%
  \providecommand\transparent[1]{%
    \errmessage{(Inkscape) Transparency is used (non-zero) for the text in Inkscape, but the package 'transparent.sty' is not loaded}%
    \renewcommand\transparent[1]{}%
  }%
  \providecommand\rotatebox[2]{#2}%
  \newcommand*\fsize{\dimexpr\f@size pt\relax}%
  \newcommand*\lineheight[1]{\fontsize{\fsize}{#1\fsize}\selectfont}%
  \ifx\svgwidth\undefined%
    \setlength{\unitlength}{420bp}%
    \ifx\svgscale\undefined%
      \relax%
    \else%
      \setlength{\unitlength}{\unitlength * \real{\svgscale}}%
    \fi%
  \else%
    \setlength{\unitlength}{\svgwidth}%
  \fi%
  \global\let\svgwidth\undefined%
  \global\let\svgscale\undefined%
  \makeatother%
  \begin{picture}(1,0.75)%
    \lineheight{1}%
    \setlength\tabcolsep{0pt}%
    \put(0,0){\includegraphics[width=\unitlength,page=1]{Real_Gam_Field.pdf}}%
    \put(0.13035714,0.05404768){\makebox(0,0)[t]{\lineheight{1.25}\smash{\begin{tabular}[t]{c}0\end{tabular}}}}%
    \put(0.22723214,0.05404768){\makebox(0,0)[t]{\lineheight{1.25}\smash{\begin{tabular}[t]{c}50\end{tabular}}}}%
    \put(0.32410714,0.05404768){\makebox(0,0)[t]{\lineheight{1.25}\smash{\begin{tabular}[t]{c}100\end{tabular}}}}%
    \put(0.42098214,0.05404768){\makebox(0,0)[t]{\lineheight{1.25}\smash{\begin{tabular}[t]{c}150\end{tabular}}}}%
    \put(0.51785714,0.05404768){\makebox(0,0)[t]{\lineheight{1.25}\smash{\begin{tabular}[t]{c}200\end{tabular}}}}%
    \put(0.61473214,0.05404768){\makebox(0,0)[t]{\lineheight{1.25}\smash{\begin{tabular}[t]{c}250\end{tabular}}}}%
    \put(0.71160714,0.05404768){\makebox(0,0)[t]{\lineheight{1.25}\smash{\begin{tabular}[t]{c}300\end{tabular}}}}%
    \put(0.80848214,0.05404768){\makebox(0,0)[t]{\lineheight{1.25}\smash{\begin{tabular}[t]{c}350\end{tabular}}}}%
    \put(0.90535714,0.05404768){\makebox(0,0)[t]{\lineheight{1.25}\smash{\begin{tabular}[t]{c}400\end{tabular}}}}%
    \put(0.5178575,0.01988089){\makebox(0,0)[t]{\lineheight{1.25}\smash{\begin{tabular}[t]{c}\scalebox{1.5}{Frequency [Hz]}\end{tabular}}}}%
    \put(0,0){\includegraphics[width=\unitlength,page=2]{Real_Gam_Field.pdf}}%
    \put(0.12083339,0.07571429){\makebox(0,0)[rt]{\lineheight{1.25}\smash{\begin{tabular}[t]{r}-3\end{tabular}}}}%
    \put(0.12083339,0.14376982){\makebox(0,0)[rt]{\lineheight{1.25}\smash{\begin{tabular}[t]{r}-2.5\end{tabular}}}}%
    \put(0.12083339,0.21182536){\makebox(0,0)[rt]{\lineheight{1.25}\smash{\begin{tabular}[t]{r}-2\end{tabular}}}}%
    \put(0.12083339,0.27988089){\makebox(0,0)[rt]{\lineheight{1.25}\smash{\begin{tabular}[t]{r}-1.5\end{tabular}}}}%
    \put(0.12083339,0.34793643){\makebox(0,0)[rt]{\lineheight{1.25}\smash{\begin{tabular}[t]{r}-1\end{tabular}}}}%
    \put(0.12083339,0.41599214){\makebox(0,0)[rt]{\lineheight{1.25}\smash{\begin{tabular}[t]{r}-0.5\end{tabular}}}}%
    \put(0.12083339,0.48404768){\makebox(0,0)[rt]{\lineheight{1.25}\smash{\begin{tabular}[t]{r}0\end{tabular}}}}%
    \put(0.12083339,0.55210321){\makebox(0,0)[rt]{\lineheight{1.25}\smash{\begin{tabular}[t]{r}0.5\end{tabular}}}}%
    \put(0.12083339,0.62015875){\makebox(0,0)[rt]{\lineheight{1.25}\smash{\begin{tabular}[t]{r}1\end{tabular}}}}%
    \put(0.12083339,0.68821429){\makebox(0,0)[rt]{\lineheight{1.25}\smash{\begin{tabular}[t]{r}1.5\end{tabular}}}}%
    \put(0.04726196,0.38839321){\rotatebox{90}{\makebox(0,0)[t]{\lineheight{1.25}\smash{\begin{tabular}[t]{c}\scalebox{1.5}{log$_{10}$ Amplitude of Response }\end{tabular}}}}}%
    \put(0,0){\includegraphics[width=\unitlength,page=3]{Real_Gam_Field.pdf}}%
  \end{picture}%
\endgroup%